%% file: 3dN=2-CY4.tex
\documentclass[11pt,a4paper]{article}
\pdfoutput=1

\usepackage{jheppub}

\usepackage{amssymb}
\usepackage{amsfonts}
\usepackage{graphicx}
\usepackage{epstopdf}
\usepackage{dcolumn}
\usepackage{amsmath}
\usepackage{latexsym,bm}
\usepackage{amsthm}
\usepackage{slashed}
\usepackage{float}
\usepackage{color}
\usepackage{url}
\usepackage{longtable}
\usepackage{subfig}
\usepackage{tikz}
\usepackage[all,cmtip]{xy}

\usepackage{multirow}
\usepackage{longtable}

\usepackage[titletoc]{appendix}
\makeatletter
\newtheorem*{rep@theorem}{\rep@title}
\newcommand{\newreptheorem}[2]{%
\newenvironment{rep#1}[1]{%
 \def\rep@title{#2 \ref{##1}}%
 \begin{rep@theorem}}%
 {\end{rep@theorem}}}
\makeatother
\newtheorem{lemma}{Lemma}[subsection]
\newreptheorem{lemma}{Lemma}
\newtheorem{theorem}[lemma]{Theorem}

\newtheorem{prop}[lemma]{Proposition}
\newtheorem{conj}[lemma]{Conjecture}
\newreptheorem{conj}{Conjecture}

\theoremstyle{definition}

\newcommand \xoverline[2][0.75]{
    \sbox{\myboxA}{$\m@th#2$}
    \setbox\myboxB\null
    \ht\myboxB=\ht\myboxA
    \dp\myboxB=\dp\myboxA
    \wd\myboxB=#1\wd\myboxA
    \sbox\myboxB{$\m@th\overline{\copy\myboxB}$}
    \setlength\mylenA{\the\wd\myboxA}
    \addtolength\mylenA{-\the\wd\myboxB}
    \ifdim\wd\myboxB<\wd\myboxA
       \rlap{\hskip 0.5\mylenA\usebox\myboxB}{\usebox\myboxA}%
    \else
        \hskip -0.5\mylenA\rlap{\usebox\myboxA}{\hskip 0.5\mylenA\usebox\myboxB}%
    \fi}
\makeatother

\newcommand{\ba}{\begin{aligned}}
\newcommand{\ea}{\end{aligned}}
\def\be{\begin{equation}}
\def\ee{\end{equation}}
\def\bsp{\begin{split}}
\def\esp{\end{split}}
\def\bea{\begin{eqnarray}}
\def\eea{\end{eqnarray}}

\def\mc{\mathcal}

\def\mb{\mathbb}
\def\mbf{\mathbf}
\def \bp{\begin{pmatrix}}
\def\ep{\end{pmatrix}}

\def\R{\mathbb{R}}
\def\N{\mathcal{N}}
\def\P{\mathbb{P}}
\def\C{\mathbb{C}}
\def\Z{\mathbb{Z}}
\def\F{\mathbb{F}}

\def\mk{\mathfrak}

\newcommand{\FTfive}{\mathcal{T}^{\rm 5d}_{X_3}}
\newcommand{\FT}{\mathcal{T}^{\rm 3d}_{X_4}}
\newcommand{\FTsing}{\mathcal{T}^{\rm 3d}_{X_{\text{4,sing}}}}

\usepackage{tikz}
\usepackage{diagbox}
\usetikzlibrary{positioning}
\usetikzlibrary{calc}
\usetikzlibrary{decorations.pathreplacing,calligraphy}

\usepackage{xstring}
\usetikzlibrary{decorations.pathmorphing} 
\usetikzlibrary{decorations.markings} 
\usetikzlibrary{snakes}
\usetikzlibrary{arrows} 
\usetikzlibrary{shapes} 
\usetikzlibrary{matrix} 
\usetikzlibrary{positioning} 
\usepackage[english]{babel} 
\usepackage[autostyle]{csquotes}

\usepackage{tikz}
\usetikzlibrary{arrows}
\usetikzlibrary{arrows.meta}
\usetikzlibrary{shapes.geometric,calc,arrows, positioning,shapes.misc,decorations.markings}
\tikzset{
  big arrow/.style={
    decoration={markings,mark=at position 1 with {\arrow[scale=2,#1]{>}}},
    postaction={decorate},
    shorten >=0.4pt},
  big arrow/.default=black}
  
\pgfdeclarelayer{edgelayer}
\pgfdeclarelayer{nodelayer}
\pgfsetlayers{edgelayer,nodelayer,main} 
\tikzstyle{none}=[inner sep=0pt] 

\tikzstyle{NodeCross}=[draw, shape=circle, cross out, inner sep=0pt, minimum size=6pt,line width=0.25mm]
\tikzstyle{Circle}=[draw, shape=circle, black,  fill=black, inner sep=0pt, minimum size=6pt]
\tikzstyle{Star}=[draw, shape=star, fill=black, star points=8, inner sep=0pt, minimum size=8pt]

\tikzstyle{DashedLine}=[-, densely dashed, line width=0.25mm]
\tikzstyle{DottedLine}=[-, dotted, line width=0.25mm]
\tikzstyle{ThickLine}=[-, line width=0.25mm]
\tikzstyle{ArrowLineRight}=[-, -{Stealth[scale=1.75]}, line width=0.1mm, scale=5]
\tikzstyle{RedLine}=[-, draw={rgb,255: red,191; green,0; blue,0}, fill=none, line width=0.25mm]
\tikzstyle{DottedRed}=[-, dotted, draw={rgb,255: red,191; green,0; blue,0}, fill=none, line width=0.25mm]
\tikzstyle{DashedLineThin}=[-, densely dashed, line width=0.125mm, fill=none, draw=black]
\tikzstyle{ArrowLineRed}=[-, -{Stealth[scale=1.75]}, draw={rgb,255: red,191; green,0; blue,0}, line width=0.1mm, scale=5]
\tikzstyle{brane}=[draw]
\tikzset{D7/.style={circle, draw=black, inner sep=0pt, fill=white, minimum size=3mm}}
\tikzset{hasse/.style={circle, fill,inner sep=2pt}}
\tikzset{flavor/.style={regular polygon,fill=white,regular polygon sides=4,inner sep=2.5pt, draw}}
\tikzset{gauge/.style={circle, draw,inner sep=2.5pt}}
\tikzset{gaugeb/.style={circle, draw,fill=black,inner sep=2.5pt}}
\tikzset{gauger/.style={circle, draw,fill=cyan,inner sep=2.5pt}}
\tikzset{gaugeg/.style={circle, draw,fill=red,inner sep=2.5pt}}
\tikzset{SUd/.style={circle, draw=black, inner sep=0pt, fill=yellow, minimum size=2mm}}
\tikzset{bd/.style={circle, draw=black, inner sep=0pt, fill=black, minimum size=2mm}}
\tikzset{wd/.style={circle, draw=black, inner sep=0pt, fill=white, minimum size=2mm}}
\tikzset{Dynkin/.style={circle, draw=black, inner sep=0pt, fill=white, minimum size=2mm}}
\tikzstyle{ligne}=[draw, thick] 
\tikzset{doublearrow/.style={ draw=black!75, color=black!75, thick, double distance=3pt, }} 

\usepackage{enumitem} 

\usepackage{physics}

\usepackage{chemfig}
\usepackage{import}

\makeatletter
\newcommand\xleftrightarrow[2][]{%
  \ext@arrow 9999{\longleftrightarrowfill@}{#1}{#2}}
\newcommand\longleftrightarrowfill@{%
  \arrowfill@\leftarrow\relbar\rightarrow}
\makeatother

\usepackage{booktabs}

\usepackage{braket}

\graphicspath{ {figs/} }

\title{3d $\mc{N}=2$ theories from M-theory on CY4 and IIB brane box}

\preprint{\today \hspace*{0.1in} }

\author[a]{Marwan Najjar}
\author[b,c]{Jiahua Tian}
\author[a,d]{Yi-Nan Wang}

\affiliation[a]{Center for High Energy Physics, Peking University, \\
Beijing 100871, China}
\affiliation[b]{School of Physics and Electronic Science, East China Normal University, \\
Shanghai 200050, China}
\affiliation[c]{School of Physics, Korea Institute for Advanced Study, \\
85 Hoegi-ro Dongdaemun-gu, Seoul 02455, Republic of Korea}
\affiliation[d]{School of Physics, \\
Peking University, Beijing 100871, China}

\emailAdd{marwan.najjar@pku.edu.cn, jtian1905@gmail.com, ynwang@pku.edu.cn}

\abstract{We study 3D $\mathcal{N}=2$ supersymmetric field theories geometrically engineered from M-theory on non-compact Calabi-Yau fourfolds (CY4). We establish a detailed dictionary between the geometry and topology of non-compact CY4 and the physics of 3D $\mathcal{N}=2$ theories in three different regimes. The first one is the Coulomb branch description when the CY4 is smooth. The second one is non-abelian gauge theory when the CY4 has a degenerate $\mathbb{P}^1$-fibration structure. The third one is the strongly coupled SCFT from a CY4 singularity. We find interesting flavor symmetry enhancements in the singular limit of CY4, as well as an interesting and previously unexplored phenomenon in 3D, termed ``flavor symmetry duality''. Many examples are analyzed with an emphasis on toric CY4s and $\mathbb{C}^4$ orbifolds with crepant resolutions. We develop a new brane box method to study the physics of Coulomb branch of 3D $\mathcal{N}=2$ theory that admits a toric construction. Via IIB/M-theory duality we find that the brane box diagram living in $\mathbb{R}^3$ can be physically realized as a configuration of intersecting 4-branes which are extended objects in 8D maximal supersymmetric theory, which is shown to be consistent via various chains of dualities. The rank, effective gauge coupling and certain hints to flavor symmetry enhancement of the 3D $\mathcal{N}=2$ theory are read off from the brane box and cross-checked against the results obtained from geometric engineering. The exotic branes in 8D maximal supersymmetric theory and the 4-string junctions thereof are shown to play a crucial role in the construction of the brane box.}

\begin{document}

\maketitle

\section{Introduction}
\import{sections/}{intro}

\section{Geometric dictionary}
\import{sections/}{dictionary}

\section{Examples from non-compact geometries}\label{examplesgeo}
\import{sections/}{examples}

\section{Brane box and 3D $\mc{N} = 2$}\label{sec:branebox}

\import{sections/}{brane}

\section{Conclusion and outlook}\label{sec:conclusion}
\import{sections/}{conclusion}

\appendix

\import{sections/}{app}


\newpage

\bibliographystyle{JHEP}
\bibliography{F-ref}

\end{document}

%% file: sections/intro.tex
In recent years, the classification and study of 5d superconformal field theories (SCFTs) has been an active subject, from either M-theory on local Calabi-Yau threefold geometries (canonical threefold singularities) or brane web constructions in IIB superstring theory~\cite{Seiberg:1996bd,Morrison:1996xf,Intriligator:1997pq,Aharony:1997ju,Aharony:1997bh,DeWolfe:1999hj,Benini:2009gi,Kim:2012gu,Bergman:2013aca,Bergman:2013koa,Zafrir:2014ywa,Bergman:2015dpa,Hayashi:2015zka,Zafrir:2015rga,Zafrir:2015uaa,Zafrir:2015ftn,Xie:2017pfl,Ferlito:2017xdq,Jefferson:2017ahm,Hayashi:2018bkd,Hayashi:2018lyv,Jefferson:2018irk,Bhardwaj:2018vuu,Closset:2018bjz,Cabrera:2018jxt,Apruzzi:2018nre,Bhardwaj:2018yhy,Apruzzi:2019vpe,Apruzzi:2019opn,Apruzzi:2019enx,Apruzzi:2019kgb,Bhardwaj:2019ngx,Bhardwaj:2019jtr,Bhardwaj:2019fzv,Bhardwaj:2019xeg,Hayashi:2019jvx,Saxena:2020ltf,Bhardwaj:2020gyu,Eckhard:2020jyr,Morrison:2020ool,Collinucci:2020jqd,Closset:2020scj,vanBeest:2020kou,Bhardwaj:2020ruf,Hubner:2020uvb,Bhardwaj:2020avz,VanBeest:2020kxw,Kim:2020hhh,Hayashi:2021pcj,Apruzzi:2021vcu,Collinucci:2021ofd,vanBeest:2021xyt,Acharya:2021jsp,Tian:2021cif,Closset:2021lwy,Kim:2021fxx,DelZotto:2022fnw,Collinucci:2022rii,Xie:2022lcm,DeMarco:2022dgh,Bourget:2023wlb,DeMarco:2023irn,Mu:2023uws}. Physical information of the 5d SCFTs such as the properties of its Coulomb branch (CB) and Higgs branch (HB), UV enhanced flavor symmetry, IR non-abelian gauge descriptions and generalized symmetries have been extensively studied via different approaches. Various partial classification results have been achieved as well.

In contrast, very few progress has been made on the direct geometric engineering of 3d $\mc{N}=2$ SUSY field theories and SCFTs $\FT$ from M-theory on local Calabi-Yau fourfolds, apart from the early works~\cite{Leung:1997tw,Diaconescu:1998ua,Gukov:1999ya} and more recent developments~\cite{Intriligator:2012ue,Jockers:2016bwi,Chen:2022vvd}. In this paper, we aim to establish a systematic  framework of this geometric approach, which represents an entirely new direction to study the rich family of 3d $\mc{N}=2$ field theories~\cite{Hanany:1996ie,Aharony:1997bx,Aharony:1997gp,Aharony:1997ju,Kapustin:1999ha,Bergman:1999na,Dorey:1999rb,Tong:2000ky,Borokhov:2002cg,Gaiotto:2007qi,Benini:2009qs,Imamura:2011su,Benini:2011cma,Jafferis:2011ns,Dimofte:2011ju,Benini:2011mf,Cecotti:2011iy,Dimofte:2011py,Closset:2012ep,Closset:2012vg,Closset:2012vp,Intriligator:2013lca,Aharony:2013dha,Aharony:2013kma,Amariti:2015yea,Closset:2019hyt,Eckhard:2019jgg,Nii:2020ikd,Sacchi:2021afk,vanBeest:2022fss,Benvenuti:2023qtv}. 

\begin{figure}
\centering
\includegraphics[height=6cm]{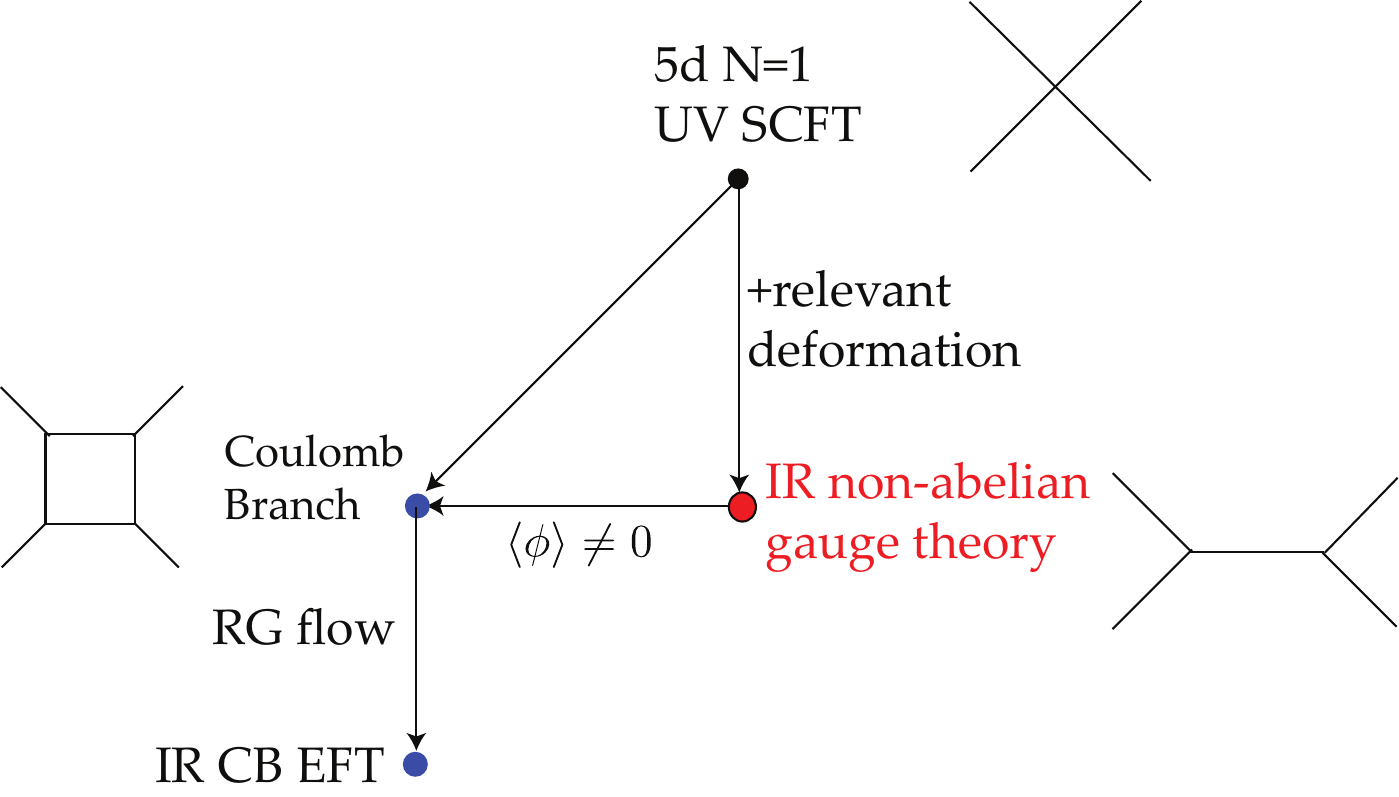}
\caption{The relation between 5d SCFTs, non-abelian gauge theory description and Coulomb branch theories.}\label{fig:5dbranches}
\end{figure}

We would like to first comment on the fundamental difference between the 5d and 3d physics picture. We plot the different phases of 5d theories in figure~\ref{fig:5dbranches}. A 5d $\mc{N}=1$ SCFT $\FTfive$ is defined as M-theory on a canonical threefold singularity, which is the full singular limit of a non-compact CY3 $X_3$. The CB deformation of $\FTfive$ is given by M-theory on the crepantly resolved $X_3$, and one obtains the IR CB effective theory after integrating out the massive modes and RG flow to the deep IR. Finally, for some cases of $X_3$, one can define a $\mb{P}^1$-fibration (ruling) structure and let all $\mb{P}^1$ fibers shrink to zero volume. From the field theory perspective, this corresponds to an IR 5d non-abelian SUSY gauge theory description of the UV SCFT $\FTfive$, which can be obtained after adding a relevant deformation to $\FTfive$ and triggering an RG flow. We have omitted the Higgs branch from the picture.

\begin{figure}
\centering

\includegraphics[height=7cm]{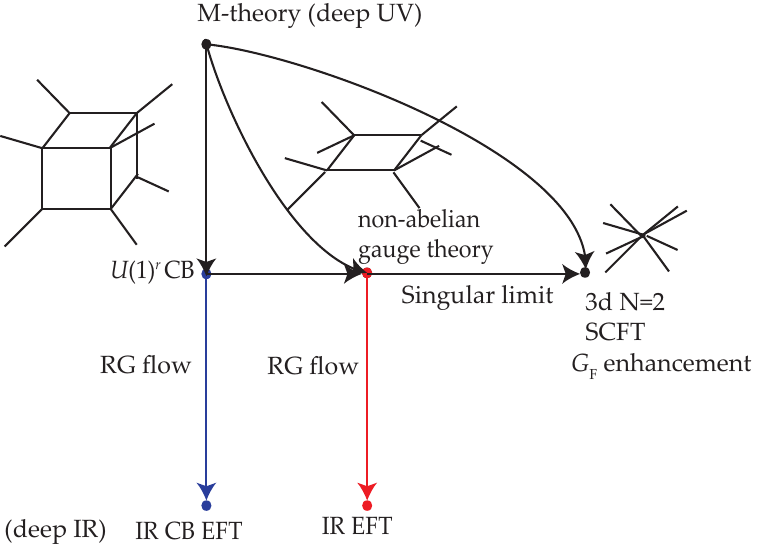}
\caption{Three geometric limits of $X_4$ and the physics of M-theory on different limits. Note that there are three different IR limits.}\label{fig:3pictures1}
\end{figure}

In the 3d case, the general physical picture is different since for a generic 3d $\mc{N}=2$ SUSY field theory, one expect a (possibly free or gapped) fixed point in the IR. We plot the different geometric limits that correspond to different physical phases in figure~\ref{fig:3pictures1}. In the picture we assume that there is no free $G_4$ flux being turned on.

Before getting into the details of $X_4$ or $\FT$ we comment on the different physical scales involved in this work. There is a deep UV where every state in $\FT$ is effectively massless though the physics at this scale is relatively unattractive to us, as by construction we would expect to recover M-theory itself in the deep UV. The physics that is interesting to us happens below the compactification scale, which is much lower than the 11D Planck scale due to that our space $X_4$ has infinite volume (analogous to \cite{Cordova:2009fg}). At this scale where gravity has been already decoupled while the gauge dynamics is still weakly coupled, $\FT$ can be studied semi-classically without caring too much about the quantum corrections. Therefore we expect that there exists a very precise match between the geometric data of $X_4$ and the physical data of $\FT$. Finally, there is a deep IR where all the massive states are integrated out in a Wilsonian manner. At this scale, the gauge dynamics becomes strongly coupled such that any semi-classical analysis fails. More importantly, the non-perturbative superpotential that was suppressed by weak gauge coupling in the UV can no longer be ignored at this scale and it does significantly modify the moduli space in the IR, sometimes completely lifts it. We will remind the reader of the scale we are at whenever it may not be clear from the context, and when it is necessary we will distinguish between different scales via denoting by $\mc{T}^{UV}_{X_4}$ the physical theory just below the compactification scale where gravity has been decoupled while the gauge dynamics is still weakly coupled, and by $\mc{T}^{IR}_{X_4}$ the physical theory in the deep IR where all the massive states have been integrated out.

Now we briefly discuss the different geometric limits in figure~\ref{fig:3pictures1}, and we present a more complete discussion in section~\ref{sec:geometric-limits}. First, the case of M-theory on fully resolved CY4 $X_4$ correspond to a CB description $\mc{T}^{UV}_{X_4}$ that is a $U(1)^r$ gauge theory coupled to charged matter fields (here $r$ is the rank of the 3d $\mc{N}=2$ theory). From this point, it can RG flow to a deep IR CB effective theory $\mc{T}^{IR}_{X_4}$. Secondly, if $X_4$ has a $\mb{P}^1$-fibration structure, one can consider a geometric limit $\widehat{X}_4$ where all the $\mb{P}^1$ fibers are shrunk to zero volume. In this case, the W-bosons in the CB description become massless and we obtain a 3d $\mc{N}=2$ non-abelian gauge theory $\mc{T}^{UV}_{\widehat{X}_4}$. Such a theory would flow to another deep IR theory $\mc{T}^{IR}_{\widehat{X}_4}$\footnote{One should keep in mind that generically the theories $\mc{T}^{UV}_{X_4}$ and $\mc{T}^{UV}_{\widehat{X}_4}$ may have a non-zero superpotential $W$, which lead to no SUSY vacua (the CB is lifted). This issue is further discussed in section~\ref{sec:superpotential}. Nonetheless we would neither compute the detailed dependence of the superpotential on all the geometric moduli fields nor compute too much details about the deep IR effective field theories. Also note that even if the CB is lifted, we still use the notion of CB to describe the deformed 3d field theory from a resolved CY4.}. Finally, if all the compact cycles in $X_4$ are shrunk to zero volume, we arrive at the singular limit $X_{\text{4,sing}}$ of $X_4$, with no geometric scale. M-theory
on $X_{\text{4,sing}}$ would give rise to a 3d $\mc{N}=2$ SCFT $\FTsing$ with $W=0$, since the geometric picture contains no scale parameter at all. In section~\ref{sec:shrinkability} we present a detailed shrinkability criteria of $X_4$ for the existence of $\FTsing$.

In section~\ref{sec:dictionary}, we establish a dictionary between the geometrical and topological data of resolved $X_4$ and the physical properties of $\FT$, including the geometric origins of gauge fields, gauge coupling constants, CB parameters and FI parameters, background gauge fields for flavor symmetries and charged particles. We discuss the BPS particle spectrum from M2-brane wrapping 2-cycles $C$ on $X_4$, whose normal bundle in $X_4$ is either $\mc{O}\oplus\mc{O}\oplus\mc{O}(-2)$ or $\mc{O}\oplus\mc{O}(-1)\oplus\mc{O}(-1)$, from the twisted compactification of either 7D or 5D SYM theory, analogous to \cite{Beasley:2008dc,Intriligator:2012ue}. We further discuss non-BPS higher modes as well in appendix~\ref{sec:non-BPS}. We also briefly discuss the schematics of superpotential and the physical effects of turning on a non-zero $G_4$ flux. In particular, after such $G_4$ flux is included, the singular limit of $X_4$ would be topologically obstructed, since the 4-cycles supporting $G_4$ flux cannot shrink to zero volume\footnote{We thank Hong Lu, Ruben Minasian and Dan Xie for illuminating discussions about this point.}. In the IR limit of the CB effective field theory, a non-zero $G_4$ flux will induce a non-zero Chern-Simons level from the dimensional reduction of 11d SUGRA action.

In section~\ref{sec:toric-CY4}, we give a more detailed physical dictionary for the cases of toric CY4 $X_4$, which is the main class of examples covered in this paper. In section~\ref{sec:fibration}, we discuss the identification of ruling structures of $X_4$ and non-abelian gauge theory descriptions of $\FT$. In section~\ref{sec:flavor}, we discuss in detail the non-abelian flavor symmetry enhancement of the 3d $\mc{N}=2$ SCFT $\FTsing$ at the singular limit $X_{\text{4,sing}}$. It can be read off from the construction the flavor Cartan subalgebra from certain linear combinations of non-compact divisors and flavor W-bosons from M2-brane wrapping certain 2-cycles (which can be non-effective). We find a generic flavor symmetry enhancement for the cases of local $\mb{P}^1\times\mc{S}$. More interestingly, in section~\ref{sec:flavor-duality} we find cases with two different possible flavor symmetry enhancements at the singular limit and we will call this phenomenon ``flavor symmetry duality''. In section~\ref{sec:higher_form_symmetry} we review the calculation of 1-form symmetry and the SymTFT from geometry, following the general recipe in~\cite{vanBeest:2022fss}.

In section~\ref{sec:local_CY4}, we discuss  many geometric and physical aspects of toric CY4 examples. In particular, we studied the case of local $\mb{P}^1\times\mb{P}^1\times\mb{P}^1$ in detail. Such a geometry has three $\mb{P}^1$-fibration structures, which lead to triality among three different $SU(2)$ gauge theory descriptions. We have also observed an interesting $\mk{su}(3)$ flavor symmetry enhancement at the singular limit via identifying flavor Cartan and flavor W-bosons on local $\mb{P}^1\times\mb{P}^1\times\mb{P}^1$ using the dictionary developed in section~\ref{sec:dictionary} and~\ref{sec:toric-CY4}. Furthermore, for this example we also discuss its 1-form symmetry, the corresponding SymTFT and the consequences of turning on $G_4$ flux. In this section we also present a number of other rank-1 examples, as well as higher-rank examples which admit $SU(N)$ gauge theory limits. In section~\ref{sec:C4_orbifolds} we initiate the study of a rich class of non-compact Calabi-Yau geometries which can be constructed by orbifolding $\mb{C}^4$ by a finite subgroup $\Gamma$ of $SU(4)$, i.e. $X_4 = \mb{C}_4/\Gamma$. While these finite groups have been already classified in \cite{HananyHe_SU4}, certain properties that are particularly relevant for geometrically engineering $\FT$ have not been studied thoroughly in literature. Via applying higher dimensional McKay correspondence \cite{ItoReid, ito2011mckay}, we calculate and tabulate those group theoretical quantities that correspond to certain discrete data of $\FT$, e.g. its gauge rank and flavor rank. In section~\ref{sec:N=4} we present a 3d $\mc{N}=4$ example of local $T^2\times\mb{P}^1\times\mb{P}^1$. This case is technically excluded by the shrinkability criteria we proposed in section~\ref{sec:shrinkability}, but it is of physical interest in its own right. We observe an $\mk{su}(2)$ geometric flavor symmetry enhancement in this case.

Besides geometric engineering of 3D $\N=2$ theory from M-theory on non-compact CY4, we develop a new \emph{brane box} method to study 3D $\N=2$ theories in section~\ref{sec:branebox}. We find that the branes in a brane box diagram are the $(p,q,r)$ 4-branes labelled by a $(p,q,r)$ charge triplet in 8D maximal supersymmetric theory obtained from type IIB on $T^2$ studied in~\cite{Leung:1997tw,Lu:1998sx,Lu:1998mr}. We emphasize that the brane boxes in this work are fundamentally different from the brane webs that were used to construct  3d $\mc{N}=2$ theories in, e.g.~\cite{Hanany:1996ie,Aharony:1997ju,Bergman:1999na,Benvenuti:2016wet,Amariti:2019rhc,Cheng:2021vtq,Yagi:2022tot}.

In section~\ref{sec:pqr4branes}, starting from a system of Kaluza-Klein (KK) monopoles in M-theory, we establish the duality between a toric Calabi-Yau 4-fold (CY4) and an intersecting brane/plane configuration in $\mathbb{R}^3$ following the method in~\cite{Leung:1997tw}. The KK-monopole system has been studied via various string dualities in the type II dual picture and in the language of 8D supersymmetric theory in detail~\cite{Lu:1998mr,Lu:1998sx,Eyras:1998hn,ortin2004gravity,Elitzur:1997zn,deBoer:2012ma}. Following~\cite{Leung:1997tw} we show in section~\ref{sec:toricCY} that the M-theory KK-monopole configuration is equivalent to a toric CY4 and it can be visually represented as a configuration of intersecting planes in $\mb{R}^3$. Motivated by the study of 5-brane webs~\cite{Aharony:1997ju, Aharony:1997bh} we find in section~\ref{sec:Dual_IIB_description} that the intersecting planes are physically realized as intersecting branes in type IIB via the IIB/M-theory duality. We then find in section~\ref{sec:8DSALG} that these branes are best understood in the framework of 8D maximal supersymmetric field theory obtained from IIB on $T^2$. In this 8D framework certain symmetry properties, e.g. an $SL(3,\mb{Z})$ modular symmetry that were manifest in the M-theory description but was obscured in the 10D IIB description, become manifest again. The branes in the brane box are shown to be the $(p,q,r)$ 4-branes studied in~\cite{Leung:1997tw,Lu:1998sx,Lu:1998mr}. Various physical properties of a consistent intersecting $(p,q,r)$ 4-brane configuration are then studied, including the $(p,q,r)$ charge conservation law at the intersections of 4-branes or of 4-strings, brane bending when a $(p,q,r)$ 4-brane meets a $(p',q',r')$ 4-brane, etc. We are then able to establish a dictionary between the diagrammatical properties of the brane box diagram and the geometric properties of the dual toric CY4 in section~\ref{sec:duality_branebox_CY4}. We note that our brane box construction is different from that dubbed the same name in \cite{Hanany:2018hlz,Hanany:1998it,Feng:1999fw}.

Having established the duality between a brane box diagram and a toric CY4 in section~\ref{sec:pqr4branes}, we devote section~\ref{3dphysicsfrombrane} to the study of the physics of a 3D $\N=2$ theory on its CB when it is constructed as an intersecting $(p,q,r)$ 4-brane system. We are able to read off the rank of CB, the effective coupling and the limits of non-abelian gauge theory enhancement and SCFT from the diagrammatical properties of the brane box without referring to the dual toric geometry. The key is to relate the properties of the theory on CB to (local) deformations of the branes in the brane box in a way similar to that in~\cite{Aharony:1997ju, Aharony:1997bh}. We cross-check the results obtained from the brane box against the results obtained from toric geometry in section~\ref{sec:examples} and find perfect match.

Motivated by the introduction of 7-branes into a 5-brane web~\cite{DeWolfe:1999hj}, in section~\ref{seccodim2branes} we extend the construction of brane boxes to include the codimension-2 branes in 8D, which are 8D 5-branes, that can be introduced into the original brane box without further breaking SUSY. We review the relevant facts about the relation between these 8D 5-branes and the exotic branes that has been extensively studied in~\cite{Hull:1994ys,ortin2004gravity,Elitzur:1997zn,Eyras:1999at,Obers:1998fb,deBoer:2010ud,deBoer:2012ma,Bergshoeff:2011qk,Bergshoeff:2011se,Kleinschmidt:2011vu,Kimura:2016yqa,Fernandez-Melgarejo:2018yxq}. We find that these 8D 5-branes also carry $(p,q,r)$ charges allowing the 4-branes and effective strings in 8D carrying the same charge to end on them properly without breaking the law of charge conservation. Furthermore, we show that a brane creation/annihilation process similar to Hanany-Witten effects~\cite{Hanany:1996ie} are realized in this extended brane box system via studying brane movements in different type IIB descriptions of the same brane box.

In the section~\ref{sec:flavsymbrane} we investigate various new diagrammatical properties of the brane box diagram extended by adding codimension-2 branes, i.e. the $(p,q,r)$ 5-branes in 8D. The notion of a global deformation of the brane box is then introduced and used to calculate the rank of the flavor symmetry of the corresponding 3D $\N=2$ theory. We then investigate the flavor symmetry enhancement of a certain class of 3D $\N=2$ theories. Such a theory is associated to a brane box diagram dual to a geometry in certain special class of toric CY4's which can be constructed via interlacing two different toric CY3's (see Appendix~\ref{app:interlacing} for an introduction to the notion of interlacing CY's). The interlacing structure of geometry can be studied via looking at the brane box at various decoupling limits in different type IIB duality frames. At certain limits one recovers the ordinary 5-brane web description with additional 7-branes and the flavor symmetries of the theories associated to these brane webs can be read off using the standard methods in~\cite{DeWolfe:1998zf, DeWolfe:1998bi,DeWolfe:1999hj,Distler:2019eky}. Going back to the brane box description via relaxing from the various limits, the flavor symmetry of the 3D theory associated to the brane box can be viewed as an interweaving of the symmetries that can be read off from the ordinary 5-brane webs obtained from looking at those limits.

%% file: sections/dictionary.tex
\label{sec:dictionary}

In this section, we consider several general aspects of 3D $\mc{N}=2$ theories $\mc{T}_{X_4}$ obtained from M-theory on a non-compact Calabi-Yau fourfold $X_4$. We present the dictionary between the geometry and topology of $X_4$ and the field theoretical quantities of 3d $\mc{N}=2$ theory. Certain subtle aspects of the dictionary will be explained by looking at concrete examples where $X_4$ can be a (smooth) toric CY4, or a crepant resolution of a CY4 singularity.

\subsection{Uncharged states from supergravity multiplet and gauge coupling}

We consider a non-compact resolved CY4 $X_4$ with a number of compact divisors $D_i$ $(i=1,\dots,r)$, non-compact divisors $S_\alpha$ $(\alpha=1,\dots,f)$ and $2b$ 5-cycles. A non-compact divisor $S_\alpha$ in $X_4$ cannot be written as a linear combination of $D_i$'s.

In the reduction of M-theory on $X_4$, the 3-form field $C_3$ in M-theory can be decomposed as
\be
C_3=\sum_{i=1}^r A_i\wedge\omega_i^{(1,1)}+\sum_{k=1}^{b} (\mc{N}_k\omega_k^{(2,1)}+\widetilde{\mc{N}}_k\bar{\omega}_k^{(2,1)})+\sum_{\alpha=1}^f B_\alpha\wedge\omega_\alpha^{(1,1),F}
\ee 
There is a dynamical $U(1)$ gauge field $A_i$ associated with each compact divisor $D_i\subset X_4$ where $\omega_i^{(1,1)}$ is the Poincar\'{e} dual $(1,1)$-form of $D_i$. Following the notations in \cite{Grimm:2010ks}, each $\mc{N}_k$ is a real scalar associated to a (2,1)-form $\omega_k^{(2,1)}$ which is the Poincar\'{e} dual of a compact 5-cycle in $X_4$. In this paper we are mostly focusing on the either toric CY4 or $\mathbb{C}^4$ orbifolds where there are no 5-cycles hence there will be no $\mc{N}_k$'s in the corresponding 3d field theory. The decomposition of other fields in the gravity multiplet can be fixed by SUSY.

Besides the dynamical gauge fields $A_i$, $C_3$ also gives rise to the background $U(1)$ 1-form gauge fields $B_\alpha$ that generate 0-form flavor symmetries. They correspond to non-compact divisors $S_\alpha$ Poincar\'{e} dual to $\omega_\alpha^{(1,1),F}$. In some cases, one can gauge the flavor symmetries by compactifying $S_\alpha$ to make $B_\alpha$ dynamical.

To compute the volume of various holomorphic cycles in $X_4$, we need to introduce the K\"{a}hler $(1,1)$-form
\be
J(X_4)=\sum_{i=1}^r a_i \omega_i^{(1,1)}+\sum_{\alpha=1}^f b_\alpha \omega_\alpha^{(1,1),F}\,,
\ee
which is Poincar\'{e} dual to
\be
J^c(X_4)=\sum_{i=1}^r a_i D_i+\sum_{\alpha=1}^f b_\alpha S_\alpha\,.
\ee
Then the volume of each compact complex curve (2-cycle) $C$ can be computed as
\be
\ba
V_C&=\int_C J\cr
&=C\cdot J^c|_{X_4}\,.
\ea
\ee
The volume of each compact complex surface (4-cycle) $\mc{S}$ can be computed as
\be
\ba
V_\mc{S}&=\frac{1}{2}\int_\mc{S} J\wedge J\cr
&=\frac{1}{2}\mc{S}\cdot J^c\cdot J^c|_{X_4}\,.
\ea
\ee
The volume of each compact divisor (6-cycle) $D$ can be computed as
\be
\ba
V_D&=\frac{1}{6}\int_D J\wedge J\wedge J\cr
&=\frac{1}{6}D\cdot J^c\cdot J^c\cdot J^c|_{X_4}\,.
\ea
\ee
The 3d gauge coupling $g_i$ of dynamical gauge field $A_i$ is computed from the dimensional reduction of the $G_4$ kinetic term in the 11D supergravity action (we only keep the terms with the field strength $F_i$ of $A_i$)\footnote{Here we omitted the higher-derivative corrections and other quantum corrections of the supergravity.}:
\be
\ba
\frac{1}{2}\int_{\mb{R}^{1,2}\times X_4}G_4\wedge\star G_4&=\frac{1}{2}\int_{\mb{R}^{1,2}}F_i\wedge\star F_i\int_{X_4}\omega_i^{(1,1)}\wedge\star\omega_i^{(1,1)}+(\dots)\cr
&=\frac{1}{2g_i^2}\int_{\mb{R}^{1,2}}F_i\wedge\star F_i+(\dots)
\ea
\ee
where
\be
\ba
\label{gauge-coupling}
\frac{1}{g_i^2}&=\int_{X_4}\omega_i^{(1,1)}\wedge\star\omega_i^{(1,1)}\cr
&=-\frac{1}{2}\int_{X_4}\omega_i^{(1,1)}\wedge\omega_i^{(1,1)}\wedge J\wedge J\cr
&=-\frac{1}{2}\int_{D_i\cdot D_i}J\wedge J\cr
&=-\frac{1}{2}D_i\cdot D_i\cdot J^c\cdot J^c\cr
&=-\text{Vol}(D_i\cdot D_i)\cr
&=\text{Vol}(-K_{D_i}|_{D_i})\,
\ea
\ee
in terms of the 11D Planck unit. Here we used the expression for Hodge star operator in CY4~\cite{Weissenbacher:2019mef}, and that Vol$(X_4)\rightarrow\infty$. Note that in the toric cases, which will be elucidated in section~\ref{sec:toric-CY4}, (\ref{gauge-coupling}) is equal to the sum of the volume of all toric divisors on $D_i$.

Now let us discuss the two real parameters. The first one is real CB parameter $\sigma_i$ which equals to the vev of the real scalar field in the  dynamical (gauge) $U(1)$ vector multiplet. The second one is the FI parameter $\xi_\alpha$ which equal to the VEV of the scalar field in the background (flavor) $U(1)$ vector multiplet. Analogous to the 5d case~\cite{Morrison:1996xf}, they come from the coefficients in the K\"ahler form:
\be
 J(X_4) 
=\sum_{i=1}^r a_i \omega_i^{(1,1)}+\sum_{\alpha=1}^f b_\alpha \omega_\alpha^{(1,1),F}\,.
\ee
\be
\label{FI-geometric}
\sigma_i={a_i}\ ,\ \xi_\alpha={b_\alpha}\,.
\ee
At a generic point on the CB, these parameters are non-vanishing.

\subsection{Charged states from M2-brane wrapping modes}
\label{sec:M2modes}

The charged BPS states\footnote{It was mentioned in \cite{Aharony:1997bx} that the total energy of a 3d particle charged under $U(1)$ gauge symmetry diverges. Nonetheless here our discussions are in the UV where the presence of light W-bosons provide a screening effect of the otherwise logarithmically decayed Coulomb interaction.} of the 3d $\mc{N}=2$ theory come from M2-branes wrapping various 2-cycles $C\subset X_4$. The charge of these BPS states under the $U(1)$ gauge (flavor) symmetry associated with $D_i$ ($S_\alpha$) can be calculated via the intersection numbers:
\be
q_i=C\cdot D_i\quad (\mathrm{or}\ C\cdot S_\alpha)\,.
\ee

In this paper we only consider the case of a single M2 (anti-M2) brane on a rational curve $C=\mb{P}^1$, analogous to the discussions in the CY3 cases~\cite{Witten:1996qb,Kachru:2018nck,Tian:2018icz}. Moreover in this section we assume that there is no $G_4$-flux on $X_4$.

The normal bundle of $C$ in $X_4$, $N_{C|X_4}$ is a rank-3 vector bundle over $C$, which can be decomposed as
\be
N_{C|X_4}=\mc{O}(d_1)\oplus\mc{O}(d_2)\oplus\mc{O}(d_3)\,.
\ee
Applying Riemann-Roch formula on $C$ given the Ricci-flatness of $X_4$, we have
\be
d_1+d_2+d_3=-2\,.
\ee
If $C$ is a complete intersection curve of three divisors $D_1$, $D_2$ and $D_3$, i.e.
\be
C=D_1\cdot D_2\cdot D_3\,,
\ee
the degrees $d_i$ can be computed as $d_i=D_i\cdot C$, i.e.
\be
\ba
d_1&=D_1^2\cdot D_2\cdot D_3\ ,\ d_2=D_1\cdot D_2^2\cdot D_3\ ,\ d_3=D_1\cdot D_2\cdot D_3^2\,.
\ea
\ee
Note that since there is at least one $d_i\geq 0$, which means that $h_0(N_{C|X_4})>0$, a $\mb{P}^1$ curve inside a Calabi-Yau fourfold is never rigid. 

To identify the particle spectrum of $\mc{T}_{X_4}$ we need to calculate the spacetime quantum numbers of the BPS states from M2-brane wrapping $C$ with different $N_{C|X_4}$. These results can be compared with those from various literature on M-theory or F-theory on CY4 \cite{Beasley:2008dc,Donagi:2008ca,Jockers:2016bwi}.

\begin{enumerate}

\item{$N_{C|X_4}=\mc{O}\oplus\mc{O}\oplus\mc{O}(-2)$.

In this case, the moduli space of $C$ is a complex surface $S$, extended in the directions of $\mc{O}\oplus\mc{O}$. The resulting subsector of $\mc{T}_{X_4}$ can be viewed as a twisted reduction of a 7D SYM on $S$.

A 7D SYM in flat spacetime preserves the global symmetry group $SO(1,6)\times SU(2)_R$. Upon reduction on a Riemannian fourfold $S$ we have the original global symmetry group branches as follows:
\begin{equation}\label{eq:branching}
    \begin{split}
        SO(1,6)\times SU(2)_R &\rightarrow SO(1,2) \times SO(4) \times SU(2)_R \\
        &\simeq SO(1,2) \times SU(2)\times SU(2) \times SU(2)_R
    \end{split}
\end{equation}
and in order to have a 3D theory with 4 supercharges we require that 4 out of the original 16 supercharges under $(\mathbf{8},\mathbf{2})$ of $SO(1,6)\times SU(2)_R$ to be preserved after the topological twist. The branching rule (\ref{eq:branching}) leads to:
\begin{equation}
    (\mathbf{8},\mathbf{2}) \rightarrow (\mathbf{2}, (\mathbf{2},\mathbf{1}))_{\pm\frac{1}{2}} + (\mathbf{2}, (\mathbf{1},\mathbf{2}))_{\pm\frac{1}{2}}
\end{equation}
where $\pm\frac{1}{2}$ is the charge of $\mathbf{2}$ under the $U(1)$ center of $SU(2)_R$. We further require $S$ to be K\"ahler hence the $SO(4)$ holonomy becomes $U(2)_L$ under which $(\mathbf{2},\mathbf{1})$ becomes $\mathbf{2}_0$ and $(\mathbf{1},\mathbf{2})$ becomes $\mathbf{1}_{+1}\oplus \mathbf{1}_{-1}$ \cite{Beasley:2008dc}. Applying the K\"ahler condition, the original branching rule becomes:
\begin{equation}
\begin{split}
    SO(1,6)\times SU(2)_R &\rightarrow SO(3)\times SO(4) \times U(1)_R \\
    (\mathbf{8},\mathbf{2}) &\rightarrow (\mathbf{2}, \mathbf{2}_0)_{\pm\frac{1}{2}} + (\mathbf{2}, \mathbf{1}_{+1}\oplus \mathbf{1}_{-1})_{\pm\frac{1}{2}}.
\end{split}
\end{equation}
After the topological twist $J = J_L + 2J_R$ we have:
\begin{equation}\label{eq:superC_branching}
\begin{split}
    SO(1,6)\times SU(2)_R &\rightarrow SO(3)\times U(2)_L \times U(1)_R \\
    (\mathbf{8},\mathbf{2}) &\rightarrow (\mathbf{2}, \mathbf{2}_{\pm 1})_{\pm\frac{1}{2}} + (\mathbf{2}, \mathbf{1}_{\pm 2})_{\pm\frac{1}{2}} + (\mathbf{2}, \mathbf{1}_0)_{\pm\frac{1}{2}}.
\end{split}
\end{equation}
The last piece of the above equation that is neutral under the $U(2)$ holonomy of $S$ represents the 4 supercharges of the 3D $\mathcal{N} = 2$ theory. From (\ref{eq:superC_branching}) we can also read off that the 3D fermions transform as:
\begin{equation}\label{eq:spec_7Dfermion}
    \chi \in \Omega^{2,0}_S,\ \widetilde{\chi} \in \Omega^{0,2}_S,\ \psi \in \Omega^{1,0}_S,\ \widetilde{\psi} \in \Omega^{0,1}_S,\ \lambda, \widetilde{\lambda} \in \Omega^{0,0}_S.
\end{equation}

Next we look at the twisted reduction of the vector field $(\mathbf{7},\mathbf{1})$. We have:
\begin{equation}
    \begin{split}
        (\mathbf{7},\mathbf{1}) &\rightarrow (\mathbf{3}, \mathbf{1}_0)_0 + (\mathbf{1}, \mathbf{2}_{\pm 1})_0
    \end{split}
\end{equation}
from which we obtain:
\begin{equation}\label{eq:spec_7Dvector}
    A_\mu \in \Omega^{0,0}_S,\ \phi \in \Omega^{1,0}_S,\ \widetilde{\phi} \in \Omega^{0,1}_S.
\end{equation}

Finally we look at the twisted reduction of the scalar field $(\mathbf{1}, \mathbf{3})$. We have:
\begin{equation}
    \begin{split}
        (\mathbf{1}, \mathbf{3}) \rightarrow (\mathbf{1}, \mathbf{1}_{\pm 2})_{\pm 2} + (\mathbf{1}, \mathbf{1}_0)_{0}.
    \end{split}
\end{equation}
from which we obtain:
\begin{equation}\label{eq:spec_7Dscalar}
    \chi \in \Omega^{2,0}_S,\ \widetilde{\chi} \in \Omega^{0,2}_S,\ \rho \in \Omega^{0,0}_S.
\end{equation}

The states in (\ref{eq:spec_7Dfermion}), (\ref{eq:spec_7Dvector}) and (\ref{eq:spec_7Dscalar}) can be packaged into the following 3D $\mc{N}=2$ multiplets:
\begin{equation}\label{eq:vector_spec}
    \begin{split}
        &V: (A_\mu, \lambda, \widetilde{\lambda}, \rho) \in \Omega^{0,0}_S, \\
        &\Psi: (\psi, \phi) \in \Omega^{1,0}_{S},\ \widetilde{\Psi}: (\widetilde{\psi}, \widetilde{\phi}) \in \Omega^{0,1}_{S} \\
        &X: (\chi, \rho) \in \Omega^{2,0}_{S},\ \widetilde{X}: (\widetilde{\chi}, \widetilde{\rho}) \in \Omega^{0,2}_{S}.
    \end{split}
\end{equation}
Therefore besides one vector multiplet $V$, there are $h^{0,1}(S)$ vector-like chiral multiplets $\Psi + \widetilde{\Psi}$ and $h^{0,2}(S)$ vector-like chiral multiplets $X+\widetilde{X}$. This result should be compared with the ``bulk'' spectrum in \cite{Beasley:2008dc, Donagi:2008ca}.

}

\item{$N_{C|X_4}=\mc{O}\oplus\mc{O}(-1)\oplus\mc{O}(-1)$.

In this case, the moduli space of $C$ is a Riemann surface $\Sigma$ extended in the direction of $\mc{O}$. The resulting 3D theory can be viewed as a twisted reduction of 5D matters on $\Sigma$ where the 5D theory is a defect theory living within the previously discussed 7D SYM.

We consider the following branching rule of 5D theory compactified on a Riemannian twofold $\Sigma$:
\begin{equation}
    SO(1,4) \times SU(2)_R \rightarrow SO(1,2) \times SO(2) \times U(1)_R.
\end{equation}
We will focus on the matter sector, i.e. the twisted reduction of the 5D hypermultiplets. Since $\Sigma$ is complex, the $SO(2)$ holonomy becomes $U(1)_L$.

Let us first look at the fermion in $\mathbf{4}$. Under the twisted reduction we have:
\begin{equation}
    (\mathbf{4},\mathbf{1}) \rightarrow \left( \mathbf{2}, +\frac{1}{2} \right)_0 + \left( \mathbf{2}, -\frac{1}{2} \right)_0.
\end{equation}
Clearly the fermions cannot be twisted due to vanishing $R$-charge. Therefore we obtain:
\begin{equation}\label{eq:spec_5Dfermion}
    \xi \in K^{1/2}_\Sigma,\ \widetilde{\xi} \in \overline{K}^{1/2}_\Sigma.
\end{equation}

We then look at the complex scalars $(\mathbf{1}, \mathbf{2})$. Upon the twist $J = J_L + J_R$ the scalars become:
\begin{equation}
    (\mathbf{1}, \mathbf{2}) \rightarrow \left(\mathbf{1}, \pm\frac{1}{2} \right)_{\pm \frac{1}{2}}.
    \label{eq:spec_5Dscalar}
\end{equation}
Therefore we obtain complex scalars:
\begin{equation}
    \eta \in K^{1/2}_\Sigma,\ \widetilde{\eta} \in \overline{K}^{1/2}_\Sigma.
\end{equation}

The states in (\ref{eq:spec_5Dfermion}) and (\ref{eq:spec_5Dscalar}) can then be packaged into the following 3D $\mc{N}=2$ multiplets:
\begin{equation}\label{eq:chiral_spec}
    \Xi: (\xi, \eta) \in K^{1/2}_\Sigma,\ \widetilde{\Xi}: (\widetilde{\xi}, \widetilde{\eta}) \in \overline{K}^{1/2}_\Sigma.
\end{equation}
Note that since $\widetilde{\Xi} \in \overline{H^0(K_\Sigma^{1/2})} = H^0(K_\Sigma^{1/2})^*$\footnote{See \cite{Beasley:2008dc} for a similar calculation in the context of 4D F-theory compactification.}, we have:
\begin{equation}
    \chi(\Xi) = h^0(K_\Sigma^{1/2}) - h^0(K_\Sigma^{1/2}) = 0.
\end{equation}
Therefore in the absence of $G_4$-flux the 3D spectrum is composed of $h^0(K_\Sigma^{1/2})$ chiral multiplets $\Xi$ and $h^0(K_\Sigma^{1/2})$ anti-chiral multiplets $\widetilde{\Xi}$. In other words, the reduction of the 5D hypermultiplet on $\Sigma$ without $G_4$-flux leads to a vector-like matter spectrum in 3D.

In particular, if the moduli space $\Sigma$ is a $\mb{P}^1$, since we know that 
\be
h^0(K_{\mb{P}^1}^{1/2})=h^0(\mc{O}(-1))=0\,,
\ee
there would not be any BPS matter chiral multiplet at all in absence of $G_4$ flux. This is consistent with the fact that Dirac operator on $\mb{P}^1\equiv S^2$ has no zero mode. For all the cases of toric $\mb{P}^1$ curves $C$ in a toric CY4 $X_4$ with $N_{C|X_4}=\mc{O}\oplus\mc{O}(-1)\oplus\mc{O}(-1)$, the previous statement applies and the matter field is absent. For the higher Eigenstates of the Dirac operator on $S^2$, we present a detailed discussion in appendix~\ref{sec:non-BPS}.

}

\end{enumerate}

For the other cases, such as $N_{C|X_4}=\mc{O}(1)\oplus\mc{O}(-1)\oplus\mc{O}(-2)$, the moduli space $\mc{M}_C$ of $C$ is more complicated, and the resulting BPS states are typically higher-spin multiplets. In the case of curves on CY4, different representatives of a given homology class $[C]\in H_2(X_4,\mb{Z})$ can have different normal bundles. We will not discuss these complicated cases in the current paper.

\subsection{Superpotential}
\label{sec:superpotential}

The superpotential $W$ plays an important role in theories with four supercharges. In this section we discuss a number of possible sources of $W$:

\begin{enumerate}
\item{Euclidean M5-brane over a divisor $D$

This type of superpotential was introduced in \cite{Witten:1996bn}. If we have
\be
\label{Rigid-divisor}
h^{0,1}(D)=h^{0,2}(D)=h^{0,3}(D)=0\,,
\ee
there is a non-perturbative superpotential generated from Euclidean M5-brane wrapping $D$, of the form
\be
\label{M5-superpotential}
W_{EM5}=T(m_a) e^{-V_D+i\phi_D}\,.
\ee
$V_D$ is the volume of $D$ and $\phi_D$ is an axion field and $T(m_a)$ is a function depending on all other moduli. In fact, $(\phi_D,\eta_D,V_D)$ form a 3d $\mc{N}=2$ linear multiplet. Note that the $V_D$ is not directly related to the $U(1)$ gauge coupling in (\ref{gauge-coupling}).

}

\item{Polynomial superpotentials among chiral multiplets

This is analogous to the Yukawa superpotential in 4d F-theory \cite{Blumenhagen:2007zk,Beasley:2008dc}. When we have $n$ curves $C_i$, and over a common point $p$ in their moduli spaces $\mc{M}_{C_i}$ they satisfy
\be
\sum_{i=1}^n (C_i)|_p=0\,,
\ee
it induces a superpotential with degree $n$:
\be
W_{Pol}=AX_1 X_2 \dots X_n
\ee
where $X_i$ is the chiral matter field from $C_i$. 

}
\item{Euclidean M2-brane over a 3-cycle

When $X_4$ has rigid 3-cycles $\gamma$, Euclidean M2-brane over it generates non-perturbative superpotentials of the form \cite{Harvey:1999as}
\be
W_{EM2}=Ae^{-V_\gamma+i\phi_\gamma}\,.
\ee
Again $A$ would be a function of other moduli fields.

}
\end{enumerate}

We comment on the lifting of 3d $\mc{N}=2$ Coulomb branch due to a non-trivial superpotential. In general when there are rigid divisors satisfying (\ref{Rigid-divisor}) in the resolved $X_4$, we would expect the non-perturbative superpotential in (\ref{M5-superpotential}) to appear. This would always happen for the cases of local toric CY4, which are the main examples in this paper. Hence for generic values of other moduli fields, we would expect that the Coulomb branch is lifted (becomes massive) by the non-zero superpotential, and there is no SUSY vacuum. Nonetheless, it is certainly possible that the prefactor $T(m_a)$ in (\ref{M5-superpotential}) would vanish for special values of moduli fields $m_a$, and hence an actual SUSY vacuum exists. We would not discuss the complications of SUSY breaking in this paper, since there is no known way to compute the prefactor $T(m_a)$ in the general geometric setups.

In general, in the absence of $G_4$-flux there will be no obstruction to achieving the geometric limit where all compact divisors along with the subvarieties thereof shrink. In fact this is guaranteed by our construction of $X_4$ such that all of its compact divisors are shrinkable, see section~\ref{sec:shrinkability}. At the end point of this limiting process there is no length scale in the geometry hence there must be no energy scales in the corresponding field theory in the UV either. Therefore the physical theory associated to the geometry at the geometric limit is indeed a CFT.

Generally a CFT living in an infinite flat spacetime has zero vacuum energy. The existence of a zero energy vacuum implies no SUSY breaking, assuming we have started with supersymmetry. Therefore we expect the CFT associated with $X_4$ at the geometric limit to be a SCFT. This in turn suggests no superpotential be present. Hence $W$ must vanish at the geometric limit in the absence of $G_4$-flux.

\subsection{$G_4$-flux}\label{sec:G4flux}

$G_4$-flux is an essential ingredient in compactification on Calabi-Yau fourfolds. It satisfies the quantization condition~\cite{Witten:1996md}
\be
\label{G4-quant}
G_4+\frac{1}{2}c_2(X_4)\in H^4(X_4,\mb{Z})\,.
\ee

We restrict ourselves to self-dual $G_4$ flux of the type (2,2) in order to preserve supersymmetry. The inclusion of $G_4$-flux changes the matter spectrum of the 3d $\mc{N}=2$ theory. In practice, we consider the Poincar\'{e} dual of $G_4$:
\be
G_4^c\in H_4(X_4,\mb{R})\,.
\ee
In order to make the total energy of the $G_4$-flux finite, it is necessary that $G_4$ is compactly supported~\cite{Gukov:1999ya}:
\be
G_4\in H^4_{cpct}(X_4,\mb{R})\,
\ee
which in turn means that its Poincar\'{e} dual $G_4^c$ should only have compact components in $H_4(X_4,\mb{R})$.

In terms of $G_4^c$ and intersection numbers, (\ref{G4-quant}) can be written as
\be
\label{G4c-quant}
(G_4^c+\frac{1}{2}c_2^c(X_4))\cdot \mc{S}\in\mb{Z}\quad\ ,\ \forall\mc{S}\in H_4(X_4,\mb{Z})\,.
\ee
Here $c_2^c(X_4)$ is the Poincar\'{e} dual 4-cycle of $c_2(X_4)$, and $\mc{S}$ is an arbitrary compact 4-cycle of $X_4$. When $X_4$ is toric we have
\be
c_2^c(X_4)=\sum_{\sigma_2=v_i v_j}D_i\cdot D_j\,,
\ee
where $\sigma_2$ are all the 2d cones of $X_4$ and $D_i$ is the toric divisor corresponding to the ray $v_i$.

Now we briefly discuss the change in the spectrum of M2-brane wrapping modes after introducing $G_4$-flux. Again we will focus on the case where the normal bundle of curve wrapped by an M2-brane is either $\mc{O}\oplus \mc{O}\oplus \mc{O}(-2)$ or $\mc{O}\oplus \mc{O}(-1)\oplus \mc{O}(-1)$.

\begin{enumerate}

\item{For the bulk matter from M2-brane wrapping a curve $C\simeq \mb{P}^1$ with $N_{C|X_4}=\mc{O}\oplus\mc{O}\oplus\mc{O}(-2)$, we suppose that the compact divisor $D$ is a $\mb{P}^1$-fibration over a complex surface $S$. Then the intersection product
\be
D\cdot G_4^c\in CH_1(S)
\ee
defines a divisor class in $S$, which further corresponds to a line bundle $L\in \mathrm{Pic}(S)$ on $S$. Now the chiral and anti-chiral spectrum are
\be
\ba
\mathrm{chiral}:\ & h^1(S,L)+h^0(S,L\otimes K_S)\cr
\mathrm{anti-chiral}:\ & h^2(S,L)+h^1(S,L\otimes K_S)
\ea
\ee

In particular, the net chirality is \cite{Beasley:2008dc}:
\be
\label{surface-chirality}
\chi=-\int_S c_1(S)c_1(L)\,.
\ee

}

\item{For the chiral matter from M2-brane wrapping a curve $C\simeq \mb{P}^1$ with $N_{C|X_4}=\mc{O}\oplus\mc{O}(-1)\oplus\mc{O}(-1)$, suppose that the moduli space of $C$ is $\mc{M}_C=\Sigma$ and we further define a complex surface 
\be
\mc{S}=C\rightarrow\Sigma\in D\,,
\ee
as the fibration of $C$ over $\Sigma$. Such $\mc{S}$ is also known as the \emph{matter surface} of $C$ in the context of 4D F-theory compactification \cite{Donagi:2008ca, Beasley:2008dc}.

The intersection product
\be
\mc{S}\cdot G_4^c\in CH_0(\Sigma)
\ee
defines a divisor class and line bundle $L\in\mathrm{Pic}(\Sigma)$ on $\Sigma$. The chiral and anti-chiral spectrum are
\be
\ba
\mathrm{chiral}:\ & h^0(\Sigma,L\otimes\sqrt{K_{\Sigma}})\cr
\mathrm{anti-chiral}:\ & h^1(\Sigma,L\otimes\sqrt{K_{\Sigma}})\,.
\ea
\ee

We can compute the net chirality
\be
\chi(C)=G_4^c\cdot \mc{S}\,.
\ee

}
\end{enumerate}

Another important consequence of non-trivial $G_4$-flux is the generation of an effective Chern-Simons term in the action from the dimensional reduction of classical 11d SUGRA action:
\be
\label{Eff-CS}
S_{CS}=\frac{1}{4\pi}\int \sum_{i,j=1}^r k_{ij} A_i\wedge F_j\,.
\ee
Here $A_i$ $(i=1,\dots,r)$ are the $U(1)$ gauge fields associated to each compact divisor $D_i$. Note that such a CS term is present in the UV theory $\mc{T}^{UV}_{X_4}$ in figure~\ref{fig:3pictures1}. The Chern-Simons level $k_{ij}$ can be computed from the $G_4$ flux as
\be
\ba
\label{CS-G4-flux}
k_{ij}&=\int_{X_4}G_4\wedge\omega_i^{(1,1)}\wedge\omega_j^{(1,1)}\cr
&=G_4^c\cdot D_i\cdot D_j\,.
\ea
\ee
When there exists a 4D F-theory lift of our 3D $\mathcal{N} = 2$ theory, it should match the 1-loop corrections of \emph{massless} fermions \cite{Grimm:2011fx}:
\be\label{eq:CSlevel_fermionloop}
k_{ij}=\frac{1}{2}\sum_f(q_f)_i(q_f)_j\mathrm{sign}\left(\sum_{k=1}^r(q_f)_k\sigma^k+m_f\right)\,
\ee
where $\sigma^k$ is the CB parameter of $U(1)_k$, $m_f$ is the classical mass of the fermion $f$ and one has to set $m_f\rightarrow 0$, and there will be effective Chern-Simons terms from integrating out massive fermions with charge $q_i(i=1,\dots,r)$ under the gauge groups $U(1)^r$ and $q_\alpha^F(\alpha=1,\dots,f)$ under the flavor Cartan $U(1)^f$ \cite{PhysRevLett.52.18, Dunne:1998qy,Intriligator:2013lca}. In absence of the parity anomaly, the Chern-Simons levels should be all quantized, i.e. $k_{ij}\in\mb{Z}$. It is clear that $k_{ij}$'s vanish when $G_4^c = 0$. This matches $k_{ij}$ calculated via (\ref{eq:CSlevel_fermionloop}) when there exists a 4D lift due to the vanishing $m_f$ and the trivial chiral index \cite{Grimm:2011fx}.

Finally, turning on $G_4$-flux induces a contribution to the superpotential, known as the Gukov-Vafa-Witten superpotential~\cite{Gukov:1999ya}
\be
W_{GVW}=\int_{X_4} G_4\wedge \Omega_4\,.
\ee
Nonetheless, since we have already required that $G_4$ has no $(0,4)$ component, the above term vanishes. 

In the cases of compact CY4, there is also the scalar potential term in 3d
\be
V(G_4)=\pi\int_{X_4}G_4\wedge *G_4
\ee
and the $D$-term superpotential
\be
D=\int_{X_4}G_4\wedge J\wedge J\,.
\ee
Nonetheless, since our $X_4$ is always non-compact, such terms will not play any role here either.

In studying various geometric limits of $X_4$, we need to be careful that $G_4$-flux cannot pass through shrinking 4-cycles. More precisely, the limit Vol$(\mc{S})\rightarrow 0$ cannot be achieved if
\be
\int_{\mc{S}}G_4=G_4^c\cdot \mc{S}\neq 0\,.
\ee
for a compact 4-cycle $\mc{S}$.

\subsection{Geometric limits and 3d $\mc{N}=2$ SCFTs}
\label{sec:geometric-limits}

We take $X_4$ to be a smooth non-compact CY4, which has a singular limit $X_{4,\text{sing}}$ when all the compact cycles in $X_4$ shrink to zero volume. We discuss the three geometric limits and the corresponding physics respectively, see Figure~\ref{fig:3pictures1}.

Note that the various limiting behaviours in Figure~\ref{fig:3pictures1} are analogous to what happens in 5D obtained from M-theory compactified on a local CY3 \cite{Apruzzi:2019vpe, Apruzzi:2019opn}. In 5D at a generic point on the Coulomb branch the IR EFT is a free $U(1)^r$ theory where $r$ is the rank of the CB. Along certain locus in the CB the theory can be enhanced to a non-Abelian gauge theory. It happens typically when the fibral $\mathbb{P}^1$ of a ruled surface in CY3 is shrunk to zero volume. In the limit where all compact cycles in the CY3 are shrunk to zero volume the theory becomes a 5D SCFT. 


The 3D theory is different from 5D theory in several aspects. One important difference is that a Chern-Simons term generated after integrating out a massive charged fermion generates a mass for the gauge field in the presence of YM Lagrangian \cite{Dunne:1998qy, DESER1982372}. Therefore the existence of massive charged fermions in the UV will drastically change the IR physics. Clearly, in the absence of massive charged fermions in UV no CS term will be generated along the RG flow, hence the IR will be a strongly coupled SYM. In the following subsections we will assume the existence of massive charged fermions in UV, which is generally true for a theory obtained from geometric engineering on a smooth variety.

\subsubsection{Generic point on the Coulomb branch} 
\label{sec:CB-limit}

We first consider M-theory on a smooth non-compact $X_4$, which leads to a $U(1)^r$ gauge theory with non-zero gauge coupling and massive charged matter fields. The rank $r$ equals to the number of irreducible compact divisors $D_i$, i.e. $r = h^{1,1}(X_4)$. We denote the $U(1)$ gauge fields by $A_1$, $\dots$, $A_r$.

Besides the compact divisors, there are also $f=b_2(X_4)-b_6(X_4)$ non-compact divisors $S_\alpha$, which correspond to non-dynamical 1-form gauge fields $B_1$, $\dots$, $B_f$. These can be interpreted as background gauge fields for the flavor symmetries, with the field contents summarized in the vector multiplet:
\be
(\sigma_\alpha,B_\alpha,D_\alpha)\,.
\ee
In particular, the VEV $\langle \sigma_\alpha \rangle := m_\alpha$ is the FI parameter for the flavor symmetry $B_\alpha$~\cite{Closset:2019hyt}.

Even in the absence of $G_4$-flux, there will be massive BPS particles from M2-brane wrapping 2-cycles $C\subset X_4$ are charged under both $U(1)^r$ and $U(1)^f$. Such states include

\begin{itemize}
\item W-bosons of the broken gauge symmetries, whose charge under $U(1)^r$ takes the form of a weight in the adjoint representation of certain enhanced non-abelian Lie algebra. $C$ has normal bundle $N_{C|X_4}=\mc{O}\oplus\mc{O}\oplus\mc{O}(-2)$.
\item Flavor W-bosons, whose charge under $U(1)^f$ takes the form of a weight in the adjoint representation of the enhanced flavor algebra $G_F$. They are also uncharged under all the gauge groups, which means $C\cdot D_i=0$, $\forall D_i$. $C$ has normal bundle $N_{C|X_4}=\mc{O}\oplus\mc{O}\oplus\mc{O}(-2)$.
\item Matter chiral multiplets from $C$ with normal bundle $N_{C|X_4}=\mc{O}\oplus\mc{O}(-1)\oplus\mc{O}(-1)$. If the moduli space of $C$ is $\mb{P}^1$ (e. g. in the toric  cases), strictly speaking there is no zero mode in absence of $G_4$ flux. Nonetheless there are still massive particles from M2-brane wrapping $C$ which will be discussed shortly.
\item Other states where $C$ has a different normal bundle than the previous cases.
\end{itemize}

Now we consider the RG flow to the IR. In the IR, the kinetic terms of $U(1)^r$ become irrelevant. However, there will be effective Chern-Simons terms from integrating out massive fermions with charge $q_i(i=1,\dots,r)$ under the gauge groups $U(1)^r$ and $q_\alpha^F(\alpha=1,\dots,f)$ under the flavor Cartan $U(1)^f$ \cite{PhysRevLett.52.18, Dunne:1998qy,Intriligator:2013lca}. The quantizations of these $U(1)$ charges are the same from the intersection number calculations. There are three types of terms arising from such integrating-out procedure: 
\begin{itemize}
    \item The CS terms from integrating out massive fermions charged under gauge symmetries:
    \be
    \label{CStermsAA}
    \left(\frac{1}{8\pi} \sum_f(q_f)_i(q_f)_j\mathrm{sign}\left(\sum_{k=1}^r(q_f)_k\sigma^k+m_f\right)\right) A_i\wedge d A_j\,,
    \ee
    \item The mixed CS terms from integrating out massive fermions charged under both gauge and flavor symmetries:
    \be
    \label{CStermsAB}
    \left(\frac{1}{8\pi}\sum_f (q_f)_i (q_f)_\alpha^F\mathrm{sign}\left(\sum_{k=1}^r(q_f)_k\sigma^k+m_f\right)\right) A_i\wedge d B_\alpha\,,
    \ee
    \item The CS terms from integrating out massive fermions charged under flavor symmetries:
    \be
    \label{CStermsBB}
    \left(\frac{1}{8\pi}\sum_f (q_f)_\alpha^F (q_f)_\beta^F\mathrm{sign}\left(\sum_{k=1}^r(q_f)_k\sigma^k+m_f\right) \right)B_\alpha\wedge d B_\beta\,.
    \ee
\end{itemize}

Here $m_f$ is the UV real mass of the particle and there is a shift term related to the CB parameters $\sigma^k$. Note that one should sum over fermions from both the M2-brane and anti-M2-brane. In terms of Lie algebra representations, it means that we need to sum over the weights $w$ and $-w$ when they are both present~\cite{Grimm:2011fx}. When $G_4=0$, the 3d $\mc{N}=2$ spectrum is generally expected to be non-chiral, which means that the numbers of chiral fermions in the weight $w$ and $-w$ are identical, and they would give rise to identical contributions to (\ref{CStermsAA}), (\ref{CStermsAB}), (\ref{CStermsBB}) but with opposite signs. In this case, we expect that the CS terms are all vanishing, which is consistent with the dimensional reduction of 11d SUGRA action (\ref{Eff-CS}). For the non-BPS states from higher eigenmodes of the Dirac operators on the moduli space of complex curves, we give a more detailed discussion in appendix~\ref{sec:non-BPS}, and the physical result of no CS terms still holds.

When there is a non-zero $G_4$-flux on $X_4$, the will be additional chiral zero-modes from M2-brane wrapping curves, as well as contributions to the CS terms depending on the $G_4$-flux:
\be
\left(\frac{1}{4\pi}G_4^c\cdot D_i\cdot D_j\right)A_i\wedge d A_j\,,
\ee
\be
\left(\frac{1}{4\pi}G_4^c\cdot D_i\cdot S_\alpha\right)A_i\wedge d B_\alpha\,,
\ee
\be
\left(\frac{1}{4\pi}G_4^c\cdot S_\alpha\cdot S_\beta\right)B_\alpha\wedge d B_\beta\,.
\ee
Moreover, after adding $G_4$-flux, there can be additional flavor symmetry rotating different new chiral modes. The charges and mixed CS term for these new flavor symmetries can be computed in an analogous way, and we would not discuss in more details.

\subsubsection{Non-abelian enhancement locus on the Coulomb branch}
\label{sec:nonab-limit}

In the cases where $X_4$ has a $\mb{P}^1$-fibration structure, $U(1)^r$ can be enhanced to certain non-abelian gauge groups $G$ when the fibral $\mb{P}^1$s shrink to zero volume. More precisely we require each compact divisor of $X_4$ is a $\mathbb{P}^1$-ruled subvariety and all the compact divisors can be glued in an uniform way (see section~\ref{sec:toric-CY4} for a detailed explanation of this point). In this geometry the M2-brane wrapping modes over the fibral $\mb{P}^1$s become the massless W-bosons whose charges are determined by its intersection numbers with the compact divisors associated to $r$ Cartan $U(1)$'s. This is a non-Abelian gauge theory limit of the $U(1)^r$ theory on the Coulomb branch.

Besides the massless W-bosons, there are also massive particles from M2-brane wrapping curves in the base directions, which are charged under both the gauge symmetry and $U(1)^f$ flavor symmetry. We will see in section~\ref{sec:toric-CY4} and section~\ref{sec:examples} that these states should be interpreted as disorder operators in 3D.

Therefore in this case, along the RG flow one first integrates out the massive fermion modes charged under both $U(1)$ and non-Abelian symmetries and gets a 3D CS theory with both $G$ and $U(1)^f$ as the gauge algebra.

Here we comment on the precise correspondence between the non-abelian gauge coupling and the geometric data of $X_4$. For this recall that from (\ref{gauge-coupling}) we have:
\begin{equation}
    \frac{1}{g_i^2} = -\text{Vol}(D_i\cdot D_i).
\end{equation}
For simplicity let us focus on one such compact divisor $D$ and study the corresponding $U(1)$ gauge coupling:
\begin{equation}
    \frac{1}{g^2} = -\text{Vol}(D\cdot D) = -\text{Vol}(K_D) = -K_D\cdot J^c\cdot J^c.
\end{equation}
Recall that the current situation where the UV theory admits a gauge theory description, $D$ is a $\mb{P}^1$-ruled variety which is topologically $C\times B$, $C\simeq \mathbb{P}^1$. Alternatively $D$ can be viewed as the projective bundle associated to a rank-2 vector bundle $E$ over $B$. Hence we have:
\begin{equation}\label{eq:K_D}
    - K_D = 2\xi - \pi^*(K_B + \det(E))
\end{equation}
where $\xi$ is the natural line bundle $O_D(1)$ on $D$ (i.e. a section of fibral $C\simeq \mb{P}^1$) and $\pi$ is the projection of $D$ to $B$. Since $\pi^*(K_B + \det(E))$ is a vertical divisor of $D$, its volume will be suppressed by the hierarchy needed for a valid gauge theory description. Therefore the gauge coupling will be set by:
\begin{equation}
    \frac{1}{g^2} = -K_D\cdot J^c\cdot J^c \simeq 2\xi \cdot J^c\cdot J^c.
\end{equation}
Since $[\xi] = [B]$, we arrive at:
\begin{equation}
    \frac{1}{g^2} \simeq 2\text{Vol}(B).
\end{equation}
Therefore, given the hierarchy $\text{Vol}(B) \gg \text{Vol}(C)$ the gauge coupling $1/g^2$ is essentially set by the volume of the base $B\subset D$ \footnote{A very similar calculation along the same line of logic was done in \cite{Halverson:2022mtc} in a slightly different context.}.

\subsubsection{3d $\mc{N}=2$ SCFT at the origin of the Coulomb branch}
\label{sec:shrinkability}

In the singular limit where all the compact divisors $D_i$ are shrunk to zero volume such that the theory is at the origin of CB, since all $D_i\cdot D_i$'s shrink with $D_i$, all $U(1)$ gauge theories discussed in section~\ref{sec:CB-limit} will become strongly coupled assuming the description (\ref{gauge-coupling}) in a smooth geometry can be extrapolated to the singular limit. Even without looking into the details of the limiting process it is natural to expect that the theory will be conformal due to the absence of all the scales in the singular geometry $X_{4,\text{sing}}$.

Now we discuss the conditions for reaching the singular limit of the smooth local CY4 $X_4$, such that the 3d $\mc{N}=2$ SCFT $\FTsing$ exists.

The first condition is the shrinkability of $X_4$. Before presenting the proposal, we provide a streamlined set of shrinkability conditions for the local CY3 case $X_3$, stated differently from \cite{Jefferson:2018irk}.

$\bullet$ Shrinkability for local CY3

\begin{enumerate}
\item For any compact curve $C=D\cdot S\subset X_3$, where $D$ is a compact divisor and $S$ is a non-compact divisor, we require that $N_{\mc{C}|D}-K_{C|D}\geq 0$. Equivalently, it means that $C\cdot D\leq 0$. The condition can exclude non-examples such as local $\mb{F}_3$. Physically, it means that in the singular limit of $X_3$ where all compact cycles have zero volume, all the non-compact divisors $S$ still have \be
\text{Vol}(-S\cdot S)=\text{Vol}(-K_S)\rightarrow\infty
\ee
to make sure that the gauge coupling of the background gauge field for the non-compact divisor $S$ satisfy $1/g_F^2\rightarrow\infty$~\cite{Morrison:1996xf}.

\item Consider a general point in the K\"{a}hler cone such that all compact curves Vol$(C)\geq 0$ and compact divisors Vol($D)\geq 0$, it is required that for any compact divisor $D\subset X_3$, we have
\be
\label{KD-vol-vanish}
\text{Vol}(-D\cdot D)\equiv \text{Vol}(-K_D)=0
\ee
if and only if all curves $C$ on $D$ have zero volume. Physically, the condition (\ref{KD-vol-vanish}) means that the gauge coupling $g$ of $U(1)$ associated to $D$ satisfies $1/g^2\rightarrow 0$~\cite{Morrison:1996xf} only when $D$ shrinks exactly to a point. The condition can exclude non-examples such as local dP$_9$ or local $T^2\times \mb{P}^1$.
\end{enumerate}

Now let us present the analogous shrinkability criterion for local CY4 $X_4$.

$\bullet$ Shrinkability for local CY4

 \begin{enumerate}
\item For any compact 4-cycle $\mc{S}=D\cdot S$, where $D$ is a compact divisor and $S$ is a non-compact divisor, we require that $N_{\mc{S}|D}-K_{\mc{S}|D}\geq 0$, or equivalently $N_{\mc{S}|D}-K_{\mc{S}|D}$ is effective on $D$.
\item Moreover, for any compact curve $C=D\cdot S_1\cdot S_2$ where $D$ is a compact divisor and $S_1$, $S_2$ are non-compact divisors, we require that $C\cdot D\leq 0$. The physical meanings of condition 1 and 2 are similar to the CY3 case. We make sure that in the singular limit of $X_4$ where all compact cycles have zero volume, all the non-compact divisors $S$ still have \be
\text{Vol}(-S\cdot S)=\text{Vol}(-K_S)\rightarrow\infty\,,
\ee
such that the gauge coupling of the background gauge field for the non-compact divisor $S$ satisfy $1/g_F^2\rightarrow\infty$.

\item Consider a general point in the K\"{a}hler cone such that all compact curves Vol$(C)\geq 0$, 4-cycles Vol$(\mc{S})\geq 0$ and compact divisors vol($D)\geq 0$, it is required that for any compact divisor $D\subset X_4$, we have
\be
\text{Vol}(D\cdot D)\equiv \text{Vol}(-K_D)=0
\ee
if and only if all 4-cycles $\mc{S}$ of $D$ have zero volume. It means that the gauge coupling $g$ of $U(1)$ associated to $D$ satisfies $1/g^2\rightarrow 0$ (\ref{gauge-coupling}) only when $D$ shrinks to a point or a line.

\end{enumerate}

We list a number of cases that are excluded by the above criteria, where $X_4$ has a $T^2$ fibration structure in which case the resulting 3D SCFT is not $\mc{N}=2$.

(a) If $X_4$ is a non-trivial $T^2$-fibration, then the geometry $X_4$ has a 4d F-theory uplift when the $T^2$-fibration shrink to zero volume. This corresponds to the geometric construction of a 4d $\mc{N}=1$ theory.

(b) If $X_4$ is a direct product of $T^2$ and a local CY3 $X_3$, then the geometry describes a 3d $\mc{N}=4$ theory with the volume of $T^2$ is finite. It uplifts to 4d $\mc{N}=2$ when the Vol$(T^2)\rightarrow 0$, where the theory is IIB on $X_3$.

(c) If $X_4$ is a direct product of (local) K3 and (local) K3, the theory is also 3d $\mc{N}=4$. 

(d) If $X_4$ is a direct product of $T^4$ and a local K3 $X_2$, the geometry describes a 3d $\mc{N}=8$ theory.

Specifically, in the cases of toric CY4, the shrinkability  conditions are equivalent to requiring the toric polytope $\mc{P}_3$ to be convex.

There is also additional criterion in the CY4 case about the $G_4$ flux. It is required that no compact 4-cycle $\mc{S}$ in $X_4$ can support any $G_4$-flux, that is for any $\mc{S}$,
\be
\int_{\mc{S}}G_4=0\,.
\ee
This means that we cannot turn on $G_4$ flux in the resolved $X_4$ as its existence obstructs the singular limit.

In the singular limit, the M2-brane wrapping modes all become massless, which include the W-bosons of the $U(1)^f$ geometric flavor symmetry as well. Hence $U(1)^f$ can be enhanced to a non-abelian flavor symmetry group (algebra) $G_F$.

\section{3d $\N=2$ theories from toric CY4}
\label{sec:toric-CY4}

In this section, we consider a smooth toric CY4 $X_4$, which has a singular limit $X_{4,\text{sing}}$ when all the compact cycles in $X_4$ shrink to zero volume. There exists a $GL(4,\mb{Z})$ transformation, such that all rays of $X_4$ take the form of $\tilde{v}_i=(v_i,1)$. We define the convex hull of $v_i$s to be the 3d polytope $\mc{P}_3$. One can compute the quadruple intersection numbers of divisors using the standard toric methods.

\subsection{Fibration structures and non-abelian gauge theory}
\label{sec:fibration}

When the local CY4 has a $\mb{P}^1$-fibration structure (ruling), one can generally consider the non-abelian gauge theory description when the $\mb{P}^1$ fiber direction is shrunk to zero volume. If there are multiple $\mb{P}^1$-fibration structures, it would imply dualities between different non-abelian gauge theories in the UV.

In the toric CY4, the $\mb{P}^1$-fibration structures can be read off from the structure of (triangulated) $\mc{P}_3$. Namely, whenever there is a set of straight parallel lines $L_i$ in $\mc{P}_3$ that pass through all the lattice points in $\mc{P}_3$ while not intersecting the boundary of $\mc{P}_3$ over non-lattice point, then one can define a $\mb{P}^1$ fibration structure where the base are along the $L_i$ direction in the toric diagram and the $\mb{P}^1$ fibers are transversal to the $L_i$ direction.

To determine the non-abelian gauge group and flavor symmetry group in the geometric limit of a particular $\mb{P}^1$-fibration, one count the number $n_i$ of internal lattice points on the parallel lines $L_i$. If these internal lattice points are in the interior of $\mc{P}_3$, they give rise to a $SU(n_i+1)$ gauge group factor. If these points are on the boundary of $\mc{P}_3$, they give rise to a $\mk{su}(n_i+1)$ flavor symmetry factor.

When there are multiple $SU(n_i+1)$ gauge group factors, one may ask whether there are bifundamental matter fields charged between them. In general, if there is no $G_4$ flux, the absence of BPS states from M2-brane wrapping $\mc{O}\oplus\mc{O}(-1)\oplus\mc{O}(-1)$ curves in the toric CY4 would imply that there are no bifundamental matter, unlike the 5d cases.

For example, in the case of local $\mb{P}^1\times\mb{P}^1\times\mb{P}^1$, from the toric diagram in figure~\ref{fig:P3_P1P1P1}
we see that there are three set of lines that satisfy the criteria above, and they represents three different $\mb{P}^1$-fibration structures. For instance if one choose the line $L_1=(0,0,1)\sim (0,0,-1)$ colored in red, then the base direction is given by $S_3$ and the fiber is given by $D\cdot S_1\cdot S_2$. There is an $SU(2)$ gauge theory description since the line $L_1$ has one interior point that corresponds to a compact divisor. For the other two $\mb{P}^1$-fibration structures colored in green and blue, they give $SU(2)$ gauge theory descriptions as well. 

The case in figure~\ref{fig:F2SU3SU22face} also has three distinct $\mb{P}^1$-fibration structures from the lines in $x$, $y$ and $z$-axes. For the $\mb{P}^1$-fibration structure giving by parallel lines in red, one can read off an $SU(3)$ gauge group and $SU(3)^2$ non-abelian flavor symmetry. For the $\mb{P}^1$-fibration structure given by the lines in green, one read off $SU(2)^2$ gauge group and $SU(2)^4$ flavor symmetry. For the $\mb{P}^1$-fibration structure in blue, one read off $SU(2)^2$ gauge group and $SU(2)^6$ flavor symmetry.

Note that we have not discussed the full flavor symmetry enhancement, which is elaborated in section~\ref{sec:flavor}.

\subsection{Flavor symmetry from toric CY4}
\label{sec:flavor}

We focus on the geometric flavor symmetry algebra $G_F$ associated to the non-compact divisors in $X_4$. The rank of $G_F$ equals to the total number of linearly independent non-compact divisors, which can also be expressed as
\be
f:=\text{rk}(G_F)=b_2(X_4)-b_6(X_4)\,.
\ee
In the case of smooth $X_4$, the 3d $\mc{N}=2$ theory is on its extended Coulomb branch such that only the maximal abelian subgroup $U(1)^f\subset G_F$ can be seen. In the singular limit $X_{4,\text{sing}}$, $U(1)^f$ can be enhanced to certain larger flavor symmetry algebra $G_F$.

In order to see the enhancement, one should first identify the linear combinations of non-compact divisors $S_1$, $\dots$, $S_f$ that generate the Cartan subalgebra of $G_F$ (the flavor Cartans). Then for the parts that enhance to non-abelian factors $S_k$, there exist flavor W-bosons wrapping curves $C_k=S_k\cdot \mc{D}$ where $\mc{D}$ is a compact 4-cycle. The flavor W-boson $C_k$ satisfy the following conditions
\begin{enumerate}
{\item $C_k$ is neutral under all $U(1)$ gauge groups generated by compact divisors.}
{\item $C_k$ has the correct normal bundle in $X_4$: $N_{C_k|X_4}=\mc{O}\oplus\mc{O}\oplus\mc{O}(-2)$. }
{\item The intersection matrix $C_k\cdot S_j$ correctly  corresponds to (the negative of) the Cartan matrix of $G_F$.}
\end{enumerate}
Such conditions provide constraints on the choices of $S_k$ and $C_k$, which may not be unique.

For a toric $X_4$, the flavor symmetry enhancement can be partially read off from the toric data. For $X_4$ with a 3d polytope $\mc{P}_3$, there are two ways to get an enhanced non-abelian flavor symmetry factor:

\begin{enumerate}
\item Each edge $\theta=v_i v_j$ on a facet of $\mc{P}_3$ contributes to a non-abelian flavor symmetry factor $\mk{su}(l^*(\theta)+1)$, where $l^*(\theta)$ is the number of interior points on the edge $\theta$.

To see this, we construct a new 3d polytope $\widetilde{\mc{P}}_3$ by adding a new vertex $v$, such that the edge $\theta$ becomes an internal edge in $\widetilde{\mc{P}}_3$. In this case, let us take the $l^*(\theta)$ internal points and its adjacent cones, which form a sub-polytope $\mc{P}_f\subset\widetilde{\mc{P}}_3$. $\mc{P}_f$ precisely describes a string of $l^*(\theta)$ $\mb{P}^1$'s fibered over a complex surface (see Figure~\ref{fig:add_vertex}). 

\begin{figure}
    \centering
    \includegraphics[width=0.45\textwidth]{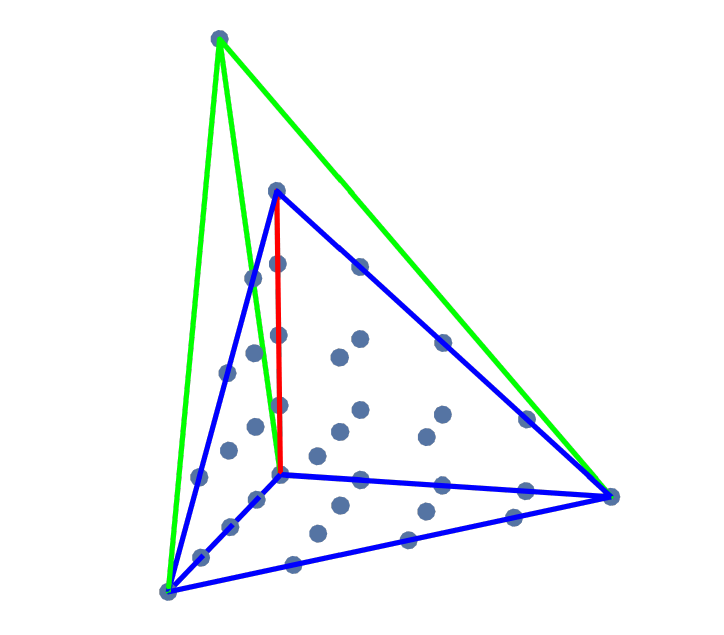}
    \caption{A new polyhedron $\widetilde{\mc{P}}_3$ is made by adding an extra vertex at the intersection of the three green edges to the original tetrahedron $\mathcal{P}_3$ bounded by the red and the blue edges. The red edge is the one that is made internal, i.e. $\theta$ by adding the extra vertex.}
    \label{fig:add_vertex}
\end{figure}

Hence the internal points of $\theta$ generate the Cartan subalgebra $U(1)^{l^*(\theta)}$ of a non-abelian gauge group $\mk{su}(l^*(\theta)+1)$, and we restore the full non-abelian gauge group when all the $\mb{P}^1$ fibers are shrunk to zero volume.

Finally, let us decouple this $\mk{su}(l^*(\theta)+1)$ gauge group by removing the vertex $v$, and $\mk{su}(l^*(\theta)+1)$ becomes a flavor symmetry in the original 3d $\mc{N}=2$ theory from $X_4$.

\item For the internal points of a 2d face $F\subset\mc{P}_3$, each of them generates a $\mk{u}(1)$ flavor symmetry, though further enhancement to a non-abelian flavor algebra may exist given a suitable triangulation of $\mc{P}_3$.

\end{enumerate}

As an example of the second case, consider the configuration of the following 2d facet $\mc{F}$:
\be
\begin{tikzpicture}
\draw[ligne,black] (1,0) -- (-1,-2);
\draw[ligne,black] (1,0) -- (0,1);
\draw[ligne,black] (0,1) -- (-1,-2);
\draw[ligne,black] (0,0) -- (0,1);
\draw[ligne,black] (0,0) -- (1,0);
\draw[ligne,black] (0,0) -- (-1,-2);
\draw[ligne,black] (0,0) -- (0,-1);
\node[bd] at (1,0) [label=right:$S_1$] {};
\node[bd] at (0,0) [label=above right:$S$] {};
\node[bd] at (0,1) [label=above:$S_2$] {};
\node[bd] at (0,-1) [label=below:$S_3$] {};
\node[bd] at (-1,-2) [label=left:$S_4$] {};
\end{tikzpicture}
\ee
We denote by $D$ the compact divisor that connects all points in $\mc{F}$. To see the enhanced $\mk{su}(3)$ flavor symmetry, we write down the flavor Cartan subalgebra generators
\be
F_1=S_1+S\ ,\ F_2=S_3\,.
\ee
The flavor W-boson associated to $F_1$ and $F_2$ are 
\be
C_1=D\cdot F_1\cdot S_1=D\cdot S\cdot S_1
\ee
and
\be
C_2=D\cdot F_2\cdot S=D\cdot S\cdot S_3\,.
\ee
We can check the intersection matrix
\be
\label{su3-noneh-Cartan}
\begin{array}{c|cc}
& F_1 & F_2\\
\hline
C_1 & -2 & 1\\
C_2 & 1 & -2
\end{array}
\ee
which is indeed the Cartan matrix of $\mk{su}(3)$.

Furthermore, we argue that there can be a  flavor symmetry enhancement in the general  cases of local $\mb{P}^1\times\mc{S}$, where $\mc{S}$ is a weak-Fano complex surface that is not $\mb{P}^2$, see the detailed examples in Section~\ref{sec:examples}. In the resolved geometry, we have a single compact divisor $D$ and a number of non-compact divisors $S$, $S_1$, $\dots$, $S_n$. Note that $D\cdot S\cong \mc{S}$, and $D\cdot S\cdot S_i$ corresponds to the curves $C_i$ on $\mc{S}$. In general, we have the following intersection numbers on $X_4$:
\be
\ba
&S^2=0\ ,\ D\cdot S\cdot S_i\cdot S_j=(C_i\cdot C_j)|_\mc{S}\ ,\ D\cdot D\cdot S\cdot S_i=(K_D\cdot C_i)|_\mc{S}\ ,\ \cr
&D\cdot D\cdot S_i\cdot S_j=-2(C_i\cdot C_j)|_\mc{S}\,.
\ea
\ee

In the toric language, $D$ has the toric coordinate $v_D=(0,0,0)$, $S$ corresponds to the toric coordinates $(0,0,1)$ and $(0,0,-1)$ and $S_i$s has toric coordinates $(v_{i,x},v_{i,y},0)$.

Now first for all the $(-2)$-curves $C_i$ on $\mc{S}$, they also give rise to flavor W-bosons $C_i=D\cdot S\cdot S_i$ in $X_4$ corresponding to flavor Cartan $F_i=S_i$ in the Dynkin diagram of $G_{F,nA}$. One can check that 
\be
D\cdot C_i=D\cdot D\cdot S\cdot S_i=(K_D\cdot C_i)|_\mc{S}=0\,.
\ee
Furthermore, whenever $\mc{S}\neq\mb{P}^2,\mb{F}_1$, there exists another flavor Cartan $F=S_\alpha-S$, where $S_\alpha$ corresponds to a rational 0-curve $C_\alpha$ on $\mc{S}$ ($(C_\alpha^2)|_\mc{S}=0$, $g(C_\alpha)=0$). The flavor W-boson for $F$ is given by $C=D\cdot (S_\alpha-S)\cdot (S_\alpha+ F_j)$. Here $ F_j$ is a flavor Cartan satisfying $F_j\cdot C_\alpha=1$, which corresponds to a $(-2)$-curve $C_j$ on $\mc{S}$. One can compute the charges
\be
\ba
C\cdot (S_\alpha-S)&=-2\cr
C_i\cdot (S_\alpha-S)&=(C_\alpha\cdot C_i)|_{\mc{S}}\cr
C\cdot F_i&=(-C_\alpha\cdot C_i-C_i\cdot C_j)|_{\mc{S}}\,.
\ea
\ee
Now let us see whether the flavor Cartans $F$ and $F_i$'s can be combined into a larger non-abelian subalgebra with the correct (negative of) Cartan matrix. First, one can compute that
\be
C_j\cdot (S_\alpha-S)=1\ ,\ C\cdot F_j=1\,,
\ee
which it seems that one can attach the node $F$ to $F_j$ in the Dynkin diagram.

Here comes a subtlety: if there are other flavor Cartans $F_k$ intersecting $F_j$ $(k\neq j)$, there would be a negative charge
\be
C\cdot F_k=(-C_k\cdot C_j)|_{\mc{S}}\neq 0\,.
\ee
Nonetheless, the problem can be remove by redefining the flavor Cartan $F_k\rightarrow F_k'=S_k+S$, and we get $C\cdot F_k'=0$ and there is no issue.

Note that in the construction for the additional flavor Cartan $F=S_\alpha-S$ and its flavor W-boson $D\cdot (S_\alpha-S)\cdot (S_\alpha+F_j)$, they are not effective in $X_4$. Nonetheless, as one reaches the full singular limit $X_{\text{4,sing}}$ of $X_4$, the flavor W-bosons all become massless and there is no obstruction to having such BPS states from non-effective curves. From another perspective, one can consider a complex structure deformation $X_4\rightarrow X_4'$ where $X_4'$ has a larger effective cone of divisors and curves such that $S_\alpha-S$ becomes effective. 

\begin{figure}
\centering
\includegraphics[height=4cm]{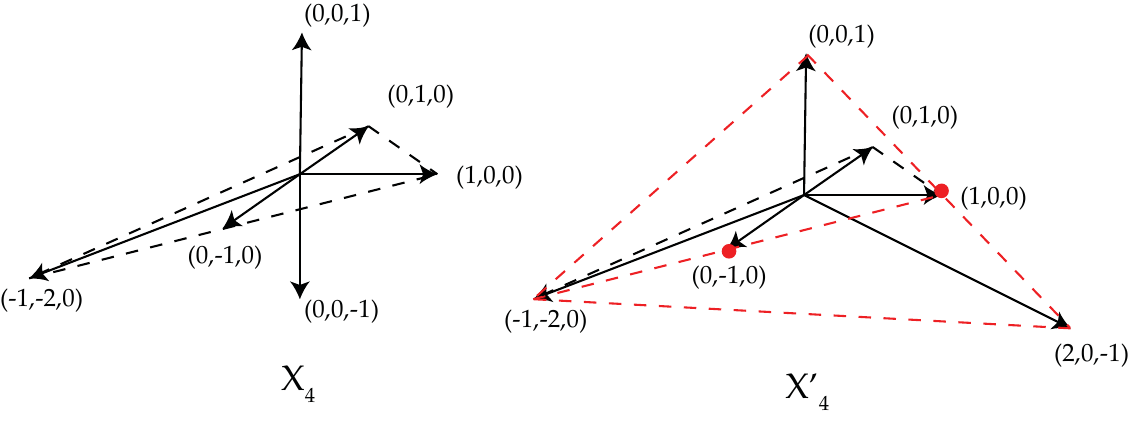}
\caption{The toric diagram of $X_4=$ local $\mb{P}^1\times\mb{F}_2$ and its deformation $X_4'$. On $X_4'$, one explicitly see the flavor W-bosons of $\mk{su}(3)$ flavor symmetry, which are not effective on $X_4$.}\label{fig:P1F2-def}
\end{figure}

As an example, we draw the toric diagram of the case of $\mc{S}=\mb{F}_2$ in figure~\ref{fig:P1F2-def}. As one can see, in the deformed $X_4'$, the red lines and points lie on the same plane, where the red points form the enhanced $\mk{su}(3)$ flavor algebra. The ray $v=(1,0,0)$ exactly corresponds to the flavor Cartan $F$ in the original geometry $X_4$. There is a subtlety that after the deformation $X_4'$ only has a single $\mb{P}^1$-fibration structure (along the $y$-axis), while $X_4$ has two $\mb{P}^1$-fibration structures. Precisely speaking, the deformation $X_4\rightarrow X_4'$ changes the physical description along the Coulomb branch, but they still give rise to the identical enhanced flavor symmetry at the singular limit $X_{\rm{X_4,sing}}$.

In summary, whether $\mc{S}$ has a 0-curve $C_\alpha$ that intersects some $(-2)$-curves, it can be assigned node in the Dynkin diagram of the enhanced flavor algebra $G_{F,nA}$. Note that the analysis works for non-toric $\mc{S}$ as well.

\subsection{Flavor symmetry dualities}
\label{sec:flavor-duality}

In the cases of 3d $\mc{N}=2$ theories from M-theory on local CY4 singularities, we observe a new phenomenon of ``flavor symmetry duality'', that follows from the existence of multiple $\mb{P}^1$-fibration structures mentioned in section~\ref{sec:fibration}.

Let us consider a 2d facet $\mc{F}$ in the toric diagram with two possible $\mb{P}^1$-fibration structures, such as
\be
\label{fig:SU3SU22face}
\includegraphics[height=4.5cm]{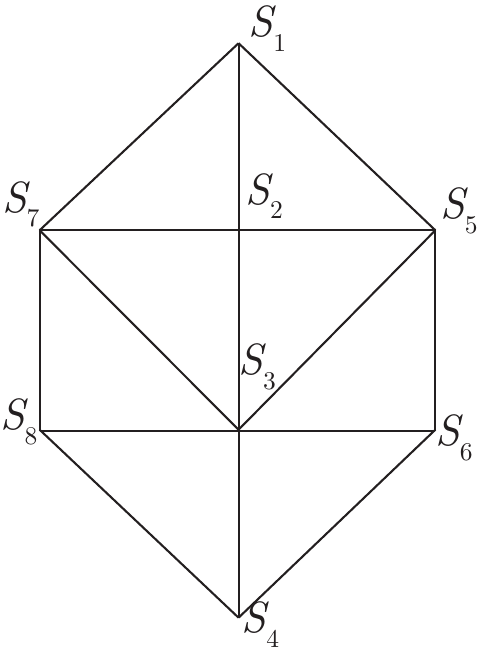}
\ee
If one considers a 5d theory from M-theory on the toric threefold $X_\mc{F}$ in the figure (\ref{fig:SU3SU22face}), it would have IR dual descriptions $SU(3)_0+2\mbf{F}$ and $SU(2)_\pi-SU(2)_\pi$~\cite{Jefferson:2018irk}. In the description $SU(3)_0+2\mbf{F}$, the vertical $\mb{P}^1$ fibers $S_2\cdot S_5$, $S_3\cdot S_5$, $S_3\cdot S_6$ are shrunk to zero volume. While in the description $SU(2)-SU(2)$, the horizontal $\mb{P}^1$ fibers $S_2\cdot S_3$, $S_3\cdot S_5$, $S_3\cdot S_7$ are shrunk to zero volume. In the singular CY3 limit, the massless states can be expressed as either $SU(3)$ or $SU(2)^2$ representations.

\begin{figure}
\centering
\includegraphics[height=6cm]{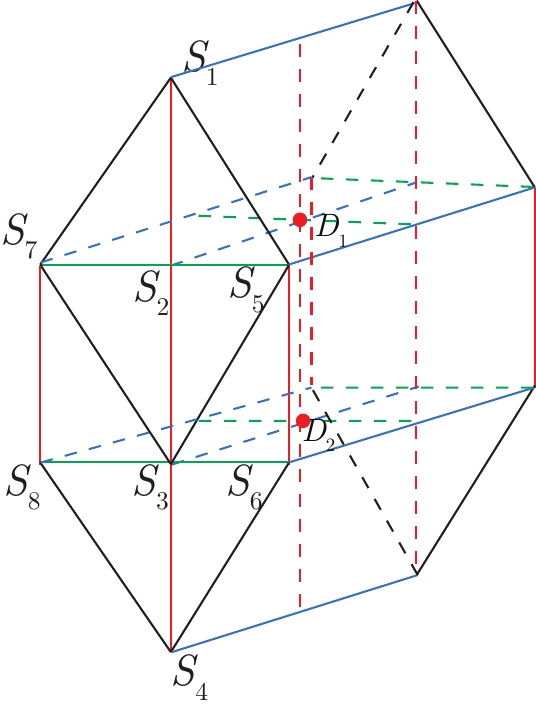}
\caption{A rank-2 example with three $\mb{P}^1$-fibration structures, and a 2d facet that demonstrates the flavor symmetry duality. The total flavor rank of the corresponding 3D theory is 18.}\label{fig:F2SU3SU22face}
\end{figure}

Now let us consider the 3d $\mc{N}=2$ case where figure~\ref{fig:F2SU3SU22face} is a 2d facet, and all divisors $S_i$ are non-compact. We assume that the non-compact CY4 $Y_\mc{F}$ also has two $\mb{P}^1$-fibration structures. Now in the limit where the vertical fibers are shrunk to zero volume, the 3d theory would have enhanced flavor symmetry $\mk{su}(3)$. On the other hand, in the limit where the horizontal fibers are shrunk to zero volume, the enhanced flavor symmetry is $\mk{su}(2)\oplus \mk{su}(2)$. Finally, in the full singular limit where all curves are shrunk to zero size, one can either choose the $\mk{su}(3)$ or $\mk{su}(2)\oplus \mk{su}(2)$ flavor symmetry enhancements. These two choices represent two different ways to arrange the physical states as non-abelian flavor Lie algebra representations, and either of them makes sence. We hence call this ``flavor symmetry duality''.

For example let us consider the resolved non-compact CY4 in figure~\ref{fig:F2SU3SU22face}.
This describes a rank-2 theory with compact divisors $D_1$ and $D_2$, and there are two copies of the facet (\ref{fig:SU3SU22face}). It has three $\mb{P}^1$-fibration structures:
\begin{enumerate}
\item when the vertical $\mb{P}^1$ fibers (in red) shrinks to zero volume, we have a $SU(3)$ gauge theory with $\mk{su}(3)$ flavor symmetry enhancement from the facet $\mc{F}$. Counting the other non-compact divisors, the enhanced flavor symmetry from toric geometry is $\mk{su}(3)^{\oplus 2}\oplus\mk{u}(1)^{\oplus 14}$.

\item When the horizontal $\mb{P}^1$ fibers (in green) shrink to zero volume, we have a $SU(2)^2$ gauge theory with $\mk{su}(2)\oplus \mk{su}(2)$ flavor symmetry enhancement from the facet $\mc{F}$. Counting the other non-compact divisors, the enhanced flavor symmetry from toric geometry is $\mk{su}(2)^{\oplus 4}\oplus\mk{u}(1)^{\oplus 14}$.

\item In the third limit where the $\mb{P}^1$ fibers along the $D_1\cdot S_2$ direction (in blue) shrink to zero volume, we have a $SU(2)^2$ gauge theory with no flavor symmetry enhancement from the facet $\mc{F}$. Counting the other non-compact divisors, the enhanced flavor symmetry from toric geometry is $\mk{su}(2)^{\oplus 6}\oplus\mk{u}(1)^{\oplus 12}$.

\end{enumerate}

Finally in the full singular limit of $X_4$, we can choose to use either of the enhanced flavor symmetry $\mk{su}(3)$ or $\mk{su}(2)\oplus\mk{su}(2)$, as there is a flavor symmetry duality between them. After we count all the contributions of the edges and faces, the flavor symmetry duality is between $G_F=\mk{su}(3)^{\oplus 2}\oplus\mk{su}(2)^{\oplus 6}\oplus\mk{u}(1)^{\oplus 8}$ and $G_F=\mk{su}(2)^{\oplus 10}\oplus\mk{u}(1)^{\oplus 8}$.

Such flavor duality appears in a variety of 3D theories which can be seen as follows. For a 2D facet $\mc{F}$ we consider both its associated local CY3 $X_\mc{F}$ and a toric CY4 $Y_\mc{F}$ whose associated 3D polyhedron $\mc{P}_3$ has $\mc{F}$ one of its 2D facets. If the 5D SCFT obtained from M-theory on $X_\mc{F}$ admits multiple IR gauge theory descriptions, then the 3D SCFT obtained from M-theory on $Y_\mc{F}$ admits an IR flavor duality. From its construction it is clear that this duality depends only on the property of $\mc{F}$ and is independent from the way it is embedded in $\mc{P}_3$\footnote{Here we do not completely exclude the possibility of a further enhancement to a larger global symmetry. Nonetheless, to confirm this one needs to do a careful full analysis of the BPS spectrum from M2-brane wrapping all the holomorphic curves on the CY4. This problem is beyond the scope of the current paper, but it is an interesting question to investigate in the future.}.

\subsection{Higher-form symmetries and SymTFT from toric CY4}\label{sec:higher_form_symmetry}

The computation of higher-form symmetries and SymTFT action from M-theory on a CY4 was developed in \cite{vanBeest:2022fss} , which we briefly recap here. To be more precisely, we focus on the ``geometric'' global symmetries that are generated by the non-compact cycles of the CY4.

We denote by $\mc{T}_{X_4}$ the 3D theory obtained from M-theory compactification on CY4 $X_4$. To compute the 1-form symmetry of $\mc{T}_{X_4}$, we need to compute the Smith normal form of the intersection matrix $\{C_i\cdot D_j\}$, where $C_i$ $(i=1,\dots,b_2(X_4))$ are compact 2-cycles on $X_4$ and $D_j$ $(j=1,\dots,r = b_6(X_4))$ are the compact divisors.

Taking the \emph{Smith normal form} of the matrix $\{C_i\cdot D_j\}$:
\be
\label{SNF-int}
SNF(\{C_i\cdot D_j\})=
\bp l_1 & 0 & \dots & 0\\
0 & l_2 & \dots & 0\\
\vdots & \vdots & \ddots & 0\\
0 & 0 & \dots & l_r\\
0 & 0 & \dots & 0\\
\vdots & \vdots & \ddots & \vdots\\
0 & 0 & \dots & 0\\
\ep
=A\cdot \{C_i\cdot D_j\} \cdot B\,,
\ee
one can read off the 1-form symmetry
\be
\Gamma=\bigoplus_{i=1}^r (\mb{Z}/l_i\mb{Z})=\bigoplus_i\mb{Z}_{l_i}\,.
\ee

In the toric cases a simpler algorithm can be applied \cite{Morrison:2020ool,Luo:2023ive}. One take the $n\times 4$ matrix where each row is the 4d toric ray $\tilde{v}_i=(v_i,1)$. It is required that $v_i$ locates on the boundary of the polytope $\mc{P}$. Such a $n\times 4$ matrix has the same Smith normal form as the one in (\ref{SNF-int}), which gives rise to the same 1-form symmetry.

We write down the generators of $\mb{Z}_{l_\alpha}$ 1-form symmetry using a linear combination of compact divisors:
\be
Z_6^\alpha=B_{\alpha j} D_j\,.
\ee

Besides the 1-form symmetries, there are also torsional $(-1)$-form symmetries, which one can compute from the intersection matrix of compact 4-cycles $\{\mc{S}_i\cdot \mc{S}_j\}$ $(i=1,\dots,b_4(X_4))$. We can use the Smith decomposition
\be
\label{SNF-int}
SNF(\{\mc{S}_i\cdot \mc{S}_j\})=
\bp m_1 & 0 & \dots & 0\\
0 & m_2 & \dots & 0\\
\vdots & \vdots & \ddots & 0\\
0 & 0 & \dots & m_r\\
0 & 0 & \dots & 0\\
\vdots & \vdots & \ddots & 0\\
0 & 0 & \dots & 0\\
\ep
=\mc{A}\cdot \{\mc{S}_i\cdot \mc{S}_j\} \cdot\mc{B}\,,
\ee
to read off such symmetry $\bigoplus_i(\mb{Z}/m_i\mb{Z})$ and the generators $Z_4^\beta=\mc{B}_{\beta j}\mc{S}_j$ for each $\mb{Z}_{m_\beta}$ factor.

The SymTFT for the 3d $\mc{N}=2$ theory can be computed following the general formula in \cite{vanBeest:2022fss}:

\be
\label{SymTFT}
\frac{S_{\rm SymTFT}}{2\pi}=-\sum_{j\alpha\beta}\Lambda_{22}^{j\alpha\beta} b^\alpha\int_{\mc{M}_4}F_2^j \cup B_2^\beta-\sum_{\alpha\beta\gamma}\frac{\Lambda_{24}^{\alpha\beta\gamma}}{2}b^\alpha\int_{\mc{M}_4}B_2^\beta \cup B_2^\gamma\,.
\ee
Here $F_2^j$ corresponds to the field strength of 0-form flavor symmetry, where $j$ labels each non-compact divisors $S_j$. $b^\alpha$ denotes the amount of torsional background flux from $G_4$ over $\mb{Z}_{m_\alpha}$ torsional 4-cocycles (which is dual to the 4-cycle $Z_4^\alpha$). $B_2^\beta$ corresponds to the background gauge field of $\mb{Z}_{l_\beta}$ 1-form symmetry, which arise from $G_4$ over torsional 2-cocycles (which is dual to the 6-cycle $Z_6^\beta$). The coefficients can be computed from the intersection number calculation
\be
\ba
\Lambda_{22}^{j\alpha\beta}&=\left[\frac{S_j\cdot Z_4^\alpha\cdot Z_6^\beta}{m_\alpha l_\beta}\right]_{\rm mod\ 1}\,,\cr
\Lambda_{24}^{\alpha\beta\gamma}&=\left[\frac{Z_4^\alpha\cdot Z_6^\beta\cdot Z_6^\gamma}{m_\alpha l_\beta l_\gamma}\right]_{\rm mod\ 1}\,.
\ea
\ee

Note that for the SymTFT to be non-trivial, one has to introduce torsional flux $b^\alpha$ that generates discrete $(-1)$-form symmetries.

%% file: sections/examples.tex
\label{sec:examples}

In this section we present concrete examples of non-compact CY4 and analyze the physics of the corresponding 3D theories.

\subsection{Toric Calabi-Yau fourfolds}\label{sec:local_CY4}

In this section we study non-compact CY4's that can be explicitly constructed as the canonical bundle of a complex 3D compact variety, whose Ricci-flatness trivially follows from $c_1 = 0$ and Yau's theorem. Many of the examples that fall into this class can be constructed torically which allows us to analyze their geometry with standard toric geometry tools. Using the dictionary developed in section~\ref{sec:dictionary} we compute both the naive and the enhanced flavor symmetry group, the gauge group and various other important physical quantities. We will also present an analysis of the possible $G_4$-flux profile on such CY4 and the resulting change in the particle spectrum in field theory. In particular, we analyze the cases of local $\mb{P}^3$, local $\mb{P}^1\times\mb{P}^2$, $\mb{P}^1\times\mb{P}^1\times\mb{P}^1$, local $\mb{P}^1\times\mb{F}_1$ and local $\mb{P}^1\times \text{dP}_2$ in full details.

\subsubsection{Local $\mb{P}^3$}

We first consider the simple case of $X_4=$ local $D$, where $D=\mb{P}^3$. It gives rise to a $U(1)$ gauge theory on the CB. There are no curves with normal bundle $\mc{O}\oplus\mc{O}\oplus\mc{O}(-2)$ or $\mc{O}\oplus\mc{O}(-1)\oplus\mc{O}(-1)$ on $\mb{P}^3$, and there is no $\mb{P}^1$-fibration structure either. Thus there is no geometric limit such that the gauge group is enhanced to $G=SU(2)$.

We define a non-compact divisor $S$ such that $D\cdot S|_D=H$ is the hyperplane class of $D$. The compact curves on $X_4$ are in the homology class $C_a=aD\cdot S^2=aH^2|_D$. We have 
\be
K_D=-4H=-4D\cdot S\,
\ee
and the intersection numbers
\be
D\cdot S^3=1\ ,\ D^2\cdot S^2=-4\ ,\ D^3\cdot S=16\ ,\ D^4=-64\,.
\ee

\paragraph{K\"{a}hler form and $U(1)$ gauge coupling}

We denote the (Poincar\'{e} dual) of K\"{a}hler form to be
\be
\ba
J^c&=xS\cr
&=-\frac{1}{4}xD\,.
\ea
\ee
We can compute
\be
\text{Vol}(C_a)=ax\ ,\ \text{Vol}(D\cdot S)=\text{Vol}(H)=\frac{1}{2}x^2\ ,\ \text{Vol}(D)=\frac{1}{6}x^3\,.
\ee

At a generic point in the K\"{a}hler moduli space, $x>0$, $D$ and any complex curve $C\subset D$ have a positive volume. Hence the 3d $\mc{N}=2$ theory $\mc{T}^{UV}_{X_4}$ is a $U(1)$ gauge theory coupled to an infinite tower of massive particle states, which are generated by M2-branes wrapping every $C\in H_2(D,\mb{Z})$. 

The gauge coupling is (\ref{gauge-coupling})
\be
\frac{1}{g^2}=\text{Vol}(-D\cdot D)=2x^2\,.
\ee

In the strong coupling limit $x\rightarrow 0$, these massive BPS states become massless, and one recovers the geometric limit of local $\mb{P}^3$ singularity $X_{\text{4,sing}}$, leading to a 3d $\mc{N}=2$ SCFT $\FTsing$. 

\paragraph{M2-brane wrapping states, 1-form symmetry and SymTFT}

For the massive BPS states in this case, let us consider M2 brane wrapping curves $C_a$ on $D=\mb{P}^3$. Its charge under the $U(1)$ gauge group is
\be
C_a\cdot D|_{X_4}=C_a\cdot K_D|_D=-4aH^3|_D=-4a\,.
\ee
Hence every massive particle BPS state has charge $-4a$ $(a\in\mb{Z})$ under the $U(1)$. We therefore expect a $\mb{Z}_4$ 1-form symmetry in both $\mc{T}^{UV}_{X_4}$ and $\FTsing$ from the Coulomb branch analysis.

For the intersection between 4-cycles, we have
\be
(D\cdot S)\cdot (D\cdot S)=-4\,,
\ee
hence there is a $\mb{Z}_4$ $(-1)$-form symmetry as well with the generator
\be
Z_4=D\cdot S\,.
\ee
Using (\ref{SymTFT}) we can compute the SymTFT action
\be
\frac{S_{\text{SymTFT}}}{2\pi}=-\frac{1}{8}b\int_{\mc{M}_4}B_2\cup B_2\,,
\ee
where $b$ is the torsional $G_4$ flux over $\mb{Z}_4$ torsional 4-cocycles and $B_2$ is the background gauge field for $\mb{Z}_4$ 1-form symmetry.

\subsubsection{Local $\mb{P}^1\times\mb{P}^2$}

Now let us consider the case with an $SU(2)$ gauge theory description. To construct such a theory one starts with a compact complex threefold $D$ which is a $\mb{P}^1$-fibration over a (weak-Fano) complex surface $\mc{S}$:
\be
D=\mb{P}^1\hookrightarrow \mc{S}
\ee
Let us denote the fibral $\mb{P}^1$ by $F$. The normal bundle of $F$ is
\be
\ba
N_{F|D}&=\mc{O}\oplus\mc{O}\,,\cr
N_{F|X_4}&=\mc{O}\oplus\mc{O}\oplus\mc{O}(-2)\,.
\ea
\ee
If $\mc{S}$ satisfies $h^{0,1}(\mc{S})=h^{0,2}(\mc{S})=0$, the M2-brane wrapping mode over $F$ is a single vector multiplet with charge
\be
F\cdot D=-2
\ee
under the $U(1)$ gauge group associated to $D$. The anti-M2-brane wrapping mode over $F$ is also a vector multiplet, whose $U(1)$ charge is +2. Hence they form the W-bosons of the $SU(2)$ gauge group. In the limit of Vol$(F)\rightarrow 0$, the W-bosons become massless and we restore the non-abelian $SU(2)$ gauge group. On the other hand, there are also M2 brane wrapping modes over curves $C\subset \mc{S}$, which correspond to massive BPS states that are different for each choice of $\mc{S}$.

In this section, we consider the case of a local $\mb{P}^1\times\mb{P}^2$, i.e. $\mc{S}=\mb{P}^2$. In this case, we write down the rays in the toric fan of CY4 $\tilde{v}_i=(v_i,1)$:
\be
v=(0,0,0)\ ,\ v_1=(1,0,0)\ ,\ v_2=(0,1,0)\ ,\ v_3=(-1,-1,0)\ ,\ v_4=(0,0,1)\ ,\ v_5=(0,0,-1)\,.
\ee
The 4d cones are
\be
\{\tilde{v} \tilde{v}_1 \tilde{v}_2 \tilde{v}_4\ ,\ \tilde{v} \tilde{v}_1 \tilde{v}_3 \tilde{v}_4\ ,\ \tilde{v} \tilde{v}_2 \tilde{v}_3 \tilde{v}_4\ ,\ \tilde{v} \tilde{v}_1 \tilde{v}_2 \tilde{v}_5\ ,\ \tilde{v} \tilde{v}_1 \tilde{v}_3 \tilde{v}_5\ ,\ \tilde{v} \tilde{v}_2 \tilde{v}_3 \tilde{v}_5\}\,.
\ee
The Picard group of $X_4$ is generated by the compact divisor $D$ associated to $\tilde{v}$, the non-compact divisor $S$ associated to $\tilde{v}_4$ and $H$ associated to $\tilde{v}_1$, see the following figure of $\mc{P}_3$:
\begin{center}
\includegraphics[height=4cm]{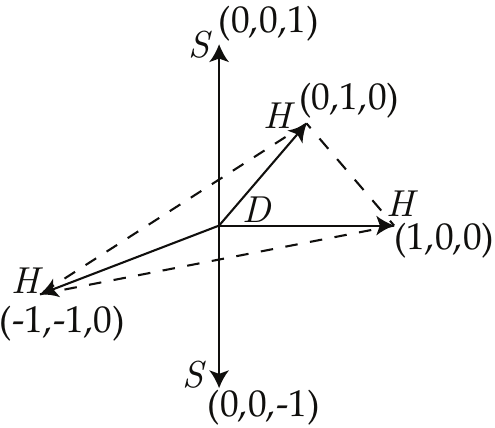}
\end{center}
The canonical class of $D$ is:
\be
K_D=(-2S-3H)|_D\,.
\ee
The triple intersection numbers on $D$ are
\be
S\cdot H^2|_D=1\ ,\ S^3|_D=S^2\cdot H|_D=H^3|_D=0\,.
\ee
The (Poincar\'{e} dual) of K\"{a}hler form is taken as
\be
\ba
\label{P1P2-kahler}
J^c&=xS+yH\cr
&=-\frac{y}{3}D+\left(x-\frac{2y}{3}\right)S\,.
\ea
\ee
We discuss the following physical aspects.

\paragraph{K\"{a}hler moduli and $U(1)$ gauge coupling} 

From (\ref{P1P2-kahler}) we can compute the volume of curves and cycles
\be
\ba
\label{P1P2-volumes}
&\text{Vol}(D\cdot S\cdot H)=y\ ,\ \text{Vol}(D\cdot H\cdot H)=x\,,\cr
&\text{Vol}(D\cdot S)=\frac{1}{2}y^2\ ,\ \text{Vol}(D\cdot H)=xy\,,\cr
&\text{Vol}(D)=\frac{1}{2}xy^2\,.
\ea
\ee
Hence if all of these volume are non-nagetive, it is required that $x>0$, $y>0$.

The $U(1)$ gauge coupling (\ref{gauge-coupling}) is
\be
\frac{1}{g^2}=\text{Vol}(-D\cdot D)=y^2+3xy\,.
\ee

\paragraph{Charged particles from M2-brane wrapping modes and 1-form symmetry}

For an M2-brane wrapping a curve in the homology class
\be
C_{ab}=D\cdot(a S\cdot H+b H^2)\,,
\ee
its charge under the $U(1)$ gauge group is
\be
C_{ab}\cdot D=(C\cdot K_D)|_D=-2b-3a\,.
\ee
For example the curve $C_{10}= D\cdot S\cdot H$ has charge 3, which breaks the $\mb{Z}_2$ 1-form symmetry of the $SU(2)$ gauge group. 

On the other hand, $C_{01}=D\cdot H\cdot H$ is the $\mb{P}^1$ fiber with normal bundle $N_{C_{01}|X_4}=\mc{O}\oplus\mc{O}\oplus\mc{O}(-2)$. 

\paragraph{$SU(2)$ gauge theory limit}

M2-brane wrapping $C_{01}$ acts as the W-boson for the $SU(2)$ gauge group, which become massless in the limit 
\be
\ba
\text{Vol}(C_{01})&=x=0\,.
\ea
\ee
In this limit if $y>0$, we get an $SU(2)$ gauge theory description with $SU(2)$ gauge coupling\
\be
\frac{1}{g_{YM}^2}=y^2\,.
\ee
Furthermore, in the limit $y\rightarrow 0$, we get a strongly coupled SCFT $\FTsing$ where all the volume of compact cycles (\ref{P1P2-volumes}) shrink to zero.

\paragraph{Flavor symmetry $G_F$ for $\FTsing$}

In this case, there is no flavor W-boson from M2-brane wrapping a curve with normal bundle $\mc{O}\oplus\mc{O}\oplus\mc{O}(-2)$, and the flavor symmetry would be $G_F=\mk{u}(1)$ for $\FTsing$ at the singular limit. Let us choose the non-compact divisor generating $G_F$ to be $S$, the charges of BPS states from M2-brane wrapping the generators of Mori cone are in table~\ref{P1P2-charges}. $Q_e$ is the gauge $U(1)$ charge and $Q_A$ is the flavor charge.

\begin{table}
\centering
\begin{tabular}{c|cc}
$U(1)$ Charge & $Q_e$ & $Q_A$\\
\hline
Divisor & $D$  & $S$\\
\hline
$C_{10}$ & $-3$ & 0\\
$C_{01}$ & $-2$ & 1
\end{tabular}
\caption{Gauge and flavor charges of M2-brane wrapping states on local $\mb{P}^1\times\mb{P}^2$.}\label{P1P2-charges}
\end{table}

\paragraph{$c_2$ and additional $G_4$ flux on the resolved $X_4$}

Now we can compute the Poincar\'{e} dual of $c_2(X_4)$, which is
\be
c_2^c(X_4)=2D\cdot S+3D\cdot H+3H^2+6S\cdot H\,.
\ee
From
\be
\frac{1}{2}c_2^c(X_4)\cdot D \cdot H\in\mb{Z}\ ,\ \frac{1}{2}c_2^c(X_4)\cdot D \cdot S\in\mb{Z}\,,
\ee
we see that a non-zero $G_4$ flux is not required.

On the other hand, if one chooses to turn on $G_4$ flux, the most general $G_4^c$ dual to compact 4-cycles satisfying (\ref{G4c-quant}) is
\be
G_4^c=mD\cdot S+nD\cdot H (m,n\in\mb{Z})\,.
\ee
We can compute the chirality for the adjoint matter charged under $SU(2)$, which comes from M2-brane wrapping the $\mb{P}^1$ fiber $D\cdot H\cdot H$. The curve $D\cdot H\cdot H$ has normal bundle $N_{D\cdot H\cdot H}=\mc{O}\oplus\mc{O}\oplus\mc{O}(-2)$
\be
\ba
\chi &=-\int_{D\cdot S} c_1(D\cdot S)c_1(L)\cr
&=D\cdot G_4^c\cdot (-3H)\cr
&=(m S+n H)\cdot K_D\cdot (-3H)\cr
&=9m+6n\,.
\ea
\ee

From the $G_4$ flux there are induced CS terms from 11d SUGRA action (\ref{Eff-CS})(\ref{CS-G4-flux})
\be
\ba
S_{CS}&=\frac{1}{4\pi}(G_4^c\cdot D\cdot D)\int A\wedge dA+\frac{1}{2\pi}(G_4^c\cdot D\cdot S)\int A\wedge dB+\frac{1}{4\pi}(G_4^c\cdot S\cdot S)\int B\wedge dB\cr
&=\frac{1}{4\pi}(9m+12n)\int A\wedge dA-\frac{3n}{2\pi}\int A\wedge dB\,.
\ea
\ee
Here $A$ is the $U(1)$ gauge field and $B$ is the $G_F=\mk{u}(1)$ background gauge field.

\subsubsection{Local $\mb{P}^1\times\mb{P}^1\times\mb{P}^1$}\label{secP1cube}

\begin{figure}[h]
\centering
\includegraphics[height=4cm]{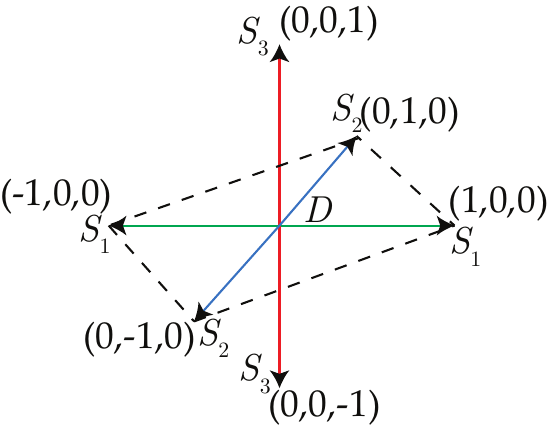}
\caption{$\mc{P}_3$ of local $\mb{P}^1\times\mb{P}^1\times\mb{P}^1$.}\label{fig:P3_P1P1P1}
\end{figure}

The case of $X_4=$ local $\mb{P}^1\times\mb{P}^1\times\mb{P}^1$ is particularly interesting because it admits multiple $\mb{P}^1$-fibration structures and a non-trivial flavor symmetry enhancement $G_F=\mk{su}(3)$. We write down the rays in the toric fan of $X_4$ $\tilde{v}_i=(v_i,1)$:
\be
\ba
&v=(0,0,0)\ ,\ v_1=(1,0,0)\ ,\ v_2=(0,1,0)\ ,\ v_3=(0,0,1)\ ,\ v_4=(-1,0,0)\ ,\ \cr
&v_5=(0,-1,0)\ ,\ v_6=(0,0,-1)\,.
\ea
\ee
The cones are
\be
\label{P1P1P1cones}
\{\tilde{v} \tilde{v}_1 \tilde{v}_2 \tilde{v}_3\ ,\ \tilde{v} \tilde{v}_1 \tilde{v}_2 \tilde{v}_6\ ,\ \tilde{v} \tilde{v}_1 \tilde{v}_3 \tilde{v}_5\ ,\ \tilde{v} \tilde{v}_1 \tilde{v}_5 \tilde{v}_6\ ,\ \tilde{v} \tilde{v}_2 \tilde{v}_3 \tilde{v}_4\ ,\ \tilde{v} \tilde{v}_2 \tilde{v}_4 \tilde{v}_6\ ,\ \tilde{v} \tilde{v}_3 \tilde{v}_4 \tilde{v}_5\ ,\ \tilde{v} \tilde{v}_4 \tilde{v}_5 \tilde{v}_6\}\,.
\ee
The Picard group of $X_4$ is generated by the divisors $D$, $S_1$, $S_2$ and $S_3$ which corresponds to the rays $\tilde{v}$, $\tilde{v}_1$, $\tilde{v}_2$ and $\tilde{v}_3$ respectively. The corresponding 3D polytope $\mc{P}_3$ is drawn in Figure~\ref{fig:P3_P1P1P1}.
The canonical class of $D$ is:
\be
K_D=(-2S_1-2S_2-2S_3)|_D\,.
\ee
The intersection numbers on $D$ are
\be
S_1\cdot S_2\cdot S_3|_D=1\ ,\ S_1^2|_D=S_2^2|_D=S_3^2|_D=0\,.
\ee
We discuss the following physical aspects of the model.

\paragraph{K\"{a}hler form and $U(1)$ gauge coupling}

We take the (Poincar\'{e} dual of) the K\"{a}hler class to be
\be
\ba
\label{P1P1P1-Kahler}
J^c&=xS_1+yS_2+zS_3\cr
&=-\frac{1}{6}(x+y+z)D+\frac{1}{3}(2x-y-z)(S_1-S_2)+\frac{1}{3}(x+y-2z)(S_2-S_3)\,.
\ea
\ee
The volume of curves are
\be
\label{P1P1P1-curveVol}
\text{Vol}(D\cdot S_1\cdot S_2)=z\ ,\ \text{Vol}(D\cdot S_2\cdot S_3)=x\ ,\ \text{Vol}(D\cdot S_1\cdot S_3)=y\,.
\ee
The criteria for all the compact cycles to have non-negative volume is $x\geq 0$, $y\geq 0$, $z\geq 0$.

The gauge coupling of $U(1)$ CB gauge group is (\ref{gauge-coupling})
\be
\ba
\label{P1P1P1-coupling}
\frac{1}{g^2}&=-\frac{1}{2}D\cdot D\cdot J^c\cdot J^c\cr
&=2(xy+xz+yz)\,.
\ea
\ee
We can see that it is equal to the sum of the volume of all the toric divisors of $D$
\be
\frac{1}{g^2}=2\text{Vol}(D\cdot S_1)+2\text{Vol}(D\cdot S_2)+2\text{Vol}(D\cdot S_3)\,.
\ee

\paragraph{Charged particles from M2-brane wrapping modes and 1-form symmetry}

For M2-brane wrapping a curve in the homology class
\be\label{eq:Cabc_P1P1P1}
C_{abc}=D\cdot(a S_1\cdot S_2+b S_2\cdot S_3+c S_1\cdot S_3)\,,
\ee
its charge under the $U(1)$ gauge group is
\be
\label{P13Cabc-charge}
C_{abc}\cdot D=-2a-2b-2c\,,
\ee
which is always a multiple of 2, hence the $\mb{Z}_2$ 1-form symmetry is preserved on the Coulomb branch. 

The Mori cone of $X_4$ is generated by $C_{100}$, $C_{010}$ and $C_{001}$, note that these curves all have normal bundle $N_{C_{100}|X_4}=N_{C_{010}|X_4}=N_{C_{001}|X_4}=\mc{O}\oplus\mc{O}\oplus\mc{O}(-2)$. Their charges under the $U(1)$ gauge group are
\be
C_{100}\cdot D=C_{010}\cdot D=C_{001}\cdot D=-2\,.
\ee
Hence M2-brane wrapping one of these curves would give rise to a vector multiplet, which can act as the W-boson for an $SU(2)$ gauge group in a certain $\mb{P}^1$ fibration limit, which will be discussed later.

\paragraph{Non-abelian flavor symmetry enhancement $G_F$}

Now let us discuss the non-abelian 0-form symmetry enhancement at the singular point $X_{\text{4,sing}}$. We first identify the generators of the flavor Cartan subalgebra $\mk{u}(1)^2$, which are
\be
\label{P1P1P1-F12}
F_1=S_1-S_2\ ,\ F_2=S_2-S_3\,.
\ee
We identify the flavor W-bosons of $F_1$ and $F_2$ to be M2-branes wrapping
\be
\label{P1P1P1-C1}
C_1=D\cdot S_3\cdot (S_1-S_2)
\ee
and
\be
\label{P1P1P1-C2}
C_2=D\cdot S_1\cdot (S_2-S_3)\,.
\ee
One can check the following requirements:
\begin{enumerate}
{\item The flavor W-bosons $C_1$ and $C_2$ are neutral under the $U(1)$ gauge symmetry generated by $D$, as one can check that $C_1\cdot D=C_2\cdot D=0$.}
{\item The flavor W-bosons $C_1$ and $C_2$ has the correct normal bundle $N_{C_1|X_4}=N_{C_2|X_4}=\mc{O}\oplus\mc{O}\oplus\mc{O}(-2)$. }
\end{enumerate}
Note that the curves $C_1$ and $C_2$ are not effective on the particular resolved geometry, they represent degrees of freedom that are decoupled to the 3d $\mc{N}=2$ SCFT in the singular limit\footnote{The story is analogous to the $SU(2)$ flavor symmetry enhancement of the 5d $SU(2)_0$ SCFT from M-theory on local $\mb{P}^1\times\mb{P}^1$, where the flavor Cartan and flavor W-boson are not effective.}. 

Now let us compute the charge of $C_1$ and $C_2$ under the flavor Cartan generators:
\be\label{su3cartan}
\begin{array}{c|cc}
& F_1 & F_2\\
\hline
C_1 & -2 & 1\\
C_2 & 1 & -2
\end{array}
\ee
As one can see, they form the Cartan matrix of an $\mk{su}(3)$ flavor algebra. Hence the 0-form flavor symmetry of the local $\mb{P}^1\times\mb{P}^1\times\mb{P}^1$ theory is actually enhanced to $G_F=\mk{su}(3)$ at the singular limit!

One can also check that the curves $C_{abc}$ with the same normal bundles form $\mk{su}(3)$ representations. Given a triple $(a,b,c)$, the curves 
\be
C_{abc}\ ,\ C_{acb}\ ,\ C_{bac}\ ,\ C_{bca}\ ,\ C_{cab}\ ,\ C_{cba}
\ee
all have the same normal bundles, which can be combined into a larger representation.

For example, let us consider the case of $(a,b,c)=(1,0,0)$. The curves $C_{100}$, $C_{010}$, $C_{001}$ have the charges $Q_e$ under the gauge Cartan and $Q_A$, $Q_B$ under the flavor Cartans as shown in table~\ref{t:BPSchargeP1cube}. Hence one can see that they form a $\mbf{3}$ representation of the enhanced flavor symmetry $\mk{su}(3)$.

For the case of $(a,b,c)=(1,1,0)$. The curves $C_{110}$, $C_{011}$, $C_{101}$ has the following charges under the flavor Cartan
\be
\begin{array}{c|cc}
& F_1 & F_2\\
\hline
C_{110} & 1 & -1\\
C_{011} & 0 & 1\\
C_{101} & -1 & 0
\end{array}
\ee
They form a $\bar{\mathbf{3}}$ representation of the enhanced flavor symmetry $\mk{su}(3)$.


\begin{table}
\centering
\begin{tabular}{c|c|c|c}
$U(1)$ Charge & $Q_e$ & $Q_A$ & $Q_B$\\
\hline
Divisor & $D$  & $F_1$ & $F_2$\\
\hline
\hline
$C_{100}$ & $-2$ & 0 & $-1$\\
$C_{010}$ & $-2$ & 1 & 0\\
$C_{001}$ & $-2$ & $-1$ & 1
\end{tabular}
\caption{U(1) gauge charge and flavor charges of BPS states from M2-brane wrapping the Mori cone generators of local $\mb{P}^1\times\mb{P}^1\times\mb{P}^1$.}\label{t:BPSchargeP1cube}
\end{table}




\paragraph{$c_2$ and additional $G_4$ flux on the resolved $X_4$}

Now let us discuss the $G_4$ flux. As $c_2(X_4)$ is an even element in $H^4(X_4,\mb{Z})$, a non-zero $G_4$ flux is not required. Nonetheless, one can still choose to add $G_4$ flux of the general form
\be
G_4^c=D\cdot(mS_1+n S_2+k S_3)\,.
\ee
The flux will induce chiral matter from M2-brane wrapping $C_{100}$, $C_{010}$ and $C_{001}$.

The moduli space of $C_{100}$ is $D\cdot S_3$, and we can compute the chirality of $C_{100}$ as
\be
\ba
\chi(C_{100})&=G_4^c\cdot D\cdot (-2S_1-2S_2)\cr
&=4m+4n+8k\,.
\ea
\ee
Similarly, the moduli space of $C_{010}$ and $C_{001}$ are $S_1$ and $S_2$ respectively, and we have
\be
\ba
\chi(C_{010})&=8m+4n+4k\cr
\chi(C_{001})&=4m+8n+4k\,.
\ea
\ee

Note that the chiralities $\chi(C_{100})$, $\chi(C_{010})$ and $\chi(C_{001})$ are different for generic choices of $m$, $n$ and $k$ which in turn seems to imply that the $SU(3)$ flavor symmetry is broken for a generic choice of $G_4$-flux since we have shown in table~\ref{t:BPSchargeP1cube} that M2-branes wrapping $(C_{100}, C_{010}, C_{001})$ form a $\mathbf{3}$ under $SU(3)$ without $G_4$-flux. This is not surprising if we look solely at the curves $C_{abc}$ since a generic choice of $G_4^c$ clearly breaks the symmetric group $S_3$ symmetry of (\ref{eq:Cabc_P1P1P1}), which one can also see from Figure~\ref{fig:P3_P1P1P1} that there is an apparent symmetry among $S_1$, $S_3$ and $S_3$ before introducing $G_4$-flux. But this is a bit puzzling if we look at M2-brane wrapping modes on $C_1$ and $C_2$ from which we obtain the 6 flavor W-bosons of the enhanced $SU(3)$ flavor symmetry. The $G_4$-flux does not give rise to non-trivial chiral indices for the states obtained from M2-brane wrapping $C_1$ and $C_2$ therefore the $SU(3)$ flavor symmetry does not seem to be broken at all from this point of view.

For this, recall that in Section~\ref{sec:G4flux} we have commented that a careless choice of $G_4$-flux may prevent certain compact 4-cycle from shrinking therefore the SCFT limit cannot be achieved. Consider for example $S = D\cdot(S_1 + S_2) \subset D$, we have $G_4^c\cdot S \neq 0$ for generic $G_4^c$. Therefore $S$, hence $D$ cannot be shrunk. But since we have shown that the flavor W-bosons are obtained from M2-brane wrapping $D\cdot S_1\cdot (S_2 - S_3)$ and $D\cdot S_3\cdot (S_1 - S_2)$ respectively, the fact that $D$ being unshrinkable implies that the enhancement to $SU(3)$ can never be realized with generic $G_4^c$. In this sense the breaking of $SU(3)$ by $G_4^c$ becomes almost tautological.

From the $G_4$ flux there are induced CS terms
\be
\ba
S_{CS}&=\frac{1}{4\pi}(8m+8n+8k)\int A\wedge dA+\frac{1}{2\pi}(2m-2n)\int A\wedge dB_1+\frac{1}{2\pi}(2n-2k)\int A\wedge dB_2\cr
&+\frac{1}{2\pi}(k+m-n)\int B_1\wedge dB_2-\frac{2k}{4\pi}\int B_1\wedge dB_1-\frac{2m}{4\pi}\int B_2\wedge dB_2\,.
\ea
\ee
Here $A$ is the $U(1)$ gauge field, $B_1$ and $B_2$ are the  background gauge fields of the flavor Cartan $F_1$ and $F_2$ of $G_F=\mk{su}(3)$.

\paragraph{SymTFT from geometry}

Although adding a free $G_4$ flux would destroy the validity of geometric singularity description, one can still discuss the addition of torsional $G_4$ flux and the resulting SymTFT action for the 3d $\mc{N}=2$ theory, as in section~\ref{sec:higher_form_symmetry}.

Let us first compute the intersection matrix between 4-cycles
\be
\{\mc{S}_i\cdot\mc{S}_j\}=
\begin{array}{c|ccc}
& D\cdot S_1 & D\cdot S_2 & D\cdot S_3\\
\hline
D\cdot S_1 & 0 & -2 & -2\\
D\cdot S_2 & -2 & 0 & -2\\
D\cdot S_3 & -2 & -2 & 0
\end{array}
\ee
Its Smith normal form is $SNF(\{\mc{S}_i\cdot\mc{S}_j\})=$diag$(2,2,4)$. The generators of $\mb{Z}_2\oplus\mb{Z}_2\oplus\mb{Z}_4$ $(-1)$-form symmetries are
\be
Z_4^1=D\cdot S_1\ ,\ Z_4^2=D\cdot S_2\ ,\ Z_4^3=D\cdot (-S_1-S_2+S_3)\,.
\ee
We can thus plug in (\ref{SymTFT}) to compute the SymTFT action
\be
\ba
\frac{S_{\rm SymTFT}}{2\pi}&=\frac{1}{2}\int_{\mc{M}_4}[(b_2-b_1) F_2^1+(b_3-b_2) F_2^2]\cup B_2+\frac{1}{4}(2b_1+2b_2-b_3)\int_{\mc{M}_4}B_2\cup B_2\,.
\ea
\ee
Here $b_i$ are the torsional flux associated to $Z_4^i$, $F_2^i$ $(i=1,2)$ are field strength of $U(1)$ background gauge fields corresponding to the flavor Cartan generators in (\ref{P1P1P1-F12}) and $B_2$ is the background gauge field for the $\mb{Z}_2$ 1-form symmetry. 

\paragraph{$\mb{P}^1$-fibration structures and the triality}

Now let us discuss the different $\mb{P}^1$-fibration structures, following section~\ref{sec:fibration}. We have three different $\mb{P}^1$-fibration structures along the red, green and blue lines in figure~\ref{fig:P3_P1P1P1}. In each of these cases when the $\mb{P}^1$-fiber shrinks to zero volume, we have an effective UV $SU(2)$ gauge theory description. 

For instance let us consider the $\mb{P}^1$-fibration structure in red, where $D\cdot S_3$ is the base while $C_{100}\equiv D\cdot S_1\cdot S_2$ is the $\mb{P}^1$ fiber. When $z=0$, the $\mb{P}^1$ fiber $C_{100}$ shrinks to zero volume. We require that in this geometric limit, $x,y>0$. In this case, the gauge coupling (\ref{P1P1P1-coupling}) becomes the $SU(2)$ gauge coupling
\be
\label{P1P1P1-SU2-coupling}
\frac{1}{g_{YM}^2}=2xy\,.
\ee

For the BPS states in table~\ref{t:BPSchargeP1cube}, $C_{100}$ becomes the massless W-boson for $SU(2)$ in this case, while $C_{010}$ and $C_{001}$ are two massive vector multiplets. For the physics interpretation, we conjecture that $C_{010}$ can be thought as the twisted dimensional reduction of the dyonic instanton in 5d $SU(2)_0$ theory~\cite{Jia:2021ikh} on $\mb{P}^1=C_{001}=D\cdot S_1\cdot S_3$. For the other state $C_{001}$, it is regarded as  the twisted dimensional reduction of the dyonic instanton in 5d $SU(2)_0$ theory on $C_{010}$ as well. These BPS states are all $(0+1)$-dimensional disorder operators in the 3d $\mc{N}=2$ $SU(2)$ field theory. 

Alternatively, we can argue that $C_{010}$ and $C_{001}$ cannot be charged local operators charged under $SU(2)$ gauge group, because they cannot be completed into a full $SU(2)$ adjoint representation (the single uncharged $U(1)$ gauge boson already formed the $SU(2)$ gauge group with the W-boson $C_{100}$). Thus in the $SU(2)$ gauge theory description, such operators can only be disorder operators.

Now let us discuss the flavor symmetry in this limit. The volume of the flavor W-bosons (\ref{P1P1P1-C1}), (\ref{P1P1P1-C2}) are
\be
\text{Vol}(F_1)=y-x\ ,\ \text{Vol}(F_2)=z-y\,.
\ee
Now we find an interesting phenomenon: when $x=y$, the volume of the flavor W-boson $F_1$ vanishes, and we have an enhanced flavor symmetry $\mk{su}(2)\oplus\mk{u}(1)$. When $x\neq y$, the flavor symmetry is still $\mk{u}(1)^{\oplus 2}$.

When going to the $\mb{P}^1$-fibration structure in green, we take Vol$(C_{010})=x=0$ and $y,z>0$ in (\ref{P1P1P1-curveVol}). the disorder operator from M2-brane wrapping $C_{010}$ becomes the massless W-boson in this limit. On the other hand, the old W-boson $C_{100}$ becomes a disorder operator. The $SU(2)$ gauge coupling is now
\be
\frac{1}{g_{YM}^2}=2yz\,.
\ee
Finally, if we take the $\mb{P}^1$-fibration structure in blue, which means that Vol$(C_{001})=y=0$ and $x,z>0$. Now M-brane wrapping the curve $C_{001}$ becomes the massless W-boson while the other $C_{ab0}$ corresponds massive disorder operators. The $SU(2)$ gauge coupling in this limit is
\be
\frac{1}{g_{YM}^2}=2xz\,.
\ee

\begin{figure}
\centering
\includegraphics[height=6cm]{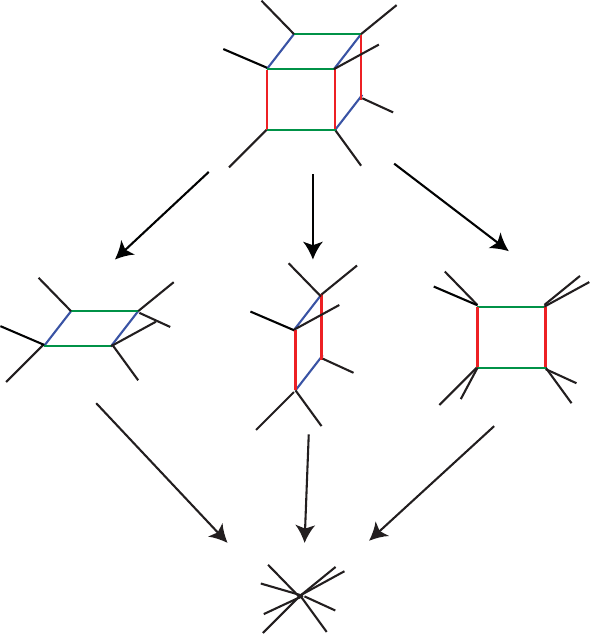}
\caption{Three different geometric limits of the local $\mb{P}^1\times\mb{P}^1\times\mb{P}^1$ geometry, where one of the $\mb{P}^1$ fibers shrink to zero volume. They represents three different $SU(2)$ gauge theory descriptions.}\label{f:P1triality}
\end{figure}

In summary, there are three $\mb{P}^1$-fibration structures, and we have a triality among the three different limits, see figure~\ref{f:P1triality}.

\subsubsection{Local $\mb{P}^1\times\mb{F}_1$}\label{sec:p1F1}

\begin{figure}[h]
\centering
\includegraphics[height=4cm]{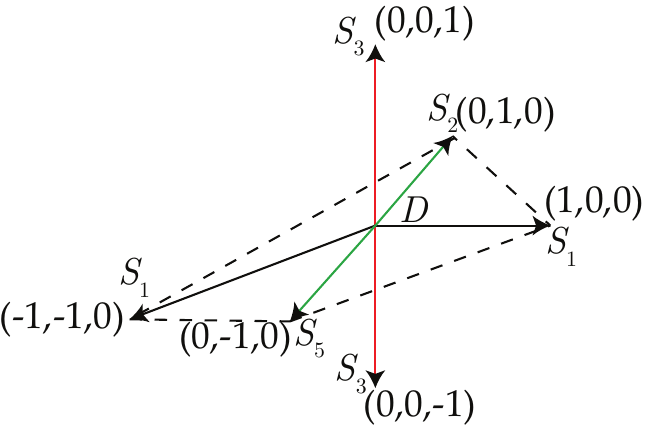}
\caption{$\mc{P}_3$ of local $\mb{P}^1\times\mb{F}_1$.}\label{fig:P3_P1F1}
\end{figure}

We write down the rays in the toric fan of CY4 $\tilde{v}_i=(v_i,1)$:
\be
\ba
&v=(0,0,0)\ ,\ v_1=(1,0,0)\ ,\ v_2=(0,1,0)\ ,\ v_3=(0,0,1)\ ,\ v_4=(-1,-1,0)\ ,\ \cr
&v_5=(0,-1,0)\ ,\ v_6=(0,0,-1)\,.
\ea
\ee
The cones are the same as the local $\mb{P}^1\times\mb{P}^1\times\mb{P}^1$ case (\ref{P1P1P1cones}).
The Picard group of $X_4$ is generated by the divisors $D$, $S_1$, $S_3$ and $S_5$ which corresponds to the rays $\tilde{v}$, $\tilde{v}_1$, $\tilde{v}_3$ and $\tilde{v}_5$ respectively, see figure~\ref{fig:P3_P1F1}.

The canonical class of $D$ is:
\be
K_D=(-3S_1-2S_3-2S_5)|_D\,.
\ee
The intersection numbers on $D$ are
\be
S_1\cdot S_3\cdot S_5|_D=1\ ,\ S_1^2|_D=S_3^2|_D=0\ ,\ S_1\cdot S_5^2|_D=0\ ,\ S_3\cdot S_5^2|_D=-1\,.
\ee
We discuss the following physical aspects of the model.

\paragraph{Charged particles from M2-brane wrapping modes and 1-form symmetry}

For M2-brane wrapping a curve in the homology class
\be
C_{abc}=D\cdot(a S_1\cdot S_3+b S_1\cdot S_5+c S_3\cdot S_5)\,,
\ee
its charge under the $U(1)$ gauge group is
\be
C_{abc}\cdot D=-2a-2b+c\,.
\ee
The 1-form symmetry is trivial.

The Mori cone generators of $X_4$ are
\begin{itemize}
\item $C_{100}=D\cdot S_1\cdot S_3$ with normal bundle $N_{C_{100}|X_4}=\mc{O}\oplus\mc{O}\oplus\mc{O}(-2)$. M2-brane wrapping $C_{100}$ gives a vector multiplet.
\item $C_{010}=D\cdot S_1\cdot S_5$ with normal bundle $N_{C_{100}|X_4}=\mc{O}\oplus\mc{O}\oplus\mc{O}(-2)$. In fact, $C_{010}$ is the $\mb{P}^1$ fiber in $D=\mb{P}^1\times\mb{F}_1$. M2-brane wrapping $C_{010}$ gives a vector multiplet as well.
\item $C_{001}=D\cdot S_3\cdot S_5$ with $N_{C_{001}|X_4}=\mc{O}\oplus\mc{O}(-1)\oplus\mc{O}(-1)$ and moduli space $\mb{P}^1$. Following the discussions in section~\ref{sec:M2modes}, when there is no $G_4$ flux, there is no BPS state (zero mode) from M2-brane wrapping $C_{001}$. 
\end{itemize}

Despite that $C_{001}=D\cdot S_3\cdot S_5$ gives no zero mode, one can consider the curve $C_{101}=D\cdot S_3\cdot S_2$ with normal bundle $\mc{O}\oplus\mc{O}(1)\oplus\mc{O}(-3)$. M2-brane wrapping such curve would give a rich set of higher-spin massive particles on the CB.

\paragraph{Non-abelian flavor symmetry enhancement $G_F$}

The non-compact divisor generating flavor Cartan of $\mk{su}(2)$ is $F_1=S_5-S_3+S_1$, with flavor W-boson
\be
\ba
C&=D\cdot (S_5-S_3+S_1)\cdot S_1\cr
&=D\cdot (S_5-S_3)\cdot S_1\,,
\ea
\ee
one can compute that $N_{C|X_4}=\mc{O}\oplus\mc{O}\oplus\mc{O}(-2)$ and $C\cdot D=0$. The non-compact divisor for the other flavor Cartan can be chosen as $F_2=S_1$\footnote{Note that there are freedom to choose $F_1$ and $F_2$. One can choose $F_1=S_5-S_3+aS_1$ $(a\in\mb{Z})$ and $F_2$ to be an arbitrary linear combination of non-compact divisors.}. Hence the enhanced flavor symmetry is $G_F=\mk{su}(2)\oplus\mk{u}(1)$.

We list the gauge charges and flavor symmetry charges under $F_1$, $F_2$ for M2-branes wrapping Mori cone generators $C_{100}$, $C_{010}$, $C_{001}$ in table~\ref{t:BPSchargeP1F1}. $Q_e$ is the gauge charge and $Q_A$, $Q_B$ are flavor charges.

\begin{table}
\centering
\begin{tabular}{c|c|c|c}
$U(1)$ Charge & $Q_e$ & $Q_A$ & $Q_B$\\
\hline
Divisor & $D$  & $F_1$ & $F_2$\\
\hline
\hline
$C_{100}$ & $-2$ & 1 & 0\\
$C_{010}$ & $-2$ & $-1$ & 0\\
$C_{001}$ & $-1$ & 0 & 1
\end{tabular}
\caption{U(1) gauge charge and flavor charges of BPS states from M2-brane wrapping the Mori cone generators of local $\mb{P}^1\times\mb{F}_1$.}\label{t:BPSchargeP1F1}
\end{table}

\paragraph{$c_2$ and additional $G_4$ flux on resolved $X_4$}

Now we consider $G_4$ flux. First we compute
\be
c_2^c(X_4)=2D\cdot S_3+3D\cdot S_1+6S_1\cdot S_3+4S_2\cdot S_3\,,
\ee
which has even intersection number with any 4-cycle on $X_4$. Hence no non-zero $G_4$-flux is required. Nonetheless we can add $G_4$-flux whose most general form is:
\be
G_4^c=mD\cdot S_1+nD\cdot S_3+kD\cdot S_5\quad (m,n,k\in\mb{Z}).
\ee

Now we compute the chirality of certain charged matter fields from M2-brane wrapping 2-cycles. For an M2-brane wrapping $C_{100}=D\cdot S_1\cdot S_3$ with normal bundle $N_{C_{100}|X_4}=\mc{O}\oplus\mc{O}\oplus\mc{O}(-2)$, its moduli space is $D\cdot S_5\sim\mb{P}^1\times\mb{P}^1$. The chirality is
\be
\chi(C_{100})=-\int_{D\cdot S_5} c_1(D\cdot S_5)c_1(L)=D\cdot G_4^c\cdot (-2S_1-2S_3)=2k+4m\,.
\ee
For an M2-brane wrapping $C_{010}=D\cdot S_1\cdot S_5$ with normal bundle $N_{C_{010}|X_4}=\mc{O}\oplus\mc{O}\oplus\mc{O}(-2)$, its moduli space is $D\cdot S_3\sim\mb{F}_1$. The chirality is
\be
\chi(C_{010})=-\int_{D\cdot S_2} c_1(D\cdot S_3)c_1(L)=D\cdot G_4^c\cdot (-3S_1-2S_5)=2k+4m+4n\,.
\ee
For an M2-brane wrapping $C_{001}=D\cdot S_3\cdot S_5$ with normal bundle $N_{C_{001}|X_4}=\mc{O}\oplus\mc{O}(-1)\oplus\mc{O}(-1)$, its moduli space is $D\cdot S_1\cdot S_5\sim\mb{P}^1$. To compute the chirality we need to integrate $G_4$ over the matter surface $D\cdot S_5$. The chirality is
\be
\chi(C_{001})=D\cdot S_5\cdot G_4^c=2k-2m+n\,.
\ee
Finally, we compute the effective Chern-Simons term from 11d SUGRA action, which is free of parity anomaly
\be
\ba
S_{CS}&=\frac{1}{4\pi}(8m+8n+4k)\int A\wedge dA+\frac{1}{2\pi}(k-3n)\int A\wedge dB_1-\frac{1}{2\pi}(2k+2n)\int A\wedge dB_2\cr
&+\frac{1}{2\pi}(k-n)\int B_1\wedge dB_2+\frac{1}{4\pi}(n-2m)\int B_1\wedge dB_1\,.
\ea
\ee
Here $A$ is the $U(1)$ gauge field, $B_1$ and $B_2$ are background gauge fields of the flavor Cartans $F_1$ and $F_2$.

\paragraph{$\mb{P}^1$-fibration structures and duality}

Let us write the Poincar\'{e} dual of the K\"{a}hler form in terms of
\be
J^c=xS_1+yS_3+zS_5\,.
\ee
The $U(1)$ gauge coupling on CB is (\ref{gauge-coupling})
\be
\ba
\frac{1}{g^2}&=-\frac{1}{2}D\cdot D\cdot J^c\cdot J^c\cr
&=2(xy+xz+yz)-z^2\,.
\ea
\ee
The volume of various curves are
\be
\text{Vol}(C_{100})=z\ ,\ \text{Vol}(C_{010})=y\ ,\ \text{Vol}(C_{001})=x-z\,.
\ee
For all the compact cycles to have a non-negative volume, we require that $x\geq z\geq 0$, $y\geq 0$.

Local $\mb{P}^1\times\mb{F}_1$ has two distinct $\mb{P}^1$ fibration structures.

If one shrink the $\mb{P}^1$ fiber $C_{010}=D\cdot S_1\cdot S_5$ along the red direction in figure~\ref{fig:P3_P1F1}, the gauge theory description is $SU(2)$ with massive matter fields from the curves $C_{a0c}$ in the base $D\cdot S_3\sim\mb{F}_1$. In this limit $y=0$, $x>z>0$, and the $SU(2)$ gauge coupling is
\be
\frac{1}{g_{YM}^2}=2xz-z^2\,.
\ee

On the other hand, if one shrink the $\mb{P}^1$ fiber $C_{100}=D\cdot S_1\cdot S_3$ along the green direction, the gauge theory description is $SU(2)$ with massive matter fields from the curves $C_{0bc}$ in the base $D\cdot S_5\sim\mb{F}_0$. In this limit $z=0$, $x,y>0$, and the $SU(2)$ gauge coupling is
\be
\frac{1}{g_{YM}^2}=2xy\,.
\ee

\subsubsection{Local $\mb{P}^1\times\mb{F}_2$}

\begin{figure}[h]
\centering
\includegraphics[height=4cm]{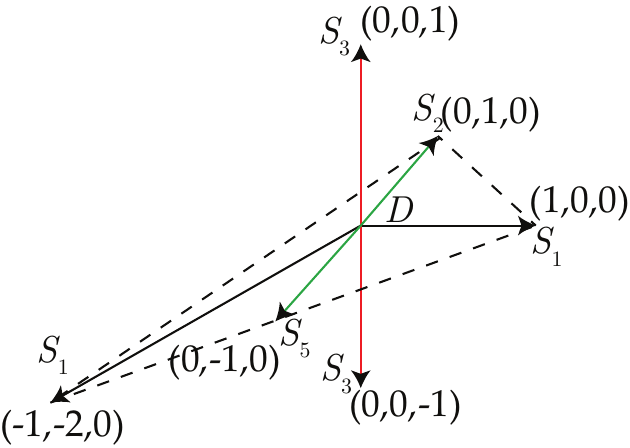}
\caption{$\mc{P}_3$ of local $\mb{P}^1\times\mb{F}_2$.}\label{fig:P3_P1F2}
\end{figure}

We write down the rays in the toric fan of CY4 $\tilde{v}_i=(v_i,1)$:
\be
\ba
&v=(0,0,0)\ ,\ v_1=(1,0,0)\ ,\ v_2=(0,1,0)\ ,\ v_3=(0,0,1)\ ,\ v_4=(-1,-2,0)\ ,\ \cr
&v_5=(0,-1,0)\ ,\ v_6=(0,0,-1)\,.
\ea
\ee
The cones are the same as the local $\mb{P}^1\times\mb{P}^1\times\mb{P}^1$ case (\ref{P1P1P1cones}).
The Picard group of $X_4$ is generated by the divisors $D$, $S_1$, $S_3$ and $S_5$ which corresponds to the rays $\tilde{v}$, $\tilde{v}_1$, $\tilde{v}_3$ and $\tilde{v}_5$ respectively, see figure~\ref{fig:P3_P1F2}.

In this case, the geometry and physics are almost exactly the same as the local $\mb{P}^1\times\mb{P}^1\times\mb{P}^1$ case, with the linear relation
\be
(S_5)|_{\mb{P}^1\times\mb{F}_2}=(S_2-S_1)_{\mb{P}^1\times\mb{P}^1\times\mb{P}^1}\,.
\ee
Hence in the case of local $\mb{P}^1\times\mb{F}_2$, $S_5$ is a flavor Cartan by itself, and $D\cdot S_5\cdot S_1$ is its flavor W-boson. In this case note that $D\cdot S_5\cdot S_1$ is an effective curve, nonetheless the M2-brane wrapping mode over it is neutral under the $U(1)$ gauge symmetry, hence it can be decoupled from the theory.

The main difference here is that there are only two $\mb{P}^1$ fibration structures, in red and green. The physical discussions of duality is similar to case of local $\mb{P}^1\times\mb{F}_2$, and we would not repeat here.

\subsubsection{Local $\mb{P}^1\times dP_2$}

\begin{figure}[h]
\centering
\includegraphics[height=4cm]{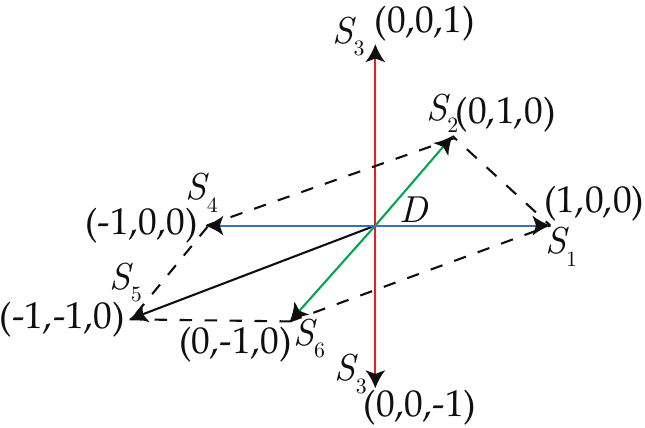}
\caption{$\mc{P}_3$ of local $\mb{P}^1\times dP_2$.}\label{fig:P3_P1dP2}
\end{figure}

The rays in the toric fan of $X_4$ are $\tilde{v}_i=(v_i,1)$
\be
\ba
&v=(0,0,0)\ ,\ v_1=(1,0,0)\ ,\ v_2=(0,1,0)\ ,\ v_3=(0,0,1)\ ,\ v_4=(-1,0,0)\ ,\ \cr
&v_5=(-1,-1,0)\ ,\ v_6=(0,-1,0)\ ,\ v_7=(0,0,-1)\,.
\ea
\ee
The Picard group of $X_4$ is generated by divisors $D$, $S_1$, $S_2$, $S_3$ and $S_5$ which corresponds to the rays $v$, $v_1$, $v_2$, $v_3$ and $v_5$ respectively, see figure~\ref{fig:P3_P1dP2}.

\be
K_D=(-2S_1-2S_2-2S_3+S_5)|_D\,.
\ee
The non-zero intersection numbers on $D$ are
\be
S_3\cdot S_1\cdot S_2|_D=1\ ,\ S_3\cdot S_5^2|_D=-1\,.
\ee

\paragraph{Charged particles from M2-brane wrapping modes and 1-form symmetry}

For the M2-brane wrapping a curve in the homology class
\be
C_{abcd}=D\cdot (aS_3\cdot S_1+bS_3\cdot S_2+cS_3\cdot S_5+dS_1\cdot S_2)\,,
\ee
its charge under the $U(1)$ gauge group is
\be
C_{abcde}\cdot D=-2a-2b-c-2d\,.
\ee
Hence there is no 1-form symmetry.

In this case, the Mori cone generators $D\cdot S_3\cdot S_4$, $D_\cdot S_3\cdot S_5$ and $D\cdot S_3\cdot S_6$ all have normal bundle $\mc{O}\oplus\mc{O}(-1)\oplus\mc{O}(-1)$, and there is no M2-brane zero mode. Nonetheless, the curves $C_{1000}=D\cdot S_3\cdot S_1$ and $C_{0100}=D\cdot S_3\cdot S_2$ have normal bundle $\mc{O}\oplus\mc{O}\oplus\mc{O}(-2)$
 and still give vector multiplets. Together wtih $C_{0001}=D\cdot S_1\cdot S_2$, they form the three possible $\mb{P}^1$ fibers in the three $\mb{P}^1$-fibration structures colored in red, green and blue in figure~\ref{fig:P3_P1dP2}.

\paragraph{Non-abelian flavor symmetry enhancement $G_F$}

The flavor Cartan subalgebra generators are
\be
\label{P1dP2-flavor}
F_1=S_1-S_2\ ,\ F_2=S_2-S_3\ ,\ F_3=S_5\,.
\ee
The flavor W-boson for $F_1$ is M2-brane wrapping
\be
C_1=D\cdot S_3\cdot (S_1-S_2)\,. 
\ee
The flavor W-boson for $F_2$ is M2-brane wrapping
\be
C_2=D\cdot S_1\cdot (S_2-S_3)\,. 
\ee
There is no flavor W-boson for $F_3$. Moreover, we can compute the intersection matrix
\be
\begin{array}{c|cc}
& F_1 & F_2\\
\hline
C_1 & -2 & 1\\
C_2 & 1 & -2
\end{array}
\ee
Hence there is a non-trivial flavor symmetry enhancement to $G_F=\mk{su}(3)\oplus\mk{u}(1)$.

The charges of BPS states under $U(1)$ gauge group and flavor Cartans are given in table~\ref{t:BPSchargeP1dP2}. $Q_e$ is the gauge charge and $Q_A$, $Q_B$, $Q_C$ are the flavor charges.

\begin{table}
\centering
\begin{tabular}{c|c|c|c|c}
$U(1)$ Charge & $Q_e$ & $Q_A$ & $Q_B$ & $Q_C$\\
\hline
Divisor & $D$  & $F_1$ & $F_2$ & $F_3$\\
\hline
\hline
$C_{1000}$ & $-2$ & $-1$ & 1 & 0\\
$C_{0100}$ & $-2$ & 1 & 0 & 0\\
$C_{0010}$ & $-1$ & 0 & 0 & $-1$ \\
$C_{0001}$ & $-2$ & 0 & $-1$ & 0
\end{tabular}
\caption{U(1) gauge charge and flavor charges of BPS states from M2-brane wrapping the curves on local $\mb{P}^1\times dP_2$.}\label{t:BPSchargeP1dP2}
\end{table}

One can see that $C_{1000}$, $C_{0100}$ and $C_{0001}$ form a $\mbf{3}_0$ representation of $G_F=\mk{su}(3)\oplus\mk{u}(1)$.

\paragraph{$c_2$ and additional $G_4$ flux on the resolved $X_4$}

Now let us consider the $G_4$ flux. First since
\be
c_2^c(X_4)=2S_3\cdot(2S_1+2S_2-S_5)+D\cdot(2S_1+2S_2+2S_3-S_5)+3S_1\cdot S_2-2S_5^2\,,
\ee
one can compute that
\be
\frac{1}{2}c_2^c(X_4)\cdot D\cdot S_1\in\mb{Z}\ ,\ \frac{1}{2}c_2^c(X_4)\cdot D\cdot S_2\in\mb{Z}\ ,\ \frac{1}{2}c_2^c(X_4)\cdot D\cdot S_3\in\mb{Z}\ ,\ \frac{1}{2}c_2^c(X_4)\cdot D\cdot S_5\in\mb{Z}\,. 
\ee
Therefore a non-vanishing $G_4$ is not required.

Nonetheless, the most general form of $G_4^c$ is
\be
\ba
G_4^c=D\cdot (nS_5+ pS_1+qS_2+kS_3)\,.
\ea
\ee
Now let us compute the chirality of charged matter fields from M2-brane wrapping specific 2-cycles.

First for M2-brane wrapping $C_{1000}=D\cdot S_1\cdot S_3$, its normal bundle is $N_{C_{1000}}=\mc{O}\oplus\mc{O}\oplus\mc{O}(-2)$. The moduli space of of $C_{1000}$ is $D\cdot S_2\sim \mb{P}^1\times\mb{P}^1$. The chirality of matter field with charge $-2$ under the $U(1)$ gauge group is
\be
\ba
\label{chiC1000}
\chi(C_{1000})&=-\int_{D\cdot S_2}c_1(D\cdot S_2)c_1(L)\cr
&=D\cdot G_4^c\cdot (-2S_3-S_1-S_4)\cr
&=4n+4p+8q+3k\,.
\ea
\ee

For M2-brane wrapping $C_{0100}=D\cdot S_2\cdot S_3$, its normal bundle is $N_{C_{0100}}=\mc{O}\oplus\mc{O}\oplus\mc{O}(-2)$. The moduli space of $C_{0100}$ is $D\cdot S_1\sim \mb{P}^1\times\mb{P}^1$. The chirality of matter field with charge $-2$ under the $U(1)$ gauge group is
\be
\ba
\label{chiC0100}
\chi(C_{0100})&=-\int_{D\cdot S_1}c_1(D\cdot S_1)c_1(L)\cr
&=D\cdot G_4^c\cdot (-2S_3-S_2-S_6)\cr
&=4n+8p+4q+3k\,.
\ea
\ee

For M2-brane wrapping $C_{0001}=D\cdot S_1\cdot S_2$, its normal bundle is $N_{C_{0001}}=\mc{O}\oplus\mc{O}\oplus\mc{O}(-2)$. The moduli space of $C_{0001}$ is $D\cdot S_3\sim dP_2$. The chirality of matter field with charge $-2$ under the $U(1)$ gauge group is
\be
\ba
\label{chiC0001}
\chi(C_{0001})&=-\int_{D\cdot S_3}c_1(D\cdot S_3)c_1(L)\cr
&=D\cdot G_4^c\cdot (-2S_1-2S_2+S_5)\cr
&=2n+4p+4q+7k\,.
\ea
\ee

Finally, for M2-brane wrapping $C_{0010}=D\cdot S_3\cdot S_5$, its normal bundle is $N_{C_{0010}}=\mc{O}\oplus\mc{O}(-1)\oplus\mc{O}(-1)$. The moduli space of $C_{0010}$ is $D\cdot S_5\cdot S_6\sim\mb{P}^1$. To compute the chirality of such matter field with charge $-1$ under the $U(1)$ gauge group, we integrate $G_4$ over the matter surface $D\cdot S_5$:
\be
\label{chiC0010}
\chi(C_{0010})=D\cdot S_5\cdot G_4^c=2n-k\,.
\ee

We compute the effective Chern-Simons term from 11d SUGRA action, which is free of parity anomaly
\be
\ba
S_{CS}&=\frac{1}{4\pi}(7k+4n+8p+8q)\int A\wedge dA+\frac{1}{2\pi}(2p-2q)\int A\wedge dB_1+\frac{1}{2\pi}(n-2k+2q)\int A\wedge dB_2\cr
&+\frac{1}{2\pi}(2n-k)\int A\wedge dB_3-\frac{2k}{4\pi}\int B_1\wedge dB_1-\frac{2p}{4\pi}\int B_2\wedge dB_2\cr
&-\frac{k}{4\pi}\int B_3\wedge dB_3+\frac{1}{2\pi}(k+p-q)\int B_1\wedge dB_2+\frac{n}{4\pi}\int B_2\wedge dB_3\,.
\ea
\ee
Here $A$ is the $U(1)$ gauge field, and $B_1$, $B_2$, $B_3$ are background gauge fields of flavor Cartans $F_1$, $F_2$, $F_3$ (\ref{P1dP2-flavor}) respectively.

\subsubsection{Local $\mb{P}^1\times dP_N(gdP_N)$ models}

According to the general rules of flavor symmetry enhancement in Section~\ref{sec:flavor}, for the general local $\mb{P}^1\times \mc{S}$ rank-1 models where $\mc{S}= dP_N(gdP_N)$, the $E_N$ flavor symmetry is enhanced to a larger flavor symmetry algebra. Namely, one look at the curve configurations on $gdP_N$ of type $E_N$, which geometrically realizes the largest flavor symmetry algebra (in 5d)~\cite{derenthal2014singular}. The $(-2)$-curves on $\mc{S}$ and the single 0-curve on $\mc{S}$ combine into a larger Dynkin diagram of the enhanced flavor symmetry algebra. From this general rule, we list the flavor enhancement at the singular limit in table~\ref{t:dPN-GF}.
\begin{table}
\centering
\begin{tabular}{c|c}
$\mc{S}$ & Enhanced $G_{F}$\\
\hline
$\mb{P}^2$ & $\mk{u}(1)$\\
$\mb{F}_1$ & $\mk{su}(2)\oplus\mk{u}(1)$\\
$\mb{F}_2$ & $\mk{su}(3)$\\
$dP_2$ & $\mk{su}(3)\oplus\mk{u}(1)$\\
$dP_3$ & $\mk{su}(4)\oplus\mk{su}(2)$\\
$dP_4$ & $\mk{su}(6)$\\
$dP_5$ & $\mk{so}(12)$\\
$dP_6$ & $E_7$ \\
$dP_7$ & $\hat{E}_7$
\end{tabular}
\caption{Flavor enhancement for the 3d $\mc{N}=2$ SCFT $\FTsing$ from M-theory on the local $\mb{P}^1\times\mc{S}$ singularities.}\label{t:dPN-GF}
\end{table}

As one can see, it is even possible to get affine flavor symmetry in this class of models.

\subsubsection{Local $\mb{P}^1\times T_N$ models}

We discuss non-compact CY4 which is a local $\mb{P}^1\times T_N$, where $T_N$ corresponds to the CY3 singularity $\mb{C}^3/(\mb{Z}_N\times\mb{Z}_N)$ that gives rise to the 5d $T_N$ theory~\cite{Eckhard:2020jyr}. 

In the resolved geometry $\widetilde{X}_4$, the non-compact divisors $S_1,\dots,S_{3N+2}$ correspond to the rays
\be
\ba
&v_1=(0,N,0)\ ,\ v_2=(1,N-1,0),\dots, v_N=(N-1,1,0)\ ,\ \cr
&v_{N+1}=(N,0,0)\ ,\ v_{N+2}=(N-1,0,0),\dots, v_{2N}=(1,0,0)\ ,\ \cr
&v_{2N+1}=(0,0,0)\ ,\ v_{2N+2}=(0,1,0),\dots, v_{3N}=(0,N-1,0)\ , \ \cr
&v_{3N+1}=(0,0,1)\ ,\ v_{3N+2}=(0,0,-1)\,.
\ea
\ee
While the compact divisors are denoted as $D_1,\dots ,D_{(N-1)(N-2)/2}$. Hence the rank $r=(N-1)(N-2)/2$.

For the flavor symmetry enhancement, we plot the configuration of flavor curves on the resolution of $\mb{C}^3/(\mb{Z}_N\times\mb{Z}_N)$ (see \cite{Eckhard:2020jyr}):
\be
\includegraphics[height=5cm]{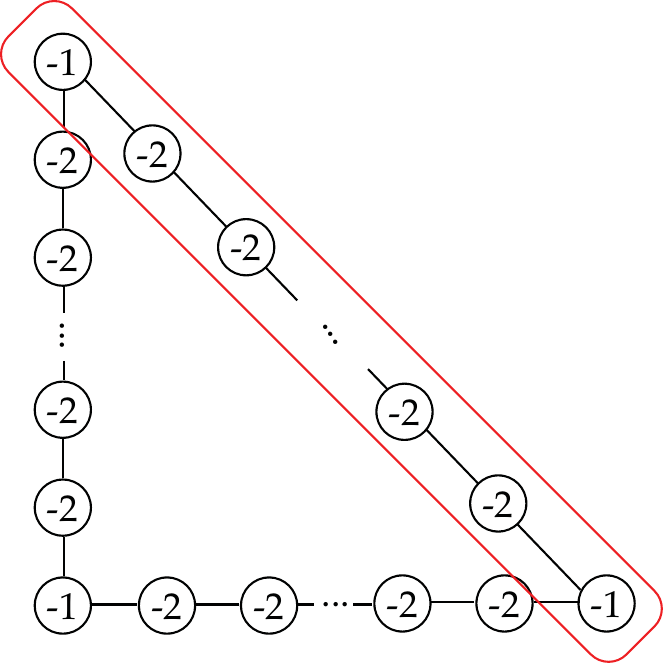}
\ee
Note that the curves in the red rectangle effectively form a 0-curve on the compact divisors, which give rise to an additional node in the Dynkin diagram of $G_{F,nA}$ following Sec.\ref{sec:flavor}. 

Hence for $N\geq 4$, the enhanced flavor symmetry is $G_{F,nA}=SU(N)\times SU(2N)$.

For $N=3$, there is a further enhancement $G_{F,nA}=E_7\supset SU(3)\times SU(6)$, because of the identification with the case of local $\mb{P}^1\times dP_6$.

\subsubsection{Higher-rank fibration over $\mb{P}^1\times\mb{P}^1$}

We can also consider toric geometries that give rise to $r>1$ theories, which are $\mb{P}^1$s fiber over a base $\mc{S}=\mb{P}^1\times\mb{P}^1$. First let us consider the geometry
\be
\label{fig:P1SU3}
\includegraphics[height=6cm]{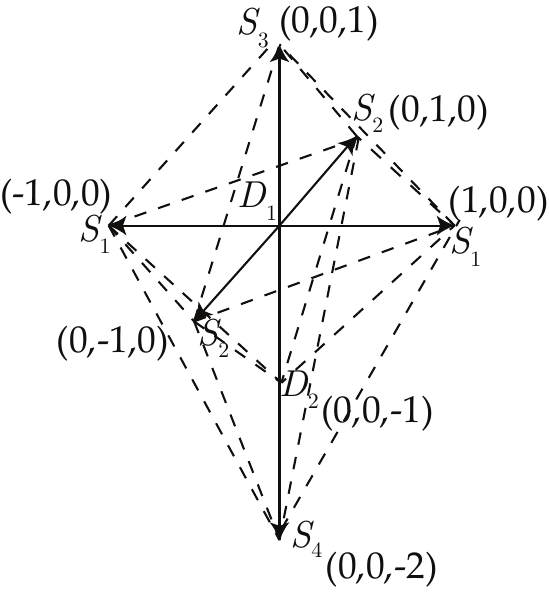}
\ee
In this case the two compact divisors are $D_1$, $D_2$ and the non-compact divisors are labeled by $S_1$, $S_2$, $S_3$, $S_4$. There is linear equivalence relation
\be
S_3=D_2+2S_4\,.
\ee
The gauge W-bosons are $C_1^g=D_1\cdot S_1\cdot S_2$, $C_2^g=D_2\cdot S_1\cdot S_2$, which satisfies the intersection matrix
\be
\begin{array}{c|cc}
& D_1 & D_2\\
\hline
C_1^g & -2 & 1\\
C_2^g & 1 & -2
\end{array}
\ee

Hence in the limit where the $\mb{P}^1$ fibers are all shrunk to zero volume limit, there is an $SU(3)$ gauge theory description.

For the flavor symmetry enhancements, we can again write down the flavor Cartans
\be
F_1=S_1-S_2\ ,\ F_2=S_2-S_3
\ee
and their corresponding flavor W-bosons
\be
C_1=D_1\cdot (S_1-S_2)\cdot S_3\ ,\ C_2=D_1\cdot (S_2-S_3)\cdot S_1\,.
\ee
From the intersection matrix
\be
\begin{array}{c|cc}
& F_1 & F_2\\
\hline
C_1 & -2 & 1\\
C_2 & 1 & -2
\end{array}
\ee
we see that the flavor symmetry enhancement to $G_F=\mk{su}(3)$ at the singular limit of $X_4$.

We can even consider higher-rank configurations by adding new rays below $S_4$ in (\ref{fig:P1SU3}). The analysis is analogous to the $SU(3)$ case, in general we will get a theory with $SU(r+1)$ gauge theory limit and flavor symmetry enhancement $G_F=\mk{su}(3)$ at the singular limit.

One can also consider non-trivial fibrations over $\mb{P}^1\times\mb{P}^1$, which we would not discuss the details here.

\subsection{$\mathbb{C}^4$ orbifolds}\label{sec:C4_orbifolds}

In this section we study a rich class of geometries obtained from orbifolding $\mathbb{C}^4$ by finite subgroups of $SU(4)$. We will denote such orbifolds by $\mathbb{C}^4/\Gamma$ for a finite subgroup $\Gamma \subset SU(4)$ and by $\widetilde{\mathbb{C}^4/\Gamma}$ its crepant resolution if there exists one. Strictly speaking $\Gamma$ is a subgroup of $SL(4,\mb{C})$ but we will not distinguish between the two and this subtlety is not important for our discussion.

The significance of such orbifolds $\mathbb{C}^n/\Gamma$, where $\Gamma$ is a finite subgroup of $SU(n)$ for general $n$, lies in the fact that $\mathbb{C}^n/\Gamma$ possesses an $SU(n)$ holonomy. Therefore SUSY field theories can be geometrically engineered from string/M-theory compactification on $\mathbb{C}^n/\Gamma$. In contrast, from a physical point of view assuming the existence of a crepant resolution for $\mathbb{C}^n/\Gamma$ is somewhat less general in our setup. When $n=4$ this assumption is equivalent to assuming the existence of a UV Coulomb branch hence making the physics more tractable. Certainly some mathematical aspects of $\mathbb{C}^n/\Gamma$ become tricky if no crepant resolution exists, but those issues are beyond the scope of the current paper.

It is well-known that a crepant resolution may not exist for any $\mathbb{C}^4/\Gamma$ and an existence theorem is yet to be established \cite{ItoReid, ito1998mckay, ito2011mckay}. In this section we will avoid this subtlety and discuss only general physical properties that a 3D theory associated to $\mathbb{C}^4/\Gamma$ must possess whenever at least one crepant resolution exists. We will leave a more detailed investigation of physical aspects of these theories and its link to various mathematical concepts to future work.


\subsubsection{Group data of $\Gamma$ and M-theory on $\widetilde{\mathbb{C}^n/\Gamma}$ or $\mathbb{C}^n/\Gamma$}

In this subsection we discuss the general relation between the group theoretical data of a finite subgroup $\Gamma$ of $SU(n)$ and the geometrical and topological data of $\widetilde{\mathbb{C}^n/\Gamma}$, assuming the existence of a crepant resolution of $\mathbb{C}^n/\Gamma$ \footnote{Here $n$ does not have to be 4, rather it can be any positive integer as long as the compactification of M-theory on $\mathbb{C}^n/\Gamma$ makes sense.}. We then relate the geometry of $\widetilde{\mathbb{C}^n/\Gamma}$ to various aspects of the lower dimensional theory from M-theory on $\widetilde{\mathbb{C}^n/\Gamma}$. Note that the geometric limit $\mathbb{C}^n/\Gamma$ can be realized by blowing down $\widetilde{\mathbb{C}^n/\Gamma}$ and certain physical properties of $\mc{T}_{\widetilde{\mathbb{C}^n/\Gamma}}$ carry over to the SCFT $\mc{T}_{\mathbb{C}^n/\Gamma}$, which will be discussed in this section in a moment. Later we will restrict to the case of interest, i.e. $n = 4$ after the discussion of the generalities, in which case we arrive at IR SCFT.

The following conjecture will play the most important role in our analysis \cite{ItoReid}:
\begin{conj}\label{conj:ItoReid}
    Let $\Gamma$ be a finite subgroup of $SL(n,\mb{C})$, $X = \mb{C}^n/\Gamma$ and $\widetilde{X}\rightarrow X$ a crepant resolution. Then there exists a basis of $H^*(\widetilde{X}, \mb{Q})$ consisting of algebraic cycles in one-to-one correspondence with conjugacy classes of $\Gamma$.
\end{conj}
Since there are no odd-dimensional cycles in $\widetilde{X}$ and $H^{2i}(\widetilde{X}, \mb{Q})$ is spanned by the algebraic cycles, given the above conjecture we must have \cite{Ito:Crepant_Resolution_Trihedral, ItoReid, Dixon:Strings_On_Orbifolds_I, Dixon:Strings_On_Orbifolds_II} :
\begin{equation}
    \chi(X) = \chi(\widetilde{X}) = \text{number of conjugacy classes of $\Gamma$}.
\end{equation}
We will denote by $\mathfrak{g}_i$, $i = 1, 2, \cdots, \chi(X)$ the conjugacy classes of $\Gamma$.

Of crucial importance is the notion of \emph{age} of a conjugacy class $\mathfrak{g}$ of $\Gamma$. We first define the order of $\mathfrak{g}$ to be the smallest integer $r$ such that $g^r = 1$, $\forall g\in\mathfrak{g}$. It is then easy to show that an arbitrary representative $g \in \mathfrak{g}$ in its natural $n$-dimensional representation can be diagonalized into the following form:
\begin{equation}
    \text{diag}(g) = \text{diag}\left( e^{2\pi i a_1/r}, e^{2\pi i a_2/r}, \cdots, e^{2\pi i a_n/r} \right).
\end{equation}
We then define the age of $\mathfrak{g}$ to be:
\begin{equation}
    \text{age}(\mathfrak{g}) = \frac{1}{r}\sum_{i=1}^n a_i
\end{equation}
Clearly, the maximal value of $\text{age}(\mathfrak{g})$ is $n-1$. The finite group $\Gamma$ admits a canonical grading by the age, $\Gamma = \cup_{i = 1}^{n-1} \Gamma_i$ where $\text{age}(\Gamma_i) = i$ \cite{ItoReid}. It is easy to see that $\Gamma_0 = \{e\}$ and $\Gamma_n = \emptyset$.

Given the conjecture~\ref{conj:ItoReid} together with the existence of an age grading of $\Gamma$, it is natural to expect that the conjugacy classes in $\Gamma_i$ give a basis of $H^{2i}(\widetilde{X}, \mb{Q})$, so that the age grading of $\Gamma$ becomes the usual grading of $H^*(\widetilde{X},\mb{Q})$. Such a correspondence has been established rigorously between the so-called \emph{junior class} $\Gamma_1$ and $H^2(\widetilde{X}, \mb{Q})$ \cite{ItoReid}:
\begin{theorem}
    Assuming the existence of a crepant resolution $\widetilde{X}\rightarrow X$, there is a canonical one-to-one correspondence between the conjugacy classes in $\Gamma_1$ and $H^2(\widetilde{X},\mb{Q})$:
    \begin{center}
        $\Gamma_1 = \{\mathfrak{g}_1^1, \cdots, \mathfrak{g}_1^{|\Gamma_1|}\} \leftrightarrow \text{exceptional prime divisors in $\widetilde{X}$} \leftrightarrow \text{basis of\ } H^2(\widetilde{X}, \mb{Q})$. 
    \end{center}
\end{theorem}
An immediate consequence of the above theorem is $b_2(\widetilde{X}) = |\Gamma_1|$. In fact a more general statement was proved in \cite{joyce} that $b_{2k}(\widetilde{X}) = |\Gamma_{k}|$. Though a detailed correspondence between the $2k$-cycles and the conjugacy classes in $\Gamma_k$ is unknown, the theorem gives a way to compute the Betti numbers without knowing the details of a resolution.

Among all the algebraic cycles in $\widetilde{X}$ we are particular interested in the divisors as they are closely related to the gauge and the flavor symmetries of a 3D theory. For this let us start with general properties of algebraic cycles in a non-compact variety. Since there are no torsional cycles in our $\widetilde{X}$, we have:
\begin{equation}
    H_i(\widetilde{X}) \cong H^i(\widetilde{X}) \cong \mb{Z}^{b_i}
\end{equation}
where the coefficient of the (co)homology group is restricted to $\mb{Z}$. From Poincar\'e-Lefschetz duality we have:
\begin{equation}
    H^i_c(\widetilde{X}) \cong H_{2n-i}(\widetilde{X}) \cong \mathbb{Z}^{b_{2n-i}}
\end{equation}
where $H^i_c(\widetilde{X})$ is the $i^{\text{th}}$ cohomology of $\widetilde{X}$ with compact support. By the same duality we have:
\begin{equation}
    H_{2n-i}(\widetilde{X}, \partial\widetilde{X}) \cong H^{2n-i}(\widetilde{X}, \partial\widetilde{X}) \cong H^{2n-i}_c(\widetilde{X}) \cong H_i(\widetilde{X}) \cong H^i(\widetilde{X}) \cong \mb{Z}^{b_i}.
\end{equation}
Applying the above results for $i = 2$, we see that there are in total $b_2$ divisors in $\widetilde{X}$, of which $b_{2n-2}$ are compact. Now applying the theorem that $b_{2k}(\widetilde{X}) = |\Gamma_k|$, we have:
\begin{equation}
    \text{number of compact divisors} = |\Gamma_{n-1}|,\ \text{number of non-compact divisors} = |\Gamma_1| - |\Gamma_{n-1}|.
\end{equation}

For the $n = 4$ cases, which are the main focus of this work, via the geometry-physics dictionary developed in section~\ref{sec:dictionary} we arrive at the following key statement:
\begin{prop}\label{prop:r_and_f}
    Denote by $\mc{T}_{\widetilde{X}}$ the 3D theory obtained from M-theory on a crepant resolution $\widetilde{X}$ of $\mathbb{C}^4/\Gamma$. Let $r$ be the rank of the gauge symmetry group of $\mc{T}_{\widetilde{X}}$ and $f$ the rank of its flavor symmetry group, we have:
    \begin{center}
        $r = |\Gamma_3|,\ f = |\Gamma_1| - |\Gamma_3|$.
    \end{center}
\end{prop}
As we do not expect $\Gamma_i$'s to change in the resolution process, they will stay the same in the geometric limit where all the compact divisors are shrunk, i.e. when the geometry is the singular $X \simeq \mathbb{C}^4/\Gamma$. Since we have shown in section~\ref{sec:dictionary} that M-theory on $X$ leads to a 3D SCFT $\mc{T}_X$ in the UV, we hence expect that the gauge rank and the flavor rank of $\mc{T}_X$ to be $r$ and $f$, the same as those of $\mc{T}_{\widetilde{X}}$.

\subsubsection{Abelian $\Gamma$}\label{sec:orbifold_ab}

In this section we discuss the geometry of $X\simeq \mb{C}^4/\Gamma$ when $\Gamma$ is an Abelian finite subgroup of $SU(4)$. We note that in this case it is easy to establish the (non-)existence of a crepant resolution $\widetilde{X}\rightarrow X$. Since in the Abelian case $X$ admits a toric construction, a crepant resolution of $X$ exists iff there exists a fine-regular-star triangulation (FRST) of the toric fan of $X$.

We will first present the analysis for the Abelian cases via a concrete example which is a natural generalization of 5D $T_n$ theory to 3D, for which reason we will call these theories $T_n^{3D}$.

Let us consider the variety $X = \mathbb{C}^4/\Gamma$ where $\Gamma = \mathbb{Z}_n\times \mathbb{Z}_n\times \mathbb{Z}_n$ whose generators in the natural 4D representation of $SU(4)$ are:
\begin{equation}
    g_1 = \begin{pmatrix}
        \omega & 0 & 0 & 0 \\
        0 & 1 & 0 & 0 \\
        0 & 0 & 1 & 0 \\
        0 & 0 & 0 & \omega^{-1}
    \end{pmatrix},\ g_2 = \begin{pmatrix}
        1 & 0 & 0 & 0 \\
        0 & \omega & 0 & 0 \\
        0 & 0 & 1 & 0 \\
        0 & 0 & 0 & \omega^{-1}
    \end{pmatrix},\ g_3 = \begin{pmatrix}
        1 & 0 & 0 & 0 \\
        0 & 1 & 0 & 0 \\
        0 & 0 & \omega & 0 \\
        0 & 0 & 0 & \omega^{-1}
    \end{pmatrix}
\end{equation}
where $\omega = e^{2\pi i/n}$. Since $\Gamma$ is Abelian we have $|\Gamma| = n^3$ therefore by Conjecture~\ref{conj:ItoReid} $\chi(X) = \chi(\widetilde{X}) = n^3$. Without much effort it is also easy to show:
\begin{equation}
    |\Gamma_1| = \frac{1}{6}(n-1)(n^2+7n+18),\ |\Gamma_2| = \frac{1}{3}(n-1)(2n^2+2n-9),\ |\Gamma_3| = \frac{1}{6}(n-3)(n-2)(n-1).
\end{equation}
From Proposition~\ref{prop:r_and_f} we have:
\begin{equation}\label{eq:r_and_f_Tn}
    r = \frac{1}{6}(n-3)(n-2)(n-1),\ f = 2n^2 - 2.
\end{equation}
As a byproduct we see that the result indeed matches the theorem that $b_6 = |\Gamma_3|$ in the $\mb{C}^4$ orbifold case.

The above results obtained from group theoretical data can be immediately checked against the toric construction. In this case the vertices of the toric fan $F_X$ of $X$ are:
\begin{equation}
    \tilde{v}_1 = (0,0,0,1),\ \tilde{v}_2 = (n,0,0,1),\ \tilde{v}_3 = (0,n,0,1),\ \tilde{v}_4 = (0,0,n,1),\ \tilde{v}_o = (0,0,0,0)
\end{equation}
and the vertices of the corresponding 3D polyhedron $\mc{P}_3$ are:
\begin{equation}
    v_1 = (0,0,0),\ v_2 = (n,0,0),\ v_3 = (0,n,0),\ v_4 = (0,0,n).
\end{equation}
It is obvious that a crepant resolution $\widetilde{X}\rightarrow X$ can be obtained by an FRST of $F_X$, or a fine-regular triangulation of $\mc{P}_3$. By our geometry-physics dictionary the rank of the gauge symmetry is equal to the number of the interior points of $\mc{P}_3$ and the rank of the flavor symmetry is equal to the number of the points living on the exterior facets of $\mc{P}_3$ subject to 4 linear constraints coming from the Calabi-Yau condition. It is then easy to see that the results obtained from toric geometry methods indeed match those in (\ref{eq:r_and_f_Tn}).

There is a third equivalent description of $X$ as a Calabi-Yau hypersurface in $\mb{C}^5$. We have:
\begin{equation}
    X = \{ x_1x_2x_3x_4 = x_5^n | (x_1,x_2,x_3,x_4,x_5)\in\mb{C}^5 \}.
\end{equation}
The singular locus of $X$ can be immediately calculated to be:
\begin{equation}
    \{ x_i = x_j = x_5 = 0,\ \forall 1 \leq i < j \leq 5 \} \subset X.
\end{equation}
Each of the above 6 singular locus is a codimension-2 subvariety of $X$ and is locally of the form $x_ix_j = x_5^n$ hence is a product of an $A_{n-1}$ singularity and $\mathbb{C}^2$. It is obvious that these singular locus are non-compact as they extend to infinity. This then matches the 6 $A_{n-1}$ singularities one gets along each of the 6 edges of $\mc{P}_3$ before triangulation, where each edge forms a length $n$ unsubdivided toric sub-fan of $F_X$ hence gives rise to an unresolved $A_{n-1}$ singularity. We see that $6n-6$ out of the total $2n^2 - 2$ flavor rank can be accounted for via this simple analysis of the codimension-2 singularity of $X$.

The codimension-2 singularities can also be read off from the group theoretical data. Recall that any representative $g$ of a conjugacy class $\mathfrak{g}$ can be diagonalized as
\begin{equation}
    \text{diag}(g) = \text{diag}\left( e^{2\pi i a_1/r}, e^{2\pi i a_2/r}, e^{2\pi i a_3/r}, e^{2\pi i a_4/r} \right).
\end{equation}
Therefore the information of the ``orbit'' of each conjugacy class $\mathfrak{g}$ can be packaged into the following vector:
\begin{equation}
    v_{\mathfrak{g}} = \frac{1}{r}(a^{\mathfrak{g}}_1,a^{\mathfrak{g}}_2,a^{\mathfrak{g}}_3,a^{\mathfrak{g}}_4).
\end{equation}
The original multiplicative binary operation of $\Gamma$ becomes a mod-$r$ additive binary operation among $v_{\mathfrak{g}}$'s. More precisely, under the homomorphism that sends the conjugacy classes of $\Gamma$ to the orbits we have:
\begin{equation}
    g_1g_2 = g_3 \longrightarrow v_{\mathfrak{g}_1} + v_{\mathfrak{g}_2} = \frac{1}{r} (a^{\mathfrak{g}_1}_1 + a^{\mathfrak{g}_2}_1, a^{\mathfrak{g}_1}_2 + a^{\mathfrak{g}_2}_2, a^{\mathfrak{g}_1}_3 + a^{\mathfrak{g}_2}_3, a^{\mathfrak{g}_1}_4 + a^{\mathfrak{g}_2}_4) \mod\ r
\end{equation}
for $g_1\in\mathfrak{g}_1$, $g_2\in\mathfrak{g}_2$. One can then use this property to find all the orbits of the conjugacy classes of $\Gamma$. Since $\Gamma$ is a finite Abelian group, all the orbits must be closed and isomorphic to $\mathbb{Z}_k$ for some $k$. Among all the $\mb{Z}_k$ orbits of $\Gamma_1$ we are particularly interested in those one whose generator can be put into the form $g = \text{diag}(e^{2\pi i/k},e^{-2\pi i/k},1,1)$ by a similar transformation in $SU(4)$. The subvariety in $X$ that is formed by quotienting $\mb{C}^4$ by such a $\mathbb{Z}_k$ action, i.e. $\mb{C}^4/\langle g\rangle$ is clearly isomorphic to $\mathbb{C}^2/\mb{Z}_k \times \mb{C}^2$ therefore must be an $A_{k-1}$ singularity living at codimension-2 in $X$.

As a concrete example let us again consider $\Gamma = \mathbb{Z}_n^3$. Among all the various orbits in $\Gamma$ there are the following six ones, denoted by $O_2$, whose generators can be put into the form $g = \text{diag}(e^{2\pi i/n},e^{-2\pi i/n},1,1)$:
\begin{equation}
    \begin{split}
        &\left\langle \frac{1}{n}(0,0,1,n-1) \right\rangle,\ \left\langle \frac{1}{n}(0,1,0,n-1) \right\rangle,\ \left\langle \frac{1}{n}(0,1,n-1,0) \right\rangle, \\
        &\left\langle \frac{1}{n}(1,0,0,n-1) \right\rangle,\ \left\langle \frac{1}{n}(1,0,n-1,0) \right\rangle,\ \left\langle \frac{1}{n}(1,n-1,0,0) \right\rangle.
    \end{split}
\end{equation}
It is clear that each $\mathbb{C}^4/\langle g\rangle$ for a $\langle g\rangle$ of the above form is isomorphic to $\mb{C}^2/\mathbb{Z}_n\times \mb{C}^2$. Therefore from a group theoretical point of view one is also able to read off the 6 $A_{n-1}$ singularities in $X$. Each conjugacy class in an orbit in $O_2$ corresponds to a non-compact divisor of $\widetilde{X}$.

This method can certainly be generalized to the cases with more general form of $g$. Actually there are orbits of conjugacy classes, which will be denoted by $O_1$, in $\Gamma = \mathbb{Z}_n^3$ generated by the elements of the form $g = \text{diag}(e^{2\pi i p/n}, e^{2\pi i q/n}, e^{2\pi i r/n}, 1)$ with positive integers $p$, $q$ and $r$ satisfying $p+q+r = n$. For the orbits in $O_1$ we have $\mb{C}^4/\langle g \rangle \simeq \mb{C}^3/\mathbb{Z}_k \times \mb{C}$ therefore it is clear that these subvarieties are also non-compact.

Here comes a subtlety. Unlike the orbits in $O_2$ in which all the conjugacy classes are in $\Gamma_1$, the conjugacy classes in an orbit in $O_1$ can be in either $\Gamma_1$ or $\Gamma_2$. For example the orbit of $g = (1,1,n-2,0)$ contains the following conjugacy classes:
\begin{equation}\label{eq:O1orbit}
    (1,1,n-2,0),\ (2,2,n-4,0),\ \cdots,\ (n-1,n-1,2,0).
\end{equation}
Here we have omitted the trivial class $(0,0,0,0)$. Clearly the last $\left\lfloor \frac{n-1}{2} \right\rfloor$ conjugacy classes in (\ref{eq:O1orbit}) are in $\Gamma_2$. But now since the last entry is always 0 we could simply apply the results for $\mathbb{C}^3$ orbifolds in \cite{Tian:2021cif}. Let us denote by $\Gamma'$ the natural action of $\Gamma = \langle g \rangle$ on the first three coordinates of $\mathbb{C}^4$. The fact that the last $\left\lfloor \frac{n-1}{2} \right\rfloor$ conjugacy classes are in $\Gamma_2$ simply means that out of the first $n - 1 - \left\lfloor \frac{n-1}{2} \right\rfloor$ conjugacy classes in (\ref{eq:O1orbit}), there are $\left\lfloor \frac{n-1}{2} \right\rfloor$ of them correspond to compact divisors in $\widetilde{\mathbb{C}^3/\Gamma'}$ and the rest of them correspond to non-compact divisors in $\widetilde{\mathbb{C}^3/\Gamma'}$ \cite{Tian:2021cif}. Nevertheless, these divisors, whether compact or non-compact in $\widetilde{\mathbb{C}^3/\Gamma'}$, all map to non-compact divisors in $\widetilde{\mathbb{C}^4/\Gamma}$ due to the Cartesian product with an extra $\mathbb{C}$. Therefore we conclude that a conjugacy class in an orbit in $O_1$ corresponds to a non-compact divisor of $\widetilde{X}$ if it is in $\Gamma_1$.

As a sanity check of our claims above about $O_1$ and $O_2$, we have:
\begin{equation}
    4\times \frac{(n-1)(n-2)}{2} + 6\times (n-1) = 2n^2 - 2 = f.
\end{equation}
Here $\frac{(n-1)(n-2)}{2}$ is the number of partitions of $n$ into three positive integers. The first term the above equation counts the contribution to the flavor rank of $T^{3D}_n$ from the orbits in $O_1$ and the second term counts that from the orbits in $O_2$.

Another consequence of the subtlety discussed above is that unlike the orbits in $O_1$ which correspond to $A_k$ singularities which are $\mb{C}^2$ orbifolds, the orbits in $O_2$ effectively correspond to $\mathbb{C}^3$ orbifolds which are considerably more complicated. But still, using the method developed in \cite{Tian:2021cif}, one can read off at least the naive flavor symmetries related to the structure of the orbits in $O_2$. One only needs to keep in mind that both the gauge symmetries and the flavor symmetries in the $\mathbb{C}^3$ orbifold case become flavor symmetries in $\mb{C}^4$ orbifold case due to the Cartesian product with an extra $\mathbb{C}$. We will not repeat the procedure in this paper and a full account of these issues will be given in our future work.

\subsubsection{Non-abelian $\Gamma$}\label{sec:orbifold_nonab}

In this section we list the physically relevant group theoretical data of the non-abelian finite subgroups $\Gamma$ of $SU(4)$. Such groups were classified in \cite{HananyHe_SU4} where many useful group data were calculated as well. However, some of the most important group theoretical quantities relevant to geometric engineering of 3D theories were never calculated in literature. Therefore we will calculate and tabulate those quantities in this work. Again we emphasize that such $\mathbb{C}^4/\Gamma$ for non-abelian $\Gamma$ may not have a crepant resolution and we will later discuss only the physical properties that $\mc{T}_X$ must have assuming the existence of a crepant resolution of $X$.

The data of primitive (``exceptional'') and imprimitive (``dihedral'') finite groups that are relevant to the physics of $\mc{T}_{\mb{C}^4/\Gamma}$ are calculated and summarized in Table~\ref{tab:primitive_group_data} and Table~\ref{tab:imprimitive_group_data}, respectively. The notations in these tables are summarized in Appendix~\ref{app:nonab_gens}. In both tables $\chi$ is the number of conjugacy classes. In particular as a sanity check we have $\chi = 1 + |\Gamma_1| + |\Gamma_2| + |\Gamma_3|$, which matches conjecture~\ref{conj:ItoReid}. Again, crucial physical data of $\mc{T}_{\mb{C}^4/\Gamma}$ can be read off as:
\begin{equation}
    r = |\Gamma_3|,\ f = |\Gamma_1| - |\Gamma_3|.
\end{equation}

\begin{table}[h]
    \centering
    \begin{tabular}{c|c|c|c|c|c|c}
         Group & Generators & Order & $\chi$ & $|\Gamma_1|$ & $|\Gamma_2|$ & $|\Gamma_3|$ \\
         \hline
         \hline
         \multicolumn{7}{c}{Primitive simple groups}\\
         \hline
         I & $F_1, F_2, F_3$ & 240 & 18 & 3 & 13 & 1 \\
         \hline
         II & $F_1, F_2', F_3'$ & 60 & 5 & 2 & 2 & 0 \\
         \hline
         III & $F_1, F_2, F_3, F_4$ & 1440 & 26 & 4 & 19 & 2 \\
         \hline
         IV & $S, T, W$ & 5040 & 16 & 2 & 13 & 0 \\
         \hline
         V & $S, T, R$ & 336 & 11 & 2 & 8 & 0 \\
         \hline
         VI & $T, C, D, V, F$ & 51840 & 34 & 8 & 24 & 1 \\
         \hline
         \multicolumn{7}{c}{Groups having simple normal primitive subgroups}\\
         \hline
         VII & $F_1, F_2, F_3, F''$ & 480 & 24 & 5 & 17 & 1 \\
         \hline
         VIII & $F_1, F_2', F_3', F'$ & 480 & 28 & 5 & 20 & 2 \\
         \hline
         IX & $F_1, F_2, F_3, F_4, F''$ & 2880 & 34 & 7 & 24 & 2 \\
         \hline
         \multicolumn{7}{c}{Groups having normal intransitive subgroups}\\
         \hline
         X & $\langle S_{SU(2)}, U^2_{SU(2)} \rangle \otimes \langle S_{SU(2)}, U^2_{SU(2)} \rangle$ & 288 & 25 & 5 & 19 & 0 \\
         \hline
         XI & $x_1\otimes x_2, x_1\otimes x_2^T, x_3\otimes x_4, x_5\otimes x_6$ & 576 & 20 & 5 & 14 & 0 \\
         \hline
         XII & $\langle S_{SU(2)}, U^2_{SU(2)} \rangle \otimes \langle S_{SU(2)}, U_{SU(2)} \rangle$ & 576 & 29 & 4 & 24 & 0 \\
         \hline
         XIII & $\langle S_{SU(2)}, U^2_{SU(2)} \rangle \otimes \langle S_{SU(2)}, V_{SU(2)}, U^2_{SU(2)} \rangle$ & 1440 & 32 & 3 & 28 & 0 \\
         \hline
         XIV & $\langle S_{SU(2)}, U_{SU(2)} \rangle \otimes \langle S_{SU(2)}, U_{SU(2)} \rangle$ & 1152 & 34 & 6 & 27 & 0 \\
         \hline
         XV & $\langle S_{SU(2)}, U_{SU(2)} \rangle \otimes \langle S_{SU(2)}, V_{SU(2)}, U^2_{SU(2)} \rangle$ & 2880 & 37 & 3 & 33 & 0 \\
         \hline
         XVI & $\langle S_{SU(2)}, V_{SU(2)}, U^2_{SU(2)} \rangle \otimes \langle S_{SU(2)}, V_{SU(2)}, U^2_{SU(2)} \rangle$ & 7200 & 41 & 4 & 36 & 0 \\
         \hline
         \multicolumn{7}{c}{Groups having X-XVI as normal primitive subgroups}\\
         \hline
         XVII & (XI), $T_1$ & 2304 & 50 & 11 & 34 & 4 \\
         \hline
         XVIII & (XI), $T_2$ & 2304 & 38 & 10 & 25 & 2 \\
         \hline
         XIX & (X), $T_1$ & 1152 & 46 & 9 & 32 & 4 \\
         \hline
         XX & (XVI), $T_1$ & 28800 & 68 & 12 & 48 & 7 \\
         \hline
         XXI & (XIV), $T_1$ & 4608 & 58 & 13 & 40 & 4 \\
         \hline
         \multicolumn{7}{c}{Groups having normal imprimitive subgroups}\\
         \hline
         XXII & (K), $T'$ & 320 & 26 & 4 & 20 & 1 \\
         \hline
         XXIII & (K), $T', R'^2$ & 640 & 28 & 7 & 19 & 1 \\
         \hline
         XXIV & (K), $T', R'$ & 1280 & 32 & 8 & 22 & 1 \\
         \hline
         XXV & (K), $T', S'B$ & 3840 & 34 & 7 & 24 & 2 \\
         \hline
         XXVI & (K), $T', BR'$ & 3840 & 26 & 6 & 18 & 1 \\
         \hline
         XXVII & (K), $T', A$ & 7680 & 50 & 11 & 35 & 3 \\
         \hline
         XXVIII & (K), $T', B$ & 7680 & 37 & 8 & 26 & 2 \\
         \hline
         XXIX & (K), $T', AB$ & 23040 & 37 & 8 & 26 & 2 \\
         \hline
         XXX & (K), $T', S'$ & 46080 & 59 & 12 & 42 & 4 \\
    \end{tabular}
    \caption{Relevant group theoretical data of primitive (``exceptional'') finite subgroups of $SU(4)$.}
    \label{tab:primitive_group_data}
\end{table}

\begin{table}[h]
    \centering
    \begin{tabular}{c|c|c|c|c|c|c|c}
        Group & Generators & Order & $n$ & $\chi$ & $|\Gamma_1|$ & $|\Gamma_2|$ & $|\Gamma_3|$ \\
        \hline
        \hline
        \multirow{7}{*}{XXXI} & \multirow{7}{*}{$\Delta_n, A_4$} & \multirow{7}{*}{$12n^3$} & 2 & 11 & 5 & 5 & 0 \\
        \cline{4-8}
        & & & 3 & 13 & 5 & 7 & 0 \\
        \cline{4-8}
        & & & 4 & 24 & 8 & 14 & 1 \\
        \cline{4-8}
        & & & 5 & 28 & 8 & 18 & 1 \\
        \cline{4-8}
        & & & 6 & 45 & 13 & 29 & 2 \\
        \cline{4-8}
        & & & 7 & 53 & 14 & 35 & 3 \\
        \cline{4-8}
        & & & \multicolumn{5}{c}{$\cdots$} \\
        \hline
        \multirow{7}{*}{XXXII} & \multirow{7}{*}{$\Delta_n, S_4$} & \multirow{7}{*}{$24n^3$} & 2 & 13 & 5 & 7 & 0 \\
        \cline{4-8}
        & & & 3 & 26 & 7 & 17 & 1 \\
        \cline{4-8}
        & & & 4 & 33 & 11 & 20 & 1 \\
        \cline{4-8}
        & & & 5 & 56 & 12 & 39 & 4 \\
        \cline{4-8}
        & & & 6 & 57 & 16 & 37 & 3 \\
        \cline{4-8}
        & & & 7 & 106 & 21 & 75 & 9 \\
        \cline{4-8}
        & & & \multicolumn{5}{c}{$\cdots$} \\
        \hline
        \multirow{7}{*}{XXXIII} & \multirow{7}{*}{$\Delta_n, D_8$} & \multirow{7}{*}{$8n^3$} & 2 & 20 & 9 & 9 & 1 \\
        \cline{4-8}
        & & & 3 & 36 & 12 & 21 & 2 \\
        \cline{4-8}
        & & & 4 & 33 & 11 & 20 & 1 \\
        \cline{4-8}
        & & & 5 & 44 & 19 & 23 & 1 \\
        \cline{4-8}
        & & & 6 & 126 & 44 & 75 & 6 \\
        \cline{4-8}
        & & & 7 & 206 & 45 & 141 & 19 \\
        \cline{4-8}
        & & & \multicolumn{5}{c}{$\cdots$} \\
        \hline
        \multirow{7}{*}{XXXIV} & \multirow{7}{*}{$\Delta_n, \mathbb{Z}_2\times \mathbb{Z}_2$} & \multirow{7}{*}{$4n^3$} & 2 & 17 & 9 & 7 & 0 \\
        \cline{4-8}
        & & & 3 & 15 & 7 & 7 & 0 \\
        \cline{4-8}
        & & & 4 & 40 & 16 & 22 & 1 \\
        \cline{4-8}
        & & & 5 & 44 & 16 & 26 & 1 \\
        \cline{4-8}
        & & & 6 & 87 & 29 & 53 & 4 \\
        \cline{4-8}
        & & & 7 & 103 & 32 & 65 & 5 \\
        \cline{4-8}
        & & & \multicolumn{5}{c}{$\cdots$} \\
        \hline
    \end{tabular}
    \caption{Relevant group theoretical data of imprimitive (``dihedral'') finite subgroups of $SU(4)$.}
    \label{tab:imprimitive_group_data}
\end{table}

Again one can study the geometry of $\mb{C}^4/\Gamma$ via looking for a concrete embedding of it into certain $\mb{C}^n$ as a CY hypersurface or a CY complete intersection. Though finding such embeddings is generally hard for the groups listed in Table~\ref{tab:primitive_group_data} and~\ref{tab:imprimitive_group_data}, we find cases which can indeed be realized as CY hypersurface in $\mb{C}^5$. For example consider the quotient of $\mathbb{C}^4/\mathbb{Z}_n^3$ by $A_4$ which is type XXXI in Table~\ref{tab:imprimitive_group_data}, which is in some sense a generalization to complex 4D of the ``dihedral'' finite group $\Delta(3n^2)\subset SU(3)$ studied in \cite{Acharya:2021jsp, Tian:2021cif}. For this case we find the invariant polynomials of $\langle \mathbb{Z}_n^3, A_4 \rangle$ to be:
\begin{equation}\label{eq:inv_polys}
    \begin{split}
        u_1 &= x_1^n + x_2^n + x_3^n + x_4^n,\ u_2 = x_1^{2n} + x_2^{2n} + x_3^{2n} + x_4^{2n},\ u_3 = x_1^{3n} + x_2^{3n} + x_3^{3n} + x_4^{3n}, \\
        u_4 &= x_1^{3n} x_2^{2n} x_3^n + x_1^n x_2^{3n} x_3^{2n} + x_1^{2n} x_2^n x_3^{3n} + x_1^{2n} x_2^{3n} x_4^n + x_1^{3n} x_3^{2n} x_4^n + x_2^{2n} x_3^{3n} x_4^n + x_1^{3n} x_2^n x_4^{2n}\\ 
        &\ \  + x_2^{3n} x_3^n x_4^{2n} + x_1^n x_3^{3n} x_4^{2n} + x_1^n x_2^{2n} x_4^{3n} + x_1^{2n} x_3^n x_4^{3n} + x_2^n x_3^{2n} x_4^{3n}, \\
        u_5 &= x_1x_2x_3x_4.
    \end{split}
\end{equation}
Since there are exactly 5 invariant polynomials, type XXXI can be expressed as a CY hypersurface in $\mathbb{C}^5$ by calculating the kernel of the following ring homomorphism:
\begin{equation}
    H: \mb{C}[u_1, u_2, u_3, u_4, u_5] \rightarrow \mb{C}[x_1, x_2, x_3, x_4]
\end{equation}
defined by (\ref{eq:inv_polys}). We find that $\text{ker}(H)$, hence the hypersurface equation in $\mb{C}^5\cong\mb{C}[u_1, u_2, u_3, u_4, u_5]$ is:
\begin{equation}
    \begin{split}
        F_{\text{XXXI}}(u_1,u_2,u_3,u_4,u_5) &= u_1^{12}-15 u_1^{10} u_2+12 u_1^9 u_3+90 u_1^8 u_2^2-144 u_1^7 u_2 u_3-246 u_1^6 u_2^3 \\
        &\ \ +60 u_1^6 u_3^2+576u_1^5 u_2^2 u_3+261 u_1^4 u_2^4-468 u_1^4 u_2 u_3^2-144 u_1^4 u_2 u_4 \\
        &\ \ -768 u_1^3 u_2^3 u_3+128 u_1^3 u_3^3+144 u_1^3 u_3 u_4-27 u_1^2 u_2^5+900 u_1^2 u_2^2 u_3^2 \\
        &\ \ +432 u_1^2 u_2^2 u_4+u_5^{2n} \left(4320 u_1^4-18144 u_1^2 u_2+13824 u_1 u_3+7776u_2^2 \right) \\
        &\ \ +u_5^n \left(-108 u_1^8+972 u_1^6 u_2-720 u_1^5 u_3-2988 u_1^4 u_2^2+4032 u_1^3 u_2 u_3 \right. \\
        &\ \ \left.+2916 u_1^2 u_2^3-1440 u_1^2 u_3^2+864 u_1^2 u_4-4464 u_1 u_2^2u_3-216 u_2^4+2016 u_2 u_3^2 \right. \\
        &\ \ \left.+1728 u_2 u_4\right) + 36u_1 u_2^4 u_3-480 u_1 u_2 u_3^3-720 u_1 u_2u_3 u_4-12 u_2^3 u_3^2 \\
        &\ \ +96 u_3^4+288 u_3^2 u_4+864u_4^2-55296 u_5^{3n}.
    \end{split}
\end{equation}
As before, the singular locus, in particular the codimension-2 singularities can be studied via calculating $\mb{C}^5/\langle F, \partial F \rangle$.

The structure of the singular locus in $\mb{C}^4/\Gamma$ can also be read off from checking the orbits of the conjugacy classes of $\Gamma$ in exactly the same way as we described in section~\ref{sec:orbifold_ab}, though we will not present details of this analysis in this work.

As a concluding remark of this section we again emphasize that though the $\mb{C}^4$ orbifolds constitute a very interesting class of geometries that can be used in geometrical engineering of 3D theories, there are many subtleties and open problems in the study of their mathematical properties. Therefore at this moment it is hard to reach a series of structured results as in \cite{Tian:2021cif}. We will come back to this issue in section~\ref{sec:conclusion}, where we will also propose several outstanding questions along this direction.

\subsection{A 3d $\mc{N}=4$ example}
\label{sec:N=4}

In this section we briefly discuss an example with enhanced $\mc{N}=4$ supersymmetry, which is given by a complex fourfold with $SU(3)$ holonomy. It does not satisfy the shrinkability conditions for a 3d $\mc{N}=2$ geometry in section~\ref{sec:shrinkability}. Let us consider a rank-1 example, where the single compact divisor is $D=T^2\times\mb{P}^1\times\mb{P}^1$. The three non-compact divisors are $S_1$, $S_2$ and $S_3$, and the Mori cone generators are
\be
\ba
&\mb{P}^1:\ D\cdot S_1\cdot S_2\cr
&\mb{P}^1:\ D\cdot S_1\cdot S_3\cr
&T^2:\ D\cdot S_2\cdot S_3\,.
\ea
\ee
The canonical divisor of $D$ is
\be
K_D=(-2S_2-2S_3)|_D\,.
\ee
The triple intersection numbers on $D$ are
\be
S_1\cdot S_2\cdot S_3|_D=1\ ,\ S_1^2=S_2^2=S_3^2=0\,.
\ee

\paragraph{K\"{a}hler form and $U(1)$ gauge coupling}

We denote the (Poincar\'{e} dual) of the K\"{a}hler form to be
\be
\ba
J^c&=xS_1+yS_2+zS_3\cr
&=-\frac{1}{4}(y+z)D+\frac{1}{2}(y-z)(S_2-S_3)+xS_1\,.
\ea
\ee
The volume of compact curves are
\be
\text{Vol}(D\cdot S_1\cdot S_2)=z\ ,\ \text{Vol}(D\cdot S_1\cdot S_3)=y\ ,\ \text{Vol}(D\cdot S_2\cdot S_3)=x\,.
\ee
The volume of compact 4-cycles are
\be
\text{Vol}(D\cdot S_1)=yz\ ,\ \text{Vol}(D\cdot S_2)=xz\ ,\ \text{Vol}(D\cdot S_3)=xy\,.
\ee
The $U(1)$ gauge coupling is given by (\ref{gauge-coupling})
\be
\ba
\label{T2P1P1-coupling}
\frac{1}{g^2}&=-\text{Vol}(D\cdot D)\cr
&=2x(y+z)\,.
\ea
\ee
Now note that when $x=0$, $y,z>0$, we have $\frac{1}{g^2}=0$, but the volume of 4-cycle $\text{Vol}(D\cdot S_1)=yz>0$, hence the shrinkability condition in section \ref{sec:shrinkability} is violated. It implies that one can reach the strong coupling point while keeping the 4-cycle $D\cdot S_1\sim\mb{P}^1\times\mb{P}^1$ at a finite volume.

\paragraph{Charged particles from M2-brane wrapping modes, 1-form symmetry and SUSY enhancement}

For M2-brane wrapping a curve in the homology class
\be
C_{abc}=D\cdot (a S_1\cdot S_2+b S_2\cdot S_3+c S_1\cdot S_3)\,,
\ee
its charge under the $U(1)$ gauge group is
\be
C_{abc}\cdot D=-2a-2c\,.
\ee
It is always a multiple of 2, hence there is a $\mb{Z}_2$ 1-form symmetry.

Now let us consider the M2-brane wrapping mode over $C_{100}=D\cdot S_1\cdot S_2$. Its moduli space is $\mc{M}_{C_{100}}=D\cdot S_3=\mb{P}^1\times T^2$, with Hodge numbers
\be
h^{0,1}(\mb{P}^1\times T^2)=h^{1,0}(\mb{P}^1\times T^2)=1\ ,\ h^{0,2}(\mb{P}^1\times T^2)=h^{2,0}(\mb{P}^1\times T^2)=0\,.
\ee

From the discussions in section~\ref{sec:M2modes}, the particle spectrum consists of a 3d $\mc{N}=2$ vector multiplet and a vector-like pair of chiral multiplets. They combine into a 3d $\mc{N}=4$ vector multiplet.

Similar statements hold for M2-brane wrapping mode over $C_{001}=D\cdot S_1\cdot S_3$ as well. Hence we expect the SUSY enhancement to $\mc{N}=4$ in this case, which is also expected from the $T^2$ structure in the geometry.

\paragraph{Non-abelian flavor symmetry enhancement $G_F$}

For the non-abelian flavor symmetry enhancement at the singular point $X_{\text 4,sing}$, we identify the flavor Cartan generators to be
\be
F_1=S_2-S_3\ ,\ F_2=S_1\,.
\ee
There is a single flavor W-boson from M2-brane wrapping
\be
C_1=D\cdot S_1\cdot (S_2-S_3)\,.
\ee
One can check that $C_1$ satisfy the conditions $C_1\cdot D=0$ and $N_{C_1|X_4}=\mc{O}\oplus\mc{O}\oplus\mc{O}(-2)$. Hence the enhanced flavor symmetry is $G_F=\mk{su}(2)\oplus\mk{u}(1)$.

\paragraph{Physical interpretations}

Since the CY4 geometry is local $T^2\times\mb{P}^1\times\mb{P}^1$, in the limit when Vol$(T^2)=x\rightarrow 0$, $y,z\rightarrow 0$ the 3d $\mc{N}=4$ theory can be thought as the dimensional reduction of 5d $SU(2)_0$ SCFT on $T^2$ and then flow to IR. Namely, the theory is the ``electric quiver theory'' EQ5 for the 5d $SU(2)_0$ theory~\cite{Closset:2020scj}, which is also called $T[SU(2)]$ in the literature. The theory is a 3d $\mc{N}=4$ $U(1)$ gauge theory coupled to 2 fundamental matter  hypermultiplets, with $\mk{su}(2)_F\oplus  \mk{su}(2)_J$ flavor symmetry in the IR~\cite{Gaiotto:2008ak}. Here $\mk{su}(2)_F$ rotates the matter hypermultiplets and $\mk{su}(2)_J$ is enhanced from $\mk{u}(1)_J$.

We attempt to identify the geometric $\mk{su}(2)$ flavor symmetry factor from CY4 with $\mk{su}(2)_J$, because $\mk{su}(2)_J$ originates from the 5d $\mk{su}(2)$ flavor symmetry that is enhanced from the 5d $\mk{u}(1)$ topological symmetry that acts on the dyonic instanton and the W-boson.

On the other hand, one can also take $x\rightarrow 0$ and $y,z>0$. The theory would be the dimensional reduction of the CB theory of 5d $SU(2)_0$ SCFT on $T^2$ and then flow to IR, nonetheless the 3d gauge coupling can still go to infinity.

%% file: sections/brane.tex
In this section, we develop a brane construction of the 3D $\mc{N}=2$ theory associated with a toric CY4. The brane configuration can be depicted as a \emph{brane box} diagram living in real 3D. We will show that the branes in the brane box are indeed the $(p,q,r)$ 4-branes studied in the context of 8D $\N=2$ superalgebra \cite{Lu:1998sx, Lu:1998mr}. Various BPS states in the brane box diagram can then be identified with the dimensional reduction of the more familiar objects in type IIB string theory.

In section~\ref{sec:pqr4branes}, we develop the language of brane box, with an emphasis on its duality with the toric diagram of the local CY4. Starting from a configuration of M-theory KK-monopoles we will be able to identify the branes in the brane box diagram as the reduction of various brany objects in IIB via certain chains of dualities.

In section~\ref{3dphysicsfrombrane}, we study the physics of a 3D $\mc{N}=2$ theory on its Coulomb branch via analyzing the properties of the corresponding brane box. In a fashion parallel to those developed in the context of 5-brane webs in \cite{Aharony:1997bh, Aharony:1997ju} we will show how to calculate the rank of CB, the effective gauge coupling and the charges of various electrically charged states using only the data of the brane box. We will also analyze various limits of the brane box diagram, in particular the limit where the theory admits a gauge theory description, and the SCFT limit. We will show a complete match between the results obtained using brane box methods and the results obtained using the toric geometry methods discussed in section~\ref{sec:toric-CY4} and~\ref{sec:examples}.

In section~\ref{seccodim2branes}, we introduce the codimension-2 branes into the brane box. These codimension-2 branes, referred to as "non-standard branes" in literature~\cite{Eyras:1999at, deBoer:2010ud, deBoer:2012ma} plays the same role as the $(p,q)$ 7-branes in a 5-brane web~\cite{DeWolfe:1999hj}. We will show via the IIB/M-theory duality that these codimension-2 branes also come from the familiar brany objects in IIB. We will see in section~\ref{sec:flavsymbrane} that the introduction of these codimension-2 branes enables us to study the flavor symmetries of the corresponding 3D $\N = 2$ theory.

\subsection{Toric CY4 and $(p,q,r)$ $4$-branes}\label{sec:pqr4branes}

In this section, we develop a brane box method to study 3D $\N=2$ theories via constructing a brane box diagram dual to a toric CY4, on which a compactification of M-theory leads to the same 3D theory given by the brane box. We will start with an M-theory KK-monopole configuration which will be shown to be equivalent to a toric CY4 in section~\ref{sec:toricCY}. In section~\ref{sec:Dual_IIB_description} we dualize the M-theory KK-monopole configuration to IIB on a torus and find that that the branes in brane box are nothing but certain familiar brany objects in IIB. In addition to the M-theory KK-monopole origin and the IIB brane origin of the brane box, we will show in section~\ref{sec:m5origin} that there is a dual M5-brane construction of the same 3D $\N=2$ theory.

\subsubsection{Toric geometry, KK-monopole and hints to a brane construction}\label{sec:toricCY}

A complex $n$-dimensional toric variety $X_n$ admits a $T^{n}$ action. For our purpose, it is illuminating to view the toric variety as a $T^n$-fibration over a real $n$-dimensional base with corners where the $T^n$-fibration can degenerate \cite{Leung:1997tw, Bouchard:Lecture,Closset:2009sv,Hori:2003ic,Kreuzer:2006ax}. Typically the physics becomes subtle near the locus where the fiber degenerates. But in this case the fiber is $T^n$ and the physics at the degeneration limit is actually well-understood.


Let us first consider the simplest case of M-theory compactification on an $S^1$ that can degenerate at several points in spacetime. After a suitable coordinate transformation one can view this as M-theory compactification on a multi-center Taub-NUT space (TN) with metric:
\begin{equation}\label{eq:TN_metric}
    ds_{TN}^2 = U^{-1}(dx^0 + \vec{\omega}\cdot d\vec{x})^2 + Ud\vec{x}^2.
\end{equation}
Here $dx^0$ is the 1-form on an $S^1$ which we will call the TN circle, $\vec{x}\in \mathbb{R}^3$, $U = 1 + \sum_I\frac{4m_{I}}{|\vec{x} - \vec{x}_I|}$ where $\vec{x}_I$'s are the coordinates of the centers of TN and $\vec{\nabla}\times \vec{\omega} = - \vec{\nabla}U$. This then defines an M-theory KK-monopole solution given by the metric:
\cite{Hull:1997kt, Townsend:1997wg, Lozano-Tellechea:2000mfy,ortin2004gravity},
\begin{equation}\label{defKK7M}
    \text{KK7M:}\qquad ds^{2} = -dt^{2} + ds^2_{(6)} + ds^{2}_{TN}
\end{equation}
where $-dt^{2} + ds^2_{(6)}$ is the flat metric of $\mb{R}^{1,6}$. A single-center KK7M can then be viewed as a 6-brane living at a point in $\mb{R}^3\subset \text{TN}$ which is indeed dual to a D6-brane in IIA upon a reduction on the TN circle \cite{Hull:1994ys, Witten_1995}.

As an illuminating warm-up example we consider M-theory geometry given by the lift of $n+1$ parallel D6-branes in IIA. Locally the geometry is given by a resolved $A_n$ singularity \cite{Acharya_2004} and the $n+1$ centers of TN can be made to be lined up on a line $l\subset \mathbb{R}^3$. To make the discussion more generalizable to the cases of interest to be addressed later, we view the 7D theory as obtained from M-theory compactified on $S^1$ which degenerates at $n+1$ distinct points on $l$. Such a configuration can readily be studied via toric methods. More precisely, the $S^1$ can be viewed as fibering over $l$ such that each $S^1$-fibration between two neighboring degeneration points on $l$ is topologically an $S^2$. Thus this fibration gives rise to in total $n$ $\mathbb{P}^1\simeq S^2$ intersecting at the middle $n-1$ degeneration points. In addition to the $n$ $\mb{P}^1$'s there is also a base $\mathbb{R}^3\backslash l \simeq \mathbb{R}^2\simeq \mathbb{C}$ which can be viewed as an $S^1$-fibration over $[0, +\infty)$ with one degeneration point at $0$. Therefore we have the toric description of a resolved $A_n$ singularity as a $T^2$-fibration over $\mathbb{R}\times \mathbb{R}_+$ with various degeneration loci. Note in particular that the toric diagram is the dual of the configuration of D6-branes or equivalently the M-theory KK-monopoles. More precisely for $n$ parallel D6-branes the corresponding toric diagram consists of $n+1$ rays out of which $n-1$ rays are internal. In summary in this case the following four constructions are equivalent:
\begin{itemize}
    \item M-theory on $S^1$ with degenerations, i.e. KK7M.
    \item M-theory on a toric CY2 whose compact part is given by an $S^1$ fibration over a real 1D base with corners represented by points.
    \item Parallel D6-branes in IIA.
\end{itemize}

As a further warm-up let us go down to 5D by considering M-theory on $T^2\simeq S^1_9\times S^1_{10}$. We will assume that $S^1_9$ is the $A$-cycle and $S^1_{10}$ is the $B$-cycle. The KK-monopole associated with the $(p,q)$-cycle of the $T^2$ fiber is dual to a $(p,q)$ 5-brane in type IIB \cite{Leung:1997tw}. This can be seen by first reducing along $S^1_9$ to get $p$ D6-branes and $q$ IIA KK-monopoles then $T$-dualizing along $S^1_{10}$ to get $p$ D5-branes and $q$ NS5-branes in IIB \cite{Eyras:1998hn}. Generalizing the above statement to a $(p,q)$ 5-brane web it is not hard to see that the brane web can be dualized to a local toric description where the $T^2$-fibration has various degeneration limits at the corners of the real 2D base \cite{Leung:1997tw, Aharony:1997bh, Aharony:1997ju, Kol:1997fv}. More precisely, the M-theory geometry is a CY3 which is the total space of the canonical bundle of the real four-dimensional space equipped with the above $T^2$-fibration structure viewed as a complex variety. The $T^2$ action together with the extra $S^1$ action rotating the bundle direction constitutes the $T^3$-action on the CY3 viewed as a toric variety.

A useful way to see how the toric description arises is as follows. For simplicity let us denote by KK7M$^{(9)}$ (KK7M$^{(10)}$) the KK-monopoles whose TN circle is $S^1_9$ ($S^1_{10}$) and centers distributed along the 8-direction (7-direction). We first perform a reduction on $S^1_9$ to get IIA such that KK7M$^{(9)}$ becomes parallel D6's distributed along the 8-direction. On the other hand, a IIA KK-monopole is carried over from KK7M$^{(10)}$ upon the reduction on $S^1_9$. A subsequent $T$-dualization along $S^1_{10}$ turns D6 to D5 while the IIA KK-monopole to NS5. In the IIB picture, it is clear that a degeneration of $S^1_9$ indicates the position of a D5 on the 8-direction while a degeneration of $S^1_{10}$ indicates the position of a NS5 on the 7-direction. More generally a $(p,q)$-cycle of $T^2\simeq S^1_9\times S^1_{10}$ degenerates along a $(p,q)$ 5-brane on $\mb{R}^2_{78}$, i.e. along a line in the $(p,q)$-direction. Therefore each $(p,q)$ 5-brane will be dual to an edge of the polytope of the dual toric fan. In summary, in this case the following three constructions are equivalent:
\begin{itemize}
    \item M-theory on $T^2$ with degenerations, i.e. KK7M's.
    \item M-theory on a toric CY3 whose compact part is given by an $T^2$ fibration over a real 2D base with corners represented by lines.
    \item $(p,q)$ 5-brane web in IIB.
\end{itemize}

Now let us generalize the above discussions to the case of real interest where M-theory is compactified on $T^3 \cong S^{1}_{10}\times S^{1}_9\times S^{1}_8$ which gives rise to 3D theories \cite{Leung:1997tw}. In this case one can define a KK7M for any 1-cycle of $T^3$, i.e. a KK7M can be associated to a $(p,q,r)$-cycle of $T^{3}$. Physically, $(p,q,r)$ indicates the winding number of the KK7M in $H_1(T^3)$ generated by:
\begin{equation}\label{cyclespqr}
    (1,0,0)\text{-cycle: }S^1_{10},\ (0,1,0)\text{-cycle: }S^1_9,\ (0,0,1)\text{-cycle: }S^1_8.
\end{equation}
Hence $(p,q,r)$ must transform as a triplet under the $SL(3,\Z)$ group acting on $T^3$. We will see that $(p,q,r)$ can also be properly viewed as the charge of the corresponding KK7M \cite{Hull:1997kt}, which we can refer to as $(p,q,r)$-KK7M or KK7M$^{(p,q,r)}$. Equivalently, we can refer to a specific KK7M brane by the direction of the associated TN cycle, e.g. KK7M$^{(1,0,0)}$ is nothing but KK7M$^{(10)}$.

To be more concrete let us take a closer look at the KK7M configuration in this case as shown in Table~\ref{originalKK7Ms}. The KK7M configuration in Table~\ref{originalKK7Ms} is equivalent to M-theory on $T^3$ with various degenerations \cite{Leung:1997tw} and it serves as the starting point of most discussions in the following sections. 
\begin{table}[H]
\begin{center}
	\begin{tabular}{c|c|c|c||c|c||c|c|c||c|c|c}
		& 0 & 1 & 2 & 3 & 4 & 5 & 6 & 7 & $S^{1}_{8}$ & $S^{1}_{9}$ & $S^{1}_{10}$\\
		\hline
		KK7M$^{(\#)}$ & $\checkmark$ & $\checkmark$ & $\checkmark$ & $\bullet$ &$\bullet$ & $\bullet$& $\checkmark$& $\checkmark$ & $\checkmark$ &$\checkmark$ & TN  \\
		\hline
		KK7M$^{(9)}$ & $\checkmark$ & $\checkmark$ & $\checkmark$ &$\bullet$ &$\bullet$ & $\checkmark$ & $\bullet$ & $\checkmark$  & $\checkmark$ & TN & $\checkmark$\\ 
		\hline
		KK7M$^{(8)}$ & $\checkmark$ & $\checkmark$ & $\checkmark$  &$\bullet$ &$\bullet$ & $\checkmark$  & $\checkmark$ & $\bullet$ & TN & $\checkmark$ & $\checkmark$ \\ 
  \hline
  \hline
   	M2$^{(1,0,0)}$ & $\checkmark$ & $\bullet$ &   $\bullet$ & $\bullet$ & $\bullet$ & $\checkmark$  & $\bullet$ & $\bullet$   &  $\bullet$ & $\bullet$ & $\checkmark$ \\ 
		\hline
 	M2$^{(0,1,0)}$ & $\checkmark$ & $\bullet$ & $\bullet$  & $\bullet$ & $\bullet$ & $\bullet$  & $\checkmark$ & $\bullet$  &  $\bullet$ & $\checkmark$ & $\bullet$ \\ 
\hline
  	M2$^{(0,0,1)}$ & $\checkmark$ & $\bullet$ & $\bullet$   & $\bullet$ & $\bullet$& $\bullet$  & $\bullet$ & $\checkmark$   &  $\checkmark$ & $\bullet$ & $\bullet$ \\ 
	\end{tabular}
	\caption{The KK7M configuration associated to M-theory on $T^3\cong S^1_8\times S^1_9\times S^1_{10}$. In this table, we also anticipate the configuration of the M2-branes, that can end on these KK monopoles, which will play an important role in the later discussions. Here, $\#$ denotes the $10^\text{th}$ direction.}
	\label{originalKK7Ms}
\end{center}
\end{table}

It is not hard to show that the KK7M configuration is indeed supersymmetric by recalling that SUSY projection operators are:
\begin{equation}\label{eq:kk7mconstraines}
    \begin{split}
        &P_{(\#)}: \ \varepsilon = \Gamma^{0126789}\, \varepsilon, \ \quad P_{(9)}: \ \varepsilon = \Gamma^{012578\#}\, \varepsilon, \  \quad P_{(8)}: \ \varepsilon = \Gamma^{012569\#}\, \varepsilon,
    \end{split}
\end{equation}
where $P_{(i)}$ with $i\in\{\#,9,8\}$ denotes the projector associated with the KK7M$^{(i)}$ monopole in the above table. In particular, the superscript indicates the location of the TN circle associated with the KK monopole. Here, $\#$ denotes the $10^\text{th}$ direction and $\Gamma^{ij\cdots k} = \Gamma^{[i}\Gamma^j\cdots \Gamma^{k]}$. It is easy to verify that each pair of projectors in (\ref{eq:kk7mconstraines}) commute therefore can be simultaneously diagonalized \cite{deRoo:1997gq,Bergshoeff:1997bh,Bergshoeff:1997kr}. Furthermore, since each projector is traceless and squares to identity, it halves the number of constant spinors. Therefore there will be in total four constant spinors left invariant on the common intersection of the three KK7M's in Table~\ref{originalKK7Ms}, i.e., on $\R^{1,2}$ spanning the 012-direction. Hence the KK7M configuration in Table~\ref{originalKK7Ms} engineers a 3D $\N=2$ theory as expected.

Note that in Table~\ref{originalKK7Ms} we have also anticipated the existence of various M2-branes that can be introduced into the KK7M system in a supersymmetric fashion\footnote{The introduction of M2-branes into the KK7M system will be discussed in more detail in section~\ref{sec:additional_KK7M}.} ~\cite{deRoo:1997gq,Bergshoeff_1997,Eyras:1998hn,Bergshoeff:1997bh,Bergshoeff:1997kr}. It is clear from the configuration in Table~\ref{originalKK7Ms} that these M2-branes stretch between the KK7M's in the 567-direction (which will be denoted by $\mb{R}^3_{567}$ from now on) and they become particles moving in $\R^{1,2}$.

Similar to the 7D and 5D cases discussed earlier, M-theory on $T^3 \cong S^1_{10}\times S^1_9\times S^1_8$, i.e. the KK7M system in Table~\ref{originalKK7Ms}, is actually equivalent to M-theory on a toric variety where the above $T^3$ degenerates along various sub-loci of certain base. To generalize the 7D and 5D cases, one must now consider a real 3D base with corners where $T^3$ degenerates and it is indeed natural to expect such base lives in $\mb{R}^3_{567}$. From the configuration in Table~\ref{originalKK7Ms} it is clear that KK7M$^{(10)/(9)/(8)}$ degenerates at points along 5/6/7-direction. We define a $(p,q,r)$-plane as a plane normal to the vector $(p,q,r)$ in $\mb{R}^3_{567}$. The statement above can be equivalently expressed as follows: a $(1,0,0)$-cycle vanishes on a $(1,0,0)$-plane, a $(0,1,0)$-cycle vanishes on a $(0,1,0)$-plane and a $(0,0,1)$-cycle vanishes on a $(0,0,1)$-plane. More generally a $(p,q,r)$-cycle of $T^3$ vanishes on a $(p,q,r)$-plane in $\mb{R}^3_{567}$. This then immediately leads to a local toric geometry such that each corner of the base in $\mb{R}^3_{567}$ is represented by a $(p,q,r)$-plane where a $(p,q,r)$-cycle of $T^3$ degenerates. In summary we have shown the equivalence of the following two constructions of 3D $\N=2$ theory:
\begin{itemize}
    \item M-theory on $T^3$ with degenerations, i.e. a KK7M configuration in Table~\ref{originalKK7Ms}.
    \item M-theory on toric CY4 whose compact part is given by $T^3$ fibration over a real 3D base with corners represented by $(p,q,r)$-planes.
\end{itemize}

Comparing the above result with the results we have summarized from the 7D and 5D cases, one may wonder if there is an equivalent brane construction of the same 3D $\N=2$. To be more precise let us denote by $\Delta_n$ the diagram dual to the polytope $\mc{P}_n$ associated to the toric fan of a CY $(n+1)$-fold. Here the vertices of $\mc{P}_n$ are obtained by deleting the last component of the rays of the toric fan of the CY $(n+1)$-fold as described in section~\ref{sec:examples}. In 7D we have seen that a parallel D6 configuration gives rise to $\Delta_1$ dual to $\mc{P}_1$ associated to the toric fan of a resolved $A_n$ singularity. In 5D we have seen that a $(p,q)$ 5-brane web gives rise to $\Delta_2$ dual to $\mc{P}_2$ associated to the toric fan of a resolved CY3. Hence in 3D it is natural to ask if there exists a physical brane system giving rise to $\Delta_3$ dual to $\mc{P}_3$ associated to the toric fan of a resolved CY4. In other words it is natural to ask if the $(p,q,r)$-planes defined above are actually physical branes in string theory. It is our task in the following sections to show that they are indeed brany objects in string theory.

\subsubsection{Dual type IIB description}\label{sec:Dual_IIB_description}

To identify the previously introduced $(p,q,r)$-planes with physical branes in string theory, in particular with branes in type IIB we will look at the following IIB/M-theory duality:
\begin{equation}\label{eqduality}
   \begin{split}
        &\text{M-theory on $S^{1}_{10}\times S^{1}_{9}\times S^{1}_{8}$} \qquad \xleftrightarrow{\text{duality}} \qquad \text{Type IIB on $\widetilde{S}^{1}_{9}\times S^{1}_{8}$}
   \end{split}
\end{equation}
Here $\widetilde{S}^{1}_{9}$ is the T-dual of $S^{1}_{9}$. In the limit of small $\widetilde{S}^{1}_{9}\times S^{1}_{8}$ the RHS of \ref{eqduality} reduces to 8D with 32 supercharges. This point of view will become useful later in section~\ref{sec:8DSALG} when we identify the $(p,q,r)$-planes with the effective 4-branes which can be seen from 8D $\N=2$ superalgebra.

To arrive at a type IIB description of the KK7M system in Table~\ref{originalKK7Ms}, we first do a KK reduction to type IIA on a circle. Any KK7M can be reduced on a circle in three different ways based on the choice of the KK circle \cite{Elitzur:1997zn,ortin2004gravity,Eyras:1999at}:
\begin{enumerate}[label=(\alph*)]\label{KK7MtoIIA}
    \item If the KK circle coincides the TN circle of the KK7M, the KK7M will be reduced to D6; 
    \item If the KK circle is on the world volume of the KK7M, the KK7M will be reduced to a KK6A; 
    \item If the KK circle is transverse to both the TN circle and the world volume of the KK7M, the KK7M will be reduced to a KK7A.
\end{enumerate}
Here KK6A is nothing but embedding the TN metric in (\ref{eq:TN_metric}) into 10D and we will postpone a full account of KK7A until section~\ref{seccodim2branes}, whose nature for now does not play any essential role in the discussions to follow. Given this freedom to choose the KK circle, we choose the KK circle to be $S^1_{10}$. Upon this KK reduction to IIA we have:
\begin{equation}
    (\text{KK7M}^{(10)},\text{KK7M}^{(9)}, \text{KK7M}^{(8)}) \rightarrow \text{(D6,KK6A,KK6A)}.
\end{equation}
Note that choosing a different KK circle corresponds to a permutation of the resulting IIA brane configuration. For instance, had we chosen the KK circle to be $S^1_9$ we would have obtained
\begin{equation}
    (\text{KK7M}^{(10)},\text{KK7M}^{(9)}, \text{KK7M}^{(8)}) \rightarrow \text{(KK6A,D6,KK6A)}
\end{equation}
which clearly permutes D6 to be the one from reducing KK7M$^{(9)}$.

Note that though there is no natural triplet of winding numbers in IIA since one circle of $T^3$ in the KK7M configuration has been used as the KK circle thus is shrunk, we can nevertheless define a $(p,q,r)$ charge for the resulting branes in the (D6,KK6A,KK6A) system by simply assigning $(1,0,0)$ to the D6, and $(0,1,0)$ and $(0,0,1)$ to the KK6A from KK7M$^{(9)}$ and KK7M$^{(8)}$ respectively.

To achieve a type IIB description we will further T-dualize the (D6,KK6A,KK6A) system. Again, there are three ways to T-dualize a given KK6A as follows \cite{Eyras:1998hn,Eyras:1999at,ortin2004gravity,Elitzur:1997zn,deBoer:2012ma}: 
\begin{enumerate}[label=(\alph*)]\label{KK6AtoIIB}
    \item T-dualize on the TN circle of KK6A leads to NS5-brane; 
    \item T-dualize on a circle in worldvolume direction of KK6A leads to KK6B;
    \item T-dualize on a circle transverse to both the TN circle and worldvolume direction of KK6A leads to an exotic brane \cite{deBoer:2010ud, deBoer:2012ma}.
\end{enumerate}
Here KK6B is again defined by embedding the TN metric in (\ref{eq:TN_metric}) in 10D. A full account of the exotic branes will be deferred until section~\ref{seccodim2branes} along with the discussion of KK7A as mentioned earlier. Again, given this freedom of choosing the T-dual circle, we choose to T-dualize on $S^1_9$. Upon this T-dualization we have:
\begin{equation}
    \text{(D6,KK6A,KK6A)} \rightarrow \text{(D5, NS5, KK6B)}.
\end{equation}

Now that we have obtained the end configuration in IIB of the KK7M in Table~\ref{originalKK7Ms}, let us study the IIB objects that the M2-branes in Table~\ref{originalKK7Ms} are dualized to under the same choice of the KK circle and the T-dual circle. This is rather easy and we simply give the results as follows:
\begin{equation}
    \text{(M2$^{(1,0,0)}$, M2$^{(0,1,0)}$, M2$^{(0,0,1)}$)} \rightarrow \text{(F1, D1, D3)}.
\end{equation}

In summary, upon a KK reduction on $S^1_{10}$ and a subsequent T-dualization on $S^1_9$ the KK7M/M2-brane configuration in Table~\ref{originalKK7Ms} becomes the brane/string configuration in IIB shown in Table~\ref{d5ns5kk6b}. Note that we have also labelled the IIB objects by a triplet of numbers which are the $(p,q,r)$ charge naturally defined in the M-theory dual frame. Though they are not defined naturally in IIB in the sense that they do not correspond to any winding numbers in a canonical way, they can nevertheless be defined in a meaningful way in the sense that these numbers label the corresponding dual objects in Table~\ref{originalKK7Ms} in M-theory without ambiguity.
\begin{table}[H]
\begin{center}
	\begin{tabular}{c|c|c|c||c|c||c|c|c||c|c}
		& 0 & 1 & 2 & 3 & 4 & 5 & 6 & 7 & $S^{1}_{8}$ & $\widetilde{S}^{1}_{9}$  \\
		\hline
		D$_{5}^{(1,0,0)}$  & $\checkmark$ & $\checkmark$ & $\checkmark$ & $\bullet$ &$\bullet$ & $\bullet$& $\checkmark$& $\checkmark$ & $\checkmark$ &$\bullet$   \\
		\hline
		NS$_{5}^{(0,1,0)}$  & $\checkmark$ & $\checkmark$ & $\checkmark$ &$\bullet$ &$\bullet$ & $\checkmark$  & $\bullet$ & $\checkmark$  & $\checkmark$ &$\bullet$  \\ 
		\hline
		KK6B$^{(0,0,1)}$  & $\checkmark$ & $\checkmark$ & $\checkmark$  & $\bullet$ &$\bullet$ & $\checkmark$  & $\checkmark$ & $\bullet$ & TN & $\checkmark$  \\ 
         \hline
        \hline
		   F$_{1}^{(1,0,0)}$     & $\checkmark$ & $\bullet$ & $\bullet$ & $\bullet$  &$\bullet$  & $\checkmark$ & $\bullet$  & $\bullet$ & $\bullet$ & $\bullet$   \\
		\hline
		   D$_{1}^{(0,1,0)}$ & $\checkmark$ & $\bullet$ & $\bullet$ & $\bullet$& $\bullet$& $\bullet$   & $\checkmark$& $\bullet$  & $\bullet$ &$\bullet$  \\ 
		\hline
		D$_{3}^{(0,0,1)}$ & $\checkmark$ & $\bullet$ & $\bullet$  & $\bullet$ & $\bullet$ & $\bullet$   & $\bullet$  & $\checkmark$   & $\checkmark$  & $\checkmark$  \\ 
	\end{tabular}
	\caption{The dual IIB configuration of the M-theory configuration in Table~\ref{originalKK7Ms}.}
	\label{d5ns5kk6b}
\end{center}
\end{table}

It follows from Table~\ref{d5ns5kk6b} that the 10D spacetime is now decomposed as:
\begin{equation}
    \R^{1,2}_{(012)} \times \R^{2}_{(34)}  \times \R^{3}_{(567)}   \times \text{T}^{2}_{(89)}.
\end{equation}
such that the original $SO(1,9)$ symmetry branches into:
\begin{equation}
    SO(1,2)\times SO(2)\times SO(3)\times SO(2). 
\end{equation}
The $\R^{1,2}_{(012)}$ is spanned by the common worldvolume of (D5,NS5,KK6B) along the 012-directions  and is the spacetime where the 3D theory lives. Due to the IIB/M-theory duality the (D5,NS5,KK6B) system must also preserve four supercharges as the KK7M system does. This implies that the (D5,NS5,KK6B) system gives rise to a 3D $\N=2$ theory. In fact, the 3D $\N=2$ superalgebra enjoys an $SO(2)_{R}$ R-symmetry, which can be naturally identified with the $SO(2)$ symmetry acting on $\R^{2}_{(34)}$ transverse to the all the branes in the (D5,NS5,KK6B) system.

Let us take a closer look at the various types of intersections between objects in Table~\ref{d5ns5kk6b}. Both D5 and NS5 intersect with the KK6B along 3 spatial dimensions. Therefore they become solitonic 3-branes as seen on the KK6B world volume. There are two 0-form fields propagating on the KK6B world volume transforming as a doublet under S-duality \cite{Eyras:1998hn}. The solitonic 3-branes are magnetically charged under this 0-form doublet hence transform non-trivially as a doublet under S-duality as well. This is a remnant of the fact that D5 and NS5 transform in the same manner under S-duality as seen on the KK6B world volume.

The D3-brane in Table~\ref{d5ns5kk6b} intersects both D5 and NS5 along 2 spatial dimensions. On the D5 world volume there propagates a 1-form field under which the above D3-brane is a magnetically charged solitonic 2-brane as seen on the D5 world volume. The intersection between D3 and NS5 can then be understood as the S-dual configuration of the intersection between D3 and D5. This S-dualization which brings D5 to NS5 can be performed via rotating the 56-plane by 90 degree \cite{Aharony:1997bh}. From Table~\ref{d5ns5kk6b} it is clear that D3 remains unchanged under this rotation as expected. This D3-brane intersects KK6B in one spatial dimension while wrapping its TN circle. On the KK6B world volume there propagates a S-self-dual 2-form \cite{Eyras:1998hn} under which the D3 is a charged effective string as seen on the KK6B world volume.

The F1 and D1-strings end on the D5 and NS5-branes in the usual sense as studied in the context of 5-brane webs \cite{Aharony:1997bh}. What is new in the configuration shown in Table~\ref{d5ns5kk6b} is the intersection of (F1,D1) with KK6B along one spatial dimension. The true nature of this intersection can be better understood from a IIA dual frame where a fundamental string can intersect with an NS5 along one spatial direction \cite{Eyras:1998hn}. Upon a T-dualization transverse to the NS5 the above IIA configuration transform to F1 intersecting KK6B along one spatial direction. Since KK6B is neutral to S-duality, a subsequent S-dualization transforms the (F1,KK6B) configuration to a (D1,KK6B) configuration.

Note that the (D5,NS5) sub-configuration in Table~\ref{d5ns5kk6b} is indeed the well-studied $(p,q)$ 5-brane web \cite{Aharony:1997bh,Aharony:1997ju}. It is natural to expect that many techniques that were applied in studying the 5-brane web can be transplanted to study the configuration in Table~\ref{d5ns5kk6b}. We will see in later sections that this is indeed the case when we discuss charge conservation, brane bending, Hanany-Witten moves, etc.

Motivated by the fact that the (D5,NS5) sub-system of (D5,NS5,KK6B) can be viewed as a $(p,q)$ 5-brane web, one may wonder if the full (D5,NS5,KK6B) system in Table~\ref{d5ns5kk6b} can be viewed as intersecting branes in $\R^{3}_{567}$. In this configuration, in contrast to the 5-brane web, each pair of the branes intersect along an edge rather than a vertex.

An important observation is that the LHS and the RHS of (\ref{eqduality}) treat the $(p,q,r)$ charges on different footings. On the M-theory side, the $(p,q,r)$ charge which labels a $(p,q,r)$-cycle in $T^3$ transform as a triplet under the $SL(3,\Z)$ modular symmetry of $T^3$. However in the dual type IIB configuration only an $SL(2,\mathbb{Z})$ symmetry is manifest under which $(p,q)$ transforms as a doublet of $SL(2,\mathbb{Z})$ whereas $r$ transforms as a singlet. Therefore in the dual type IIB picture there is no canonical way to treat the $(p,q,r)$ charge in a totally symmetric manner. Therefore it is favored to have a formulation that treats the $(p,q,r)$ charge symmetrically as in the M-theory dual frame.

There is also a related problem in the context of brane dynamics. It is well-known that a D5 bends when it meets an NS5 due to charge conservation~\cite{Aharony:1997ju, Aharony:1997bh}. The D5 and NS5 in Table~\ref{d5ns5kk6b} bend for the same reason in the background of KK6B's. But it remains unknown if and how a KK6B bends when it meets a D5 or an NS5 or vice versa, an investigation of which certainly requires a closer look at the dynamics of these objects. This asymmetry in brane bending when D5, NS5 and KK6B meet in pairs is also a reflection of the asymmetry in the $(p,q,r)$ charge discussed in the previous paragraph.

As we will discuss in greater detail in the next section, the full story is better understood in the context of maximal supersymmetric theory in 8D, where the $(p,q,r)$ charge is treated symmetrically and the brane dynamics is clearer hence the brane bending problem can be solved.

\subsubsection{Branes from 8D superalgebra}\label{sec:8DSALG}

Motivated by the RHS of the duality in (\ref{eqduality}), we now study extended objects in 8D maximal supersymmetric theory obtained from IIB on $T^2$. We will see in a moment that the D5, NS5 and KK6B descend to various $(p,q,r)$ 4-branes in this limit studied in \cite{Lu:1998mr, Lu:1998sx}.

To see the existence of those extended objects we look at type IIB superalgebra in $\mb{R}^{1,7}\times T^{2} \cong \mb{R}^{1,7}\times \widetilde{S}^1_9\times S^1_8$ spacetime:
\begin{equation}\label{z4z5zkk}
\begin{split}
      \{Q^{\alpha},Q^{\beta}\}^{\mb{R}^{1,7}\times \widetilde{S}^{1}_{9}\times S^{1}_{8}} \  \supset \   &(\Gamma^{M_{1}\cdots M_{4}8}C^{-1}) \, Z^{(p,q)}_{M_{1}\cdots M_{4}8} 
         \\
         &+ \  (\Gamma^{M_{1}\cdots M_{4}9}C^{-1}) \ Z^{(r)}_{M_{1}\cdots M_{4}98}\,k^{8}
\end{split}
\end{equation}
where the tilde over $\widetilde{S}^1_9$ means that the corresponding circle is the one on which a T-dualization is performed when obtaining IIB from IIA, and $k^{8}$ represents a vector in the direction of the TN circle of KK6B. Here the 5-form charges $Z^{(p,q)}_{M_{1}\cdots M_{4}8}$ and $Z^{(r)}_{M_{1}\cdots M_{4}98}\,k^{8}$ are appended with $p$, $q$ and $r$ labels, meaning that the corresponding branes are D5, NS5 and KK6B respectively~\cite{Hull:1997kt,Townsend:1997wg,ortin2004gravity}. Moreover, they are the winding numbers of the dual KK7M's upon running the duality chain (\ref{eqduality}) from RHS to LHS.

Reducing (\ref{z4z5zkk}) on $T^2 \cong \widetilde{S}^{1}_{9}\times S^{1}_{8}$ by taking the small volume limit of $T^2$ we get the superalgebra of maximally supersymmetric 8D theory. The reduction can be done by eliminating the $\hat{M}$ and $M_{5}$ directions along with the $k^{M_{5}}$ vector from (\ref{z4z5zkk}). Consequently, all the 5-forms in (\ref{z4z5zkk}) are reduced to 4-forms labelled by the same $(p,q,r)$ charges. After the reduction, the genuine 8D superalgebra contains the following term:
\begin{equation}\label{zpqr}
\{Q^{\alpha},Q^{\beta}\}^{\text{8D}} \  \supset \ \  (\Gamma^{M_{1}\cdots M_{4}}C^{-1}) \   Z^{(p,q,r)}_{M_{1}\cdots M_{4}}.
\end{equation}
Clearly there will be extended objects charged under the ``central charge'' $Z_{M_1M_2M_3M_4}^{(p,q,r)}$ labelled by $(p,q,r)$. It is also clear that in (\ref{zpqr}) we have recovered the symmetry between the $(p,q,r)$ charges that is manifest in the KK7M configuration in Table~\ref{originalKK7Ms} but is not manifest in the 10D IIB description discussed in the previous section or the 10D superalgebra in (\ref{z4z5zkk}). Actually one can identify the $SL(3,\Z)$ modular group of the M-theory $T^3$ with the $SL(3,\Z)$ subgroup of the full $SL(3,\Z)\times SL(2,\Z)$ U-duality group of the 8D maximal supersymmetric theory~\cite{Hull:1994ys,hawking1981superspace,Alonso-Alberca:2000wkg}.

The set of all 4-branes are in $(\textbf{3},\textbf{2})$ representation of the U-duality group in the sense that they form a doublet of two triplets under $SL(3,\Z)\times SL(2,\Z)$~\cite{Lu:1998sx}. The $(p,q,r)$ 4-branes that are relevant to our discussion form one of these two triplets. Therefore all the discussions to follow will be made in one particular $SL(2,\Z)$ duality frame, while a global $SL(2,\mb{Z})$ transformation takes the triplet in one $SL(2,\mb{Z})$ duality frame to a triplet in some other $SL(2,\mb{Z})$ duality frame.

The 4-brane configuration in 8D shown in Table~\ref{4braneconfiguration} is obtained from the (D5,NS5,KK6B) configuration in Table~\ref{d5ns5kk6b} by simply eliminating the 8 and the 9-directions. Since the number of supercharges does not change in the procedure we will arrive at a 3D $\mathcal{N}=2$ theory living along the common directions of the $(p,q,r)$ 4-branes. Applying the same $T^2$-reduction to the (F1,D1,D3) system in Table~\ref{d5ns5kk6b}, we obtain the effective $(p,q,r)$ strings in 8D $\N=2$ which transform as $(\textbf{3},\textbf{1})$ under the U-duality group~\cite{Lu:1998mr,Lu:1998sx}. Moreover, it has been worked out in \cite{Lu:1998mr} that a $(p,q,r)$ string in 8D is indeed a bound state of $p$ F1-strings ($(1,0,0)$ string), $q$ D1-strings ($(0,1,0)$ string) and $r$ D3-branes ($(0,0,1)$ string). The configuration of these effective strings are also presented in Table~\ref{4braneconfiguration}.

Motivated by the studies of $(p,q)$-strings and $(p,q)$ branes, one may expect to see charge conservation, brane bending and other aspects of physics in the context 4-brane (4-string) junctions. We will now discuss $(p,q,r)$ charge conservation and 4-brane bending in turns.

\begin{table}[H]
\begin{center}
	\begin{tabular}{c|c|c|c||c|c||c|c|c}
		& 0 & 1 & 2 & 3 & 4 & 5 & 6 & 7   \\
		\hline
		(1,0,0) 4-brane & $\checkmark$ & $\checkmark$ & $\checkmark$ & $\bullet$ & $\bullet$& $\bullet$ & $\checkmark$ &$\checkmark$   \\
		\hline
		(0,1,0) 4-brane & $\checkmark$ & $\checkmark$ & $\checkmark$  &$\bullet$ & $\bullet$  & $\checkmark$   & $\bullet$ &$\checkmark$  \\ 
		\hline
		(0,0,1) 4-brane & $\checkmark$ & $\checkmark$ & $\checkmark$   &$\bullet$ & $\bullet$  & $\checkmark$    &$\checkmark$  & $\bullet$  \\
  	\hline
    \hline
		   (1,0,0)-string    & $\checkmark$ & $\bullet$ & $\bullet$  &$\bullet$  & $\bullet$ & $\checkmark$    & $\bullet$ & $\bullet$   \\
		\hline
		   (0,1,0)-string  & $\checkmark$ & $\bullet$ & $\bullet$   & $\bullet$& $\bullet$   & $\bullet$     & $\checkmark$ &$\bullet$  \\ 
		\hline
		(0,0,1)-string  & $\checkmark$ & $\bullet$ & $\bullet$     & $\bullet$ & $\bullet$   & $\bullet$      & $\bullet$  & $\checkmark$  \\
	\end{tabular}
	\caption{The $(p,q,r)$ 4-brane configuration of the 8D $\mathcal{N}=2$ theory that corresponds to that of Table \ref{d5ns5kk6b} in the described limit above.}
	\label{4braneconfiguration}
\end{center}
\end{table}

\paragraph{Stability and charge conservation}

A stable intersecting brane configuration can be achieved by imposing the force-free condition, i.e. vanishing tension along the edge of intersection. Though the general form of the 4-brane tension formula can be found in \cite{Lu:1998sx}, we will limit our discussions of $(\textbf{3},\textbf{2})$ 4-branes to a fixed $SL(2,\mathbb{Z})$ dual frame as discussed earlier in this section. Given such a choice the tension formula can be expressed as follows:
\begin{equation}\label{eq:4brane_tension}
    \Vec{\tau}_{p,q,r} = e^{\phi_{0}/2- \varphi_{0}/2\sqrt{3}}\, (p-\chi_{30}q - \chi_{20}r)\,\hat{e}_{5} + e^{- \phi_{0}/2-\varphi_{0}/2\sqrt{3}}  \,(q-\chi_{10}r)\,\hat{e}_{6} + e^{\varphi_{0}/\sqrt{3}}\,r\,\hat{e}_{7}.
\end{equation}
Here the tension is expressed as a vector in an $\R^{3}$ which in our setup is $\mb{R}^3_{567}$ which can be seen from Table~\ref{4braneconfiguration} and $\phi_{0}$, $\varphi_{0}$, and $\{\chi_{i0}\}$, $i=1,2,3$ represent the asymptotic VEVs of the two dilaton fields and the three axions, respectively. These scalar fields can be understood as coordinates on the coset $SL(3,\R)/SO(3)$ part of the U-manifold~\cite{Kumar:1996zx,Lu:1998sx}.

It is easy to see from (\ref{eq:4brane_tension}) that the orientation of a $(p,q,r)$ 4-brane in $\R^3_{567}$ is determined by both the VEVs of the scalar fields and the $(p,q,r)$ charge. For simplicity and without loss of generality, we set the VEVs of the scalar fields to zero so that the orientation of a 4-brane is fully determined by its $(p,q,r)$ charge. More precisely, it can be seen from (\ref{eq:4brane_tension}) that a $(p,q,r)$ 4-brane has its tension along $p\hat{e}_5+q\hat{e}_6+r\hat{e}_7$ after setting the VEVs of all the scalars fields to 0. Therefore visually one can draw a $(p,q,r)$ 4-brane as a plane normal to the $(p,q,r)$-direction in $\mb{R}^3_{567}$. This makes the 4-branes exactly the objects we have been looking for as physical realizations of the 4-planes discussed in section~\ref{sec:toricCY}. Unless said otherwise we will stick to the assumption that the VEVs of all the scalar fields vanish.

Given (\ref{eq:4brane_tension}), the force-free condition can be expressed as follows:
\begin{equation}\label{eq:force-free}
    \sum_{i}\,\Vec{\tau}_{(p_{i},q_{i},r_{i})} = 0.
\end{equation}
applied at each edge where the branes intersect. The crucial observation here is that (\ref{eq:4brane_tension}) depends multi-linearly on $p$, $q$ and $r$. This immediately implies that the force-free condition (\ref{eq:force-free}) at an edge is equivalent to the following \emph{static condition}:
\begin{equation}\label{eq:static_condition}
    \quad \sum_{i}\,p_{i} = \sum_{i}\,q_{i} = \sum_{i}\,r_{i}  = 0.
\end{equation}
The above equation can immediately be recognized as charge conservation at an edge where the branes intersect.

Along the same line of arguments, for $(p,q,r)$-strings whose tension is \cite{Lu:1998mr}:
\begin{equation}\label{Tofstrings}
    \vec{\tau}_{(p,q,r)} = (p - \chi_{30}\,q - \chi_{20}\,r) \, \hat{e}_{5} + e^{-\phi_{0}} (q - \chi_{10}\,r) \hat{e}_{6} + e^{-\phi_{0}/2 + \sqrt{3}\varphi_{0}/2}\,r \,\hat{e}_{7},
\end{equation}
the force-free condition $\sum_i\vec{\tau}_i$ for a set of intersecting strings is again equivalent to the static condition~\cite{Lu:1998mr}. Since (\ref{Tofstrings}) is again multi-linear in $p$, $q$ and $r$, it implies charge conservation at each point where the $(p,q,r)$-strings meet.

It is easy to see from Table~\ref{4braneconfiguration} that a $(1,0,0)/(0,1,0)/(0,0,1)$-string can end on a $(1,0,0)/(0,1,0)/(0,0,1)$ 4-brane, respectively. More generally a $(p,q,r)$ string can end on a $(p,q,r)$ 4-brane and looks like a charged point particle moving on the 4-brane worldvolume. Moreover, from (\ref{Tofstrings}) after setting the VEVs of all the scalar fields to 0, the direction of a $(p,q,r)$ string should be identified with the normal direction of a $(p,q,r)$ 4-brane. Therefore a $(p,q,r)$-string can end perpendicularly on a $(p,q,r)$ 4-brane in $\mb{R}^3_{567}$.

\begin{figure}[H]
\centering{
\includegraphics[scale=0.5]{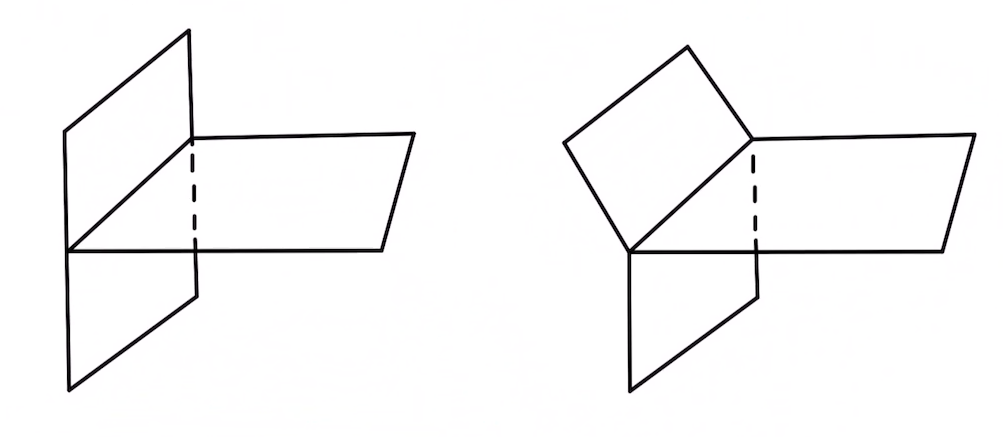}
}
\caption{The left-hand side is the brane configuration before bending. The right-hand side represents brane bending subject to charge conservation that is well-defined in the 8D limit.}
\label{4braneintersectbend}
\end{figure}

\paragraph{4-branes bending} Now let us discuss brane bending in our brane box setup similar to the phenomenon studied in the context of 5-brane webs~\cite{Aharony:1997ju, Aharony:1997bh}. Without loss of generality we consider a ${(0,0,1)}$ 4-brane ending on a ${(1,0,0)}$ 4-brane as drawn in Figure~\ref{4braneintersectbend}. Each 4-brane appears as a solitonic 3-brane on the world volume of the other as can be seen from Table~\ref{4braneconfiguration}. The solitonic 3-brane can in turn be viewed as a point particle living in the one-dimensional space transverse to the world volume of the solitonic 3-brane and carrying a charge under the world volume $U(1)$ gauge field. In the case at hand a $(0,0,1)$ 4-brane can be viewed as a particle living in the 7-direction. The dynamics of such a charged particle can be described by solving the following Laplace equation analogous to the case in \cite{Aharony:1997ju,Aharony:1997bh}:
\begin{equation}\label{eqforbending}
        \nabla^{2}_{1+1}\,\widetilde{x} \,=\, \delta(x) \Rightarrow \widetilde{x} = Q\,(x + |x|)
\end{equation}
where $Q$ is a certain electric charge. Here $(\widetilde{x},x)$ parameterizes the plane in which a 4-brane ends on another 4-brane, which is the case at hand is $\mb{R}^2_{57}$. (\ref{eqforbending}) then implies that the $(1,0,0)$ 4-brane bends to a $(1,0,1)$ 4-brane when it meets a $(0,0,1)$ 4-brane. In particular, we see that the branes bend in a way that charge must be preserved along an edge they meet.

The above brane intersection configuration can immediately be generalized to any configuration of three 4-branes meeting along an edge via an $SL(3,\mathbb{Z})$ transformation. The branes bend in a way that respects charge conservation, i.e. along each edge we must have (\ref{eq:static_condition}). It is not hard to see that any complicated configuration of brane intersection can be deformed to a set of ``simple'' configurations that each edge is formed by exactly three 4-branes meeting each other. In this sense the configuration in Figure~\ref{4braneintersectbend} is the ``building block'' of all possible brane intersection configurations. Therefore we have shown that brane bending is also a critical physical phenomenon in our brane box setup and is a consequence of $(p,q,r)$ charge conservation.

\subsubsection{Duality between brane box and toric CY4}\label{sec:duality_branebox_CY4}

Now we have found a physical realization of the intersecting planes in $\mb{R}^3_{567}$ discussed in section~\ref{sec:toricCY} as an intersecting 4-brane box system in the 8D $\N=2$ theory subject to $(p,q,r)$ charge conservation. Such a brane box construction of a 3D $\N=2$ theory is equivalent to geometrically engineering the same theory via M-theory on a dual toric CY4. More precisely, as we have discussed in section~\ref{sec:toricCY}, the brane box living in $\mathbb{R}^3_{567}$ is dual to the 3D polyhedron $\mc{P}_3$ associated to the toric fan of a CY4. Everything we have discussed for the $(p,q,r)$-planes defined in section~\ref{sec:toricCY} can be immediately applied to the $(p,q,r)$ 4-branes discussed in section~\ref{sec:Dual_IIB_description} and~\ref{sec:8DSALG}. The $(p,q,r)$ triplet, which in section~\ref{sec:toricCY} labels merely the normal direction of a plane in $\mathbb{R}^3_{567}$, now acquires a very concrete physical meaning as the charge triplet of a 4-brane.

\begin{figure}[H]
\centering{
\includegraphics[scale=0.45]{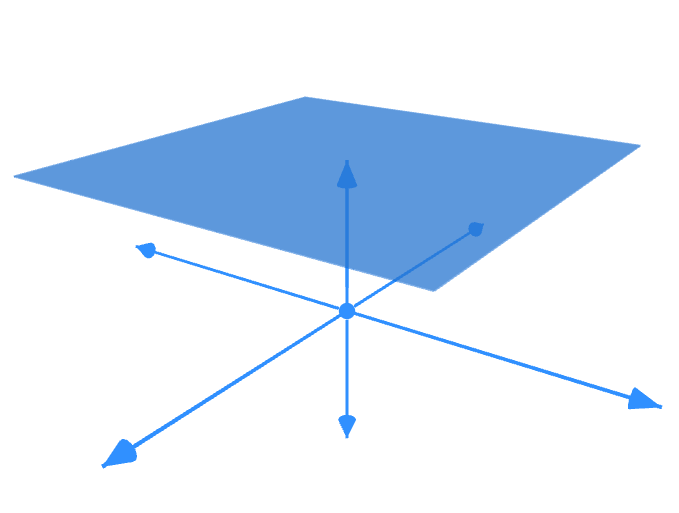}
}
\caption{The skeleton of the 3D polyhedron associated to local $\P^{1}\times\P^{1}\times\P^{1}$ and a $(0,0,1)$ 4-brane dual to the edge pointing upwards.}
\label{toricandbrane}
\end{figure}

\begin{figure}[H]
\centering{
\includegraphics[scale=0.4]{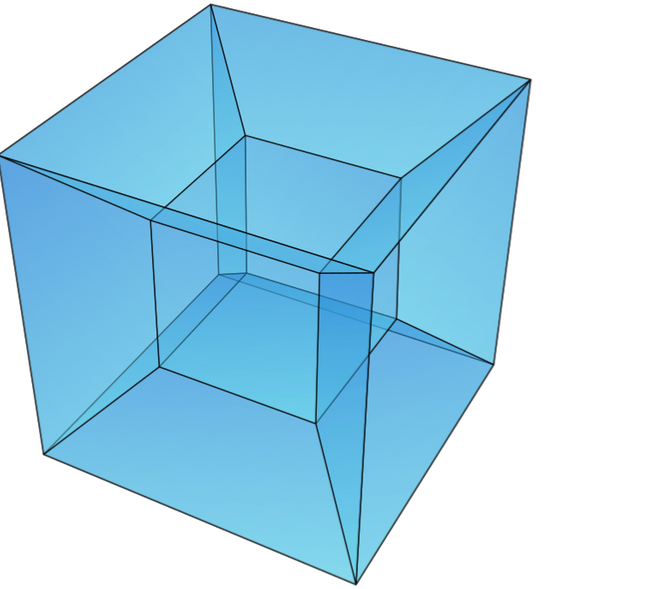}
}
\caption{The $(p,q,r)$ 4-brane configuration in $\mb{R}^3$ that is dual to the 3D polyhedron associated to local $\P^{1}\times\P^{1}\times\P^{1}$ as shown in Figure~\ref{fig:P3_P1P1P1}.}
\label{p1p1p1brane}
\end{figure}

As a concrete illustration, in Figure~\ref{toricandbrane} we show the skeleton of the 3D polyhedron $\mc{P}_3$ associated to the toric fan of local $\mb{P}^1\times \mb{P}^1\times \mb{P}^1$. A $(0,0,1)$ 4-brane is dual to the edge pointing upwards, which is exactly the normal direction of a $(0,0,1)$-plane. By dualizing the whole $\mc{P}_3$ we arrive at the brane box diagram presented in Figure~\ref{p1p1p1brane}.

Similar to the dictionary between the properties of brane web and CY3 geometry provided in \cite{Kol:1998cf}, we summarize in Table~\ref{4branetoricduality} a dictionary between the properties of brane box and CY4 geometry.
\begin{table}[H]
\centering
 \begin{tabular}{||c |c ||} 
 \hline
 $(p,q,r)$ 4-brane box & toric CY4  \\ [0.5ex]
 \hline\hline

 brane box & moment map \\ 
 \hline
 $(p,q,r)$ 4-brane & locus of degenerating $(p,q,r)$-cycle of $T^3$ \\ \hline

finite cell & compact divisor \\ \hline

semi-infinite cell & non-compact divisor \\ \hline

 semi-infinite 4-brane & normal direction to compact divisor \\
 \hline

 Charge conservation & Calabi-Yau condition \\ \hline
 \hline
 \end{tabular}
 \caption{A dictionary between toric CY4 and $(p,q,r)$ 4-brane box.}
 \label{4branetoricduality}
\end{table}

The dictionary needs some explanation since we have been mostly discussing the duality between a brane box and a 3D polyhedron $\mc{P}_3$ associated to the toric fan of a CY4, rather than the CY4 itself. The first and the second lines in Table~\ref{4branetoricduality} are evident from the construction of a brane box diagram as intersecting planes in $\mb{R}^3$ with a $T^3$-fibration over it as discussed in section~\ref{sec:toricCY} (cf.~\cite{Kol:1998cf}). The third and the fourth lines in Table~\ref{4branetoricduality} are not hard to see from basics of toric geometry \cite{Bouchard:Lecture, Hori:2003ic, cox2011toric}. Another way to see the third line is to note that even though the $T^3$ fiber does not degenerate inside a finite cell, the subvariety $D$ which is the $T^3$-fibration over that finite cell is indeed a locus of degeneration of a 1-cycle $S \simeq T^4/T^3$ of the toric CY4 (cf.~\cite{Kol:1998cf}). This $S$ is fibering over an $\mathbb{R}_+$ such that they together form topologically a cone and the codimension-1 subvariety $D$ lives at the tip of that cone. Therefore $D$ must be a compact divisor of the CY4 viewed as a cone such that its compact part lives at the tip.

The last line is very important in the dictionary. In geometry the charge conservation, i.e. the static condition (\ref{eq:static_condition}) translates to a simple consistency condition that a $\left( \sum_i p_i, \sum_i q_i, \sum_i r_i \right)$-cycle must vanish on the locus where the the loci of vanishing $(p_i, q_i, r_i)$-cycles meet. We denote by $D$ the compact part of the local geometry. Recall that Calabi-Yau condition means that the normal bundle to $D$ is the canonical bundle of $D$, i.e. $N_D = K_D$. This then implies that the $T^3$-action on $N_D$ must act the same way as it acts on $K_D$. Therefore the vanishing $(p,q,r)$-cycle in $T^3$ near $N_D$ must match that near $K_D$. Now consider an intersection between a semi-infinite 4-brane and a set $S_f$ of finite 4-branes such that local geometry is described by the restriction of $K_D$ to the loci of branes in $S_f$. The degenerating $(p,q,r)$-cycle on $K_D$ in this local patch is then given by $\left(\sum_{i\in S_f} p_i, \sum_{i\in S_f} q_i, \sum_{i\in S_f} r_i\right)$. By charge conservation this is exactly the vanishing cycle on the semi-infinite 4-brane intersecting branes in $S_f$. As suggested by the next to last line in Table~\ref{4branetoricduality} this semi-infinite 4-brane corresponds to $N_D$ restricted to the same local patch. Therefore we have $N_D = K_D$, hence $c_1(X_4) = 0$ in each local patch in the same manner as in~\cite{Leung:1997tw, Kol:1998cf}.

\subsubsection{Realization of 4-branes in terms of M5-branes}\label{sec:m5origin}

The $(p,q,r)$ 4-branes with the configuration shown in Table \ref{4braneconfiguration} can be dualized to a system of M5-branes. This is similar to the dualization of the $(p,q)$ 5-branes to M5-branes \cite{Kol:1997fv,Brandhuber:1997ua, Witten:1997sc, Benini:2009gi}. We consider the configuration of three M5-branes each wrapping one $S^1$ of $S^{1}_{8}\times S^{1}_{9}\times S^{1}_{10}$ as shown in Table \ref{m5branesto4branes}. It is easy to show that this configuration preserves $1/8$ of 32 supercharges \cite{deRoo:1997gq,Bergshoeff:1996rn}.

To show the equivalence of the M5-brane configuration in Table~\ref{m5branesto4branes} and the 4-brane configuration in Table~\ref{4braneconfiguration}, we first reduce along $S^{1}_{10}$ which results leads to (D4,NS5,NS5) in IIA. We then T-dualize along $S^{1}_{8}$, which leads to (D5,NS5,KK6B) in IIB which is indeed the configuration shown in Table~\ref{d5ns5kk6b}. A subsequent $T^2$-reduction on $S^1_8\times S^1_9$ then reproduces the 4-brane configuration in Table~\ref{4braneconfiguration}.

In summary, we see that there are two equivalent M-theory origins of the 8D 4-brane configuration shown in Table~\ref{4braneconfiguration}: one in terms of the M-theory KK-monopoles shown in Table~\ref{originalKK7Ms}, the other in terms of the M5-branes shown in Table \ref{m5branesto4branes}. For us the former description was discussed in more detail since it is relatively more directly related to the duality between the toric CY4 and the brane box. We expect the latter plays an important role for other purposes, e.g. constructing an SW-like geometry as in \cite{Witten:1997sc, Benini:2009gi}, which we will not discuss in this work.

\begin{table}[H]
\begin{center}
	\begin{tabular}{c|c|c|c||c|c||c|c|c||c|c|c}
		& 0 & 1 & 2 & 3 & 4 & 5 & 6 & 7 & $S^{1}_{8}$ & $S^{1}_{9}$ & $S^{1}_{10}$\\
		\hline
		M5$^{(1,0,0)}$ & $\checkmark$ & $\checkmark$ & $\checkmark$ & $\bullet$  &$\bullet$ & $\bullet$& $\checkmark$& $\checkmark$  & $\bullet$ &$\bullet$ & $\checkmark$  \\
		\hline
		M5$^{(0,1,0)}$ & $\checkmark$ & $\checkmark$ & $\checkmark$ & $\bullet$ &$\bullet$ & $\checkmark$  & $\bullet$ & $\checkmark$  & $\checkmark$ &$\bullet$ & $\bullet$\\ 
		\hline
		M5$^{(0,0,1)}$ & $\checkmark$ & $\checkmark$ & $\checkmark$  & $\bullet$ & $\bullet$ & $\checkmark$  & $\checkmark$ & $\bullet$  &$\bullet$  & $\checkmark$ & $\bullet$ \\ 
	\end{tabular}
	\caption{The M5-brane configuration dual to the 4-brane configuration in Table~\ref{4braneconfiguration}.}
	\label{m5branesto4branes}
\end{center}
\end{table}

\subsection{Coulomb branch physics from 4-branes box}\label{3dphysicsfrombrane}

In this section we discuss the Coulomb branch (CB) physics of a 3D $\mc{N}=2$ theory associated to a 4-brane box. The rank of CB, the effective gauge coupling of a gauge theory description at various limits of the brane box diagram, and the charges of the electrically charged states can all be read off from the brane box without referring to the dual toric geometry. The results from the brane box perspective are compared and matched with those from a dual toric geometric perspective using the toric-brane-box dictionary developed in section~\ref{sec:pqr4branes}. In section~\ref{sec:CBgauge}, we will also exemplify our method via a concrete example, which is the brane box dual to the local $\mb{P}^1\times \mb{P}^1\times \mb{P}^1$ discussed in section~\ref{secP1cube}.

In section~\ref{sec:effective_strings}, we will briefly discuss the effective strings in a 3D $\mc{N}=2$ theory from the brane box perspective. In particular we will identify the origin of these effective strings in string/M-theory.

\subsubsection{Coulomb branch, non-abelian enhancement and SCFT limit}\label{sec:CBgauge}

M-theory compactified on a smooth toric CY4 yields leads to a 3D $\N=2$ $U(1)^r$ gauge theory in the UV, where $r$ is the rank of CB and we have seen that such theory can be constructed by a 4-brane box where all compact cells have non-vanishing volume in $\mathbb{R}^3_{567}$. Given the duality between the 4-brane box and dual toric geometry discussed in section~\ref{sec:toricCY}, it is clear that $r$ must be equal to the number of finite cells in a 4-brane box diagram.

From the perspective of the brane box without referring to the toric-brane-box duality, the rank of CB can be calculated via counting the number of local deformations of the brane box. Here by a local deformation we mean a movement of a finite 4-brane along its normal direction in $\mathbb{R}^3_{567}$. It is clear that each internal (finite) facet can move along its normal direction therefore provides one real degree of freedom. Each (internal) edge is an intersection of at most three facets hence it provides one constraint. However, the constraints themselves are not independent since a vertex is at the intersection of four edges. Therefore each vertex represents one redundant constraint. A final constraint comes from the fact that the brane box is connected. Denote by $\#C$, $\#F$, $\#E$ and $\#V$ the number of finite cells, finite faces, finite edges and vertices of a brane box respectively, from the above discussion we have:
\begin{equation}\label{eq:rankCB}
    \text{rank(CB)} = \#(\text{local deformations}) = \#F - \#E + \#V - 1.
\end{equation}

It is not hard to see that the above result matches what we have obtained from the toric-brane-box duality due to the following fact:
\begin{equation}\label{eq:rankCB_euler}
    \#V-\#E+\#F-\#C = 1.
\end{equation}
The above equation is readily recognized as Euler's formula applied to the compact part of the brane box, which can be viewed as a polygonization of a solid ball whose Euler characteristic is 1.

Having read off the rank of CB from the brane box, we go on to study the $U(1)$ gauge coupling of the theory on the CB using a method similar to that in \cite{Aharony:1997bh}. We will focus on the effective gauge coupling in the UV where the theory is semi-classical and admits a Lagrangian description as follows \cite{deBoer:1997kr,Strassler:2003qg}:
\begin{equation}\label{eq:U(1)_Lagrangian}
    S \supset \frac{1}{g^2} \int d^3x\ \left( \frac{1}{2} \partial_\mu \phi \partial^\mu \phi -\frac{1}{4} F_{\mu\nu}F^{\mu\nu} \right)
\end{equation}
where $\phi$ is the scalar component of the vector multiplet and without loss of generality we focus on one particular $U(1)$. Given the above Lagrangian it is clear that the calculation of the effective gauge coupling amounts to a calculation of the metric of CB, i.e. $1/g^2 = g_{11}^{CB}$. Note that here since we have focused on a particular $U(1)$ the metric becomes that of a real 1D moduli space.

\begin{figure}[H]
\centering
\includegraphics[width=0.35\textwidth]{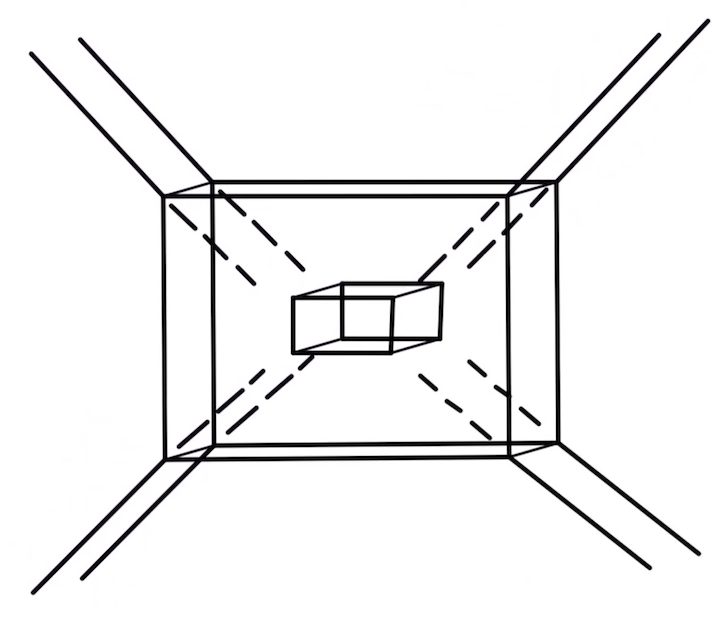}\qquad \qquad \qquad
\includegraphics[width=0.36\textwidth]{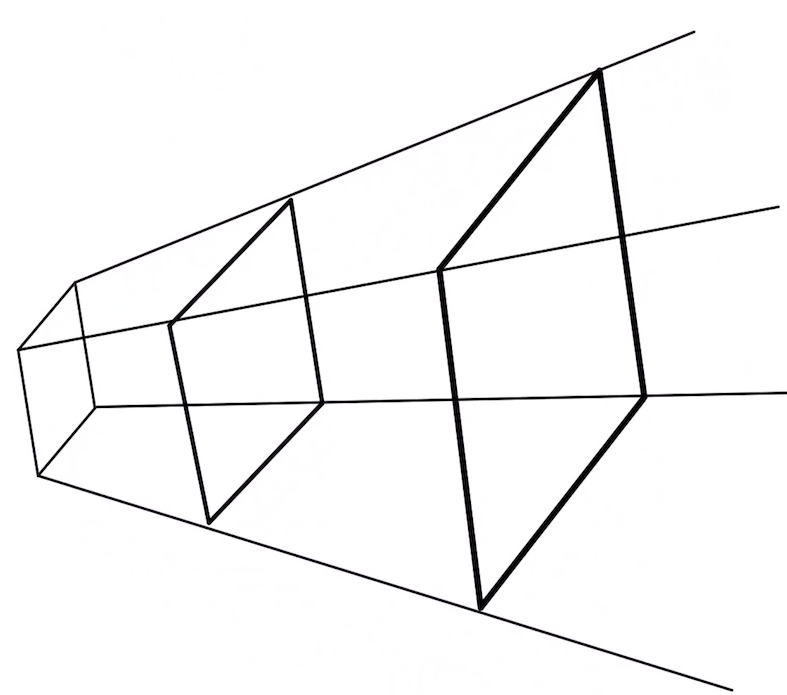}
\caption{The figure illustrates the local deformation of the brane box. The left-hand side shows the local deformation on the brane box dual to the local geometry over $\P^{1}\times \P^{1}\times \P^{1}$. The right-hand side shows the behaviour of each finite 4-brane, involved in the brane box, under the deformation.}
\label{braneboxdeform}
\end{figure}

From the brane box perspective, it is clear that the a change in the effective gauge coupling of the $U(1)$ associated to a compact cell $C$ must get contribution from the movements of all the finite 4-branes that are facets of $C$, i.e. local deformation. One way to see this is that moving one such facet along its normal direction inevitably causes movements of other facets along their normal directions if all the other finite and semi-infinite 4-branes of the brane box that are not facets of $C$ stay still (see Figure.~\ref{braneboxdeform} for an illustration of such a movement). It is also clear that the positions of the branes along its normal direction must be governed by the CB parameter $\phi$ hence their movements governed by a change of $\phi$. Therefore the movements of all the facets of $C$ must be taken into consideration if there is a change in $\phi$.

Now consider an infinitesimal change in the CB parameter $\phi \rightarrow \phi + \delta\phi$. Given the argument in the above paragraph, each facet of $C$ must be shifted by an amount proportional to $\delta\phi$ at lowest order. We will denote by $\delta_i \delta\phi$ the shift of the $i^{\text{th}}$ facet of $C$ along its normal direction. If the mass of the $i^{\text{th}}$ facet is $m_i$, the contribution of its shift along its normal direction to the change of kinetic energy will be:
\begin{equation}
    \delta_i KE = \frac{1}{2} m_{i} (\delta_i\delta\phi)^2.
\end{equation}
Clearly $m_i$ is proportional to the area $A_i$ of the $i^{\text{th}}$ facet of $C$, i.e. $m_{i} = z_i A_i$. To avoid clustering symbols we will simply absorb $z_i$ into the $\delta_i$, hence the above equation becomes:
\begin{equation}
    \delta_i KE = \frac{1}{2} A_i (\delta_i\delta\phi)^2.
\end{equation}
Now looking at (\ref{eq:U(1)_Lagrangian}) we see that at the lowest order we have:
\begin{equation}
    \delta KE = \sum_i \delta_i KE \propto \frac{1}{g^2} (\delta\phi)^2
\end{equation}
where the summation is taken over all the facets of $C$. Therefore we have:
\begin{equation}\label{eq:U1_coupling_brane}
    \frac{1}{g^2} = \sum_i \frac{1}{2} A_i(\delta_i)^2.
\end{equation}
Here again we have absorbed the proportionality constant into a redefinition of $\delta_i$'s. Note that this result is similar to that in \cite{Aharony:1997bh} in the context of 5-brane webs. This is a remnant of the fact that the real scalar in either 3D or 5D can be viewed as the compact component of a reduction of a vector in one higher dimension on a circle, hence the coefficient of the kinetic energy of that real scalar must be the same as the $U(1)$ coupling of the gauge kinetic term.

The expression of effective gauge coupling in (\ref{eq:U1_coupling_brane}) is also reassuring our calculation in (\ref{gauge-coupling}) from the geometric perspective. Recall that for one $U(1)$ we have:
\begin{equation}
    \frac{1}{g^2} = -\text{Vol}(D\cdot D) = -\text{Vol}(K_D).
\end{equation}
Therefore in the toric case which is dual to a brane box construction the calculation of gauge coupling becomes calculating the sum of the volumes of all the toric divisors of $D$. Based on the toric geometry/brane box duality discussed in section~\ref{sec:toricCY}, $D$ is dual to a compact cell $C$ while each toric divisor of $D$ is dual to a facet of $C$. Therefore the two perspectives are matched as:
\begin{equation}
    \frac{1}{2} \sum_i A_i (\delta_i)^2 = \frac{1}{g^2} = \sum_i \text{Vol}(S_i)
\end{equation}
where the summation on the LHS is taken over all the facets of $C$ while the summation on the RHS is taken over all the toric divisors $S_i\subset D$.

In the brane constructions for field theories, the massive states in the Coulomb branch are given by strings, string junctions and various branes stretching between different branes \cite{Aharony:1997bh}. The masses of the particles arising from strings or string junctions are directly proportional to the lengths of these strings and their spacetime quantum properties are determined by the details of their intersections with the branes.

For us the most important physical data of a string, besides its mass, is its electric charge under the Cartan $U(1)$ gauge group on CB. The Cartan $U(1)$ associated to a finite cell $C$ of the brane box must arise as a combination of all $U(1)$'s on the world volume of the finite 4-branes bounding $C$. A similar situation was studied in the context of 5-brane webs in \cite{Aharony:1997bh}. In brane box this is a consequence of the bounding finite 4-branes not being free to move along their normal directions independently as discussed earlier. As the CB parameter reflects the positions of the finite 4-branes in $\mathbb{R}^3_{567}$, it then must be a combination of the position parameters of all the finite 4-branes bounding $C$.

This fact reflects itself in determining the 3D electric charge of string denoted by $Q_{e}$. The calculation of the electric charge of a particle associated to a string or a string junction amounts to counting its endpoints on the finite 4-branes of the brane box. More precisely we have:
\begin{equation}\label{eq:Qe_string}
    Q_{e} = - N_{b},
\end{equation}
where $N_{b}$ is the number of endpoints of the string on the finite 4-branes. Note that the normalization in the above equation differs from the one presented in \cite{Aharony:1997bh}.

\begin{figure}[H]
\centering
\includegraphics[width=0.2\textwidth]{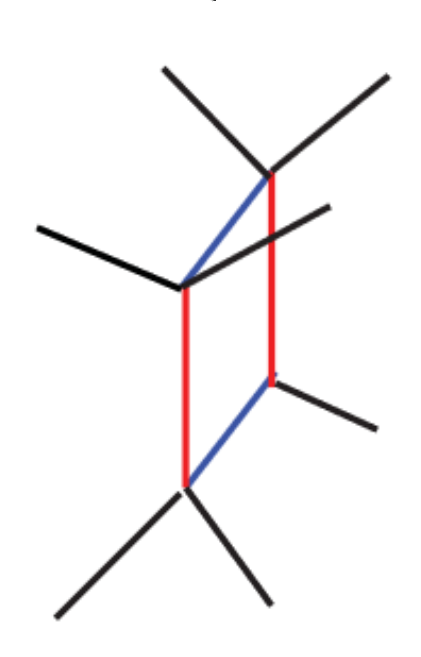}\qquad \qquad \qquad
\includegraphics[width=0.23\textwidth]{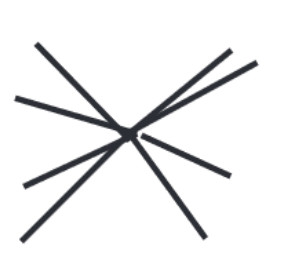}
\caption{On the left-hand side, the brane box when the two $(1,0,0)$ 4-branes coincide, which depicts the $SU(2)$ gauge theory limit. On the right-hand side, the SCFT fixed point where no scale is present in the theory.}
\label{squeezed_P1P1P1}
\end{figure}

\paragraph{Non-Abelian and SCFT limits} It is interesting to look at various limits of the brane box which are dual to tuning the K\"ahler parameters of the dual toric variety. In particular we are interested in the limit where a (non-abelian) gauge theory description is valid. For simplicity let us focus on one finite cell $C$ of the brane box bounded by two parallel $(1,0,0)$ 4-branes, two parallel $(0,1,0)$ 4-branes and two parallel $(0,0,1)$ 4-branes. In this case due to the hierarchy $\text{Vol}(B) \gg \text{Vol}(\mb{P}^1)$ discussed in section~\ref{sec:geometric-limits}, there will be a light W-boson from M2-brane wrapping $\mb{P}^1$ with small $\text{Vol}(\mb{P}^1)$. In our brane box picture this W-boson, as we will see shortly, is given by a string stretching between the $(1,0,0)$ 4-branes whose mass is proportional to the its length $l$, i.e. the distance between the two parallel $(1,0,0)$ 4-branes in $\mathbb{R}^3_{567}$. Such small $l$ will greatly suppress the area of branes that are parallel to $(1,0,0)$-direction while leaving the area of $(1,0,0)$ 4-branes unsuppressed. Therefore, according to (\ref{eq:U1_coupling_brane}) the inverse coupling $1/g^2$ will be dominated by the area of the $(1,0,0)$ 4-branes of $C$. Approximately we have:
\begin{equation}
    \frac{1}{g^2} \simeq KA_{(1,0,0)}
\end{equation}
where $K = \left(\delta_{(1,0,0)}\right)^2$. One can then squeeze down the $(1,0,0)$ direction to arrive at:
\begin{equation}
    \frac{1}{g_0^2} = K\widetilde{A}_{(1,0,0)}
\end{equation}
where $\widetilde{A}_{(1,0,0)}$ is the area of the $(1,0,0)$ 4-brane at the $l = 0$ limit where the theory enhances to an $SU(2)$ gauge theory. The resulting configuration is illustrated on the left of Figure~\ref{squeezed_P1P1P1}. The whole process is under well-control as long as the theory remains weakly-coupled, i.e. $A_{(1,0,0)}$ remains large. According to the discussion in section~\ref{sec:CB-limit} and~\ref{sec:nonab-limit}, we must have:
\begin{equation}
    A_{(1,0,0)} \propto \text{Vol}(B).
\end{equation}
This is indeed the case as can be seen from the ruling structure of the divisor $D$ dual to $C$, using the method discussed in section~\ref{sec:toric-CY4}.

Having arrived at a non-abelian enhancement point, one may ask whether the brane box can be further collapsed down to a singular limit corresponding to an SCFT as illustrated on the right of Figure~\ref{squeezed_P1P1P1}. Obviously such collapsing cannot be done via any local deformations, i.e. the movements of the finite $(1,0,0)$ 4-branes along $(1,0,0)$ direction. Rather one has to move the semi-infinite 4-branes along their normal directions to achieve the singular limit. In particular one has to tune the strings stretching between the squeezed $(0,1,0)$ 4-branes (the red lines on the left of Figure~\ref{squeezed_P1P1P1} and the squeezed $(0,0,1)$ 4-branes (the blue lines on the left of Figure~\ref{squeezed_P1P1P1} massless, which corresponds to tuning the mass parameters of the gauge theory.

We see that this also matches what we learn from the dual toric geometry picture. Since the fibral $\mb{P}^1$ has been shrunk at $l = 0$, the remaining electrically charged massive states must be from M2-brane wrapping curves in $B$, which are dual to the colored lines on the left of Figure~\ref{squeezed_P1P1P1} due to the toric-brane-box duality introduced in section~\ref{sec:toricCY}. As we shrink $B$ to achieve the strongly coupled limit $1/g^2\sim \text{Vol}(B) \rightarrow 0$, the curves in $B$ will be shrunk along with $B$ hence the masses of various M2-brane wrapping modes are tuned as well. Therefore we see that the singular limit can be achieved via local and global deformations of the brane box.

Next let us look at a concrete example to see how the CB physics is read off a brane box.

\paragraph{The Local $\P^{1}\times \P^{1}\times \P^{1}$ example} We analyze a concrete brane box example drawn in Figure \ref{p1p1p1brane}, which is dual to local $\P^{1}\times \P^{1}\times \P^{1}$. It is immediate to see that the brane configuration in Figure~\ref{p1p1p1brane} has $\#C=1$, $\#F=6$, $\#E=12$, and $\#V=8$ which match (\ref{eq:rankCB}) and~\ref{eq:rankCB_euler}. It also matches our analysis from geometry in section~\ref{secP1cube} that the rank of CB of this theory is one.

The strings stretching between the two parallel $(1,0,0)$ 4-branes, the two parallel $(0,1,0)$ 4-branes and the two parallel $(0,0,1)$ 4-branes correspond to charged massive BPS states. The masses of these particles in the weakly coupled region are given by:
\begin{equation}\label{masstension}
    m_{1} \sim l_{5}\,T_{(1,0,0)}, \qquad  m_{2} \sim l_{6}\,T_{(0,1,0)}, \qquad  m_{3} \sim l_{7}\,T_{(0,0,1)}.
\end{equation}
Here, $l_{a},\,a=5,6,7$ indicates the finite length between the pairs 4-branes that bound the central finite cell of the brane box and $T_{(p,q,r)}$ is the $(p,q,r)$-string tension. As have been discussed previously, if $A_{(1,0,0)} \gg A_{(0,1,0)}$ and $A_{(1,0,0)} \gg A_{(0,0,1)}$ as $l_5 \ll 1$ it is natural to identify the string stretching between the two parallel $(1,0,0)$ 4-branes with the massive W-boson of the broken $SU(2)$ gauge theory charged under the Cartan $U(1)$ on CB. Hence $m_1$ is identified with the mass of the W-boson. The VEV of the scalar component $\sigma$ of the CB $U(1)$ vector multiplet is proportional to $m_{W}$, i.e. $\expval{\sigma} = m_1/g$. We emphasize that there is no canonical choice of which $(p,q,r)$-string should be identified with the W-boson. The choice we have made above is nevertheless a convenient one. The $U(1)$ gauge theory is then enhanced to $SU(2)$ in the limit $\expval{\sigma} \sim m_1 = 0$. Equivalently we take the limit $l_5 \rightarrow 0$. In this limit, the two $(1,0,0)$ 4-branes coincide as drawn on the left of Figure~\ref{squeezed_P1P1P1}.

Applying (\ref{eq:U1_coupling_brane}) to this example, we have
\begin{equation}\label{CBcouplingbrane}
    \frac{1}{g^{2}} \ \sim (a l_{5}l_{6} + b l_{6}l_{7} + c l_{5}l_{7})
\end{equation}
for some positive proportionality constants $a$, $b$ and $c$. This matches (\ref{P1P1P1-coupling}) from geometry. If we had required $l_5 \ll 1$ hence naturally identified the $(1,0,0)$-string with the W-boson, we would have $1/g^2 \sim l_6l_7$ dominated by the area of the finite $(1,0,0)$ 4-brane.

Note that though we have claimed that certain strings give rise to the W-bosons that are electrically charged under the Cartan $U(1)$, we have not yet honestly determined their electric charges. To determine the electric charges of the stringy states we apply (\ref{eq:Qe_string}). It is easy to see that the states in (\ref{masstension}) all possess the same electric charge $(-2)$. We see that this result matches those listed in Table~\ref{t:BPSchargeP1cube}. Note that a W-boson is indeed a charge $-2$ (or $+2$ with a reversed orientation) particle under the Cartan $U(1)$ and our identification of the mass $m_1$ particle with the W-boson does match this fact.

We then need to look at the $-2$ charged states associated with the remaining two massive strings whose masses are $m_2$ and $m_3$ in the limit $m_1 \ll m_2$ and $m_1 \ll m_3$. In the weakly coupled limit, the theory can be viewed as a familiar $(p,q)$ 5-brane web living in $\mb{R}^2_{67}$ giving rise to a 5D $\mathcal{N}=1$ field theory living on the D5-brane worldvolume spanning in the 67-direction. In the $l_5\rightarrow 0$ limit, the state of mass $m_{2}$ which corresponds to the $(0,1,0)$-string coming from the D1-string stretching between the NS5's in IIB, is dissolved into the $S^{1}_{r}$-wrapped D5-brane. This then must be a dyonic instanton of the 5D $SU(2)_{0}$ theory living on the D5-brane world volume \cite{Aharony:1997bh,Douglas:1995bn,Witten:1995gx,Douglas:1996uz}. Therefore in the 3D $\N=2$ $SU(2)$ gauge theory this $(0,1,0)$-string corresponds to the dimensional reduction of the 5D dyonic instanton.

To identify the state associated with the mass $m_{3}$, we choose another duality frame in which the $(0,1,0)$ 4-brane corresponds to the $S^{1}_{10}$-wrapped D5-brane. This can be achieved from the KK7M configuration (see Table \ref{originalKK7Ms}) by reducing along $S^{1}_{9}$ then T-dualizing along $S^{1}_{8}$. Upon this chain of dualizations, the original (D5,NS5,KK6B) type IIB system becomes (KK6B,D5,NS5). In this frame, the $(0,1,0)$-string corresponds to the gauge theory W-boson, while the $(0,0,1)$-string corresponds to a dissolved D1-brane in the D5 worldvolume. The discussion in the previous paragraph can then be applied in this situation without modification, and the $(0,0,1)$-string must be a dyonic instanton of the 5D $SU(2)_{0}$ gauge theory in this new duality frame.

In summary, the states in (\ref{masstension}) all have electric charge $(-2)$ under the Cartan $U(1)$, though depending on a specific choice of a type IIB frame only one of them can naturally be identified as the light W-boson in a gauge theory description. Consequently, at an $SU(2)$ enhancement point, only one such state becomes massless hence part of the $SU(2)$ vector multiplet, while the remaining two states should be identified as disorder operators. The above discussion from the brane box perspective agrees with that in section \ref{secP1cube} from a geometric perspective.

The strong coupling limit of the 3D theory, which occurs at the origin of the UV CB, can be achieved by taking $\text{Vol}(C)\rightarrow 0$, or equivalently:
\begin{equation}
    l_5 \rightarrow 0,\ l_6 \rightarrow 0,\ l_7\rightarrow 0.
\end{equation}
Such a limit can of course be achieved in various ways. In particular, it can be achieved by shrinking $l_6$ and $l_7$ starting from a gauge theory phase where $l_5 \ll l_{6,7}$ in the UV. In this process, the finite face bounded by the coloured edges in the LHS of Figure~\ref{squeezed_P1P1P1} shrinks to zero size as illustrated in the RHS of the same figure. It is clear that the $\text{Vol}(C)\rightarrow 0$ limit there are no scale parameters of the brane box hence we expect that the end-point brane box of the limiting process gives rise to a conformal theory. It is also obvious that the end-point brane box shown on the right of Figure \ref{fig:3pictures1} corresponds to the singular geometric limit of the corresponding toric CY4 discussed in section~\ref{sec:geometric-limits}.

To flow from the UV picture described above to deep IR we take the opposite limit where the volume of the central finite cell goes to infinity, i.e. $\text{Vol}(C)\rightarrow \infty$. In this limit, the inverse of the coupling constant of the abelian gauge theory diverges:
\begin{equation}
    \frac{1}{g^{2}} \sim \sum_{i} A_{i}  \rightarrow \infty
\end{equation}
where $A_i$ is the area of the $i^\text{th}$ 4-brane bounding $C$. Hence, the effective theory in the deep IR is weakly coupled. Furthermore, the masses of the electrically charged BPS states become infinite as can be seen from (\ref{masstension}). Hence these states should be integrated out in a Wilsonian fashion in order to have an IR effective description.

\subsubsection{3D $\N=2$ effective strings}\label{sec:effective_strings}

In addition to the charged particles, in 3D $\N=2$ theories there are also massive effective strings whose existence can be inferred from the 3d $\mc{N}=2$ superalgebra \cite{Ferrara:1997tx}:
\begin{equation}
  \begin{split}
        &\{Q_{\alpha}^{a},Q_{\beta}^{b}\} = (\gamma^{\mu}C)_{\alpha\beta}\,P_{\mu}\,\delta^{ab}\,+\,(\gamma^{\mu}C)_{\alpha\beta}\,Z_{\mu}^{(ab)} + C_{\alpha\beta}\,Z^{[ab]},
  \end{split}
\end{equation}
with the constraint $Z_{\mu_{1}\,\cdots\mu_{p}}^{ab}\delta_{ab}=0$ for $a,b = 1,2$. Here the appearance of $Z_{\mu}^{ab}$ indicates the existence of two types of half-BPS effective strings in 3D.

The M-theory origins of these 3D effective strings are M5-brane and exotic $5^{3}$ brane in the configuration shown in Table~\ref{m553tostrings}. It is easy to check that these two M-theory branes give rise to the same supersymmetry projection operator therefore they can co-exist as a supersymmetric bound-state \cite{Kimura:2016xzd}.

From Table~\ref{m553tostrings} we see that these M-theory branes span all of $\mathbb{R}^3_{567}$ while are transverse to 89\#-direction, hence they must be singlets under the $SL(3,\Z)$ subgroup of the full U-duality group. Such transformation property remains unchanged after they are dualized to 5-branes in IIB on $T^2$. Hence, to distinguish between these 5-branes and the 5-branes that will be discussed in the next section which transform non-trivially under $SL(3,\mathbb{Z})$, we refer to such branes as singlet $(p,q,r)$ 5-branes in IIB on $T^2$. The duality chain from M-theory on $T^3$ to IIB on $T^2$ is as follows:
\begin{equation}\label{M553tononpqr5branes}
    \begin{split}
        &\text{M5} \xrightarrow[]{\text{KK on 10}} \text{NS5}  \xrightarrow[]{\text{T along 9}} \text{KK6B} \xrightarrow[]{\text{Reduce along 89}}\text{$(p,q,r)$-singlet 5-brane}
        \\
        &5^{3} \, \xrightarrow[]{\text{KK on 10}}\,5_{2}^{2} \xrightarrow[]{\text{T along 9}} \,\,\text{KK6B} \xrightarrow[]{\text{Reduce along 89}} \text{$(p,q,r)$-singlet 5-brane}.
    \end{split}
\end{equation}

There are two notable facts about these $(p,q,r)$-singlet 5-branes we want to mention before the end of this section. The first is that though they transform trivially under $SL(3,\mathbb{Z})$, they transform non-trivially as a doublet under the $SL(2,\Z)$ subgroup of the full U-duality group \cite{Bergshoeff:2011se, deBoer:2012ma}. The second is that the tension of these 5-branes are proportional to $\text{Vol}(C)$ since the corresponding singlet 5-branes wrap the whole compact cell $C$ of the brane box in $\mathbb{R}^3_{567}$. Note that they also wrap the non-compact 3 direction therefore they actually become infinitely heavy stringy defects in 3D.

\begin{table}[H]
\begin{center}
	\begin{tabular}{c|c|c|c||c|c||c|c|c||c|c|c}
		& 0 & 1 & 2 & 3 & 4 & 5 & 6 & 7 & $S^{1}_{8}$ & $S^{1}_{9}$ & $S^{1}_{10}$\\
		\hline
 	M$_{5}$ & $\checkmark$ & $\checkmark$ &   $\bullet$ & $\checkmark$ & $\bullet$ & $\checkmark$ & $\checkmark$ & $\checkmark$ &  $\bullet$ & $\bullet$ & $\bullet$ \\ 
		\hline
 	$5^{3}$ & $\checkmark$ & $\checkmark$ &   $\bullet$ & $\checkmark$ & $\bullet$ & $\checkmark$ & $\checkmark$ & $\checkmark$ &  $\bullet$ & $\bullet$ & $\bullet$ \\
	\end{tabular}
	\caption{The M-theory exotic branes that lead to effective 3D strings. }
	\label{m553tostrings}
\end{center}
\end{table}

\subsection{Codimension two branes}\label{seccodim2branes}

Motivated by the role played by 7-branes in a 5-brane web, we consider codimension-2 branes in brane box construction. The codimension-2 branes will be 5-branes in 8D. By the construction of the 8D theory as IIB on $T^2$, it is natural to expect that these codimension-2 branes are certain compactifications of various branes in IIB. We find that this is indeed the case. More precisely we find that the IIB origins of these codimension-2 branes in 8D can be 7-branes, 5-branes, KK6B and exotic branes, which will be discussed in greater detail in this section. The 8D 5-branes transform in various representations of the U-duality group $SL(3,\Z) \times SL(2,\Z)$. These 8D 5-branes play a similar role as IIB 7-branes play in an ordinary 5-brane web. 

In this section, we discuss the M-theory origin and the dual IIB description of these 8D 5-branes. Moreover, we will study the relation between the 8D 5-branes and the $(p,q,r)$ strings and 4-branes studied in the previous sections. We show that there exists a brane creation/annihilation process which is related to the movements of the branes in the brane box similar to Hanany-Witten moves~\cite{Hanany:1996ie}.

\subsubsection{Adding KK-monopoles}\label{sec:additional_KK7M}

We shall require that the introduction of these codimension-2 branes to the original brane box system does not further break supersymmetry. This requirement can be fulfilled by examining the products of the SUSY projection operators $P_{(i)}$ in (\ref{eq:kk7mconstraines}). By doing so, the resulting constraints will provide insights into the types of additional extended objects that can be introduced to our $(p,q,r)$ 4-brane system. Importantly, this approach guarantees that the new system preserves the same amount of supercharges, given that the new constraints are constructed from the original ones. Thus we have,     
    \begin{alignat}{2}\label{PtimesP}
       P_{(\#)(9)} &= P_{(\#)}\,P_{(9)} &&: \quad \varepsilon = - \Gamma^{569\#}\, \varepsilon,
        \notag\\
        P_{(9)(8)} &= P_{(9)}\,P_{(8)} &&: \quad \varepsilon = - \Gamma^{6789}\,\varepsilon,
        \\
          P_{(\#)(8)} &= P_{(\#)}\,P_{(8)} &&: \quad \varepsilon = - \Gamma^{578\#}\,\varepsilon,
        \notag\\
       P &= P_{(\#)}P_{(9)}P_{(8)} &&: \quad \varepsilon = - \Gamma^{012}\,\varepsilon. \notag
    \end{alignat}
The interpretation of these new conditions is as follows: each condition in the first three lines above results in two Taub-NUT solutions, offering flexibility in choosing the TN cycle. For instance, the Taub-NUT space associated with $\Gamma^{569\#}$ can have its TN circle along the $10^{\text{th}}$ or the $9^{\text{th}}$ direction. Equivalently, this results in a pair of dual KK7M monopoles occupying the same worldvolume directions, with each being associated with one choice of the TN circle. The additional KK7M branes are detailed in Table~\ref{addkk7m12}.

Note that the set ${P_{(i)(j)}}$, with the new index $(i)(j)$ runs over $\{(\#)(9),(9)(8),(\#)(8)\}$, is closed under multiplication:
\begin{equation}\label{closeproduct}
P_{(i)(j)} \cdot P_{(j)(k)} = P_{(i)(k)}.
\end{equation}
The order of $i$ and $j$ in the new index $(i)(j)$ is unimportant, given that the SUSY projection operators commute, i.e., $P_{(\#)(9)}=P_{(9)(\#)}$. The above equation implies that only two pairs of the new KK7M are independent. In the following, we will refer to the new KK7M monopoles by their constraints, i.e. we take $P_{(i)(j)}$ to represent the corresponding pair of the new KK7M-monopoles. The advantage of this notation is that it encodes the corresponding TN cycles associated with each pair. Specifically, $P_{(i)(j)}$ denotes the $(p,q,r)$ charges of each pair, where $(i)(j)\in \{(\#)(9),(9)(8),(\#)(8)\}$.  In this context, the first entry in $(i)(j)$ signifies the charge of the first brane in the pair, while the second entry represents that of the second brane. For instance, $P_{(\#)(9)}$ indicates the pair of $(1,0,0)$-KK7M and $(0,1,0)$-KK7M sharing the same worldvolume direction, respectively.

(\ref{closeproduct}) also implies that the codimension-two brane system, when considered in isolation of any 8D 4-branes, the system preserves 8 real supercharges out of the 32 supercharges of the 8D type II supergravity.

Regarding the interpretation of the above equation, the last line is interpreted as an M2-brane with a worldvolume along the $(x^{0},x^{1},x^{2})$ directions. Further discussion of this additional M2-brane will not be pursued.

At this point, we can demonstrate that the M2-branes appearing in Table \ref{originalKK7Ms} do not break additional supercharges. The approach involves multiplying the $P_{(i)(j)}$ constraints by $(-\Gamma^{0}\Gamma^{0})$. Focusing solely on the Dirac matrices in the constraints, we obtain
\begin{equation}
\begin{split}
         - \Gamma^{569\#} &\rightarrow \Gamma^{05\#}\ \Gamma^{069},
         \\
         - \Gamma^{569\#} &\rightarrow \Gamma^{05\#}\ \Gamma^{078},
              \\
         - \Gamma^{569\#} &\rightarrow \Gamma^{069}\ \Gamma^{078}
\end{split}
\end{equation}
Each of the 3-form Dirac matrices above is associated with the worldvolume of the M2-branes. In particular, our set of M2-branes are associated with the Dirac matrices: $\{\Gamma^{05\#}$, $\Gamma^{069}$, $\Gamma^{078}\}$. This reflects their spacetime location and coincides with that of Table \ref{originalKK7Ms}. Once again, since these conditions are constructed from the original constraints in (\ref{eq:kk7mconstraines}), the presence of these M2-branes does not break additional supercharges. The $(p,q,r)$ charges that labelling the M2-branes of the same table then inherited from the $P_{(i)(j)}$ constrains. For instance, the first line of the above equation indicates the existence of M2$^{(1,0,0)}$ and M2$^{(0,1,0)}$, respectively.

\begin{table}[H]
\begin{center}
	\begin{tabular}{c|c|c|c||c|c||c|c|c||c|c|c}
		& 0 & 1 & 2 & 3 & 4 & 5 & 6 & 7 & $S^{1}_{8}$ & $S^{1}_{9}$ & $S^{1}_{10}$ \\
		\hline
        \hline
 KK7M & $\checkmark$ & $\checkmark$ & $\checkmark$ & $\checkmark$ &$\checkmark$ & $\bullet$  & $\bullet$ & $\checkmark$  & $\checkmark$ & $\bullet$& TN\\ 
		\hline
KK7M & $\checkmark$ & $\checkmark$ & $\checkmark$ & $\checkmark$ &$\checkmark$& $\bullet$ & $\bullet$ & $\checkmark$   & $\checkmark$ &  TN & $\bullet$ \\ 
        \hline
        \hline
KK7M & $\checkmark$ & $\checkmark$ & $\checkmark$ & $\checkmark$ &$\checkmark$ & $\checkmark$  & $\bullet$ & $\bullet$  & $\bullet$ & TN& $\checkmark$ \\ 
		\hline
KK7M & $\checkmark$ & $\checkmark$ & $\checkmark$ & $\checkmark$ & $\checkmark$ & $\checkmark$ & $\bullet$& $\bullet$    & TN & $\bullet$& $\checkmark$\\ 
        \hline
        \hline
KK7M & $\checkmark$ & $\checkmark$ & $\checkmark$ & $\checkmark$ &$\checkmark$& $\bullet$  & $\checkmark$ & $\bullet$  & $\bullet$ & $\checkmark$& TN \\ 
		\hline
KK7M & $\checkmark$ & $\checkmark$ & $\checkmark$ & $\checkmark$ &$\checkmark$& $\bullet$  & $\checkmark$ & $\bullet$  & TN & $\checkmark$& $\bullet$ \\
	\end{tabular}
	\caption{The additional KK7M monopoles. The first two lines are due to $P_{(\#)(9)}$, the second two lines are due to $P_{(9)(8)}$, and the last two lines are due to $P_{(\#)(8)}$. We refer to each pair by their constrain $P_{(i)(j)}$.}
	\label{addkk7m12}
\end{center}
\end{table}

\subsubsection{Type IIB description and exotic branes}

In the following analysis, we will examine the dual type IIB description of the new KK7M monopoles, denoted by $P_{(i)(j)}$. The initial step involves following the rule in section \ref{KK7MtoIIA} to obtain the type IIA description. We perform a reduction along the M-theory circle, chosen to be along the $10^{\text{th}}$ direction as before. After the reduction, each pair $P_{(i)(j)}$ results in:
\begin{equation}
    \begin{split}
        P_{(\#)(9)} &\xrightarrow{\text{reduction on $S^{1}_{10}$}} \text{(D6, KK7A)},
        \\
         P_{(9)(8)} &\xrightarrow{\text{reduction on $S^{1}_{10}$}} \text{(KK6A, KK6A)},
         \\
          P_{(\#)(8)} &\xrightarrow{\text{reduction on $S^{1}_{10}$}} \text{(D6, KK7A)}.
    \end{split}
\end{equation}
The KK7A brane arises through an M-theory reduction of KK7M that is transverse to its worldvolume and the associated TN cycle. To understand the nature of such a brane, we will consider its mass formula following the notation of \cite{ortin2004gravity}. Let us wrap the KK7M on a $T^{6}$ torus along its worldvolume directions except the time direction, and perform the reduction \cite{Elitzur:1997zn,ortin2004gravity,Eyras:1999at}: 
\begin{equation}\label{kk7mtokk7a}
    M_{\text{KK7M}} = \frac{R_{1}\cdots R_{6}\,(R_{TN})^{2}}{(l_{p}^{(11)})^{9}} \ \rightarrow\   \frac{R_{1}\cdots R_{6}\,(R_{TN})^{2}}{g^{3}_{s}(l_{s})^{9}} = M_{\text{KK7A}}.
\end{equation}
In the reduction above we have used the fact that $l_{p}^{(11)}=g_{s}^{1/3}l_{s}$. Here, $R_{1}\cdots R_{6}$ represent the radii along the $T^{6}$ torus, while $R_{TN}$ denotes the TN cycle associated with the KK7M brane. In addition, $l_{p}^{(11)}$ denotes the 11D Planck scale, while $g_{s}$ and $l_{s}$ are the string coupling and the string length, respectively. From the mass formula of the KK7A above, one concludes that the KK7A is the exotic $6_{3}^{1}$ brane \cite{deBoer:2010ud,deBoer:2012ma}. Very briefly, the mass formula of a generic brane $b_{n}^{c}$, after wrapping its spatial directions along $T^{b}$ torus, can be written as \cite{deBoer:2010ud,deBoer:2012ma,Obers:1998fb,Eyras:1999at},
\begin{equation}
    M_{b_{n}^{c}} = \frac{R_{i_{1}}\cdots R_{i_{b}}\ (R_{j_{1}}\cdots R_{j_{c}})^{2}}{g_{s}^{n}\,l_{s}^{b+2c+1}}.
\end{equation}
Here, $\{R_{i_{1}},\cdots, R_{i_{b}}\}$ represent the radii along the $T^{b}$ torus, while $\{R_{j_{1}},\cdots, R_{j_{c}}\}$ corresponds to the radii along the transverse cycles to the brane's worldvolume. In this notation, e.g. an F1-string is denoted by $1_0$, a D$p$-brane is denoted by $p_1$, a NS5-brane in either IIA or IIB is denoted by $5_2$, KK6A or KK6B denoted as $5_{2}^{1}$ etc. There are more generic exotic branes which are irrelevant to our discussion. Hence, exotic branes refer to $b_{n}^{c}$ with either non-trivial $c$ and/or $n\geq 3$.

The second step in this analysis is to T-dualize to the type IIB frame. Let us take the T-duality circle along the $9^{\text{th}}$ direction. Then by following the rules presented in section \ref{KK6AtoIIB}, we arrive at 
\begin{equation}\label{add5BIIB}
    \begin{split}
        P_{(\#)(9)} &\xrightarrow{\text{reduction on $S^{1}_{10}$}} \text{(D6, KK7A)} \xrightarrow{\text{T-duality on $S^{1}_{9}$}} \text{(D7, NS7)} ,
        \\
         P_{(9)(8)} &\xrightarrow{\text{reduction on $S^{1}_{10}$}} \text{(KK6A, KK6A)} \xrightarrow{\text{T-duality on $S^{1}_{9}$}} \text{(NS5, $5_{2}^{2}$)} ,
         \\
          P_{(\#)(8)} &\xrightarrow{\text{reduction on $S^{1}_{10}$}} \text{(D6, KK7A)} \xrightarrow{\text{T-duality on $S^{1}_{9}$}} \text{(D5, $5_{3}^{2}$)} .
    \end{split}
\end{equation}
The exotic branes mentioned in section \ref{KK6AtoIIB} are the $5_{2}^{2}$ and $5_{3}^{2}$ branes. To arrive at this result, we check the mass formula for KK6A and KK7A under the mentioned T-duality, respectively. Recall, the $T$-duality transformation along $R$:
\begin{equation}
    R \rightarrow \frac{l_s^2}{R},\qquad g_s \rightarrow \frac{l_s}{R}\, g_s.
\end{equation}
Thus we have, 
\begin{equation}
    \begin{split}
        M_{\text{KK6A}} &= \frac{R_{1}\cdots R_{5}\,(R_{8})^{2}}{g_{s}^{2}\,l_{2}^{8}} \ \xrightarrow{\text{along}\,S^{1}_{9}}\  \frac{R_{1}\cdots R_{5}\,(R_{8}R_{9})^{2}}{g_{s}^{2}\,l_{2}^{10}} = M_{5_{2}^{2}}, 
        \\
        M_{\text{KK7A}} &= \frac{R_{1}\cdots R_{4}R_{6}R_{9}\,(R_{8})^{2}}{g_{s}^{3}\,l_{2}^{9}} \  \xrightarrow{\text{along}\,S^{1}_{9}}\  \frac{R_{1}\cdots R_{4}R_{6}\,(R_{8}R_{9})^{2}}{g_{s}^{3}\,l_{2}^{10}} = M_{5_{3}^{2}}.
    \end{split}
\end{equation}
Hence, we have justified (\ref{add5BIIB}). In addition, following the above discussion and that around (\ref{kk7mtokk7a}) one can verify the rules we had in section \ref{sec:Dual_IIB_description}.

\begin{figure}[H]
\begin{center}
\begin{tikzpicture}
   \newdimen\R
   \R=2.1cm
   \draw (0:\R) \foreach \x in {60,120,...,360} {  -- (\x:\R) };
   \foreach \x/\l/\p in
     { 60/{$5_{3}^{2}$}/above,
      120/{NS7}/above,
      180/{$5_{2}^{2}$}/left,
      240/{D5}/below,
      300/{D7}/below,
      360/{NS5}/right
     }
     \node[inner sep=1.5pt,circle,draw,fill,label={\p:\l}] at (\x:\R) {};
     \draw (0,-0.2) node[anchor=south] {$\bullet$};
      \draw (0.2,-0.2) node[anchor=south] {$\bullet$};
      \draw (5,-2.2) node[anchor=south] {$g_{s}^{-1}$};
      \draw (5,-0.2) node[anchor=south] {$g_{s}^{-2}$};
      \draw (5,1.72) node[anchor=south] {$g_{s}^{-3}$};
\end{tikzpicture}
\caption{The supersymmetric exotic branes transform in the adjoint representation of $SL(3,\Z)$; in particular, they correspond to the six non-zero weights. The D7 and the NS7 are understood to wrap the $T^{2} = \widetilde{S}^{1}_{9}\times S^{1}_{8}$ torus. The behavior of their tension according to the string coupling $g_{s}$ is shown explicitly in the diagram. The two zero weights correspond to non-supersymmetric branes \cite{Bergshoeff:2011qk,Bergshoeff:2011se,Kleinschmidt:2011vu}.}
\label{hexagonal}
\end{center}
\end{figure}
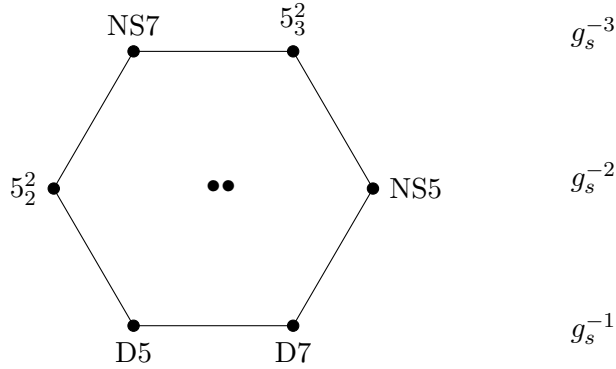

In the 8D limit, where we compactify the type IIB theory on $T^{2}=\widetilde{S}^{1}_{9}\times S^{1}_{8}$, we obtain six types of codimension-two branes, i.e. 8D 5-branes. These 8D 5-branes are the descendants of the three pairs that appear in (\ref{add5BIIB}). The worldvolume directions of the 8D 5-branes are given in Table \ref{table5branes}, which are named after their type IIB origin. The six types of 8D 5-branes fit into the $(\mbf{8},\mbf{1})$ representation of the $SL(3,\Z)\times SL(2,\Z)$ U-duality group \cite{Bergshoeff:2011qk,Bergshoeff:2011se,Kleinschmidt:2011vu}, which is illustrated in Figure \ref{hexagonal}.
Moreover, the codimension-two branes described in (\ref{add5BIIB}) along with that appear in (\ref{M553tononpqr5branes}) align with the classification presented in \cite{deBoer:2012ma}.

Note that, the exotic pairing presented in (\ref{add5BIIB}) is accomplished by associating each brane with its negative root. For instance, the NS5 and NS7 are aligned with the simple roots of the algebra. Consequently, there exist two distinct $SL(2,\Z)$ subgroups corresponding to these simple roots. Within each of these $SL(2,\Z)$ subgroups, the pairs (D7,NS7) and (NS5,$5{2}^{2}$) undergo transformations in the adjoint representation, specifically as $2 \subset \textbf{3}$. This observation explains why pairs like $(\text{D}{5},5^{2}_{3})$ are not considered independent, as inferred from (\ref{closeproduct}).

\begin{table}[H]
\begin{center}
	\begin{tabular}{c|c|c|c||c|c||c|c|c}
		& 0 & 1 & 2 & 3 & 4 & 5 & 6 & 7   \\
  \hline\hline 
		$\widetilde{S}^{1}_{9}\times S^{1}_{8}$-wrapped (D7,NS7)    & $\checkmark$ & $\checkmark$ & $\checkmark$  &$\checkmark$  & $\checkmark$ & $\bullet$    & $\bullet$ & $\checkmark$   \\
		\hline
		   (NS5,$5_{2}^{2}$)  & $\checkmark$ & $\checkmark$ & $\checkmark$   & $\checkmark$& $\checkmark$   & $\checkmark$     & $\bullet$ &$\bullet$  \\ 
		\hline
		(D5,$5_{3}^{2}$)  & $\checkmark$ & $\checkmark$ & $\checkmark$     & $\checkmark$ & $\checkmark$   & $\bullet$      & $\checkmark$  & $\bullet$  \\
	\end{tabular}
	\caption{The 8-dimensional codimension-two branes dual to the additional KK7M system given in Table \ref{addkk7m12}. }
	\label{table5branes}
\end{center}
\end{table}

\subsubsection{The 8D 5-branes}

The codimension-two branes should carry $(p,q,r)$ charges. One reason is that they are the descendent of the $P_{(i)(j)}$ KK7M-monopoles which carry $(p,q,r)$-charges through the $(i)(j)$ index as defined above. Hence, each type IIB pair that appears in (\ref{add5BIIB}) carries a $(p,q,r)$ charge according to $P_{(i)(j)}$. The first entry in $(i)(j)$ represents the charge of the first brane in the dual type IIB pair, while the second entry signifies that of the second brane in the dual type IIB pair. Hence certain codimension-two branes have the same $(p,q,r)$ charges which defines an equivalence class between these objects as 
\begin{equation}\label{pqr5_IIB}
    \begin{split}
        &(1,0,0)\ \text{5-brane: ($\widetilde{S}^{1}_{9}\times S^{1}_{8}$-wrapped D7, D5)}, \\
        &(0,1,0)\ \text{5-brane: ($\widetilde{S}^{1}_{9}\times S^{1}_{8}$-wrapped NS7, NS5)}, \\
        &(0,0,1)\ \text{5-brane: ($5_3^2, 5_2^2$)}.
    \end{split}
\end{equation}
The above pairs of 8D 5-branes appear as lines in the 3D $\mathbb{R}^{3}_{(567)}$ space. However, their $(p,q,r)$ charges are not enough to determine their location in the 3D space. In particular, those branes with the same $(p,q,r)$ charges are not in the same direction, as is evident from Table \ref{table5branes}. On general grounds, a line in 3D Euclidean space can be defined as the line at the intersection of two planes. For a $(p,q,r)$ 5-brane we can associate a plane defined as $\{P_{(p,q,r)}: px_5 + qx_6 + rx_7 = c\} \subset \mathbb{R}^3_{(567)}$. The other plane that is needed to locate a given 5-brane can be provided by the 5-brane's partner in the pairs appearing in (\ref{add5BIIB}). In the above dual frame, a $P_{(p,q,r)} \cap P_{(p',q',r')}$ determine the location of two 5-branes as:
\begin{equation}\label{5branesinplanes}
    \begin{split}
       P_{(1,0,0)} \cap P_{(0,1,0)} &: \text{$\widetilde{S}^{1}_{9}\times S^{1}_{8}$-wrapped (D7, NS7)},
       \\
       P_{(1,0,0)} \cap P_{(0,0,1)} &: \text{(D5, $5_{3}^{2}$)},
       \\
       P_{(0,1,0)} \cap P_{(0,0,1)} &: \text{(NS5, $5_{2}^{2}$)}.
    \end{split}
\end{equation}
Locating exactly the pairs we have in (\ref{add5BIIB}). This suggests that we can supplement the $(p,q,r)$ charge of a given 5-brane by the charge of its partner in the sense of (\ref{add5BIIB}) and (\ref{5branesinplanes}) as $(p,q,r)_{(p',q',r')}$. Meaning that a $(p,q,r)_{(p',q',r')}$ 5-brane is the brane with a $(p,q,r)$ charge, while the $(p',q',r')$ is the charge of its partner. For instance, a $(1,0,0)_{(0,0,1)}$ 5-brane refers to a D5 brane whose partner is the $5_{3}^{2}$ brane and it follows that these two brane live in the same direction.

Now we move to discuss the 8D 4-branes that can end on 5-brane. In the following, we will be referring to these branes by their type IIB origin up to a $T^{2}$ compactification. Let us focus on the sub-configuration involving the $(1,0,0)$ and $(0,1,0)$ 4-branes. In type IIB these are the D5$^{(1,0,0)}$ and NS5$^{(0,1,0)}$ branes, as illustrated in Table \ref{d5ns5kk6b}, corresponding to $(p,q)$ 5-brane as discussed earlier. Note the subscript on the D5 and the NS5 branes refer to their $(p,q,r)$ charges. To this sub-system, the additional 8D 5-branes are given by the first pair that appears in (\ref{add5BIIB}), the usual (D7,NS7), wrapping the compactification torus $\widetilde{S}^{1}_{9}\times S^{1}_{8}$, with $(p,q,r)$ charges that can be read from (\ref{pqr5_IIB}). Drawing from the generalized $(p,q)$ brane web, defined in situations where we have a $(p,q)$ 5-brane ends on a $(p,q)$ 7-brane, then it follows that the $(1,0,0)$ and $(0,1,0)$ 4-branes ends on the $\widetilde{S}^{1}_{9}\times S^{1}_{8}$-wrapped D7 and NS7, respectively. 

As discussed around (\ref{KK6AtoIIB}), we could have T-dualize along the $S^{1}_{8}$ cycle instead of $S^{1}_{9}$. In this dual frame, the type IIB branes that originate the $(1,0,0)$ and $(0,0,1)$ 4-branes are one again a different (D5$^{(1,0,0)}$,NS5$^{(0,0,1)}$) sub-system. In addition, in the chosen frame, the third pair in (\ref{add5BIIB}), becomes $S^{1}_{9}\times \widetilde{S}^{1}_{8}$-wrapped (D7,NS7) with $(1,0,0)$ and $(0,0,1)$ charges, respectively. Thus, we have new $(p,r)$ 5-branes that can end on new $(p,r)$ 7-branes in the usual sense. This implies that in the original frame, the $(1,0,0)$ and $(0,0,1)$ 4-branes ends on the D5 and $5_{3}^{2}$, respectively. 

On top of the above two different type IIB frames, there are four more different frames depending on the choice of the M-theory cycle and the T-dual cycle. In all such frames, the above logic holds in determining which 8D 4-brane ends on which 8D 5-brane. Thus, one concludes that
\begin{equation}
    \text{a} \   (p,q,r) \ \text{4-brane ends on a}\, (p,q,r) \ \text{5-brane},
\end{equation}
where the $(p,q,r)$ 5-branes are given in (\ref{pqr5_IIB}). Using the type IIB description in the (D5,NS5,KK6B) frame\footnote{To distinguish between different type IIB frames it is enough to refer to the ordered triplet of D5, NS5, and KK6B.}, then it follows 
\begin{equation}
    \begin{split}
       S^{1}_{8}-\text{wrapped}\,\text{D}5 \ &\ \text{ends on} \  \  \widetilde{S}^{1}_{9}\times S^{1}_{8}-\text{wrapped}\,\text{D7} \ \text{and} \ \text{D5},
       \\
       S^{1}_{8}-\text{wrapped}\,\text{NS}5 \ &\ \text{ends on}\   \  \widetilde{S}^{1}_{9}\times S^{1}_{8}-\text{wrapped}\,\text{NS7} \ \text{and} \ \text{NS5},
       \\
        \widetilde{S}^{1}_{9}-\text{wrapped}\,\text{KK6B}\ &\ \text{ends on}\  \  5^{2}_{2} \ \text{and} \ 5^{2}_{3}.
    \end{split}
\end{equation}

The strings that can end on the 8D 5-brane can be discussed in a similar way. In the usual $(p,q)$ 5-brane web \cite{Aharony:1997bh}, a $(p,q)$ string ends on a $(p,q)$ 5-brane and thus it ends on the associated $(p,q)$ 7-brane. Applying this rule to the logic above, i.e. to each frame where we can talk about a sub-configuration the resemble the generalized $(p,q)$ brane, then we conclude the following:
\begin{equation}
     \text{a $(p,q,r)$ string ends on a $(p,q,r)$ 5-brane}.
\end{equation}
In the type IIB (D5,NS5,KK6B) frame, we have
\begin{equation}
    \begin{split}
       \text{F}1 \ &\text{ends on} \ \  \widetilde{S}^{1}_{9}\times S^{1}_{8}-\text{wrapped}\,\text{D7} \ \text{and} \ \text{D5},
       \\
       \text{D}1 \ &\text{ends on} \ \  \widetilde{S}^{1}_{9}\times S^{1}_{8}-\text{wrapped}\,\text{NS7} \ \text{and} \ \text{NS5},
       \\
        \widetilde{S}^{1}_{9}\times S^{1}_{8}-\text{wrapped}\,\text{D3} \ &\text{ends on} \ \  5^{2}_{2} \ \text{and} \ 5^{2}_{3}.
    \end{split}
\end{equation}

\subsubsection{Hanany-Witten effect}

\begin{figure}[H]
\centering{
\includegraphics[scale=0.67]{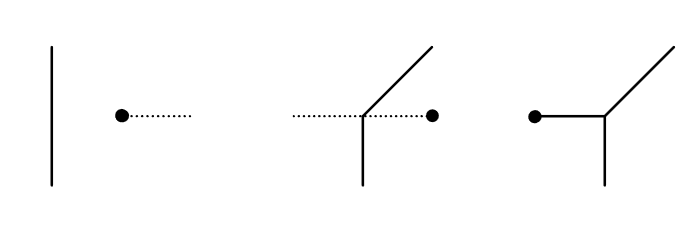}
}
\caption{The figure shows the 2D Hanany-Witten effect. From left to right one reads: (i) A codimension-two brane is depicted as a point in a plane, accompanied by a branch cut, alongside another brane represented as a line. (ii) Shifting the branch cut and enforcing charge conservation at the intersection point lead to a tilting of the branes based on their charges. (iii) Moving the codimension-two brane along its branch cut results in the creation of a new brane with the same charge as the codimension-two brane.}
\label{2Dhwmove}
\end{figure}

The Hanany-Witten (HW) transition is a phenomenon associated with the creation and annihilation of branes in string theory \cite{Hanany:1996ie}. Figure \ref{2Dhwmove} illustrates the HW transition involving a codimension-two brane and another brane represented as a line. For example, this scenario is encountered in generalized $(p,q)$ brane webs, where $(p,q)$ 5-branes terminate on $(p,q)$ 7-branes in type IIB theory. The HW transition serves as a crucial tool in brane construction, revealing various properties of the system.

The brane configuration of our interest is given in Table \ref{4braneconfiguration} and Table \ref{table5branes}, which involves a collection of $(p,q,r)$ 4-branes ending on $(p,q,r)$ 5-branes as described above. In such a system, we observe two different HW transitions involved: (i) when considering only the sub-system of $(p,q,r)$ 5-branes, and (ii)  concerning that between 4-branes and 5-branes. It should be noted that in the presence of 5-branes the axion fields are non-trivial and non-constant, so strings, 4-branes, and 5-branes can bend. This is the same feature observed in the generalized brane, when one includes $(p,q)$ 7-branes to $(p,q)$ 5-branes webs. 

Let us start with the first case concerning $(p,q,r)$ 5-branes as a sub-system. Since a full SL$(3,\Z)$ monodromy is missing in the literature, we shall consider a particular SL$(2,\Z)$ subgroup. Let us take the SL$(2,\Z)$ subgroup generated by the roots associated with the (D7,NS7) branes as in Figure \ref{hexagonal}. In the $(5,6)$ plane, these branes appear as points. Then according to charge $(p,q,r)$ assignment, the branes that constitute the $(p,q)$ brane web would be the (D5,NS5). This is nothing but the well-studied generalized $(p,q)$ brane web. In another example, we take the SL$(2,\Z)$ subgroup generated by the roots that correspond to the (NS5,$5_{2}^{2}$) branes as appear in Figure \ref{hexagonal}. In this case, the (NS5,$5_{2}^{2}$) branes appear as points in the $(6,7)$-plane of the space $\mathbb{R}^{3}_{(567)}$ as can be read from Table \ref{table5branes}. The (NS5,$5_{2}^{2}$) branes appear as a point and define $(q,r)$ codimension two 5-brane, and in particular, they would play the role of the usual $(p,q)$ 7-branes when considering the generalized $(p,q)$ 5-brane webs. This can be backed up by considering another IIB dual frame in which the (NS5,$5_{2}^{2}$) happens to be the favorite (D7,NS7) brane. In particular, we could achieve this by reducing M-theory along the $S^{1}_{q}$ in the $9^{\text{th}}$ direction and then dualize along $S^{1}_{r}$. This results in the (D7,NS7) pair with $(q,r)$ charges. Choosing the above $SL(2,\Z)$ subgroup, then our generalized $(q,r)$ web consists of the codimension-two (NS5,$5_{2}^{2}$) branes and the (NS7,$5_{3}^{2}$) branes. In the different IIB dual frame specified above, the (NS7,$5_{3}^{2}$) branes become the favoured (D5,NS5) with $q$ and $r$ charge, respectively. In either example, the standard HW transition is well-defined and depicted in Figure \ref{2Dhwmove}.

For the second case, looking at the sub-brane-system of the brane box obtained by forgetting the 7-direction, one gets an ordinary $(p,q)$ 5-brane on the 56-plane. In that case an extra 5-brane is created in the process of moving a 7-brane across a 5-brane when the branch cut generated by the 7-brane is not parallel to the 5-brane, i.e. when the charge of 7-brane is different from the 5-brane. Now recall that after all the brane web is a subsystem of the brane box, it is natural to expect that the rule governing the brane creation/annihilation in HW moves in brane web still holds in the brane box case. In other words we expect an extra 4-brane is created when a $(p,q,0)$ 5-brane moves across a $(p',q',0)$ 4-brane when $(p,q) \neq (p',q')$. More generally an extra 4-brane is created when a $(p,q,r)$ 5-brane moves across a $(p',q',r')$ 4-brane when $(p,q,r) \neq (p',q',r')$.

\begin{figure}[H]
\centering{
\includegraphics[scale=0.47]{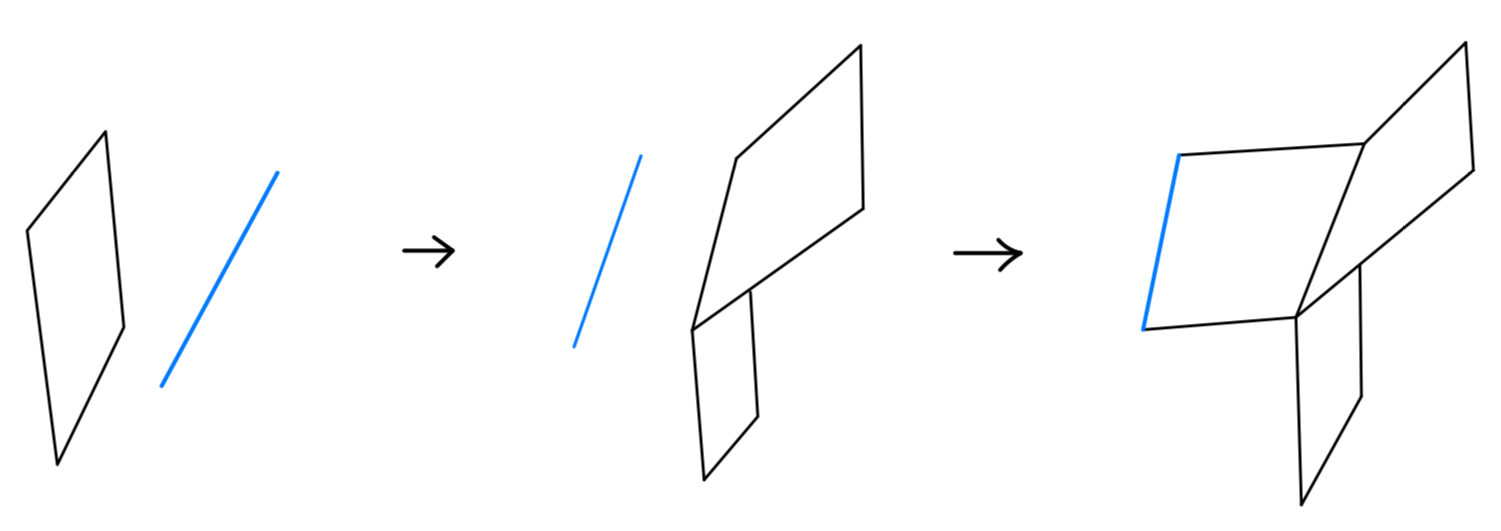}
}
\caption{The figure illustrates the 3D version of the Hanany-Witten effect. The blue line represents a 5-brane, whereas the "sheets" represent 4-branes.}
\label{3Dhwmove}
\end{figure}

As we have done above, we take an $SL(2,\Z)$ subgroup associated with a specific root fo the SL$(3,\Z)$. The choice determines the associated 8D 5-branes and we restrict ourselves to the plane in $\mathbb{R}^{3}_{(567)}$ where the 5-branes appear as points. Once again, we take the $T^{2}_{qr}$-wrapped (D7,NS7) codimension-two branes which are points in the $(5,6)$ plane. The difference between the current setup and the one considered above is that now we add to the chosen 8D 5-branes above 4-branes instead of other 5-branes. The 4-branes that we add should appear as lines in that plane, for the example above these are lines in the $(5,6)$ plane. Hence, we add $(1,0,0)$ and $(0,1,0)$ 4-brane. In the type IIB language these branes are $S^{1}_{r}$-wrapped D5 and NS5, respectively. Thus we have arrived, once again, at the flavored generalized $(p,q)$ brane web. This time however from 8D 5-branes and 4-branes instead of only 8D 5-branes. In this system, the usual HW move is well-defined. When picturing this sub-configuration in the $\mathbb{R}^{3}_{(567)}$ space rather than a plane, the HW is still valid. In particular, we have arrived at a notion of HW in the 3D space, referred to as 3D HW, between $(1,0,0)_{(0,1,0)}$ and $(0,1,0)_{(1,0,0)}$ 5-branes and $(1,0,0)$ and $(0,1,0)$ 4-branes. The notion of 3D HW is depicted in Figure \ref{3Dhwmove}. As we have learned so far, we can choose different type IIB dual frames such that we arrive at the favored generalized 5-brane web but with different charge assignments, i.e. it comes with a pair of charge out of the three $(p,q,r)$ charges. Hence, the notion of the 3D HW is defined to these sub-configurations as well. With the above argument, the 3D HW is well-defined between the following branes:
\begin{equation}
    \begin{split}
        &(1,0,0)_{(0,1,0)} \,\text{and}\, (0,1,0)_{(1,0,0)} \, \text{5-branes} \quad \text{with\quad $(1,0,0)$ and $(0,1,0)$ 4-branes,}
        \\
        &(1,0,0)_{(0,0,1)} \,\text{and}\, (0,0,1)_{(1,0,0)} \, \text{5-branes} \quad \text{with\quad $(1,0,0)$ and $(0,0,1)$ 4-branes,}
           \\
        &(0,1,0)_{(0,0,1)} \,\text{and}\, (0,0,1)_{(0,1,0)} \, \text{5-branes} \quad \text{with\quad $(0,1,0)$ and $(0,0,1)$ 4-branes.}
    \end{split}
\end{equation}

\subsection{Flavor symmetry from the brane box}\label{sec:flavsymbrane}

In this section we study the flavor symmetry of the field theory associated to a brane box. We will first determine the rank of the flavor symmetry. We then show how the charges of various stringy states are determined using only the data of the brane box. The determination of the charges will further enable us to study the enhanced flavor symmetry at the SCFT point from the brane box perspective. We will illustrate the method first via the concrete example local $\mathbb{P}^1\times \mathbb{P}^1\times \mathbb{P}^1$ then we go on to discuss more general cases.

\subsubsection{Flavor rank from brane box}

We first determine the rank of the flavor symmetry of the 3D theory associated to a brane box. Generally, analogous to the consideration of flavor symmetry of a 5-brane web \cite{Aharony:1997bh}, the rank of the global symmetry $G_{\text{global}}$ must be counted by the number of global deformations of a brane box. We denote by $F_X$ the number of semi-infinite faces. Naively there will be $F_X$ global deformations each of which corresponds to a deformation of one semi-infinite brane along its normal direction. Apparently, there are 3 redundant movements of the semi-infinite faces associated with overall translations of the brane box in $\mathbb{R}^3_{567}$ hence should not be counted as genuine global deformations. An external edge which is the intersection of three semi-infinite faces must remain pointing in the same direction therefore each semi-infinite edge provides a constraint for the global deformations. Not all the constraints are independent since all the semi-infinite edges, after all the finite cells are shrunk, must form a 3D skeleton so that when all but one of the external edges are determined the remaining one edge is already fixed without ambiguity due to charge conservation at each vertex. Therefore one of the constraints is redundant. Denote by $E_X$ the number of semi-infinite edges, we have:
\begin{equation}\label{rkglobal}
    rk_{F} := \text{rank}(G_{\text{global}}) = \#(\text{Global deformations}) = F_X - E_X - 2.
\end{equation}

For example, in the local $\mathbb{P}^1\times \mathbb{P}^1\times \mathbb{P}^1$ case (\ref{rkglobal}) gives $rk_{F} = 12 - 8 - 2 = 2$. As a slightly more complicated example, e.g., when $X_4$ is local $\mathbb{P}^1\times dP_2$ we have $rk_{F} = 15 - 10 - 2 = 3$.

Alternatively, due to the duality between the brane box and the dual toric diagram, the rank of the global symmetry group can also be calculated as:
\begin{equation}\label{rkglobal1}
    rk_{F} = C_{X} - 4,
\end{equation}
where $C_X$ is the number of semi-infinite cells in the brane box. Clearly, this is because each semi-infinite cell is dual to a non-compact divisor of $X_4$, i.e. an integral point living on an external facet of $\mc{P}_3$, and there are four linear equivalence relations among the non-compact divisors.

Having obtained two ways to calculate the rank of the global symmetry group, we must show that they are indeed equivalent. For this, it is clear that the consistency between (\ref{rkglobal}) from the brane-box perspective and (\ref{rkglobal1}) from the dual toric-geometric perspective requires:
\begin{equation}\label{eq:flavor_from_two_perspects}
    E_X - F_X + C_X = 2.
\end{equation}
To see this we will first shrink the brane box so the all the compact cells are squeezed to a point. This is a valid limiting process since we have assumed the existence of a singular limit (see section~\ref{sec:geometric-limits}. It is obvious that the numbers $E_X$, $F_X$ and $C_X$ remain unchanged in this process and all the edges meet at the origin of $\mb{R}^3$ in the singular limit. We then formally consider a one point compactification of $\mb{R}^3$ to $S^3$. Though the numbers $E_X$, $F_X$ and $C_X$ still remain unchanged after the one point compactification, all the edges meet at the extra point at infinity in $\mb{R}^3$. Therefore we have effectively achieved a polygonization of $\mathbb{S}^3$ in which there are two vertices where all of the edges meet. Hence we have:
\begin{equation}
    2 - E_X + F_X - C_X = \chi(S^3) = 0.
\end{equation}
Therefore (\ref{eq:flavor_from_two_perspects}) holds as a consequence of Euler's formula and the two perspectives indeed match.

\subsubsection{Charges of stringy states: local $\P^{1}\times\P^{1}\times\P^{1}$ example}

In this section we consider the charges of various stringy states in the local $\P^{1}\times\P^{1}\times\P^{1}$ example. The toric variety of local $\P^{1}\times\P^{1}\times\P^{1}$ as illustrated in Figure \ref{toricandbrane} can be seen as an interlacing of two local $\F_{0}$,
\begin{equation}
 \text{Local}\,( \ \mathbb{P}^{1} \times \P^{1}\times \P^{1} \ \cong \  \F_{0}|_{(xy;z=0)} \ \cap \ \F_{0}|_{(yz;x=0)}\,).
\end{equation}
The interlacing point of view on certain toric varieties is discussed in appendix \ref{app:interlacing}. To each of the interlaced varieties, there is an associated generalized $(a,b)$ brane web, with $(a,b)$ can be $(p,q)$, $(q,r)$, or $(p,r)$. This can be achieved by the following logic:\\
(i) Take the suitable type IIB frame that reflects the $(a,b)$ charge of interest. In other words, in such an IIB frame the system would be described by D5, NS5, and KK6B where the charges $(a,b)$ assigned for the sub-configuration of (D5,NS5) branes.\\
(ii) Decouple the KK6B monopoles and any bound state of the $(p,q,r)$ 4-branes in which there is a KK6B monopole. Thus one arrives at the usual (D5,NS5) brane web, with $(a,b)$ charges. The field theory on the worldvolume of the system is 5D $\N=1$ in such a limit. In the decoupling limit, the brane web is now in the 10D IIB supergravity rather than 8D type II supergravity. \\
(iii) To the above $(a,b)$ 5-brane web we can add (D7, NS7) branes, i.e. codimension two branes in the sense discussed in the previous section. Again, this can be achieved through a suitable IIB frame for the augmented four and five branes in the brane box followed by the decoupling procedure. At this point, we can apply the rules for obtaining the flavor symmetry through the codimension two branes, i.e. 7-branes, of the generalized $(a,b)$ brane web following \cite{DeWolfe:1998zf, DeWolfe:1998bi,DeWolfe:1999hj,Distler:2019eky}.  

\begin{figure}[H]
\centering{
\includegraphics[scale=0.45]{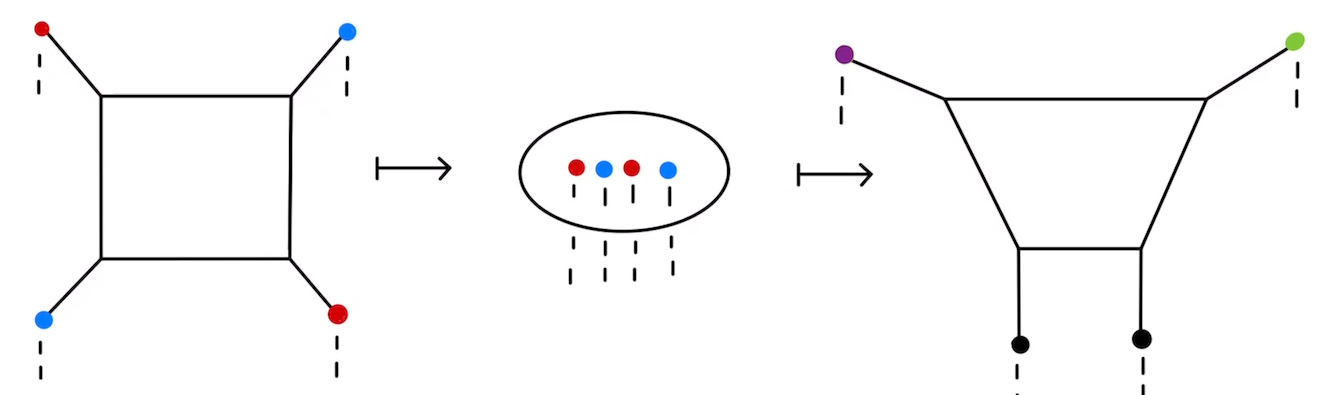}
}
\caption{The graph on the LHS is the brane web that corresponds to the local $\F_{0}$ geometry. The blue 7-branes are $(1,1)$ 7-branes and denoted by $C$, while the red ones are $(1,-1)$ 7-branes and denoted by $B$. The figure is subject to (\ref{eqf0tof2}) and the related discussion around it.}
\label{f0tof2}
\end{figure}

In this example, for each of the local $\F_{0}$ there is a generalized brane web as depicted in the LHS of Figure \ref{f0tof2}. The brane web of the first $\F_{0}$ can be obtained by decoupling the KK6B in the type IIB frame where we have (D5,NS5,KK6B). Thus we arrive at a $(p,q)$ brane web. The brane web of the second $\F_{0}$ can be taken to be in the duality frame of the (KK6B,D5,NS5) system. After decoupling the KK6B we arrive at a $(q,r)$ brane web.

In either case, the $(a,b)$ generalized brane web carries $\mk{su}(2)$ Lie algebra when considering the associated codimension-two brane \cite{DeWolfe:1999hj,Taki:2014pba}. The Lie algebra can be obtained through the procedure given in Figure \ref{f0tof2}. The first step, one makes use of the HW effect to pull all $(p,q)$ 7-branes inside the face of the 5-brane web. The second step, involve rearranging the 7-branes, which involve passing through a branch-cut and experience the associated monodromy, then applying an $SL(2,\Z)$ transformation \cite{Taki:2014pba}. Namely, we have
\begin{equation}\label{eqf0tof2}
    BCBC \, \rightarrow  \, X_{[-1,3]}BBC \, \xrightarrow{T} \, X_{[2,-1]}X_{[0,1]}X_{[0,1]}X_{[2,1]}. 
\end{equation}
Here, $(p,q)$ 7-branes are denoted by $X_{[p,q]}$, $B = X_{[1,-1]}$, $C=X_{[1,1]}$, and $T \in SL(2,\Z)$ which is given as
\begin{equation}
  T =\begin{pmatrix}
        1 & 0 \\
        1 & 1
    \end{pmatrix}.
\end{equation}
This implies that the global symmetry is realized through the semi-finite $(0,1)$ 5-branes (and 7-branes), with the flavor W-bosons arising from finite D1-strings stretched between these branes. Notably, the D1-string remains neutral under the $U(1)$ gauge group as it does not intersect any brane involved in the internal face. Consequently, the brane web dual to the local $\F_{0}$ inherently carries a well-defined flavor $\mathfrak{su}(2)$ Lie algebra. In the intermediate step of (\ref{eqf0tof2}), the $B = X_{[1,-1]}$ 7-branes generates the $\mk{su}(2)$ flavor algebra and the finite $(1,-1)$-string gives the flavor W-bosons.

In each duality frame, the same story above holds but with a different pair of the $(p,q,r)$ charges. When reconsidering the KK6B brane back to the system, i.e. examining the full 4-brane box, we conclude that the flavor algebra should be at least 
\begin{equation}\label{mingfp1cube}
    \mk{g}_{F} \ \simeq \ \mk{su}(2)\oplus \mk{u}(1).
\end{equation}
The reader may anticipate that the flavor algebra at this stage may be given as a direct sum of $\mk{su}(2)\oplus \mk{su}(2)$. After all, the two $\mk{su}(2)$ factors come from two pairs of the $B$ 5-branes with $(1,-1,0)$ and $(0,1,-1)$ charges. However, as the toric CY4 has an interlacing structure so do the two prescribed brane webs above. In other words, the charges associated with the $B$ 5-branes are interlaced as we will see below. This also implies that the two $\mk{su}(2)$ factors are intertwined. Hence, it is not necessarily a direct sum. In the following, we will show that the flavor symmetry gets enhanced to $\mk{su}(3)$ due to the non-trivial intertwining.

\paragraph{More on the local $\P^{1}\times\P^{1}\times\P^{1}$ example} 

The UV BPS states of this example, which are presented in (\ref{masstension}), carry electric charge under the $U(1)$ Coulomb gauge theory as we have seen in section \ref{sec:CBgauge}. In addition, they are charged under the $U(1)_{A}\times U(1)_{B}$ global symmetries of the configuration as we will demonstrate below. Let us take the following triplet to represent the charges of these abelian (gauge and flavor) symmetries,
\begin{equation}
    (Q_{e},\,Q_{A},\,Q_{B}).
\end{equation}
We assume that all of the above charges can be measured by the number of endpoints, boundaries which we denote as $N_{b}$, of the strings on certain 4-branes or/and 5-branes of the brane box. This is certainly true for the electric charge, as we have seen in section \ref{sec:CBgauge} and as considered in \cite{Aharony:1997bh}. To establish this measurement, we initially assign a charge $Q_{(p,q,r)}$ to each set of finite or semi-finite $(p,q,r)$ 4-branes (or 5-branes) present in the brane box. We demand that $Q_{(p,q,r)} = N_{b}^{(p,q,r)}/2$, where $N_{b}^{(p,q,r)}$ signifies the number of boundaries associated with a $(p,q,r)$ string, or a string junction, ending on a $(p,q,r)$ 4-brane (or 5-branes). In other words, $Q_{(p,q,r)}$ measures the number of end points of a $(p',q',r')$-string on a $(p,q,r)$ 4-brane. In the context of our example, the brane box features three distinct $Q_{(p,q,r)}$ charges, each associated with the specific type of $(p,q,r)$ 4-branes that are present. Namely we have $\{Q_{(1,0,0)},Q_{(0,1,0)},Q_{(0,0,1)}\}$. Consequently, the states described in (\ref{masstension}) carry a triplet of $Q_{(p,q,r)}$ charges:
\begin{equation}\label{P1cubestringQ}
\begin{array}{c|ccc}
& Q_{(1,0,0)} & Q_{(0,1,0)} & Q_{(0,0,1)}\\
\hline
T_{(1,0,0)} & 1 & 0 & 0\\
T_{(0,1,0)} & 0 & 1 & 0\\
T_{(0,0,1)} & 0 & 0 & 1 
\end{array}
\end{equation}
Here, $T_{(p,q,r)}$ refer to a $(p,q,r)$-string. The charges $(Q_{e},Q_{A},Q_{B})$ are given as linear combination of the above $Q_{(p,q,r)}$ charges. While the linear combination can be arbitrary, we take a specific choice that aligns with the prescribed duality mentioned earlier. More comments on this point would be present later. In this context, we choose the following:
\begin{equation}\label{QofP1cube}
    \begin{split}
     Q_{e} &= -2\,( Q_{(1,0,0)} + Q_{(0,1,0)} + Q_{(0,0,1)}),
        \\
     Q_{A} &=  Q_{(1,0,0)} - Q_{(0,1,0)},
          \\
     Q_{B} &=  Q_{(0,1,0)} - Q_{(0,0,1)}.
    \end{split}
\end{equation}
Once again, the above charges are seen as the generators of the corresponding $U(1)$ symmetries. In particular, $Q_{A}$ and $Q_{B}$ correspond to the flavor Cartan of the $\mk{g}_{F}$ algebra which is given above. The form of the $Q_{A}$ and $Q_{B}$ charges suggests that the two Cartans intertwined as discussed after (\ref{mingfp1cube}). In the following, we show that we indeed get an enhanced $\mk{su}(3)$ flavor algebra.

From the definition of the above charges, the finite $(p,q,r)$-strings within the brane box have the following charges 
\begin{equation}\label{p1cubechargedstates}
\begin{array}{c|ccc}
& Q_{e} & Q_{A} & Q_{B}\\
\hline
T_{(1,0,0)} & -2 & 1 & 0\\
T_{(0,1,0)} & -2 & -1 & 1\\
T_{(0,0,1)} & -2 & 0 & -1 
\end{array}
\end{equation}
Note that, (i) we arrive at the same result as in Table \ref{t:BPSchargeP1cube}. (ii) The choice we have made in determining the $Q_{A}$ and $Q_{B}$ charges reflects the choice we have to make in (\ref{P1P1P1-F12}) regarding the generators of flavor Cartan. In addition, we note that the above strings transform as a triplet of an $\mk{su}(3)$ rather than fundamentals of two districts $\mk{su}(2)$ Lie algebras. This implies that an enhancement of the above $\mk{g}_{F}$ may occur at the SCFT fixed point.

Let us explore the idea that the $Q_{A}$ and $Q_{B}$ charges can be seen as the charges associated with 5-branes, in the sense defined above. That we take,
\begin{equation}
\begin{split}
        Q_{A} &\xleftrightarrow{\text{associated with}} (1,-1,0)\, \text{5-branes},
        \\
        Q_{B} &\xleftrightarrow{\text{associated with}} (0,1,-1)\, \text{5-branes}.
\end{split}
\end{equation}
In particular, we achieve this identification when we pull these 5-branes to the interior region of the brane box. Note that in the appropriate decoupling limit, these 5-branes are nothing but the $B$ 7-branes with appropriate $(a,b)$ charges in the sense of (\ref{eqf0tof2}). In such a case, the finite strings, namely $(1,-1,0)$ and $(0,1,-1)$ strings, which are stretched between the two different $B$ 7-branes, would possibly, give rise to flavor W-bosons of the two $\mk{su}(2)$ algebras factor above. This can be achieved as these strings are neutral under the $U(1)$ gauge group. In particular, none of them touch any finite branes that constitute the brane box.

The above 5-branes and the associated strings can be seen as bound states of the ones corresponding to the $Q_{(p,q,r)}$ charges. In particular, we can write $(1,-1,0) = (1,0,0) \oplus (-1) (0,1,0)$, and $(0,1,-1) = (1,0,0) \oplus (-1)  (0,0,1)$ for the above 5-branes and strings ending on them. Let us apply the above definition of $Q_{e}$ on these bound states, denoted as $Q_{e}(p,q,r)$,
\begin{equation}
    Q_{e}(1,-1,0) = -2(1/2-1/2) =0, \qquad Q_{e}(0,1,-1) = -2(1/2-1/2) =0.
\end{equation}
Confirming that these strings are neutral under the gauge group. 
Moreover, we can apply the above $Q_{A}$ and $Q_{B}$ charges to the flavor W-boson as well, which results in 
\begin{equation}\label{su30fp1cube}
    \begin{split}
        &Q_{A}(1,-1,0) = 1+1=2,\qquad Q_{B}(1,-1,0)=-1,
        \\
         &Q_{A}(0,1,-1) = 0-1=-1,\qquad Q_{B}(0,1,-1)=1+1=2.
    \end{split}
\end{equation}
Forming the Cartan matrix of an $\mk{su}(3)$ algebra and agreeing with (\ref{su3cartan}). Furthermore, the charged states in (\ref{p1cubechargedstates}) form the fundamental, triplet, representation under the flavor algebra. We conclude that at the SCFT limit the flavor symmetry enhances to $\mk{su}(3)$. In this case, we observe that the two factors of the $\mk{su}(2)$ Lie algebra interlaced to for $\mk{su}(3)$ Lie algebra due to (\ref{su30fp1cube}). Put differently, the interlacing nature of our geometry results in the interlacing of the two 5D topological $\mathfrak{su}(2)$ algebras, giving rise to an enhanced $\mathfrak{su}(3)$ flavor algebra in the 3D $\mathcal{N}=2$ SCFT limit.


\subsubsection{Charges of stringy states: general consideration and more examples}

Let us describe the logic behind determining the non-abelian flavor symmetry for a class of toric CY4. The class is given as the local toric variety of local $\P^{1}\times\mathcal{S}$, where $\mathcal{S}$ as a weak-Fano complex surface such that there are at least two opposite non-compact divisors on a straight line. In particular, we take $\mathcal{S}$ to be one of the following set
\begin{equation}\label{localSeq}
    \{\F_{0},\F_{1},\F_{2},dP_{2},dP_{3},dP_{4},dP_{5},dP_{6},dP_{7}\}.
\end{equation}
The prescribed geometry can always be seen as an interlacing toric CY4, i.e.  
\begin{equation}
 \text{Local}\,( \ \mathbb{P}^{1} \times \mathcal{S} \ \cong \  \F_{0}|_{(xz;y=0)} \ \cap \ \mathcal{S}|_{(xy;z=0)}\,).
\end{equation}
Here, $(x,y,z)$ are axes in the 3-dimensional toric lattice $\Z^{3}$. The above geometry always has a dual 4-brane box description. From the brane box in CB, we can determine the flavor symmetry's rank through (\ref{rkglobal1}), or equivalently through (\ref{rkglobal}).

Due to the interlacing nature of the geometry, we apply the logic we followed in the $\P^{1}\times\P^{1}\times\P^{1}$ example. Namely, we consider a KK6B decoupling limit in two different type-IIB frames such that one arrives at two different $(a,b)$ brane webs. In each dual frame, the codimension two branes are the usual $(a,b)$ 7-branes, i.e. bound states of (D7,NS7). Then one reads the non-abelian flavor symmetry of each brane web independently through the 7-branes as explained in \cite{DeWolfe:1998zf, DeWolfe:1998bi,DeWolfe:1999hj,Distler:2019eky}. One brane web would be dual to the CY3 geometry of local $\F_{0}$, resulting in $\mk{su}(2)$ flavor symmetry. The other brane web is dual to that of local $\mathcal{S}$, resulting in $\mk{g}(\mathcal{S})$ flavor symmetry.

The flavor symmetry of the 4-brane box, when recovering the KK6B branes, is then seen as the intertwining between the two flavor symmetry factors above. The order of taking these $(a,b)$ brane webs matters to us, that we can take the brane web associated with the $\F_{0}$ factor first or that with $\mathcal{S}$. If we take that associated with the $\F_{0}$ factor first, then the flavor symmetry, at least, given as
\begin{equation}
  \mk{g}_{F}^{(1)} \ \simeq \ \mk{su}(2)\oplus \mk{u}(1)^{\text{rank}(\mathcal{S})}.
\end{equation}
However, if we reverse the order we arrive, at least, at
\begin{equation}
    \mk{g}_{F}^{(2)} \ \simeq \ \mk{u}(1)\oplus \mk{g}(\mathcal{S}).
\end{equation}
Having two ways of representing the flavor algebra reflects the fact that it comes from intertwined $\mk{su}(2)$ and $\mk{g}(\mathcal{S})$ algebras. Since after all, the first factor can be seen to come from the local $\F_{0}$ factor, while the other is due to the local $\mathcal{S}$. However, due to the lack of such a mathematical notion, we choose to write the flavor symmetry in two different representations as above. One should be able to work with either one which sometimes depends on the chosen example. According to \cite{DeWolfe:1998zf, DeWolfe:1998bi,DeWolfe:1999hj,Kimura:2016yqa,Distler:2019eky}, $\mk{g}(\mathcal{S})$ can be one of the following Lie algebras,
\begin{equation}
    \{\mk{su}(2),\mk{u}(1),\mk{su}(2), \mk{su}(2)\oplus \mk{u}(1) ,\mk{su}(3)\oplus \mk{su}(2), \mk{su}(5),\mk{so}(10), \mk{e}_{6}, \mk{e}_{7} \}.
\end{equation}
The order respecting that of the set in (\ref{localSeq}).

When examining the complete 4-brane box, the 7-branes become 5-branes with appropriate $(p,q,r)$ charges that are inherited from the decoupling limit. To illustrate, if we consider a $(p,q) = (0,1)$ 7-brane, it transforms into a $(0,1,0)$ 5-brane. Each of these $(p,q,r)$ 5-branes is assigned a $Q_{(p,q,r)}$ charge, denoted as $Q^{I}_{(p,q,r)}$ with $I=1,\cdots, rk_{F}$, following the approach outlined in the preceding subsection. These 5-branes may also be referred to as $(p,q,r)_{Q^{I}}$ 5-branes, and the associated strings ending on them as $(p,q,r)_{Q^{I}}$-strings.

As we have seen in (\ref{QofP1cube}), these $Q^{I}_{(p,q,r)}$ charges would be determined through linear combinations of (basis) $Q_{(p,q,r)}$ charges associated with finite $(p,q,r)$ 4-branes, and semi-finite 4-branes that ends on a sub-set of the above $(p,q,r)$ 5-branes. For example, if we have a $(1,0,0)$ 5-brane in the interior region of the brane box, then after pulling it outside it creates a $(1,0,0)$ 4-brane. Then the associated $Q_{(1,0,0)}^{(f)}$ charge should be one of the basis $Q_{(p,q,r)}$ charges. The superscript $(f)$ indicates that the charge is due to a semi-finite brane rather than a finite one in the brane box. Recall that, the $Q_{(p,q,r)}$ charges are measured through the number of boundaries of a given string, or string junction, on such 4-branes. In particular, we take $Q_{(p,q,r)} = N^{(p,q,r)}/2$.

The $(p,q,r)_{Q_{I}}$ 5-branes can be seen as the generator of the flavor symmetry once they are pulled in the interior region of the brane box. The $(p,q,r)_{Q_{I}}$-strings that end on them give rise to the flavor W-bosons. These strings should be neutral under all abelian gauge groups of the CB. Otherwise, the corresponding $(p,q,r)_{Q_{I}}$ 5-brane gives rise only to a $\mk{u}(1)$ factor in the flavor algebra. Furthermore, applying the $Q_{I}$ charges to the set of all $(p,q,r)_{Q_{I}}$-strings, i.e. flavor strings, as we have done in (\ref{su30fp1cube}), one arrives at the Cartan matrix of the, possibly enhanced, flavor symmetry. 

Let us work with the $\mk{g}^{(2)}_{F}$ given above. Then, in the case where flavor symmetry enhancement took place, the $\mk{u}(1)$ factor is upgraded to a nod in the Dynkin diagram and interlaces with the $\mk{g}(\mathcal{S})$ factor in an appropriate way that depends on specific examples. Following the above arguments, the flavor symmetry enhancement would agree with the geometrical side, which is summarized in Table \ref{t:dPN-GF}. Next, we present a few examples of the matter.

\begin{figure}[H]
\centering{
\includegraphics[scale=0.4]{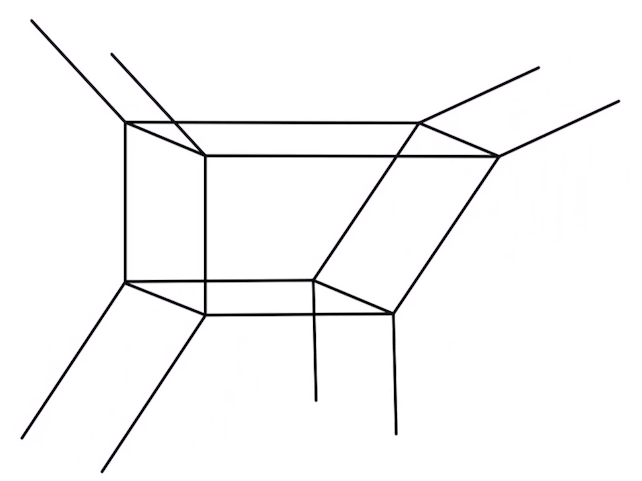}
}
\caption{The brane box dual to the local $\P^{1}\times \F_{1}$ toric CY4. Note, we have rotated our system such that the normal direction to the $(1,0,0)$ 4-brane points upward in the z-direction, i.e. the $(1,0,0)$ 4-branes are the top and the bottom facets of the finite cell.}
\label{P1F1branebox}
\end{figure}

\paragraph{The brane box dual to local $\P^{1}\times\F_{1}$:} The brane box for this example is illustrated in Figure \ref{P1F1branebox}. The dimension of the CB is one as there is only one compact cell. From the figure one also reads that there are two ways to arrive at $SU(2)$ gauge theory enhancement, i.e. two $SU(2)$ phases. In addition, we have three finite $(p,q,r)$ strings. Two of which are identical to the previous example, namely $T_{(1,0,0)}$ and $T_{(0,0,1)}$. In the appropriate type IIB dual and the gauge theory limit, these strings give rise to W-bosons and non-local order operators. The third string of charge $(0,1,0)$ connects one finite and one semi-finite $(0,1,0)$ 4-branes. In the KK6B-decoupling limit, the $(0,1,0)$-strings represent a D1-string dissolved in the D5-brane implying that we have a hypermultiplet (HM) that carries a 5D instanton charge \cite{Aharony:1997bh,Douglas:1995bn,Witten:1995gx,Douglas:1996uz}. Nonetheless, from the discussions of section~\ref{sec:M2modes}, the string states would correspond to M2-brane wrapping a curve $C=\mb{P}^1$ with normal bundle $N_{C|X_4}=\mc{O}\oplus\mc{O}(-1)\oplus\mc{O}(-1)$. In absence of $G_4$ flux, there is no BPS state from this string.

The flavor symmetry can, initially, be read from the interlacing geometry of this example, 
\begin{equation}
 \text{Local}\,( \ \mathbb{P}^{1} \times \F_{1} \ \cong \  \F_{0}|_{(xz;y=0)} \ \cap \ \F_{1}|_{(xy;z=0)}\,),
\end{equation}
which would be $\mk{g}_{F} = \mk{su}(2)\oplus\mk{u}(1)$. The rank of the flavor symmetry equals two, as we had in the previous example, which we take to be $U(1)_{A}\times U(1)_{B}$ at the CB. The $Q_{(p,q,r)}$ charges associated with the finite $(p,q,r)$ 4-branes are
\begin{equation}
    \{Q_{(1,0,0)}, Q_{(0,1,0)}, Q_{(1,1,0)}, Q_{(0,0,1)}\}.
\end{equation}
In the following, we limit ourselves to $(p,q,r)$ strings rather than string junction. It follows that none of these strings is charged under $Q_{(1,1,0)}$, so it will be dropped in the following discussion. We could write $(Q_{e},Q_{A},Q_{B})$ in terms of the above charges as the following, 
\begin{equation}
    \begin{split}
     Q_{e} &= -2\,( Q_{(1,0,0)} + Q_{(0,1,0)} + Q_{(0,0,1)}),
        \\
     Q_{A} &=  Q_{(1,0,0)} - Q_{(0,0,1)},
          \\
     Q_{B} &= 2Q_{(0,1,0)}.
    \end{split}
\end{equation}
Applying these charges on the finite $(p,q,r)$ string results in 
\begin{equation}
\begin{array}{c|ccc}
& Q_{e} & Q_{A} & Q_{B}\\
\hline
T_{(1,0,0)} & -2 & 1 & 0\\
T_{(0,1,0)} & -1 & 0 & 1\\
T_{(0,0,1)} & -2 & -1 & 0 
\end{array}
\end{equation}
We note that the states associated with $T_{(1,0,0)}$ and $T_{(0,0,1)}$ strings transform as $\textbf{2}_{0}$ under $\mk{su}(2)\oplus\mk{u}(1)$ algebra. Where as the states correspond to the $T_{0,1,0}$ string transform in $\textbf{1}_{1}$.

The bound-states of 5-brane (as well as the flavor W-bosons strings) that correspond to the $Q_{A}$ and $Q_{B}$ charges can be identified as $(1,0,-1)$ and $(0,1,0)$ 5-branes (strings), respectively. After pulling the 5-branes into the interior region of the brane box, we apply the $Q_{e}$ charge on these strings to get:
\begin{equation}
    Q_{e}(1,0,-1) =0, \qquad Q_{e}(0,1,0) = -1,
\end{equation}
which indicates that the flavor $(1,0,-1)$-string is neutral under the $U(1)$ gauge group. Hence, $U(1)_{A}$ can be seen as the Cartan of an $SU(2)$ flavor symmetry. However, the would-be $(0,0,1)$ flavor string is charged under the CB gauge group. This implies that $U(1)_{B}$ can not enhance to a flavor $SU(2)$ symmetry. All in all, the flavor symmetry algebra associated with this example is $\mk{su}(2)\oplus\mk{u}(1)$ with no further enhancement. Another check can be performed by applying the $Q_{A}, Q_{B}$ charges on the flavor strings, which confirm the aforementioned result.

\begin{figure}[H]
\centering{
\includegraphics[scale=0.38]{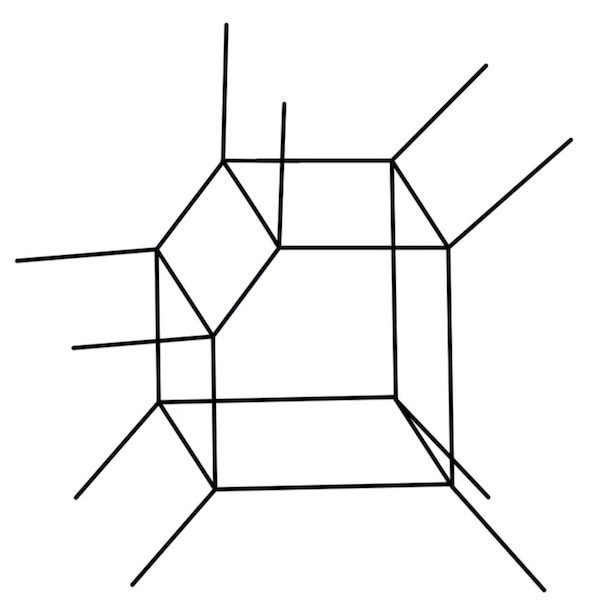}
}
\caption{The brane box dual to the local $\P^{1}\times dP_{2}$ toric CY4. Note that, we made a convention where the $(1,0,0)$ 4-branes are the top and the bottom facets of the finite cell.}
\label{p1dp2branebox}
\end{figure}

\paragraph{The local $\P^{1}\times dP_{2}$ example:} In this case the interlaced geometry is described as, 
\begin{equation}
 \text{Local}\,( \ \mathbb{P}^{1} \times \mathcal{S} \ \cong \  \F_{0}|_{(xz;y=0)} \ \cap \ dP_{2}|_{(xy;z=0)}\,).
\end{equation}
The bran box diagram is illustrated in Figure \ref{p1dp2branebox}. The flavor algebra reads from the brane web of each CY3 factor are $\mk{su}(2)$ and $\mk{su}(2)\oplus \mk{u}(1)$, respectively. Again, the flavor symmetry is seen as intertwining between these two factors. The number of finite strings within the 4-brane box is (5). In addition, there is one "strip" along the worldvolume of the $(1,1,0)$ 4-brane, which is inherited from the brane web of the local $dP_{2}$ example \cite{Aharony:1997bh}. Namely, we have 
\begin{equation}\label{stringsofdp2}
\{T_{(1,0,0)},T_{(0,1,0)},T_{(0,0,1)},T_{(1,0,0)}^{(f)},T_{(0,1,0)}^{(f)},T_{(1,-1,0)}^{\text{strip}}\}.
\end{equation}
This implies that there should be at least one string that is charged under the Cartan's of the two $\mk{su}(2)$ factors above as we will demonstrate below. The independent $Q_{(p,q,r)}$ charges that correspond to the finite and flavor 4-branes are
\begin{equation}
   \{ Q_{(1,0,0)}, Q_{(0,1,0)}, Q_{(0,0,1)}, Q_{(1,-1,0)},Q_{(1,0,0)}^{(f)}, Q_{(0,1,0)}^{(f)}\}.
\end{equation}
Three of the above strings we encountered in (\ref{P1cubestringQ}) and charged under the first three of the above charges in the same way. Two of the remaining strings connect the semi-finite and finite $(1,0,0)$ 4-branes and the semi-finite and finite $(0,1,0)$ 4-branes. The strip, which live along the worldvolume of the $(1,1,0)$ 4-brane can be taken to be charged under $Q_{(0,1,0)}$ and $Q_{(1,0,0)}^{(f)}$ only. This was a bit too quick. Let us consider the 5-branes that are involved in finding the flavor symmetry. First, we take the $(1,0,0)$ 5-brane to be associated with the $\mk{u}(1)$ factor. Once the $(1,0,0)$ 5-brane is pulled into the interior of the brane box, the $Q^{(f)}_{(0,1,0)}$ charge disappears as the associated 4-brane becomes $(1,-1,0)$ 4-brane due to charge conservation. In this case, we recover the brane box of the $\P^{1}\times\P^{1}\times\P^{1}$ example with an extra $(1,0,0)$ 5-brane live in the interior region. Hence, the results of the former example apply here. In particular, the two $\mk{su}(2)$ factors of the flavor symmetry are generated by pair of $(1,-1,0)$ 5-branes and $(0,1,-1)$ 5-branes, which get interlaced due to (\ref{su30fp1cube}). The enhanced flavor symmetry is then
\begin{equation}
    \mk{g}_{\text{enhanc.}} \ = \ \mk{su}(3) \oplus \mk{u}(1). 
\end{equation}

Nevertheless, let us work out the representation of our string in (\ref{stringsofdp2}) under the charges of the global symmetries. The electric charge $Q_{e}$ and the $Q_{I}$ charges that correspond to the Cartan of the $\mk{su}(2)\oplus\mk{su}(2)\oplus \mk{u}(1)$ can be taken as
\begin{equation}
    \begin{split}
        Q_{e} &= -2(Q_{(1,0,0)} + Q_{(0,1,0)} + Q_{(0,0,1)} + Q_{(1,-1,0)})
        \\
        Q_{A} &= Q_{(1,0,0)} - Q_{(0,1,0)} - Q_{(1,0,0)}^{(f)} + Q_{(0,1,0)}^{(f)},
        \\
        Q_{B} &= Q_{(0,1,0)} - Q_{(0,0,1)} - Q_{(0,1,0)}^{(f)},
        \\
        Q_{C} &= - 2(Q_{(1,0,0)}^{(f)} + Q_{(0,1,0)}^{(f)}).
    \end{split}
\end{equation}
Here, the charges are taken to reflect Figure \ref{p1dp2branebox} without any 5-brane being pulled in. 
When applying the $Q_{I}$ charges on the above $T^{i}_{(p,q,r)}$ strings one arrives at
\begin{equation}
\begin{array}{c|cccc}
&Q_{e} & Q_{A} & Q_{B} & Q_{C}\\
\hline
T^{1}_{(1,0,0)} & -2& 1 & 0 & 0\\
T^{2}_{(0,1,0)} & -2& -1 & 1 & 0\\
T^{3}_{(0,0,1)} & -2&0 & -1 &0\\
\hline
T^{(f)}_{(1,0,0)} &-1 &0 &0 &-1\\
T^{(f)}_{(0,1,0)} &-1 &0 &0 &-1\\
T^{\text{strip}}_{(1,-1,0)} &-1 &0 &0 &-1
\end{array}
\end{equation}
The first three strings transform in the fundamental of the $\mk{su}(3)$ Lie algebra and neutral under the abelian factor, i.e. $\textbf{3}_{0}$. Whereas the last string is trivial under the $\mk{su}(3)$ and charged under the abelian factor. This provides evidence for the above conclusion regarding the enhancement of flavor symmetry.  

Now, let us find the charges of the flavor strings when dragging all relevant 5-brane into the interior region. That we have
\begin{equation}
\begin{array}{c|ccc}
& Q_{A} & Q_{B} & Q_{C}\\
\hline
T^{1}_{(1,-1,0)} & 2 & -1 & 0\\
T^{2}_{(0,1,-1)} & -1 & 2 & 0\\
T^{3}_{(0,0,1)} & 0 & 0 &1 
\end{array}
\end{equation}
The sub $2\times 2$ matrix above represents the Cartan matrix of $\mk{su}(3)$ Lie algebra. Whereas the last entry is understood to correspond to a $\mk{u}(1)$ factor in the flavor symmetry. It follows that the enhanced flavor symmetry for this example is $\mk{su}(3)\oplus \mk{u}(1)$.

\paragraph{An observation on the local $\P^{1}\times dP_{2}$ example:}

\begin{figure}[H]
\centering{
\includegraphics[scale=0.5]{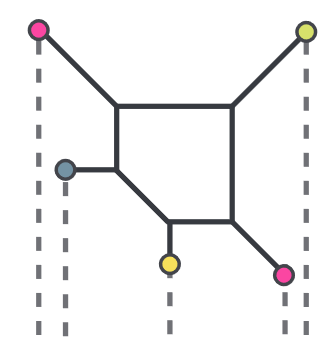} \qquad  \qquad \qquad\includegraphics[scale=0.5]{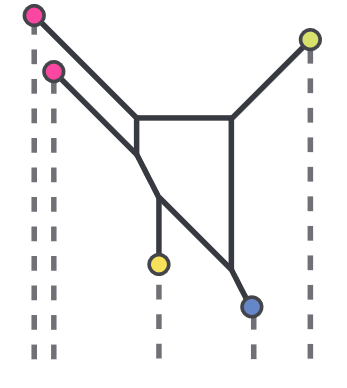}
}
\caption{On the left-hand side, the standard brane web of the local $dP_{2}$. On the right-hand side, we have the deformed web of the same geometry. The figures are due to \cite{Taki:2014pba}.}
\label{webdp2}
\end{figure}

It is worth noting that in the previous examples the $(1,-1,0)$ and $(0,1,-1)$ 5-branes played crucial roles. In the decoupling limit within the appropriate type IIB frame, these branes manifest as the $B$ 7-branes carrying $(1,-1)$ charges. In the interlacing perspective, one of these $B$ branes, specifically the $(0,1,-1)$ 5-branes, originates from the $\F_{0}$ factor, as ascribed in (\ref{eqf0tof2}). The presence of the other $B$ brane can be realized whenever we have a deformation of the brane web, involving the rearrangement of the 7-branes within the face of the generalized brane. This restructuring leads to the emergence of $B$ branes that contribute, at least in part, to the flavor symmetry. Following \cite{Taki:2014pba}, the deformation for the current example is presented in Figure \ref{webdp2}. As it is apparent from the brane web, the two parallel $B$ branes are responsible for the $\mk{su}(2)$ factor of the 5D $\N=1$ flavor symmetry. The two sets of the $B$ branes of the $(a,b)$ generalized brane web get interlaced in the sense of (\ref{su30fp1cube}), and enhance to $\mk{su}(3)$. This observation is radially applied to the next example.

\begin{figure}[H]
\centering{
\includegraphics[scale=0.55]{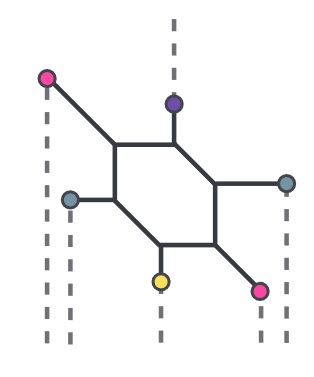} \qquad  \qquad \qquad\includegraphics[scale=0.55]{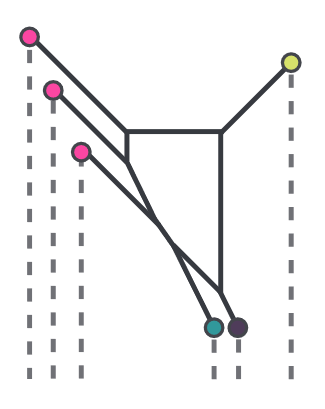}
}
\caption{On the left-hand side, the standard brane web of the local $dP_{3}$. On the right-hand side, we have the deformed web of the same geometry. The figures are due to \cite{Taki:2014pba}.}
\label{webdp3}
\end{figure}

\begin{figure}[H]
\centering{
\includegraphics[scale=0.45]{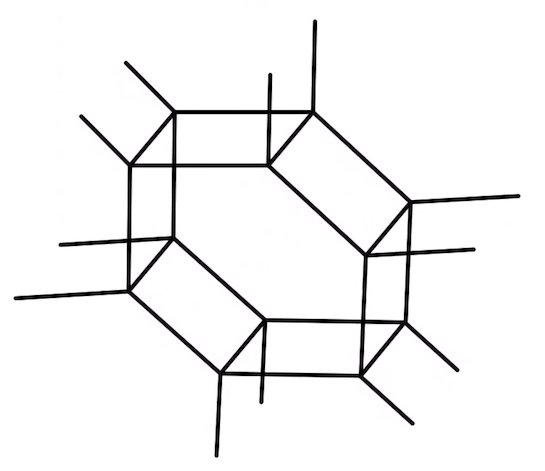}
}
\caption{The brane box dual to the local $\P^{1}\times dP_{3}$ toric CY4. Note that, we made a convention where the $(1,0,0)$ 4-branes are the top and the bottom facets of the finite cell.}
\label{p1dp3branebox}
\end{figure}

\paragraph{The local $\P^{1}\times dP_{3}$ example:} The brane box of this example is given in Figure \ref{p1dp3branebox}. The interlacing geometry is given as 
\begin{equation}
 \text{Local}\,( \ \mathbb{P}^{1} \times dP_{3} \ \cong \  \F_{0}|_{(xz;y=0)} \ \cap \ dP_{3}|_{(xy;z=0)}\,),
\end{equation}
The flavor algebra reads from the brane web of each CY3 factor are $\mk{su}(2)$ and $\mk{su}(3)\oplus \mk{su}(2)$, respectively. 
The standard brane web of the local $dP_{3}$ is given on the LHS of Figure \ref{webdp3}. We note that we can be deformed to the one given on the RHS of the same figure. Following \cite{Taki:2014pba}, the deformation is achieved by pulling the 7-branes inside the face of these and rearranging these branes. Rearranging 7-branes, as we have seen in the $\P^{1}\times\P^{1}\times\P^{1}$ example, involves branes going through the monodromy of other branes and experiencing their monodromies. The last step is to pull these branes out of the face. In this deformation, i.e. phase, the $\mk{su}(3)$ flavor symmetry is realized through $B$ 7-branes. In the brane box settings, we can take these codimension-two $B$ branes to be $(1,-1,0)$ 5-branes. These branes come with two copies of the $Q_{A}$ charge as given in (\ref{QofP1cube}).

The other part of the above interlacing geometry is the local $\F_{0}$ and as we have seen above the $\mk{su}(2)$ flavor symmetry can also be generated by $B$ 7-branes in the decoupling limit. When considering the brane box, these $B$ branes become $(0,1,-1)$ 5-branes. Which comes with one copy of the $Q_{B}$ charge as defined in (\ref{QofP1cube}).

The enhancement then occurs between the $Q_{B}$ charge and one of the $Q_{A}$ charges above, in the sense of (\ref{su30fp1cube}). Thus one arrives at
\begin{equation}
    \mk{g}_{\text{enhanc.}} \ = \ \mk{su}(4)\oplus \mk{su}(2).
\end{equation}

%% file: sections/conclusion.tex
In this paper, we set the foundations of studying 3d $\mc{N}=2$ field theories from a pure M-theory geometric engineering approach on local CY4 $X_4$. In particular we focused on the cases of toric $X_4$ which admits a dual $(p,q,r)$-brane box description in 8D maximal supersymmetric theory. We have also presented a preliminary study of M-theory compactification on $\mb{C}^4$ orbifolds.

We discussed many physical aspects of $\FT$ that are related to the geometrical and topological data of $X_4$ and/or to the diagrammatical properties of the dual brane box. We highlight the main contributions we find particularly interesting and are less-explored in previous literature:

\begin{itemize}
\item We studied the coupling $1/g^2$ of $\FT$ when it admits a gauge theory description on CB from both the geometry of $X_4$ and the dual brane box construction when $X_4$ is toric.

\item We studied the BPS spectrum of $\FT$ via investigating various M2-brane wrapping modes on complex curves $C\subset X_4$ whose normal bundle in $X_4$ is either $O\oplus O\oplus O(-2)$.

\item We studied non-abelian flavor symmetry enhancement to $G_F$ at the singular limit of $X_4$ via identifying the flavor Cartan divisors and flavor W-bosons on the resolved CY4.

\item We found that the $\mb{P}^1$-fibration structure (ruling) and non-abelian gauge enhancements can be studied via the brane box diagram and checked against the results obtained from geometric engineering. In certain cases with multiple $\mb{P}^1$-fibration structures, we also observed flavor symmetry duality from different non-abelian flavor symmetry enhancements. 

\item We studied higher-form symmetries and different global structures encoded in the SymTFT of $\FT$.

\item We studied physical effects of adding $G_4$ flux in the resolved $X_4$. Nonetheless a non-zero $G_4$ flux would generically obstruct the singular limit.

\item As a concrete example, we found that the singular limit of local $\mb{P}^1\times\mb{P}^1\times\mb{P}^1$ lead to a 3d $\mc{N}=2$ SCFT with enhanced flavor symmetry $G_F=\mk{su}(3)$ and $\mb{Z}_2$ 1-form symmetry. This example was also studied in much detail in our brane box language. We discussed the geometry and physics of other local $\mb{P}^1\times\mc{S}$ examples and a few higher-rank examples as well.

\end{itemize}

There are many aspects of $\FT$ that could be studied more clearly in the future following the frameworks developed in this paper. We point out the following aspects that we think are directly relevant to the geometric and brane box method in this paper:

\begin{itemize}
\item The 3d $\mc{N}=2$ superpotential $W$ is treated very crudely in this paper, as we only argue that $W=0$ at the singular limit while it is generically non-zero in a resolved geometry. It would be crucial to calculate the moduli dependence of superpotential in a more precise way in order to have a more complete description of $\FT$.

\item Motivated by the 5d cases where the HB is related to either the deformations of a singularity~\cite{Closset:2020scj} or the deformations of a brane web~\cite{Cabrera:2018jxt}, we expect to study the HB of 3d $\mc{N}=2$ field theories using either the M-theory geometric description or via a brane box diagram. In the examples we presented in the paper, there are no chiral matter fields when the  $G_4$ flux is absent. For the strongly coupled SCFT fixed point, one expect to see the Higgs branch information from the deformation (complex structure moduli space) of the singularity. Nonetheless, for CY4 the K\"{a}hler moduli space mixes with the complex structure moduli space, which makes the analysis much more difficult. It would also be extremely interesting to see if the brane box method we develop in this paper can be generalized to the study of HB of $\FT$.

\item It will be very interesting to learn how to realize known rich family of 3d $\mc{N}=2$ dualities using the method we develop in this paper. For instance, the 3d $\mc{N}=2$ SQED-XYZ mirror symmetry involves the HB coordinates, hence a thorough study of the HB physics closely related to the aforementioned point is necessary.

\item One can study the cases where $X_4$ is elliptically fibered hence $\FT$ can be naturally uplifted to a 4d $\mc{N}=1$ theory (SCFT). The geometric engineering of such theories in 4d F-theory is also largely unexplored, except for early results~\cite{Morrison:2016nrt,Apruzzi:2018oge}. Our framework will provide a new view point.

\item The toric examples considered in this paper, apart from the abelian quotients of $\C^{4}$, are all local $\P^{1}\times \mathcal{S}$ for certain complex surface $\mc{S}$. From the interlacing point of view these examples can be viewed as
\begin{equation}
 \text{Local}\,( \ \mathbb{P}^{1} \times \mathcal{S} \ \cong \  \F_{0}|_{(xz;y=0)} \ \cap \ \mathcal{S}|_{(xy;z=0)}\,).
\end{equation}
However, there exists a larger set of examples yet to be discussed which are of the following structure
\begin{equation}
 \text{Local}\,(  \  \mathcal{S}'|_{(xy;z=0)} \ \cap \ \mathcal{S}|_{(yz;x=0)}\,),
\end{equation}
with $\mathcal{S}$ and $\mathcal{S}'$ given in (\ref{localSeq}).

\item It was made clear in~\cite{Acharya:2021jsp, Tian:2021cif} that there is a relation between the McKay quiver of a finite subgroup $\Gamma$ of $SU(3)$ and the 1-form symmetry of a 5D theory obtained from M-theory on $\mb{C}^3/\Gamma$. It will be interesting to see if there is a similar relation between the McKay quiver and the 1-form symmetry of a 3D theory when $\Gamma$ is a finite subgroup of $SU(4)$. This requires a detailed study of the relation between a McKay quiver and the intersection pairing of the corresponding resolved $\mb{C}^4$ orbifold when a crepant resolution exists.

\item The M5-brane realization in section~\ref{sec:m5origin} of the KK7M configuration in Table~\ref{originalKK7Ms} seems to suggest the existence of another dual construction of 3D $\N=2$ theories via reducing 6D $\N=(2,0)$ on a real 3-manifold similar to those studied in~\cite{Dimofte:2011ju, Dimofte:2011py}. It will be interesting to see if our brane box diagram encodes the geometrical and topological data of certain real 3-manifolds.

\end{itemize}

\section*{Acknowledgement}

We thank Bobby Acharya, Shi Cheng, Cyril Closset, Xin Gao, Amihay Hanany, Chiung Hwang, Yukari Ito, Qiang Jia, Hee-Cheol Kim, Sung-Soo Kim, Osama Khlaif, Kimyeong Lee, Linfeng Li, Hong Lu, Ruben Minasian, Sakura Schafer-Nameki, Benjamin Sung, Xin Wang, Dan Xie and Yi Zhang for discussions. JT would like to thank Ying Zhang for her love and support. JT would like to thank the organizers of ``Seventh Annual Meeting of Simons Collaboration on Special Holonomy in Geometry, Analysis, and Physics'' and ``Special Holonomy: Progress and Open Problems 2023'' for the invitation and hospitality. YNW would like to thank the organizers of ``2023 East Asia Joint Workshop on Fields and Strings'' at Xi'an for the invitation and hospitality. JT and YNW would like to thank the organizers of ``McKay correspondence, Tilting theory and related topics'' at IPMU for the invitation and hospitality. JT is supported in part by a grant from the Simons Foundation (\#488569, Bobby Acharya). MN and YNW are supported by National Natural Science Foundation of China under Grant No. 12175004, by Peking University under startup Grant No. 7100603667 and by Young Elite Scientists Sponsorship Program by CAST (2022QNRC001, 2023QNRC001).

%% file: sections/app.tex
\section{Non-BPS spectrum on the Coulomb branch}
\label{sec:non-BPS}

We give a more detailed analysis on the article spectrum on the CB (see Figure~\ref{fig:3pictures1}), including the non-BPS states. It is natural to assume that the UV theory admits a weakly coupled gauge theory description in the UV with light charged particles from M2-brane wrapping curves $C$ intersecting non-trivially with the Cartan $U(1)$'s. To have a valid gauge theory description we will also assume that such a curve $C$ is fibered adiabatically over a base $B$ hence there is a hierarchy $\text{Vol}(B) \gg \text{Vol}(C)$. More precisely when $N_{C|X_4} = \mc{O}\oplus \mc{O}\oplus \mc{O}(-2)$ the base is a surface $S$ along the direction of $\mc{O}\oplus \mc{O}$ and when $N_{C|X_4} = \mc{O}\oplus \mc{O}(-1)\oplus O(-1)$ the base is a curve $\Sigma$ along the direction of $\mc{O}$ (c.f. section~\ref{sec:M2modes}). Given such a hierarchy one can view the UV 3D theory $\mc{T}^{UV}_{X_4}$ as the result of either a twisted compactification of a 7D SQCD on $S$ or a twisted compactification of a 5D SQCD on $\Sigma$, depending on the normal bundle of $C$. For convenience we set the Planck unit to be 1, hence $\text{Vol}(B)$, which is effectively the compactification scale, can be set to be approximately 1 while $m = \text{Vol}(C) \ll 1$ where $m$ is the mass of the light charged particle.

Now let us take a closer look at the origin of the fermions in lower dimension from compactifying certain massive fermion in a higher dimensional QFT. Note that from now on to facilitate our discussion we will assume the existence of a gauge theory description with possible non-abelian enhancements in certain region of CB of the theory, which in general does not have to be case. The implication to geometry of this assumption will be made clear in a moment. We will assume that the massive fermions in higher dimension are charged and originate geometrically from M2-brane wrapping a curve $C$ in certain background geometry $K$. A massive fermion $\psi^{(n+k)}$ living in $(n+k)$D spacetime satisfies the Dirac equation in $(n+k)$D:
\begin{equation}
    i\slashed{D}^{(n+k)} \psi^{(n+k)} = m \psi^{(n+k)}.
\end{equation}
We consider the decomposition of the original $(n+k)$D spacetime into an $n$-dimensional $X$ and a $k$-dimensional $Y$ such that $\text{Vol}(X) \gg \text{Vol}(Y) \gg \text{Vol}(\mb{P}^1)$. The Dirac operator on the spacetime and the state $\psi^{(n+k)}(x,y)$ decompose as:
\begin{equation}\label{eq:dirac_decomposition}
    \begin{split}
        \slashed{D}^{(n+k)} &= \slashed{D}^{X} + \slashed{D}^{Y}, \\
        \psi^{(n+k)}(x,y) &= \sum_i \psi_i^{X}(x) \phi_i^{Y}(y)
    \end{split}
\end{equation}
where $(x,y)\in X\times Y$ and $\phi^Y_i$ is a orthonormal complete set of solutions of the following eigenvalue problem on $Y$:
\begin{equation}
    i\slashed{D}^{Y}\phi_i^{Y}(y) = \lambda_i \phi_i^{Y}(y).
\end{equation}
Applying the Dirac operator in (\ref{eq:dirac_decomposition}) on the decomposition of $\psi^{(n+k)}$, we have:
\begin{equation}
    \sum_i\left(i \slashed{D}^{X} \psi_i^{X}(x) + \lambda_i \right) \phi_i^{Y}(y) = m \sum_i \psi^{X}_i(x) \phi^{Y}_i(y).
\end{equation}
Then the orthogonality of the eigenmodes $\phi_i^{Y}(y)$ implies:
\begin{equation}\label{eq:3Dfermion_mass}
    i \slashed{D}^{X}\psi^{X}_i = (m - \lambda_i)\psi^{X}_i
\end{equation}
In this work we take $X\cong \mb{R}^{1,2}$ therefore $\psi^{X}_i$ is a 3D fermion in $\mc{T}^{UV}_{X_4}$ with mass $|m - \lambda_i|$ and spin $\frac{1}{2}\text{sign}(m - \lambda_i)$ \cite{Witten_FermionPathIntegral}. In this case $Y$ is either a K\"ahler twofold $S$ or a Riemann surface $\Sigma$ depending on whether the decomposition is from the bulk matters in 7D or the localized matters in 5D, which in turn is determined by the normal bundle of $C$ in $X_4$, where $X_4$ is a CY4 viewed as $K$ adiabatically fibered over $Y$. In other words the CY4 geometry is either $K_2\hookrightarrow X_4 \rightarrow S$ or $K_3\hookrightarrow X_4 \rightarrow \Sigma$.

Now let us check the massive fermion spectrum in $\mc{T}^{UV}_{X_4}$. We first consider the case $K_2\hookrightarrow X_4 \rightarrow S$ in which case the fermion zero modes are shown in (\ref{eq:vector_spec}). Despite they are zero modes of the Dirac operator on $S$, the corresponding 3D fermions are massive due to non-zero $m$ in (\ref{eq:3Dfermion_mass}). Thus these massive fermions lead to the following terms in the 3D UV Lagrangian:
\begin{equation}
    \mc{L}^{UV} \supset 2h^{0,0}m \widetilde{\lambda}\lambda + 2h^{0,1}m \widetilde{\psi}\psi + 2h^{0,2}m \widetilde{\chi}\chi
\end{equation}
where all fermions have the same charge $q$ under the same Cartan $U(1)$. Integrating out the above massive fermions leads to the IR CS term \cite{Dunne:1998qy}:
\begin{equation}
    S^{IR} \supset \frac{-iq^2}{8\pi}\left(2h^{0,0}+2h^{0,1}+2h^{0,2}\right) \int d^3x\ AdA
\end{equation}
for each charged massive vector multiplet in 7D and $A$ is the $U(1)$ gauge field under which the massive fermions are charged. From (\ref{eq:3Dfermion_mass}) there seems to be a contribution from non-zero eigenmodes of the Dirac operator on $S$ as well. For that, we note that the eigenmodes of the Dirac operator always appear in pairs except for the zero modes, i.e. for each $\lambda_i \neq 0$ there will be another eigenmode with eigenvalue $-\lambda_i$. Also, in general, one would expect the first positive eigenvalue $\lambda_1$ will be of order $1/\text{Vol}(S)$. Due to the hierarchy $\text{Vol}(S) \sim 1 \gg \text{Vol}(C) = m$ we have $\lambda_1 \gg m$, therefore $m-\lambda_i \sim -\lambda_i$ for all eigenvalues. Hence in $\mc{L}^{UV}$ fermion fields with mass $|\lambda_i|$ must all appear in pairs of opposite spins. After being integrated out, these massive fermions give rise to cancelling IR CS terms.

We then consider the case $K_3\hookrightarrow X_4 \rightarrow \Sigma$ in which case the fermion zero modes are shown in (\ref{eq:chiral_spec}). In particular, we are interested in the case $\Sigma\simeq \mb{P}^1$ where there are no zero modes of the Dirac operator on $\Sigma$. According to (\ref{eq:3Dfermion_mass}) one has to look at the non-zero modes of the Dirac operator to get the massive fermions in $\mc{T}^{UV}_{X_4}$. Viewing $\mb{P}^1$ as a Riemannian sphere $S^2$ one can actually show that the eigenmodes come in pairs and the lowest positive eigenvalue is $1/\text{Vol}(S^2)$ \cite{Abrikosov:2002jr}. Therefore in this case, since $\text{Vol}(S) \sim 1 \gg \text{Vol}(C) = m$, we again have $\lambda_1 \gg m$ hence $m-\lambda_i \sim -\lambda_i$ for all eigenvalues. Following the argument in the above paragraph we see that all the induced IR CS terms associated to integrating out these massive charged fermions cancel.

There is one subtle case where $C$ is actually a curve in $B$ and such examples will be presented in section~\ref{sec:examples}. In this case the hierarchy $\text{Vol}(S) \gg \text{Vol}(C)$ is no longer valid. In this case it makes little sense to assume that $X_4$ is adiabatically fibered hence the ansatz (\ref{eq:dirac_decomposition}) of the twisted compactification may fail. Nevertheless, if we insist its validity, we will still reach the same conclusion since in this case we have $\text{Vol}(B) \sim \text{Vol}(C) \ll 1$ and the first positive eigenvalue of the Dirac operator $\lambda_1 \sim \text{Vol}(B) \gg 1 \gg \text{Vol}(C) \sim m$. Therefore we still have $m-\lambda_i \sim -\lambda_i$ for all eigenvalues. By the argument in the above two cases we will see no IR CS terms induced from these massive fermions.

Since the induced IR CS terms are from integrating out massive charged fields running in the loop, not necessarily massive charged spin-$\frac{1}{2}$ fermions (for concrete examples in the context of 5D theory see \cite{Bonetti:2013_OneLoopCS5D}), one may wonder what happens if charged higher spin states are present in 7D/5D and a subsequent twisted compactification to 3D leads to charged higher spin fields in $\mc{T}^{UV}_{X_4}$. In such cases one then has to analyze the spectrum of the spin-$\frac{n}{2}$ Dirac operator $\slashed{D}_{\frac{n}{2}}$ on $B$. But assuming that the non-zero eigenmodes of $\slashed{D}_{\frac{n}{2}}$ appear in pairs and first positive eigenvalue of $\slashed{D}_{\frac{n}{2}}$ is still of the order $1/\text{Vol}(B)$, we will still have $|m_{n/2}| \ll |\lambda_i|$ for all eigenvalues given the hierarchy $\text{Vol}(B) \sim 1 \gg \text{Vol}(C)$. Then the same argument as above leads to cancelling induced IR CS terms unless there exist zero modes of $\slashed{D}_{\frac{n}{2}}$\footnote{Note that the non-zero eigenstates of $\slashed{D}_{\frac{n}{2}}$ with eigenvalue $\pm\lambda$ pair up, because for any eigenmode $\phi^{(n)}$ satisfying $i\slashed{D}_{\frac{n}{2}}\phi^{(n)}=\lambda\phi^{(n)}$, there exists another eigenmode $\gamma\phi^{(n)}$ satisfying $i\slashed{D}_{\frac{n}{2}}\gamma\phi^{(n)}=-\lambda\gamma\phi^{(n)}$, due to $\{\slashed{D}_{\frac{n}{2}},\gamma\}=0$. Here $\gamma$ is the ``$\gamma^5$ matrix'' in either 2D or 4D in Euclidean signature. For a detailed discussion of this issue see e.g. \cite{GSW:Vol2}.}. Due to the lack of general results on the spectrum of higher spin Dirac operators on K\"ahler manifolds, we will refrain from discussing the details of the consequences of having massive charged higher spin fields in $\mc{T}^{UV}_{X_4}$.



\section{Interlacing of toric Calabi-Yau's}\label{app:interlacing}

\import{sections/}{interlacing}

\section{Generators of the non-abelian finite subgroups of $SU(4)$}\label{app:nonab_gens}

In this section we list the generators that generate the non-abelian finite subgroups of $SU(4)$ listed in Table~\ref{tab:primitive_group_data} and Table~\ref{tab:imprimitive_group_data}, which have already been fully worked out in \cite{HananyHe_SU4}.

The constants that will appear in the generators are defined as follows:
\begin{equation}
    \begin{split}
        &w = e^{2\pi i/3},\ \beta = e^{2\pi i/7}, \\
        &p = \beta + \beta^2 + \beta^4,\ q = \beta^3 + \beta^5 + \beta^6,\ s = \beta^2 + \beta^5,\ t = \beta^3 + \beta^4,\ u = \beta + \beta^6
    \end{split}
\end{equation}

The generators of primitive subgroups are listed as follows in (\ref{eq:primitive_1}), (\ref{eq:primitive_2}), (\ref{eq:primitive_3}), (\ref{eq:primitive_4}) and (\ref{eq:primitive_5}). The (K) in Table~\ref{tab:primitive_group_data} is the group generated by $A_1$, $A_2$, $A_3$ and $A_4$ in (\ref{eq:primitive_5}).
\begin{equation}\label{eq:primitive_1}
    \begin{split}
        &F_1 = \begin{pmatrix}
            1 & 0 & 0 & 0 \\
            0 & 1 & 0 & 0 \\
            0 & 0 & w & 0 \\
            0 & 0 & 0 & w^2
        \end{pmatrix},\ F_2 = \frac{1}{\sqrt{3}}\begin{pmatrix}
            1 & 0 & 0 & \sqrt{2} \\
            0 & -1 & \sqrt{2} & 0 \\
            0 & \sqrt{2} & 1 & 0 \\
            \sqrt{2} & 0 & 0 & -1
        \end{pmatrix},\ F_3 = \begin{pmatrix}
            \frac{\sqrt{3}}{2} & \frac{1}{2} & 0 & 0 \\
            \frac{1}{2} & -\frac{\sqrt{3}}{2} & 0 & 0 \\
            0 & 0 & 0 & 1 \\
            0 & 0 & 1 & 0
        \end{pmatrix}, \\
        &F'_2 = \frac{1}{3}\begin{pmatrix}
            3 & 0 & 0 & 0 \\
            0 & -1 & 2 & 2 \\
            0 & 2 & -1 & 2 \\
            0 & 2 & 2 & -1
        \end{pmatrix},\ F'_3 = \frac{1}{4}\begin{pmatrix}
            -1 & \sqrt{15} & 0 & 0 \\
            \sqrt{15} & 1 & 0 & 0 \\
            0 & 0 & 0 & 4 \\
            0 & 0 & 4 & 0
        \end{pmatrix},\ F_4 = \begin{pmatrix}
            0 & 1 & 0 & 0 \\
            1 & 0 & 0 & 0 \\
            0 & 0 & 0 & -1 \\
            0 & 0 & -1 & 0
        \end{pmatrix}, \\
        &S = \begin{pmatrix}
            1 & 0 & 0 & 0 \\
            0 & \beta & 0 & 0 \\
            0 & 0 & \beta^4 & 0 \\
            0 & 0 & 0 & \beta^2
        \end{pmatrix},\ T = \begin{pmatrix}
            1 & 0 & 0 & 0 \\
            0 & 0 & 1 & 0 \\
            0 & 0 & 0 & 1 \\
            0 & 1 & 0 & 0
        \end{pmatrix},\ W = \frac{1}{\sqrt{7}i}\begin{pmatrix}
            p^2 & 1 & 1 & 1 \\
            1 & -q & -p & -p \\
            1 & -p & -q & -p \\
            1 & -p & -p & -q
        \end{pmatrix}, \\
        &R = \frac{1}{\sqrt{7}}\begin{pmatrix}
            1 & 1 & 1 & 1 \\
            2 & s & t & u \\
            2 & t & u & s \\
            2 & u & s & t
        \end{pmatrix},\ C = \begin{pmatrix}
            1 & 0 & 0 & 0 \\
            0 & 1 & 0 & 0 \\
            0 & 0 & w & 0 \\
            0 & 0 & 0 & w^2
        \end{pmatrix},\ D = \begin{pmatrix}
            w & 0 & 0 & 0 \\
            0 & w & 0 & 0 \\
            0 & 0 & w & 0 \\
            0 & 0 & 0 & 1
        \end{pmatrix}, \\
        &V = \frac{1}{\sqrt{3}i}\begin{pmatrix}
            \sqrt{3}i & 0 & 0 & 0 \\
            0 & 1 & 1 & 1 \\
            0 & 1 & w & w^2 \\
            0 & 1 & w^2 & w
        \end{pmatrix},\ F = \begin{pmatrix}
            0 & 0 & -1 & 0 \\
            0 & 1 & 0 & 0 \\
            -1 & 0 & 0 & 0 \\
            0 & 0 & 0 & -1
        \end{pmatrix}.
    \end{split}
\end{equation}

\begin{equation}\label{eq:primitive_2}
    \begin{split}
        &S_{SU(2)} = \frac{1}{2}\begin{pmatrix}
            -1+i & -1+i \\
            1+i & -1-i
        \end{pmatrix},\ U_{SU(2)} = \frac{1}{\sqrt{2}}\begin{pmatrix}
            1+i & 0 \\
            0 & 1-i
        \end{pmatrix}, \\
        &V_{SU(2)} = \begin{pmatrix}
            \frac{i}{2} & \frac{1-\sqrt{5}}{4} - \frac{1+\sqrt{5}}{4}i \\
            -\frac{1-\sqrt{5}}{4} - \frac{1+\sqrt{5}}{4}i & -\frac{i}{2}
        \end{pmatrix},\ x_1 = \frac{1}{\sqrt{2}}\begin{pmatrix}
            1 & 1 \\
            i & -i
        \end{pmatrix},\ x_2 = \frac{1}{\sqrt{2}}\begin{pmatrix}
            i & i \\
            -1 & 1
        \end{pmatrix}, \\
        &x_3 = \frac{1}{\sqrt{2}}\begin{pmatrix}
            -1 & -1 \\
            -1 & 1
        \end{pmatrix},\ x_4 = \frac{1}{\sqrt{2}}\begin{pmatrix}
            i & 1 \\
            1 & i
        \end{pmatrix},\ x_5 = \frac{1}{\sqrt{2}}\begin{pmatrix}
            1 & -1 \\
            -i & -i
        \end{pmatrix},\ x_6 = \frac{1}{\sqrt{2}}\begin{pmatrix}
            i & -i \\
            1 & 1
        \end{pmatrix}.
    \end{split}
\end{equation}

\begin{equation}\label{eq:primitive_3}
    T_1 = \frac{1+i}{\sqrt{2}}\begin{pmatrix}
        1 & 0 & 0 & 0 \\
        0 & 0 & 1 & 0 \\
        0 & 1 & 0 & 0 \\
        0 & 0 & 0 & 1
    \end{pmatrix},\ T_2 = \begin{pmatrix}
        1 & 0 & 0 & 0 \\
        0 & 0 & 1 & 0 \\
        0 & i & 0 & 0 \\
        0 & 0 & 0 & i
    \end{pmatrix}.
\end{equation}

\begin{equation}\label{eq:primitive_4}
    \begin{split}
        &A = \frac{1+i}{\sqrt{2}}\begin{pmatrix}
            1 & 0 & 0 & 0 \\
            0 & i & 0 & 0 \\
            0 & 0 & i & 0 \\
            0 & 0 & 0 & 1
        \end{pmatrix},\ B = \frac{1+i}{\sqrt{2}}\begin{pmatrix}
            1 & 0 & 0 & 0 \\
            0 & 1 & 0 & 0 \\
            0 & 0 & 1 & 0 \\
            0 & 0 & 0 & -1
        \end{pmatrix},\ S' = \frac{1+i}{\sqrt{2}}\begin{pmatrix}
            i & 0 & 0 & 0 \\
            0 & i & 0 & 0 \\
            0 & 0 & 1 & 0 \\
            0 & 0 & 0 & 1
        \end{pmatrix}, \\
        &T' = \frac{1+i}{2}\begin{pmatrix}
            -i & 0 & 0 & i \\
            0 & 1 & 1 & 0 \\
            1 & 0 & 0 & 1 \\
            0 & -i & i & 0
        \end{pmatrix},\ R' = \frac{1}{\sqrt{2}}\begin{pmatrix}
            1 & i & 0 & 0 \\
            i & 1 & 0 & 0 \\
            0 & 0 & i & 1 \\
            0 & 0 & -1 & -i
        \end{pmatrix}.
    \end{split}
\end{equation}

\begin{equation}\label{eq:primitive_5}
    \begin{split}
        &A_1 = \begin{pmatrix}
            1 & 0 & 0 & 0 \\
            0 & 1 & 0 & 0 \\
            0 & 0 & -1 & 0 \\
            0 & 0 & 0 & -1
        \end{pmatrix},\ A_2 = \begin{pmatrix}
            1 & 0 & 0 & 0 \\
            0 & -1 & 0 & 0 \\
            0 & 0 & -1 & 0 \\
            0 & 0 & 0 & 1
        \end{pmatrix},\ A_3 = \begin{pmatrix}
            0 & 1 & 0 & 0 \\
            1 & 0 & 0 & 0 \\
            0 & 0 & 0 & 1 \\
            0 & 0 & 1 & 0
        \end{pmatrix},\ A_4 = \begin{pmatrix}
            0 & 0 & 1 & 0 \\
            0 & 0 & 0 & 1 \\
            1 & 0 & 0 & 0 \\
            0 & 1 & 0 & 0
        \end{pmatrix}.
    \end{split}
\end{equation}

The group $\Delta_n$ in Table~\ref{tab:imprimitive_group_data} is generated by:
\begin{equation}
    \begin{pmatrix}
        w^i & 0 & 0 & 0\\
        0 & w^j & 0 & 0 \\
        0 & 0 & w^k & 0 \\
        0 & 0 & 0 & w^{-i-j-k}
    \end{pmatrix}
\end{equation}
where $w = e^{2\pi i/n}$ for $i,j,k = 1,\cdots, n$.

%% file: sections/interlacing.tex
A class of Calabi-Yau 4-fold, CY$_{4}$, toric diagrams can be realized in terms of 2-fold or 3-fold toric Calabi-Yau's. One way to obtain such a construction can be done by following \cite{Franco:2020avj}, where a notion of "product" between CY$_{n}$ and CY$_{m}$ is defined according to the following rule,
\begin{equation}\label{cycy=cy}
    \text{CY}_{m+2}\ \otimes\ \text{CY}_{n+2}  \ = \ \text{CY}_{m+n+3}
\end{equation}

\begin{figure}[H]
\centering{
\includegraphics[scale=0.55]{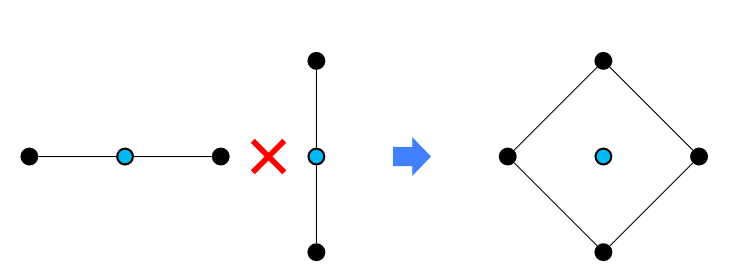}
}
\caption{The product of two toric $\C^{2}/\Z_{2}$ spaces along the middle divisor to form the geometry of CY$_{3}$ of local $\F_{0}$, as discussed in \cite{Franco:2020avj}.}
\label{producttwocy2tof0}
\end{figure}

The product is done over a toric divisor; that is, we identify a toric divisor in both of these toric spaces. In other words, the higher dimensional divisor is an interlaced geometry of two lower dimensional ones along an identified divisor belonging to each; see Figure \ref{producttwocy2tof0} for an example. In this notion, we usually need to complete (add-in) toric divisors within the boundaries of the new higher dimensional toric geometry \cite{Franco:2020avj}. In this picture, a toric CY$_{3}$ can be seen as a product of two toric CY$_{2}$. To be specific, we take the example where a product of two $\C^{2}/\Z_{2}$ is taken to form the CY$_{3}$ that is the local geometry of the zeroth Hirzebruch surfaces, $\F_{0}$, as we have in Figure \ref{producttwocy2tof0}. In this example, the geometrical engineering of 5D $\N=1$ E$_{1}$ Seiberg theory, in M-theory, can be seen as coming from two 7D field theories obtained by the reducing M-theory on different $\C^{2}/\Z_{2}$ spaces. In general, we could not obtain every 5D $\N=1$ theory in the way, e.g. $\Tilde{E}_{1}$ theory. This notion is related to the discussion in \cite{Acharya:2023bth}, where one can obtain the 5D T$_{N}$ theories from 7D theories constructed as M-theory on $\C^{2}/\Z_{N}$.  

\begin{figure}[H]
\centering{
\includegraphics[scale=0.45]{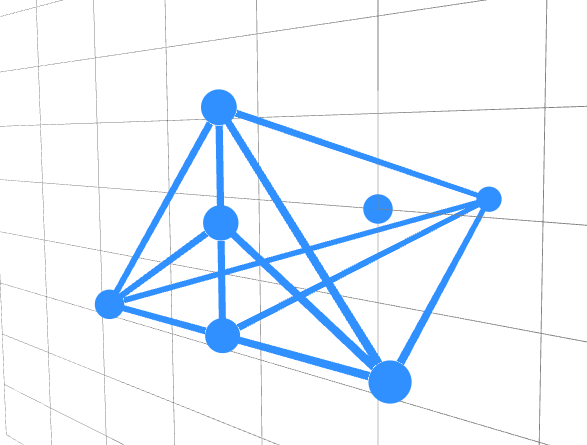}
}
\caption{The toric CY$_{4}$ geometry obtained as the interlacing of $\C^{2}/\Z_{2}$ with the CY$_{3}$ of $\F_{2}$, \href{https://www.math3d.org/ZIAHjL8o6}{the 3D model} is given by the link.}
\label{f2p1}
\end{figure}

\begin{figure}[H]
\centering{
\includegraphics[scale=0.45]{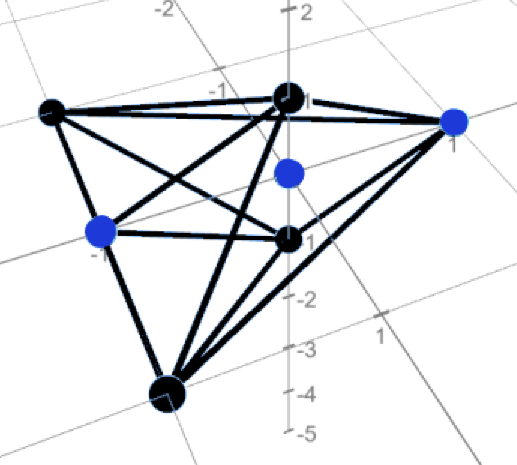}
}
\caption{The local CY$_{3}$ of $\F_{2}$ surface along the (xy)-plane "intersecting" with the Local CY$_{3}$ of $\F_{0}$ surface along the (xz)-plane. The following link directs you to \href{ https://www.math3d.org/CRI8JUtfa}{the 3D toric diagram}.}
\label{f2f0}
\end{figure}

\begin{figure}[H]
\centering{
\includegraphics[scale=0.6]{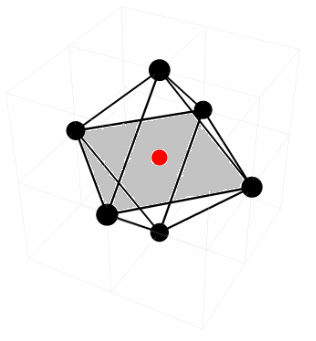}
\includegraphics[scale=0.5]{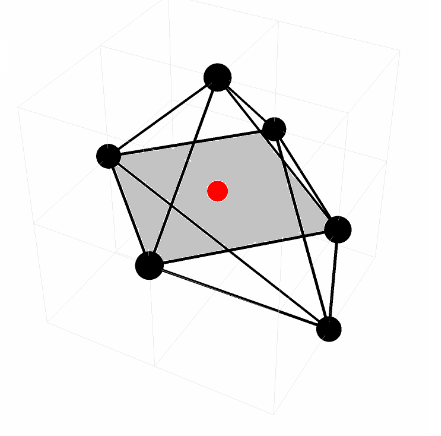}
\includegraphics[scale=0.3]{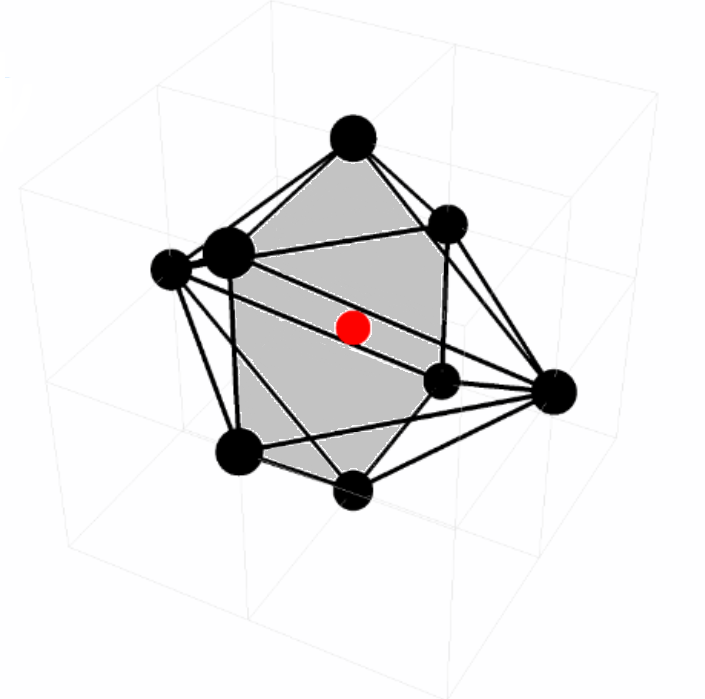}
}
\caption{More examples on the interlacing construction. From LHS to the RHS, we have the local geometry of interlacing $\F_{0}$ with $\F_{0}$, $\F_{1}$, and $dP_{3}$, respectively. }
\label{p1f0f1dp3}
\end{figure}

\begin{figure}[H]
\centering{
\includegraphics[scale=0.45]{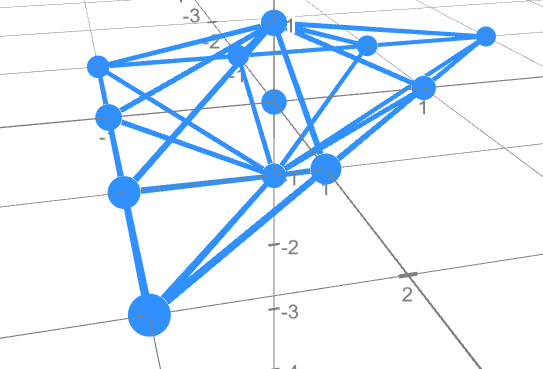}
}
\caption{The toric CY$_{4}$ geometry obtained by intersecting $\C^{3}/\Z_{3}\times\Z_{3}$ with that of local geometry of $\F_{0}$. The link gives \href{ https://www.math3d.org/Q8KuBk4dx}{the 3D model}.}
\label{T3p1}
\end{figure}

To obtain a CY$_{4}$ according to (\ref{cycy=cy}), we need to have a product between toric CY$_{2}$ and toric CY$_{3}$. For example, Figure \ref{f2p1} shows that the local geometry of $\P^{1}\times \F_{2}$ can be obtained as the interlacing of $\C^{2}/\Z_{2}$ with the CY$_{3}$ of $\F_{2}$. However, one could generalise the notion of interlacing discussed above. In particular, one can take it along a collection of divisors rather than one divisor. In this generalised notion, a toric CY$_{4}$ can be obtained through two toric CY$_{3}$'s. We emphasise that only a class of toric CY$_{4}$ can be obtained this way, not a generic toric CY$_{4}$. For example, consider the toric CY$_{3}$'s which are the local geometry of $\F_{2}$ and that of $\F_{0}$ as in Figure \ref{f2f0}. From the 3D toric space, these geometries live in orthogonal 2D subspaces, and the two geometries interlaced along three divisors, as shown in the graph. Thus, we arrive at the following statement,
\begin{equation}
 \text{Local}\,( \ \P^{1} \times \F_{2} \ \cong \  \F_{0}|_{(xz;y=0)}  \ \cap \ \F_{2}|_{(xy;z=0)}\,).
\end{equation}
More examples are presented in Figure \ref{p1f0f1dp3} and Figure \ref{T3p1}.

\bigskip

